\providecommand{\pgfsyspdfmark}[3]{}
\documentclass{aa}  

\pdfoutput=1

\usepackage{graphicx}
\usepackage{lscape}
\usepackage{longtable}
\usepackage{graphicx}
\usepackage{times}
\usepackage{graphicx}
\usepackage{xspace}
\usepackage{epsfig}
\usepackage{natbib}
\usepackage{rotating}
\usepackage{dcolumn}
\usepackage{xcolor}
\usepackage{txfonts}
\usepackage{cleveref}

\newcommand{\Td}    {T_\mathrm{d}}

\newcommand{\mo}    {$M_{\sun}$}
\newcommand{\cmt}   {cm$^{-3}$}
\begin{document}

   \title{Early phases in the stellar and substellar formation and evolution\thanks{Tables 3, 4, 5, 6 and 7 are only available in electronic form at the CDS via anonymous ftp to cdsarc.u-strasbg.fr (130.79.128.5) or via http://cdsweb.u-strasbg.fr/cgi-bin/qcat?J/A+A/}}

   \subtitle{Infrared and submillimeter data in the Barnard 30 dark cloud}

\author{D. Barrado\inst{1}
          \and
          I. de Gregorio Monsalvo
          \inst{2,3}
          \and
          N. Hu\'elamo
          \inst{1}
          \and
          M. Morales-Calder\'on
          \inst{1}
          \and
          A. Bayo
          \inst{4,5}
          \and
          A. Palau
          \inst{6}
          \and
          M.T. Ruiz
          \inst{7}
          \and
          P. Rivi\`ere-Marichalar
          \inst{8}
          \and
          H. Bouy
          \inst{1}
          \and
          \'O. Morata
          \inst{9}
          \and
          J.R. Stauffer
          \inst{10}
          \and
          C. Eiroa
          \inst{11, 12}
          \and
          A. Noriega-Crespo 
          \inst{13}
}

   \offprints{D. Barrado}

\institute{
  Depto. Astrof\'{\i}sica, Centro de Astrobiolog\'{\i}a (INTA-CSIC),  ESAC campus,
   Camino Bajo del Castillo s/n,
  E-28692 Villanueva de la Ca\~nada, Spain  \email{barrado@cab.inta-csic.es}
         \and
             European Southern Observatory, Alonso de C\'ordova 3107, Vitacura, Santiago, Chile.
             \and
             Joint ALMA Observatory, Alonso de C\'ordova 3107, Vitacura, Santiago, Chile
              \and
             Departamento de F\'{\i}sica y Astronom\'{\i}a, Facultad de Ciencias, Universidad
             de Valpara\'{\i}so, Av. Gran Breta\~na 1111, 5030 Casilla, Valpara\'{\i}so, Chile
         \and
             ICM nucleus on protoplanetary disks, Universidad de Valpara\'iso, Av. Gran Breta\~na 1111, Valpara\'iso, Chile
         \and
             Instituto de Radioastronom\'{\i}a y Astrof\'{\i}sica, Universidad Nacional Aut\'onoma de M\'exico, P.O. Box 3-72, 58090 Morelia, Michoac\'an, M\'exico
         \and
             Depto. de Astronom\'{\i}a, Universidad de Chile, Camino del Observatorio 1515, Santiago, Chile.
         \and
             European Space Astronomy Centre (ESA), Camino Bajo del Castillo s/n, 28692 Villanueva de la Ca\~nada, Madrid, Spain 
         \and
             Institute of Astronomy and Astrophysics, Academia Sinica, 11F of AS/NTU Astronomy-Mathematics Building, No.1, Sec. 4, Roosevelt Rd, Taipei 10617, Taiwan
             \and
             Spitzer Science Center, California Institute of Technology, Pasadena, CA 91125, USA.
         \and
             Depto. F\'{\i}sica Te\'orica, Fac. de Ciencias, Universidad Aut\'onoma de Madrid, Campus Cantoblanco, 28049 Madrid,  Spain.
         \and
             Unidad Asociada UAM-CAB/CSIC, Madrid, Spain.
         \and
             Space Telescope Science Institute, 3700 San Martin Dr., Baltimore, MD 21218, USA.
}

   \date{Received ; accepted }

  \abstract
 {} 
   {The early evolutionary stage of brown dwarfs are not very well characterized, specially during the embedded 
phase. Our goal is to gain insight into 
the dominant formation mechanism of very low-mass objects and brown dwarfs.
   }
   {We have conducted deep observations at 870 $\mu$m obtained with the
 LABOCA bolometer at the APEX telescope in order to identify young submillimeter sources in the 
Barnard 30 dark cloud. We have complemented these data  with multi-wavelength observations from
 the optical to the far-IR and compiled complete spectral energy distributions (SEDs) in order to 
identify the counterparts, characterize
the sources and to assess their membership to the association and stellar or substellar status based on the
available  photometric information. 
   }
   {We have identified  34 submillimeter sources and a substantial number of possible and probable 
Barnard 30 members within each individual APEX/LABOCA beam. They can be classified in three 
distinct groups. First, 15 out of these 34 have a clear optical or IR counterpart to the submm peak and nine of them  are 
potential proto-BDs candidates. Moreover, a substantial number of them could be multiple  systems. 
 A second group of 13 sources comprises candidate members with significant infrared excesses located away from the central submm emission. All of them include  brown dwarf candidates, some displaying IR excess, but their association with submm emission is unclear.  In addition, we have found six starless cores and, based on the total dust mass estimate, three might be pre-substellar (or pre-BDs) cores. Finally, the complete characterization of our APEX/LABOCA sources, focusing on those detected at 24 and/or 70 $\mu$m, indicates that in our sample of 34 submm sources there are, at least: two WTTs, four CTTs, five young stellar objects (YSOs), eight proto-BD candidates (with another three dubious cases), and one Very Low Luminosity objects (VeLLO).}
   {Our findings provide additional evidence concerning the brown dwarf formation mechanism, which seems to be a downsized
     version of the stellar formation.}  

   \keywords{circumstellar matter -- stars: formation -- stars: low-mass, brown dwarfs
             -- stars: pre-main sequence – infrared: stars -- 
             open clusters and associations: individual: Barnard 30 dark cloud
               }

   \maketitle
%

%
\section{Introduction \label{sec:intro}}
%


The Barnard 30 dark cloud (B30, hereafter) is located at the rim of the Lambda Orionis star forming region 
 (\citealt{Murdin77.1}),  a complex structure  at about 400 pc, which gives shape
to  the head of Orion.
Although bright, massive stars usually dominate stellar associations, swarms of low-mass
 objects, much less conspicuous, can be found inside them.
Our aim is to unveil this population 
and to focus on the early phases of low-mass evolution by analyzing 
this  moderately nearby and very interesting dark cloud, whose star formation and shape 
seems to be triggered and/or controlled by the O8 III $\lambda$ Orionis star
(actually, a binary, \citealt{Bouy2009-LOri})
and perhaps by a
supernova event which took place a few Myr ago (see \citealt{Maddalena1987-LOSFR, Cunha1996-LOSFR-SN, Dolan2002.1}).

Although the age of B30 is not very well constrained 
(it is normally given as 2-3 Myr), 
it is very likely significantly younger than the central cluster, Collinder 69
(C69 hereafter) 
 whose nominal age is 5-8 Myr
(\citealt{Barrado2004.3, Bayo2011.1}). 
 We note, however, that an older value has been proposed for C69 
(10-12 Myr, \citealt{Bell2013.1, Bayo2011.1}).
While there are no Class~I objects in C69  (\citealt{Barrado2007.1}),  
the ratio of Class I to Class II objects   in B30 is 1:2  (\citealt{Morales2008-PhD},
based on a Spitzer/IRAC color-color diagram (CCD), \citealt{Allen2004-IRAC-ColorsYSO}).
This  schematic  classification is described in \citet{Lada87.1} and \citet{Adams87.1}  and it is 
based on the spectral energy distribution (SED) and interpreted in an evolutionary scenario.
Therefore, B30 is an excellent hunting ground to find embedded stars and, if deep 
infrared surveys are conducted, substellar objects
(i.e., brown dwarfs or BDs, quasi stellar objects unable to burn hydrogen at any evolutionary stage.  For solar metallicity, the mass is lower than
$\sim0.072$~\mo).

\cite{Dolan1999.1, Dolan2001.1, Dolan2002.1} and \cite{Koenig2015_C69_SOri_YSO_WISE}  carried out an extensive 
search of young, low-mass stars in the Lambda Orionis star
forming region, including optical imaging and medium resolution
spectroscopy. Their initial photometric search was centered around the  
central, massive star  $\lambda$ Orionis (the C69 cluster), and near  the dark clouds B30 
and Barnard 35, located northwest and east from C69, respectively. 
In the first case, their photometric survey and subsequent spectroscopic confirmation (via radial velocity and lithium detection)
allowed the identification of  266 stellar members in an area around
these three groups. Among them,  63 can be considered members of the B30 dark cloud.
In the case of \cite{Koenig2015_C69_SOri_YSO_WISE}, they identified 544 photometric stellar candidates in a 200 sq.deg area.
A subsample of 175 was confirmed spectroscopically either by them or by previous studies, with a pollution
rate less than 24\%.Regarding the central cluster, \cite{Bayo2011.1} includes  172 spectroscopically confirmed
members for C69, and a significant number of them are very well inside the brown dwarf domain.
Now, our focus  is the significantly younger B30 cloud.

As shown by the aforementioned studies, the B30 complex is in a very early stage of
 star formation, and becomes an excellent testbed to search for the most embedded low-mass
 objects and brown dwarfs. Recent studies suggest that the particular environmental 
conditions of a star forming cloud may play an important role in determining the outcome of
 very low-mass objects and BDs in that cloud 
\citep[e.g.,][]{Levine2006,Scholz2013,Polychroni2013,Drass2016-BimodasIMF-ONC}. 
In low-mass star forming regions,
 which typically form objects in groups or loose associations, the main BD formation mechanisms
 are turbulent fragmentation \citep[][]{Padoan2004-OriginBD,Hennebelle2008},
 and ejection from multiple protostellar systems and/or fragmented disks
\citep[][]{Reipurth2001,Bate2002,Matzner2005,Whitworth2006}. 
However, in the surroundings of high-mass stars, where objects typically form more closely packed,
 there are additional plausible mechanisms, namely photo-evaporation of cores near massive stars 
\citep[e.g.,][]{Hester1996,Whitworth2004}, and gravitational fragmentation of dense
 filaments formed in a nascent cluster \citep[e.g.,][]{Bonnell2008,Bate2012}, 
suggesting that these clustered regions bathed by the feedback from high-mass stars could 
show higher BD-to-star ratios. From the observational point of view,
the  study of the most embedded and youngest BDs
 requires the use of the millimeter or submillimeter regime, where they emit the 
bulk of their energy, as they 
are dominated by cold envelopes. Thus, it is necessary to conduct systematic multi-wavelength 
searches for BDs in clouds forming low-mass stars and also in clouds forming high-mass stars 
in closely packed environments.

  Here we will mainly deal with four types of young objects.
A young stellar object (YSO) contains an accreting hydrostatic core inside, whose
final mass (after accretion and infall is finished) will be above 75 $M_{jup}$ (i.e., will form a star).
In the literature it is possible to find restricted definitions, so a YSO would be in the
Class 0 or I stages. A more flexible definition includes all the evolutionary stages that
the hydrostatic core takes before reaching the main sequence (i.e., Class  0, I, II and III).
A proto-BD also contains an embedded accreting hydrostatic core inside, whose final mass
(after accretion and infall is finished) will be in the range 13-75 $M_{jup}$. Thus, they are analogs to the
Class 0/I protostars but in the substellar regime.
An early phase would be a pre-BD, which is a  dense core which will form a brown dwarf in a future
but did not form an hydrostatic core yet (i.e., it is an analog to the  "starless" cores for higher masses).
Finally, a  VeLLO (Very Low Luminosity objects) is a dense source with an accreting hydrostatic core inside,
whose internal luminosity ($L_{int}$) is equal or below 0.1 $L_\odot$.
We note that all proto-BDs are VeLLOs, but not all VeLLOs are proto-BDs.

Several works have been focused on the search for the youngest embedded 
brown dwarfs (proto-BDs),  in different star forming regions.
A number of VeLLOs have been identified as potential
proto-BDs candidates (\citealt{Young2004-L1014Core, diFrancesco2007-LowMassCores, Dunham2008-Protostars}). 
The first hydrostatic cores (FHCs), predicted by \cite{Larson1969-CollapseProtoStar},
 might be classified as proto-BD candidates, 
since they are characterized by very low internal luminosities and very low masses, SEDs peaking at $\sim$100 $\mu$m, and presence of low velocity outflows 
\citep[e.g.][]{Machida2008-Outflow}. To date, several FHCs have been identified 
\citep[e.g.][]{Onishi1999-CondensationTaurus, Belloche2006-NGC1333,Belloche2006-ChaMMS1, 
Chen2010-HydrostaticCore,Chen2012-HydrostaticCore}.
However, the envelope masses of FHCs are too large so it seems unlikely that they will remain substellar
(see Table 4 in \citealt{Palau2014-protoBD-IC348}).
Finally, observational studies using a multi-wavelength approach have provided several 
proto-BD candidates in different clouds and star forming regions 
(\citealt{Barrado2009-ProtoBD-Taurus, 
Lee2009-VeLLO,Lee2013-ProtoBD, Kauffmann2011-VeLLO, Andre2012-PreBD, Palau2012-ProtoBD-Taurus,
Palau2014-protoBD-IC348, Morata2015-ProtoBD-Taurus, Liu2016-LOri-ClassO-protoBD-Planck, Riaz2016-ProtoBD-Serpens, deGregorio2016-ProtoBD-ChaII,
Huelamo2017-ALMA-B30}).
Despite these efforts, the number of proto-BD candidates is still very small and the characterization incomplete.

In this work we present the results regarding a search of very young low--mass stars and BDs 
in one of the bright rims of the B30 dark cloud.
The study is based on deep observations at 870 $\mu$m obtained with the LABOCA bolometer array, 
installed on the Atacama Pathfinder EXperiment 
(APEX\footnote{This work is partially based on observations with the APEX telescope.
 APEX  is a collaboration between the Max-Plank-Institute fur Radioastronomie, 
 the European Southern Observatory, and the Onsala Space Observatory})  telescope,
complemented  with multi-wavelength observations from the optical to the far-infrared.
The goal of the study  is to gain insight into 
the dominant formation mechanism of very low-mass objects and BDs in this particular region of the Lambda Orionis star forming
region.
The data are described in Sect.~\ref{sec:data}, with a detailed description of the
submillimeter imaging (subsect.~\ref{sub:LABOCAmapping}). The
master catalog for all possible counterparts to our B30 submillimeter sources 
is described in  subsect.~\ref{sub:crosscorrelation}.
In Sect.~\ref{sec:properties} we explain how we select cluster candidates from this list
 and discuss some individual cases in more detail.
 Sect.~\ref{sec:discussion}  summarizes the findings of this paper. A follow-up study based on ALMA observations
 is presented in \cite{Huelamo2017-ALMA-B30}.
 
%
\section{Observations and archival data\label{sec:data}}
%

This work is focused on submm observations obtained with APEX/LABOCA at 870 $\mu$m. 
The region mapped by LABOCA (Fig. \ref{FOV_LABOCA}a) was selected based on the Spitzer nebulosity, as can be seen in 
Fig. \ref{FOV_LABOCA}b, where we display a Spitzer/MIPS image at 24 $\mu$m together with the APEX/LABOCA detections.
This area contains a significant number of Class II objects \citep{Morales2008-PhD, Bayo2009-PhD}.
To complement the submm data,  we have  used both observations obtained
by our group and data retrieved from public archives covering a broad wavelength range.
These complementary observations are described in the following subsections, while their main characteristics 
(e.g. limiting magnitude, beam size, pointing accuracy) are summarized in Table~\ref{tabInstruments}.

%
\subsection{APEX/LABOCA mapping at 870 $\mu$m \label{sub:LABOCAmapping}}
%

Continuum observations at 870$\mu$m were carried out using APEX/LABOCA.
The field of view (FOV) of the array is 11.4 arcmin, 
and the angular resolution of each beam is 18.6$\pm$1 arcsec.
APEX pointing errors, as quoted by the 
observatory\footnote{http://www.apex-telescope.org/telescope/}, are 2 arcsec,
although  our observations have smaller errors (see below).

Our data were acquired on 2008 October 09-10 within the Chilean program 082.F-0001B,
 under excellent weather conditions (zenith opacity values of ~0.22 at 870$\mu$m).
 Observations were performed using a spiral raster mapping centered at  
$\alpha$ = 05$^{h}$31$^{m}$18.96$^{s}$, $\delta$ = +12$^{\rm o}$07$'$16.75$''$(J2000.0). 
This observing mode consist of a set of spirals with radii between 2 and 3\,arcmin at 
a combination of nine and four raster positions separated by 60 arcsec in azimuth and elevation,
  with an integration time of 40 seconds per spiral. This mode provides a fully sampled and
 homogeneously covered map in an area of  15$\times$15 arcmin. 
The total covered area has a quasi-circular shape with about 37 arcmin diameter. It represents about
a 20-25\% of the total projected size of the dark cloud, as seen by IRAS.

The total on source integration time was $\sim$2.7 hours. Calibration was performed
 using observations of Uranus as well as N2071R as secondary calibrator. The absolute 
flux calibration uncertainty is estimated to be $\sim$ 8$\%$. 
The  telescope pointing was checked every hour toward the source J0530+135,
finding an  rms pointing accuracy of $\sim$1 arcsec, smaller than the nominal pointing accuracy (2 arcsec) 
and the range quoted by \cite{Weiss2009-APEX}, 3--4 arcsec,
which was estimated  with a Monte Carlo simulation. 

We reduced the data using the BoA and MiniCRUSH software packages \citep[see][]{Kovacs2008.1}. 
The pre-processing steps consisted of flagging dead or cross-talking channels, frames with 
too high telescope accelerations and with unsuitable mapping speed, as well as temperature 
drift correction using two blind bolometers. 
The data reduction process includes flat-fielding, opacity correction, calibration, correlated
 noise removal (atmospherics fluctuations seen by the whole array, as well as electronic noise originated in 
 groups of detector channels), and de-spiking.  Every scan was visually inspected to identify and discard corrupted data.
We used an optimized data processing to recover faint sources, which smoothed the map to a final angular resolution of 27.6 arcsec.

In total, we have detected 34 sources above a 4 $\sigma$ detection threshold.
 The coordinates and the APEX/LABOCA fluxes at 870 $\mu$m are listed in
 Table~\ref{tabLABOCADetections}.
 The APEX/LABOCA image of Barnard 30 is shown in  Fig.~\ref{FOV_LABOCA}a.

%
\subsection{CAHA/Omega2000 near-IR\label{sub:O2000}}
%

Deep near-IR data were acquired with the Omega2000 instrument at the CAHA 3.5m telescope (Calar Alto Observatory in Spain\footnote{Based on observations collected at the Centro Astronomico Hispano Aleman (CAHA) at Calar Alto, operated jointly by the Max-Planck-Institut fur Astronomie and the Instituto de Astrofisica de Andalucia (CSIC)}), 
an imager with a 15x15 arcmin FOV. We collected a $\sim$1$^\circ$$\times$1$^\circ$ grid in the  $J$-band image, 
with a total ten minutes exposure per pointing. 
The data were obtained in December 2007 (a large mosaic in the 
$J$ filter) and August 30th, 2011 (deep images in $JHKs$ centered in the LABOCA pointing).
 
The initial data reduction was carried out with the "jitter" package from
  the eclipse library, version 5.0.0
 (\citealt{Devillard1999-jitter,Devillard1997-eclipse}), and involved
 flat field correction, bad pixel masking and frame combination. Typically, ten frames were
 available to combine, and we let jitter reject two to four frames per image to improve the final quality.
 The astrometry calibration was applied using Aladin with 2MASS as the reference catalog.
 Photometry was extracted using sextractor to detect the sources and to compute the aperture
 and PSF photometry for each detected object. An individual calibration was applied to each 
image, given the availability of 2MASS photometry. The process involves only zero point calibration,
 and we use  well detected sources, with 2MASS magnitude not larger than 15 and quality 
flag A. The zero point error typically goes from 0.05 to 0.09 mag.

Due to their depth, our near-IR images include a very large number of detections. Based on the count number 
(see Fig. 2 for an example), the completeness and limiting magnitudes are 
$J_{compl}$$\sim$18.75 mag,  and $J_{lim}$$\sim$20.5 mag in the case of the mosaic collected during December 2007.
For the observations collected four years later, we have derived
$J_{compl}$$\sim$19.25 mag, $H_{compl}$$\sim$18.25 mag, $Ks_{compl}$$\sim$16.25 mag;
and $J_{lim}$$\sim$21.50 mag, $H_{lim}$$\sim$20.50 mag, $Ks_{lim}$$\sim$18.25 mag.
The complete information is listed in Table \ref{tabInstruments}.

%
\subsection{WFC optical imaging\label{sub:WFC}}
%

 \cite{Morales2008-PhD} provides a detailed description of all of 
the optical and mid-IR observational data we have obtained for the Lambda Orionis star forming region 
(including the  B30 dark cloud) and our basic
 data reduction procedures. Here we provide just a brief summary.

 The optical survey on the Barnard~30 association was performed on 2005 December 24-26 with the 2.5m Isaac Newton Telescope at El Roque de los Muchachos Observatory (La Palma, Spain)
 and the Wide Field Camera\footnote{Based on observations made through the Isaac Newton Groups' Wide Field Camera Survey Programme with the Isaac Newton Telescope operated on the island of La Palma by the Isaac Newton Group in the Spanish Observatorio del Roque de los Muchachos of the Instituto de Astrofisica de Canarias}.
 Sloan-gunn $r$ and $i$ filters were used and 4 different pointings were observed covering
 the whole stellar cluster centered in Barnard~30.  Short and long exposure times where used,
 in the first case, we exposed for 5~sec for each filter whereas the long exposure
 times were 600 and 3$\times$400~sec for sloan-gunn $i$ and $r$ respectively.

The weather during the run was clear and the mean seeing as measured in the images is 1.5 arcsec.
  The reduction and photometry was performed with the CASU INT Wide Field Survey pipeline 
\citep[see][for more information on the pipeline]{Irwin2001.1}. We used an aperture of five pixels,
 designed to match median seeing of our survey data and we have used the transformations 
provided by the ING to go from sloan-gunn magnitudes to Landolt (Johnson-Cousin's). The final
completeness magnitudes are $R_{compl}=22.0$ mag and $I_{compl}=20.5$ mag.

%
\subsection{Spitzer/IRAC and MIPS data\label{sub:Spitzer}}
%

Our Spitzer\footnote{This work is based in part on observations made with the Spitzer Space Telescope, which is operated by the Jet Propulsion Laboratory, California Institute of Technology under a contract with NASA.}
data were collected in 2005 October 22nd and 2006 March 29th with IRAC, 
and 2006 March 2nd with MIPS as part of a General Observer cycle 2 program (PID:20339). 
The two epochs of IRAC imaging were obtained in mapping mode with individual exposures of
 12 seconds frame time and three dithers at each map step. 
The high dynamic range option, which provides additional shorter (0.6
 second frametime) exposures interleaved with the longer exposures, was used.
Frames from the two epochs were co-added into one single deeper mosaic at each 
of the four bandpasses using the MOPEX package \citep{Makovoz05.1}. 
The region with data in the four IRAC bandpasses covers $\sim$0.6$^\circ\times$1.2$^\circ$ 
and is centered at 05:31:22.77 +12:14:52.35. 
The mean integration time for the deep mosaics is 62.4~sec.
 MIPS data were obtained with fast rate scan mode and a total effective 
integration time per point on the sky of $\sim$15 seconds. 
The mosaic covered an area of 1$^\circ\times$2.4$^\circ$ centered around 05:30:23.17 +11:55:52.83. 
Since there were no visible artifacts in the pipeline mosaic for MIPS 24~$\mu$m we used it as
 our starting point to extract the photometry.

Both source extraction and aperture photometry were done with IRAF. 
 We used apertures of three pixels (3.66 arcsec) radius in the case of IRAC,
 and the sky was computed using a circular annulus four pixels wide, starting
 at a radius three pixels away from the center. For the MIPS data at 24~ $\mu$m,
 we used a 5.31 pixels (13 arcsec) aperture and a sky annulus from 
8.16~pixels (20 arcsec) to 13.06~pixels (32 arcsec). Aperture
correction was applied in both cases.
Details are summarized in Table \ref{tabInstruments}
and are fully reported in \cite{Morales2008-PhD}.

 Since there is some structured nebulosity in the region and in order to verify that our detections are real we have visually inspected the location of every source in a median filtered mosaic (which has part of the diffuse emission filtered). We have also performed a psf fitting to make sure that each detection is consistent with a point source instead of just part of the nebulosity. To do that we used Spitzer/APEX tool within MOPEX to filter our mosaic with a high band pass filter. That removes the majority of the nebulosity. Then we searched that image for point sources and fitted a PSF to them. Thus, when we were able to subtract a psf from the image and the filtered image showed a point source we kept those detections as real. 

 In some cases, the visual inspection of the filtered image indicated that a 24 $\mu$m  source might not be real or could be strongly affected by the  inhomogeneous extended emission. These values have been flagged with an asterisk in Table \ref{tabPhotometry} and, although some of these detections might be real, the listed values should be used with some caveats.

The LABOCA FOV is not fully covered by the Spitzer/MIPS  image at 70~$\mu$m (M2 band), 
which presents stripes with no data every 3 arcmin all along the north-to-south direction.
 We extracted the M2 photometry in all the LABOCA detections covered by this filter, 
and the fluxes are listed in Table~\ref{tabLABOCADetections}. 
For the sources with MIPS M2 emission 
peaking inside the beam, we measured the flux inside an aperture of ~30 arcsec
 of diameter (the LABOCA beam) centered on the core, and subtracted the
 background emission assessed in a nearby non-emitting region of the same area. 
For the sources with extended M2 emission we did not subtract the background 
as this cannot be easily assessed, and adopted this value as a conservative upper limit. The completeness and limiting magnitudes are:
[3.6]$_{compl}$=17.5 mag,
[4.5]$_{compl}$=17.0 mag,
[5.8]$_{compl}$=15.0 mag,
[8.0]$_{compl}$=14.5 mag,
[24.0]$_{compl}$=8.5 mag; and 
[3.6]$_{lim}$=18.25 mag,
[4.5]$_{lim}$=17.75 mag,
[5.8]$_{lim}$=15.50 mag,
[8.0]$_{lim}$=14.75 mag,
[24.0]$_{lim}$=9.25 mag, when selecting objects with magnitudes with errors smaller than 0.15 mag
(Table \ref{tabInstruments}).

%
\subsection{Public databases\label{sub:Publi}}
%

We have also searched for photometric data in public archives. Among others,
we have taken advantage of the tool implemented within Virtual Observatory SED Analyzer 
(VOSA\footnote{http://svo2.cab.inta-csic.es/svo/theory/vosa4}, \citealt{Bayo2008.1} and 
 Bayo et al. 2018, in prep.),
 under the Virtual Observatory protocols, 
and developed by the Spanish node\footnote{http://svo.cab.inta-csic.es/main/index.php}.
See details below.

\subsubsection{Akari mid- and far-IR data \label{subsub:AKARI}}
%

The Japanese satellite Akari has provided an All-sky catalog in the mid- and
 far-IR by using two different instruments: IRC and FIS.
 Updated information can be found at the mission 
webpage\footnote{http://www.ir.isas.jaxa.jp/AKARI/Publications/guideline.html\#REFS}.

In the first case, 
 comprehensive information about IRC can be found in
\cite{Ishihara2010.1}, and should be complemented with the mission release notes.
Akari has provided fluxes in the S9W and L18W filters down to 50 mJy and 90 mJy --IRC Release We note p. 29--
 (50 mJy and 120 mJy nominal, 5$\sigma$ detections). The effective beam
size is 5.5 and 5.7 arcsec, respectively 
 and sources within 7 arcsec of each other are considered the same. 
Regarding the accuracy of the pointing (IRC Release Note p. 19), the latest value is
 $0.765\pm0.574$ arcsec. 
Moreover, 95\% of the Akari/IRC detections are closer than 
 2arcsec to the 2MASS counterparts (75\% are closer than 1 arcsec).

In the case of the FIS instrument \citep{Kawada2007.1}, there are four different wide channels, namely
N60 ($65~\mu$m), WIDE-S ($90~\mu$m), WIDE-L ($140~\mu$m), and N160 ($160\; \mu$m). 
The    5$\sigma$ limiting fluxes in  survey mode are 2.4, 0.55, 1.4 and 6.3 Jy. 
Effective size of the point spread function of AKARI FIS in FWHM is
 estimated to be $37 \pm 1$ arcsec, $39 \pm 1$ arcsec, $58 \pm 3$ arcsec, and $61 \pm 4$ arcsec at N60, WIDE-S, WIDE-L, 
and N160.
Finally, the pointing accuracy is 3.8 arcsec in RA ($\sim$ cross-scan) and 4.8 arcsec in DEC ($\sim$ inscan), 
as stated in the FIS Release Note p. 27.

\subsubsection{WISE mid-IR photometry\label{subsub:WISE}}
%

We have also made use of the Wide-field Infrared Survey Explorer (WISE; \citealt{Wright2010.1})
and its latest version
(AllWISE Data Release, November 13 2013, \citealt{Cutri2013-WISE}).
WISE has mapped the sky at 3.4, 4.6, 12, and 22 $\mu$m (W1, W2, W3, W4) with an angular resolution of 6.1, 6.4, 6.5 \& 12.0 arcsec
  (FWHM psf) in the four bands, achieving 5$\sigma$  sensitivities better than 0.08, 0.11, 1 and 6 mJy in unconfused regions for point sources.
The All-Sky Release includes all data taken during the WISE full cryogenic mission phase (7 January 2010 to 6 August 2010) and its
postcryogenic phase (\citealt{Mainzer2011-NeoWISE}).
In addition, we ignored values with no errors, and also 
removed some data after a visual inspection, based on the significant source confusion.

\subsubsection{Optical data from Dolan \& Mathieu\label{subsub:DMdata}}
%

We used the published data from \cite{Dolan1999.1, Dolan2002.1}, which includes optical photometry in three
bands down to completeness and detection  limits, in the B30 area, of
$V_{compl}$=18.5, 
$R_{compl}$=17.5,
$I_{compl}$=17.0 mag; and
$V_{lim}$=21.0, 
$R_{lim}$=21.0 and
$I_{lim}$=19.0 mag.
The data were calibrated with standards stars from \cite{Landolt1992.1} and, therefore, 
the  $R$ and $I$ photometry is tied to the Cousins system (\citealt{Cousins1976.1}, also called as Kron-Cousins) 
and the $V$ to the \cite{Johnson1963-PhotometricSystems} \citep[see also][]{Landolt1973.1}.

We note that several objects spectroscopically studied by \cite{Dolan1999.1} are located within the APEX/LABOCA FOV, namely:
DM114, 115, 118, 121, 125, 127, 131, 135, 136, 142, 143, 149, 152, 158, 164, 169, and 172. They are characterized by radial velocity 
compatible with membership to the Lambda Orionis star forming region and a lithium detection, a clear indication of youth 
for late spectral type stars. Only two among this group are close to our APEX/LABOCA sources, namely DM115 --inside LB33-- and DM142 --within LB20--
(see subsections \ref{sub:groupC} and \ref{sub:groupB}, respectively).
The others are not detected at 870 $\mu$m with APEX/LABOCA.

\subsubsection{IRAS data\label{subsub:IRAS}}
%

Two of our APEX/LABOCA detections have been identified with IRAS sources, namely
B30-LB19 (IRAS05286$+$1203) and B30-LB32 (IRAS05293$+$1207),
 using a 25 arcsec radius search.
 The fluxes are listed in Table \ref{tabLABOCADetections}. 

\subsubsection{2MASS near-IR photometry\label{subsub:2MASS}}
%

The  All-Sky 2\,MASS catalog \citep{Cutri03.1, Skrutskie2006.1} has completeness and limiting magnitudes, 
in the central region of the B30 dark cloud,  of
$J_{compl}$$\sim$16.65 mag, $H_{compl}$$\sim$15.95 mag, $Ks_{compl}$$\sim$15.45 mag;
and $J_{lim}$$\sim$18.95 mag, $H_{lim}$$\sim$18.05 mag, $Ks_{lim}$$\sim$17.45 mag.
They have been used, among other things, to calibrate and complement our deep near-IR data.

\subsubsection{Absorption map based on 2MASS\label{subsub:Av}}
%

 We have followed the approach in \cite{Cambresy1997.1} to derive an extinction map of the B30
 sky area via an adaptive-step star count method. We used as base for the method the
2MASS all-sky point-source catalog \citep{Skrutskie2006.1} as in \cite{Lopez2013.2}. This technique,
 based in local density of stars as compared to an unobscured field at the same galactic latitude 
(in our case we use Collinder 69 as comparison field given its low and homogeneous extinction properties),
 provides an average resolution of 1.5\arcmin $\,$ across a region centered at 05:29:58.50 +12:01:52.0 with 1.5 degree radius.
The results, for each APEX/LABOCA detection, are listed in
Table~\ref{tabLABOCADetections}
and the conversion from A$_{\rm J}$ to A$_{\rm V}$ follows \cite{Fitzpatrick1999-Extinction}.

%
\subsection{Cross-correlation and identifications\label{sub:crosscorrelation}}
%

We have carried out a careful cross-correlation of the data taking into account both the errors in the pointing,
 which are not negligible in the case of APEX/LABOCA, and the beam size. 
The relatively wide beam sizes are
  very important when cross-matching detected sources at other wavelengths for
  Akari/FIS, Spitzer/MIPS, Wise (particularly W3 and W4) and APEX/LABOCA, and still
  significant for WISE W1 and W2, Akari/IRC, and Spitzer/IRAC.

The beam size of APEX/LABOCA has an angular size of $\sim$27.6 arcsec, so multiple identifications with optical, 
near- and mid-IR sources are possible.
Since the grid pixel is about $4\time4$ arcsec, 
when searching for possible counterparts, 
we have assumed a 5 arcsec angular distance as the optimal counterpart search radius
 (at least for the non-extended sources).
 We note that some stars could have moved from their birth site, so the position of YSO could differ from the center of the submillimeter emission. \cite{Jorgensen2008-SF-OphPerseus} have studied the correlation between MIPS and SCUBA data in Perseus and Ophiuchus and concluded that no significant dispersion has taken place if the formation is recent. In any case, we cannot completely rule out the possibility of a real ofset due to a drift.
In addition, we have visually inspected all possible cross-matches and in
 some cases rejected the WISE W3 and W4 due to confusion, as stated before. The results, including all possible
identifications and the magnitudes at each filter, are listed in 
Table~\ref{tabPhotometry}  and Table~\ref{tabPhotometryWISE} 
--near- and mid-IR photometry,  
Table~\ref{tabPhotometryOpt2M} --optical plus 2MASS, and 
Table~\ref{tabAkari} --Akari data.

%
\subsection{Classification of the  LABOCA submm sources and the counterparts\label{sub:ClassificationCounterparts}}
%

We have cross-correlated the data from LABOCA with our catalogs of optical/IR sources.
Based on the availability of data at  70 $\mu$m as a primary indicator and 24 $\mu$m as a secondary,
we have classified the cross-matches into three groups, as listed in Table \ref{tabLABOCADetections}:
\begin{itemize}
\item Group A.- Sources detected at 70 $\mu$m.
  Depending whether they  also have been detected at 24 $\mu$m they would belong to subgroup A1 or A2
  (with or without measured flux  at 24 $\mu$m, respectively).
\item Group B.- Undetected sources at 70 $\mu$m (i.e., upper limits with Spitzer/MIPS M2).
  As in the previous case, we have differentiate whether  they  also have been detected at 24 $\mu$m:
  subgroups B1 ad B2 correspond to those with or without measured flux  at 24 $\mu$m
\item Group C.- LABOCA sources outside the MIPS M2 FOV (i.e., no information at 70 $\mu$m, see subsection \ref{sub:Spitzer}).
  Some have been detected at 24 $\mu$m (subgroup C1) whereas others only display upper limits (subgroup C2). 
\end{itemize}

This classification appears in Table~\ref{tabLABOCADetections}. We have also included information regarding
whether we classify the submm source as YSO (regardless of whether it is stellar or substellar), there are counterparts with IR excesses
within the beam or the source seems to be a starless core;
whether there is a potential substellar candidate (sometimes more than one) within the whole APEX/LABOCA beam;
and whether there is a counterpart within 5 arcsec from the APEX/LABOCA emission peak (and whether it could be a brown dwarf).
A detailed discussion can be found in Appendix~\ref{appendix}.

%
\section{Properties of the submm APEX/LABOCA sources\label{sec:properties}}
%

%
\subsection{Individual submm sources and a estimate of the emiting mass \label{sub:EnvelopeMass}}
%

We show the 870 $\mu$m LABOCA emission of the B30 dark cloud,
together with the detected submm sources, in  Fig.~\ref{FOV_LABOCA}a. On top of the image we display the
H$\alpha$ nebular emission  (Virginia Tech Spectral line Survey with 6 arcmin resolution, \citealt{Finkbeiner2003-HalphaImage}).
Most of the detected sources are located in a ``valley'' in between the H$\alpha$ peaks.
Fig.~\ref{FOV_LABOCA}b correspond to a  Spitzer/MIPS1
image at 24~$\mu$m in colorscale, where the submm LABOCA sources have been 
overplotted  (see subsection \ref{sub:LABOCAmapping} for details).
The 24~$\mu$m image traces the hot dust at the border of the pillar 
facing the star $\lambda$ Ori, where the ionization front from the O-type star is interacting with B30.
Another bright region at 24~$\mu$m is found about 4~arcmin to the north of this sharp edge, 
where several point-like sources surrounded by extended emission are located. 

Our LABOCA map at 870 $\mu$m, tracing mainly the emission of cold dust, reveals tens of cores, some of 
them showing extensions and substructure, which follow the filaments seen in the MIPS image at 24 $\mu$m, 
and which are mainly distributed in two parts of the cloud. 
The first group of cores lie slightly north of the edge of the HII region, and are arranged in filaments 
elongated in the east-west direction, while the second group matches well the 24~$\mu$m bright emission 
$4'$ north of the cloud border and follow a rather north-south direction. 

We identified a total of 34 cores above the 4-$\sigma$ level (see their labels in Fig. \ref{FOV_LABOCA}),
and their flux and envelope mass (see below)
 are provided in Table~\ref{tabLABOCADetections}. Near the edge of the HII region, sources B30-LB21 to B30-LB24, and B30-LB29 to B30-LB31 follow
 remarkably well the 24 $\mu$m emission tracing the border of the pillar. In this region the two strongest sources are B30-LB19, 
which is associated with a 24 $\mu$m point source, and  B30-LB27, which is more extended and fainter in the infrared. 
A ridge of faint sources (B30-LB20, B30-LB25, B30-LB26) joins B30-LB19 and B30-LB27. The other group of submillimeter sources located
 farther to the north are also forming filamentary structures. For example, sources B30-LB05 to B30-LB07 are arranged in one 
single north-south filament, and sources B30-LB08, B30-LB09 and B30-LB12, are forming a chain of cores along the southeast-northwest direction.
 Finally, sources B30-LB02 to B30-LB04, still $4'$ farther to the north, seem to be the continuation of the B30-LB05/B30-LB07 filament. 
The chain of cores B30-LB02/B30-LB07  is dark in the infrared, 
and are associated with only very weak 24 $\mu$m emission,
 contrary to the filamentary structure of cores B30-LB08/B30-LB12, which is associated with bright 24 $\mu$m emission.

We estimated the envelope masses in the following way.
The total mass $M$ of gas and dust from thermal continuum emission, assuming that the emission is optically thin, is:

\begin{equation}
M=\frac{S_\nu D^2}{B_\nu(\Td)\kappa_\nu},
\end{equation}
where $S_\nu$ is the flux density at the frequency $\nu$, $D$ is the distance to the Sun, $B_\nu(\Td)$ is the Planck function at the dust temperature $\Td$, and $\kappa_\nu$ is the absorption coefficient per unit of total (gas+dust) mass density. Writing Eq.~(1) in practical units:

\begin{equation}
\left[\frac{M}{M_{\odot}}\right]=3.25\times
\frac{e^{0.048\,\nu/\Td}-1}{\nu^3 \kappa_\nu}\times
\left[\frac{S_\nu}{\mathrm{Jy}}\right]
\left[\frac{D}{\mathrm{pc}}\right]^{2},
\end{equation}
where $\Td$ is in K, $\nu$ is in GHz, and $\kappa_\nu$ is in cm$^2$g$^{-1}$. 
For the absorption coefficient at 870~$\mu$m or 345~GHz we used the interpolated value from the 
tables of \cite{Ossenkopf94.2}, for the case of thin ice mantles and density of 10$^6$~\cmt: 0.0175~cm$^2$\,g$^{-1}$.
  Since we do not know the evolutionary status  of the detected cores, we have adopted an average dust temperature
of 15\,K for both pre-stellar and protostellar cores \citep[see e.g.,][]{SanchezMonge2013-DenseCores}.

The total masses estimated from the 870~$\mu$m continuum emission range 
from 0.043 up to 0.166~\mo ~(or 0.279 $M_\odot$ if the extended emission is taken into account).
 The  sources with largest masses, above 0.1~$M_\odot$, are B30-LB01, B30-LB19, B30-LB27, B30-LB30,  and B30-LB32, if we only consider the peak. 
In addition, when the extended emission is taken into account, this limit is also surpassed by  B30-LB06, B30-LB08, B30-LB13, B30-LB21, and B30-LB25.
Thus, the submm sources with dust masses below 0.1  $M_\odot$ are:  B30-LB02, B30-LB03, B30-LB04, B30-LB05,
B30-LB07, B30-LB09, B30-LB10, B30-LB11, B30-LB12, B30-LB14, B30-LB15, B30-LB16, B30-LB17, B30-LB18,
B30-LB20, B30-LB22, B30-LB23, B30-LB24, B30-LB26, B30-LB28, B30-LB29, B30-LB31, B30-LB33, and B30-LB34.

%
\subsection{Evolutionary status of the optical/IR counterparts to the submm sources\label{sub:EvolutionaryStatus}}
%

After the compilation of our photometric catalog of possible counterparts of the APEX/LABOCA
sources (Table~\ref{tabLABOCADetections}),
 our first step has been to identify possible infrared excesses based on Spitzer/IRAC data.
 Figure \ref{fig_CCD_CMD_IRAC} displays a CCD -- panel a -- and several
 color-magnitude diagrams (CMD) --panels b, c, d, and e-- for the
 counterparts identified within the 34 APEX/LABOCA sources. The figure also includes data corresponding to the somewhat older  C69 cluster, 
which also belongs to the same star forming region (5 Myr for C69 versus 1-3 Myr  for B30).
The data come from \citet{Barrado2007.1}, \citet{Barrado2011.1} and \citet{Bayo2011.1}.
We have also added information (plotted on the figure) regarding to  members  in Taurus (\citealt{Luhman2006.2}) and
Serpens (\citealt{Harvey2007-Serpens_YSO-IRAC-MIPS}), which is 
about 1-2 Myr. 
We have included  several extra-galactic samples from \citet{Sacchi09.1}: we show the AGN1/QSO1/broad lines and
AGN2/obscured QSO/narrow lines as well as resolved galaxies
and emission line galaxies
 (which include AGN and star forming galaxies), respectively. We have been able to establish whether our B30 counterparts are likely members or not (see below), and our classifications are indicated in the figure.

Figure~\ref{fig_CCD_CMD_IRAC}  clearly shows that a significant number of our counterparts might have circum-(sub)stellar
disk or envelopes. Therefore, some might be classified as Class~II or Class~I objects, or something in-between. There are
a number of disk-less objects (Class III), and some of them might even belong to the association,
 in an analogous situation to what happens in Taurus (simultaneous presence of members in the Class~I, II and III phases, see
panel a). However, it is also clear, 
from the comparison with resolved and emission line galaxies and AGN from \citet{Sacchi09.1},  
that our sample can be polluted by a significant number of extra-galactic sources,
 specially  for the faint end (Spitzer/IRAC I1$\ge$15 mag). Therefore, in order to select bona-fide candidate members, 
we have carried out a detailed comparison using our wealth of data.
These figures  (see e.g., Figure~4) suggest that the pollution by extra-galactic sources should be reduced,
since the loci of most of our candidates in the CMDs are not coincident with the population of galaxies. In fact,
confirmed members in Serpens
from \cite{Harvey2007-Serpens_YSO-IRAC-MIPS} are closer to these pollutants. Moreover, the  number of
extra-galactic sources (\citealt{Surace2004-SWIRE}, \citealt{Sacchi09.1}) within the APEX/LABOCA beam should
be around 1-2.5 galaxies, and 0.07-0.03 within the central 5 arcsec (i.e., an improbable chance alignment between an
extra-galactic center and the submm source).

Both   \citet{Harvey06.1} and \citet{Gutermuth2008.1} discuss the nature of Spitzer sources based on CCDs and CMDs built with IRAC and MIPS data, namely [(I2-I3),(I3-I4)], [(I1-I3),(I2-I4)], [I2,(I2-I4)].
We have used their criteria and  verified that a small number of our possible counterparts fall in the regions
where extra-galactic sources are located. We note that young substellar objects could be located in the same areas
in the CCD and CMD \citep[see the case of 
an excellent proto-brown dwarf in Taurus][]{Barrado2009-ProtoBD-Taurus,Palau2012-ProtoBD-Taurus}. 
We have classified
these objects as possible extra-galactic, and added the tag G to the membership criteria listed in Table \ref{tabmembership}.

Moreover, for the handful also having optical photometry in the $I$ band, we have checked their position in an
 optical/near-IR/mid-IR CCD, following \citet{Bouy2009-LOri}, in order to reveal the presence of quasars. Only two of them 
appear likely to be quasars, and this information has been added to Table \ref{tabmembership}.
Obviously, additional analysis is required to assess the membership of all these counterparts to the B30 association.

%
%

\subsection{Membership of the optical/IR counterparts to the submm sources\label{sub:membership}}

Several additional CMDs, which combine near- and mid-IR photometry, can be found in 
Fig.~\ref{fig_CCD_other}. This figure is similar to Fig.~\ref{fig_CCD_CMD_IRAC}, but in it we have included several BT-Settl
isochrones from the Lyon group \citep{Allard2012.1} namely those corresponding to ages of 3  and 20 Myr, 
in order to establish whether the counterparts are young or old.
Without taking into account the interstellar and inter association reddening (as well as that produced by the object itself),
these diagrams suggest the presence of a very large number of very low-mass objects.
 However, as stated before, we expect a significant
rate of pollutants and the reddening effect should be taken into account (it will be dealt with later on).

For each object and CMD, when the photometry is available, we have assigned a qualitative membership tag,
 namely Y --younger than 20 Myr, Y? --between 20 and 50 Myr,   
N? --between 50 and 10,000 Myr,  and N --older than the 10,000 Myr isochrone.
These last two isochrones are not represented in the figure for clarity.
 We note that the interstellar reddening vector is practically parallel to the isochrones (indicated each panel), 
so it has not relevant effect in this classification.

Moreover, in order to consider all the photometric information simultaneously,
we have derived the effective temperature and bolometric luminosities of all possible counterparts 
by using VOSA \citep{Bayo2008.1}. Since the current version
 of VOSA includes the possibility of adjusting the reddening, we have produced two sets of $T_{eff}$ and $L_{Bol}$: 
i) by fixing $A_v=0.322$ mag --an average value for the Lambda Orionis  star forming region-- and 
ii) by letting it be variable. We always assumed a distance of 400 pc, solar metallicity ($[Fe/H]$=0.00)
and $log g$=3.5 dex. Results are listed in Table~\ref{tabmembership}. We selected BT-Settl models from \citet{Allard2012.1}
and two Hertzsprung-Russell Diagrams are represented in Fig.~\ref{HRD}. A similar exercise to the one carried out with the CMD,
 regarding membership and qualitative tags,
 has been executed using the effective temperature and the bolometric luminosity. However, in this case 
objects located between the 20 and the 10,000 Myr isochrones have been labeled as Y?, whereas 
 objects looking like older than the universe (below the 10,000 Myr isochrone) has been assigned 
the tag N, since they are either extra-galactic or the assumed distance is incorrect 
(and they cannot be related to the B30 cloud).

All these membership quality tags, when combined with the information regarding the number
 of photometric points in the SED, the SED shape and the evolutionary 
status (Class I, II and III, based on the IRAC data), and the possibility of being extra-galactic sources, 
 have been used to produce a final 
 membership classification (Table~\ref{tabmembership}).
 In the end, we have been able to provide a final membership classification:
 probable and possible members (tags Y and Y?) and 
possible and probable non-members (N? and N). They are represented in 
Figs.~\ref{fig_CCD_CMD_IRAC} and  \ref{fig_CCD_other},
whereas  Fig.~\ref{HRD} differentiates probable and possible members with solid and open green circles. 
We note that for some counterparts, we have not been able to elucidate whether they belong or not to the association due
to the lack of enough photometric information. These objects are listed with a ``--'' tag.

\subsection{Bolometric luminosity and temperature\label{sub:LbolTbol}}

By using our massive multi-wavelength database,
which includes photometry from 0.5 to 870 $\mu$m,
 we have estimated the bolometric temperature and luminosity 
(see details in \citealt{Palau2012-ProtoBD-Taurus}).
 The results are listed in
 Table~\ref{TABbolometric}   and
 displayed in Figs. \ref{HR_TbolLbol_Evolution_A}
 and \ref{HR_TbolLbol_Evolution_C},
 where we have included several samples 
of young stars and VeLLOs (\citealt{Chen1995-Tbol-TaurusOph};
\citealt{Young2005-SignaturesYSO}; 
\citealt{Dunham2008-Protostars};
\citealt{Bayo2011.1}; 
 \citealt{Tobin2016_Multiplicity_Protostars_Perseus}). 
 Our Barnard 30 targets appear as green, cyan and red symbols
 (depending on the figure and the classification, see subsection \ref{sub:ClassificationCounterparts}),
 where the size increases with the number of data-points 
(i.e., being more reliable those with a bigger size).
  We have only included those objects detected at 24 and/or 70 $\mu$m.
 They display significant IR excesses, so the T$_{bol}$ and $L_{bol}$ departs from our previous
 estimate of  T$_{eff}$ and $L_{bol}$ based on VOSA (the photospheric emission).
 Our B30 objects
 have been classified, in most  cases, as Class I  or Class I/II based on the IRAC data. We note that the 
flux at 870 $\mu$m has been assigned in general to the ``a'' components when there are multiple identifications
(see Appendix).
 Thus, we have studied in detail 15 objects detected at 24 $\mu$m which are within 11 APEX/LABOCA sources and another eight objects in four different LABOCA beams which display a emission at 70 $\mu$m but they do not have a detected counterpart at 24 $\mu$m.

Figures \ref{HR_TbolLbol_Evolution_A}
and \ref{HR_TbolLbol_Evolution_C}
strongly suggest that we have uncovered a few examples of
 very young, low-massive objects.
 They would be objects in the evolutionary stage between Class 0 and Class I
in the classification scheme of  \citet{Lada87.1} and \citet{Adams87.1}. 
If we take into account the $L_{int}$, as estimated from the flux at 70 $\mu$m,
one object seems to follow fully in the VeLLO category
 (LB19d, \citealt{Huelamo2017-ALMA-B30}).
In addition, there is a sample  of Class~I objects with bolometric luminosities below 0.1  $L_\odot$ and T$_{bol}$
in the range 70-700 K.
 Unfortunately, some of them have less than ten data-points in the SED
 and therefore they are not very well characterized.
 In any case, since there is also another subsample located
 in the Class II or III are, as defined by  \cite{Young2005-SignaturesYSO} and \cite{Dunham2008-Protostars}, it
 seems that there is an evolutionary sequence from Class 0/I very low-mass objects, probably substellar, down to 
the pre-main sequence. 
This  trend persists even if the fluxes at 870 $\mu$m (and 70 $\mu$m when available)
are included for these counterparts. As a matter of fact, although the
$L_{bol}$ increases,  T$_{bol}$ is shifted toward cooler values and  new candidate VeLLOs appear
(if, indeed, they are the origin of the submm emission).
Regardless of this assumption, what we might be contemplating in the Barnard 30 dark cloud is the unveiling of the brown
 dwarf population. In next subsection we explore this possibility by having a look at the individual objects.
 In any case, the 20 objects clearly detected at 24 and/or  70 $\mu$m have been classified as: two WTTs, four CTTs, five YSOs, eight proto-BD candidates and one VeLLO. Another three objects have emission at 24 $\mu$m strongly affected by the extended photometry, although they seem to be substellar and the can be classified as proto-BDs candidates.

%

%

%
\subsection{Starless cores?: LABOCA sources without unambiguous young optical/IR counterparts\label{sub:StarlessCores}}
%

In the case of six LABOCA sources, namely 
B30-LB01, B30-LB06, B30-LB13, B30-LB15, B30-LB16 and B30-LB34,  there is not any optical/IR counterpart detected within 5 arcsec.
Some include several counterparts detected between 5 and 27 arcsec which have been
classified as probable or possible cluster members, but they do not display any obvious excess.
In these six cases, we have considered the possibility that these submm sources are starless cores in B30,
and  we have  investigated their nature. We note that all of them lack data at 70 $\mu$m, either because there is no
positive detection or because the LABOCA source is outside the MIPS M2 FOV.

B30-LB15 and B30-LB16 have envelope masses around 0.05~\mo. The marginal detection of these 
sources prevents us from drawing any further conclusions regarding their nature. Each of them contain
only one candidate member (counterparts \#a in both cases, see Fig. \ref{FCh_groupC1}) 
and their SED (Fig. \ref{SED_groupC1})  are typical of a Class III, disk-less object,
although in the case of B30-LB16a it seems substellar.
 Thus, no convincing optical/IR counterpart can be assigned to these cores.

 B30-LB34 has been only detected with APEX/LABOCA. It can be considered as a starless core.
 We note, however, that it is associated to B30-LB33
 (Fig. \ref{FCh_groupC1}), a class II star. Thus, if  physically associated, they would conform a
 pair in very different evolutionary stage.

B30-LB01, B30-LB06, and B30-LB13 have relatively large submillimeter flux densities, and thus their 
estimated envelope masses are among the largest, between 0.108 and 0.260 M$_\odot$ (or 0.077 and 0.166 M$_\odot$   if we only take
the peak intensity for the extended sources). Given that these masses are
 around the substellar regime ($\sim$0.072 M$_\odot$), and the known fact that not the entire mass of a starless
 core ends up to form the star, 
these are excellent candidates to pre-substellar (pre-BD) cores, that is, cores which will form brown dwarfs in a future but did not form an hydrostatic core yet 
 (e.g., \citealt{Palau2012-ProtoBD-Taurus, Andre2012-PreBD}).
Their submm emission is also extended in these three cases, with sizes of 22, 49 and 23 arcsec, 
respectively.
In order to elucidate their true pre-BD nature, it is necessary to conduct observations of molecular 
lines, to assess whether these cores are indeed gravitationally bound. If confirmed, they would
 provide strong evidence favoring the formation of brown-dwarfs in isolation, and thus their 
formation from the gravoturbulent-fragmentation scenario.

%
\section{General discussion and conclusions\label{sec:discussion}}

The analysis of the LABOCA data of B30 has revealed the detection of 34 submm sources. 
15 of them have a clear optical/IR counterpart within 5 arcsec (Table \ref{tabLABOCADetections}), 
have been classified as a probable or possible cluster members and they display a significant infrared excess.
Among them,  six have been detected at 24 $\mu$m with MIPS M1. Another two are do not have emission at this band but have been detected at 70 $\mu$m with MIPS M2.
Therefore, we can conclude
that the excesses are well established and the sources are well characterized.
Nine out of the 15 might be substellar objects based on the mass of their envelopes and the shape of their
SED. Another one would be at the borderline between stars and brown dwarfs. In any case, the final
mass would depend on the actual reddening due both to the dark cloud and the object itself (when the 
submillimeter emission is extended). 

Two submm sources in this group  are visual binaries, in the sense that there are two
optical/IR counterparts which have been classified as probable/possible members,
display infrared excesses and are located  within the central 5 arcsec of the LABOCA beam.  
Our current spatial resolution does not allow to assign it to any component. One system, namely B30-LB19, 
has a primary component (counterpart \#a) at the borderline between stars and brown dwarfs, while the more massive component,
further away from the center of the submillimeter emission, would be stellar. Even further,
 but still inside the 27.56 arcsec of the APEX/LABOCA beam, there are several possible/probable members, classified as 
Class I sources, which could have a substellar nature.
Therefore, we may be dealing with a clump  
which is  giving birth to several very low-mass stars and/or brown dwarfs.
The other APEX/LABOCA binary, namely B30-LB25, includes counterparts \#a and \#e.

The second group is composed by those submm sources (13) with no optical/IR counterpart within 5 arcsec but which include sources
within the APEX/LABOCA beam displaying excesses
(labeled excess in Table \ref{tabLABOCADetections}).
All 13 sources contain within the submm beam at least one optical/IR counterpart which can be classified as BD if membership
is confirmed.
Eight out of 13 have envelope masses below the substellar limit and, in fact,
the individual SEDs are characteristic of brown dwarfs, if members. In any case, due to the number of possible and probable members
identified within the submm beam, we cannot unambiguously assign the source of the submmillimeter
emission to any individual object.
Nine have been detected at 24 $\mu$m (although three of these detections are problematic due to the extended nebulosity and it is possible that some among these three are not real).  

Finally, six LABOCA sources might be starless cores, with three of them having masses near
the substellar limit and thus are pre-BD core candidates. An additional follow-up of molecular
line observations is required to confirm their dynamical status and their nature.

Regarding multiplicity, it seems there is no obvious difference between the submm sources with a counterpart closer than 5 arcsec and without it. 
The first one contains multiple possible and/or probable members as counterparts
within the APEX/LABOCA beam in 13 cases out of the 15 sources. The ratio is 11 to 13 for the second group.
 However, the six starless cores might be genuinely different, since the
ratio is three out of six. However, the low number statistics might be playing a role.

Based on the availability of data at 70 and 24 $\mu$m, we have discussed the properties and evolutionary stage of our
submm sources in Barnard 30.
Overall, the most relevant
object is B30-LB19d. With L$_{int}$=0.094 $L_\odot$, as estimated from the flux at 70 $\mu$m,
it is the best Class I BD identified so far in this star forming region. \cite{Huelamo2017-ALMA-B30} has suggested that it could be associated with a bipolar nebula.
  Six counterparts, namely B30-LB11a, B30-LB12d, B30-LB22b, B30-LB22j, B30-LB30g and  B30-LB31c, have been detected at 24 $\mu$m and the have classified as proto-BD candidates.
 In addition, three objects associated to B30-LB29 (\#a, \#b and \#f) and the two counterparts related to B30-LB23 (counterparts \#a and \#c) are very interesting proto-BD candidates, since these two LABOCA sources have been detected at 70 $\mu$m,   However, the problematic photometry at 24 $\mu$m or lack of it poses some significant caveats regarding their evolutionary status.
 Finally, we have identified four sources at 24 $\mu$m which are located outside the MIPS M2 FoV (LB12a, LB18a,  LB32a and LB33a), so there is no information regarding the emission at 70 $\mu$m. They seem to be of stellar nature. B30-LB14, which includes the counterparts \#a and \#b, and has been detected at 70 $\mu$m, seems to be of stellar nature.
We note that \cite{Liu2016-LOri-ClassO-protoBD-Planck} have found Class 0 protostar and a proto-BD candidate West of our study.
Our  pre- and proto-BD candidates seem to be fainter and/or cooler than this candidate, called G192S,
based on the comparison between the SEDs (see their Figure 26).

We can compare our results in B30 with those from two regions
with very different properties: Chamaeleon II (Cha~II, see \citealt{deGregorio2016-ProtoBD-ChaII})
and the B213-L1495 clouds in Taurus 
\citep[B213 hereafter][]{Palau2012-ProtoBD-Taurus, Morata2015-ProtoBD-Taurus}.
 Unlike B30, these
are low-mass star forming regions that lack massive stars, and
therefore do not suffer from the strong and ionizing winds associated to 
hot, massive objects.

In the case of Cha~II we followed a similar strategy as in B30, and
surveyed a region of $\sim$ 34'$\times$34'
with APEX/LABOCA, detecting a total of 15 submm sources (the rms was
4\,mJy). Two of them were classified as Class~I stellar objects,
one appeared to be a proto-BD candidate, and 12 were assigned the status of starless
cores. Five (out of the 12) were good pre-BD candidates although kinematic
information is needed to assess if they are, indeed, gravitationally bound.

In the case of B213, we have performed a more exhaustive study.
We first isolated a sample of 12 proto-BDs based
on Spitzer and near-IR data \citep{Barrado2009-ProtoBD-Taurus}. We studied these candidates
with different submm, mm and cm continuum observations, together with several
gas molecules like  CO, $^{13}$CO and N$_2$H+ data \citep[for details see][]{Palau2012-ProtoBD-Taurus}.
As a result, we have confirmed two very good proto-BD candidates and one pre-substellar core candidate.
All the sources from these two works focusing on B213 were also observed with the JVLA, and we
reported the presence of thermal radiojets in four of them \citep{Morata2015-ProtoBD-Taurus}.

We have also performed  a similar detailed study to unveil the nature
of all the selected candidates in B30 (\citealt{Huelamo2017-ALMA-B30}). One important goal is to understand if the overall
properties of the region (e.g., fraction of proto-BD and pre-BD cores)
are different due to the presence of the hot star $\lambda$~Ori when compared with other regions with a
different environment.

As a final, general conclusion,
 the complete characterization of our APEX/LABOCA sources, focusing on those detected at 24 and/or 70 $\mu$m, indicates that in our sample of 34 submm sources there are, at least: two WTTs, four CTTs, five YSO, eight proto-BD candidates (with another three dubious cases due to the extended emission at 24 $\mu$m) and one VeLLO.
In addition, we have identified a substantial number of proto-BD and substellar
 starless cores in the young Barnard 30 dark cloud, located at about 400 pc.
 They are prime targets for further follow-up and characterization, as  we have done in Taurus.
 In general, high spatial
 resolution observations (with e.g., ALMA, \citealt{Huelamo2017-ALMA-B30},
 and other submm and mm arrays), both in the continuum and in specific lines, are
 necessary to assign the correct counterpart to the submm sources,  
to confirm these proto-BD and pre-BD core candidates and to reveal their characteristics and formation mechanism.


\begin{acknowledgements}
 We thank the  Calar Alto Observatory staff for their excellent work taking the near-IR data under 
 the Service Mode program. A special thanks to Rosario Lorente Balanza for her help with the Akari data.
 We do appreciate the big role the anonymous referee has played in this paper. Although we did not always agree, we thank her/him for the significant contribution.
 We also thank Almudena Alonso-Herrero for her help with the MIPS photometry.
 This research has been funded by the Spanish grants 
 AYA 2014-55840-P, and  AYA2014-57369-C3-3-P, ESP2015-65712-C5-1-R, and ESP2017-87676-C5-1-R.
 AP and MTR acknowledge financial support from UNAM-DGAPA-PAPIIT IA102815 grant, M\'exico, and from
 CATA (PB06) CONICYT, Chile, respectively.
 AB was financed by  proyecto Fondecyt Iniciaci\'on 11140572 and the Millennium Science Initiative 
(Chilean Ministry of Economy), through grant Nucleus P10-022-F.
It makes use of VOSA, developed under the Spanish Virtual Observatory project
 supported from the Spanish MICINN through grant AyA2008-02156, and
 of the SIMBAD database, operated at CDS, Strasbourg, France.
 This publication makes use of data products from the Wide-field Infrared Survey Explorer, 
which is a joint project of the University of California, Los Angeles, and the Jet Propulsion Laboratory/California Institute of Technology, 
funded by the National Aeronautics and Space Administration. 
The paper was finished during a stay at ALMA and ESO headquarters in Santiago de Chile by DB and NH, which were
supported by ALMA \& ESO and by the BBVA foundation, respectively, and another at Universidad de Valparaiso by DB, supported by proyect Proyecto Fondecyt de Iniciaci\'on 11140572 (Chile).
\end{acknowledgements}        

\small
%
\bibliographystyle{aa} 
\bibliography{00_bibliography}


%
%

\begin{figure*}[ht]
\center
\includegraphics[width=0.60\textwidth,scale=0.50]{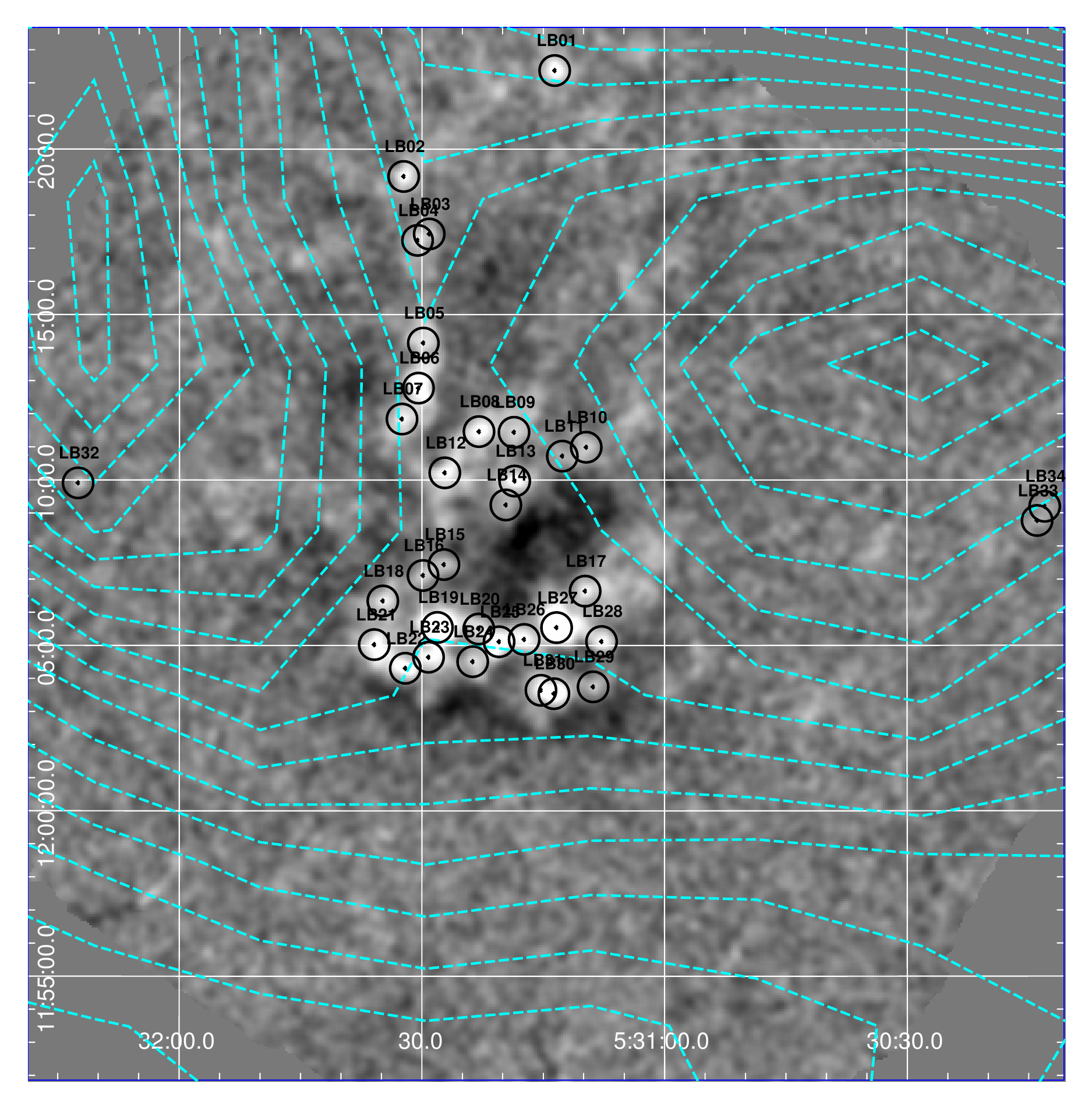} 
\includegraphics[width=0.60\textwidth,scale=0.50]{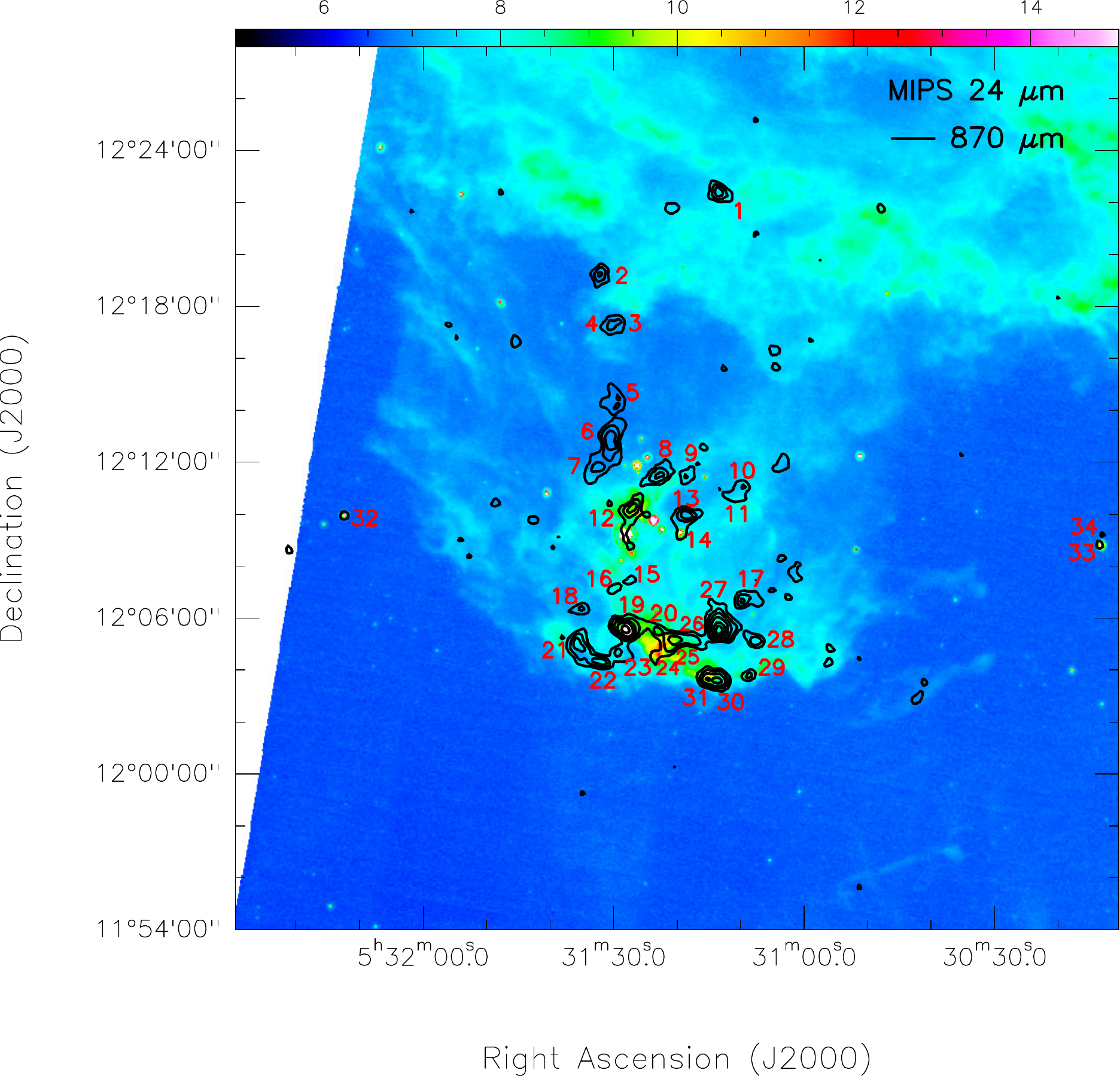} 
\caption{\label{FOV_LABOCA} 
{\bf Top.-}
 Our APEX/LABOCA image, with the 34 sources at 870 $\mu$m. Black open circles correspond
 to the LABOCA identified sources, with a 27.6 arcsec beam size. The total size of the map is about 38 arcmin in diameter.
The cyan lines represent the H$\alpha$ emission (Virginia Tech Spectral line Survey with 6 arcmin resolution, \citealt{Finkbeiner2003-HalphaImage}).
{\bf Bottom.-}
Color scale: MIPS image at 24 $\mu$m of B30. 
Black contours correspond to 3, 5, 7, 9, 12, and 15 times the rms noise of the  870 $\mu$m  emission  (which varies across the image
and it increases toward the edge of the map).
The 34 submillimeter identified sources are marked with red labels.
}
\end{figure*}

\newpage
\clearpage

\begin{figure}
\center
\includegraphics[width=\textwidth,scale=0.35]{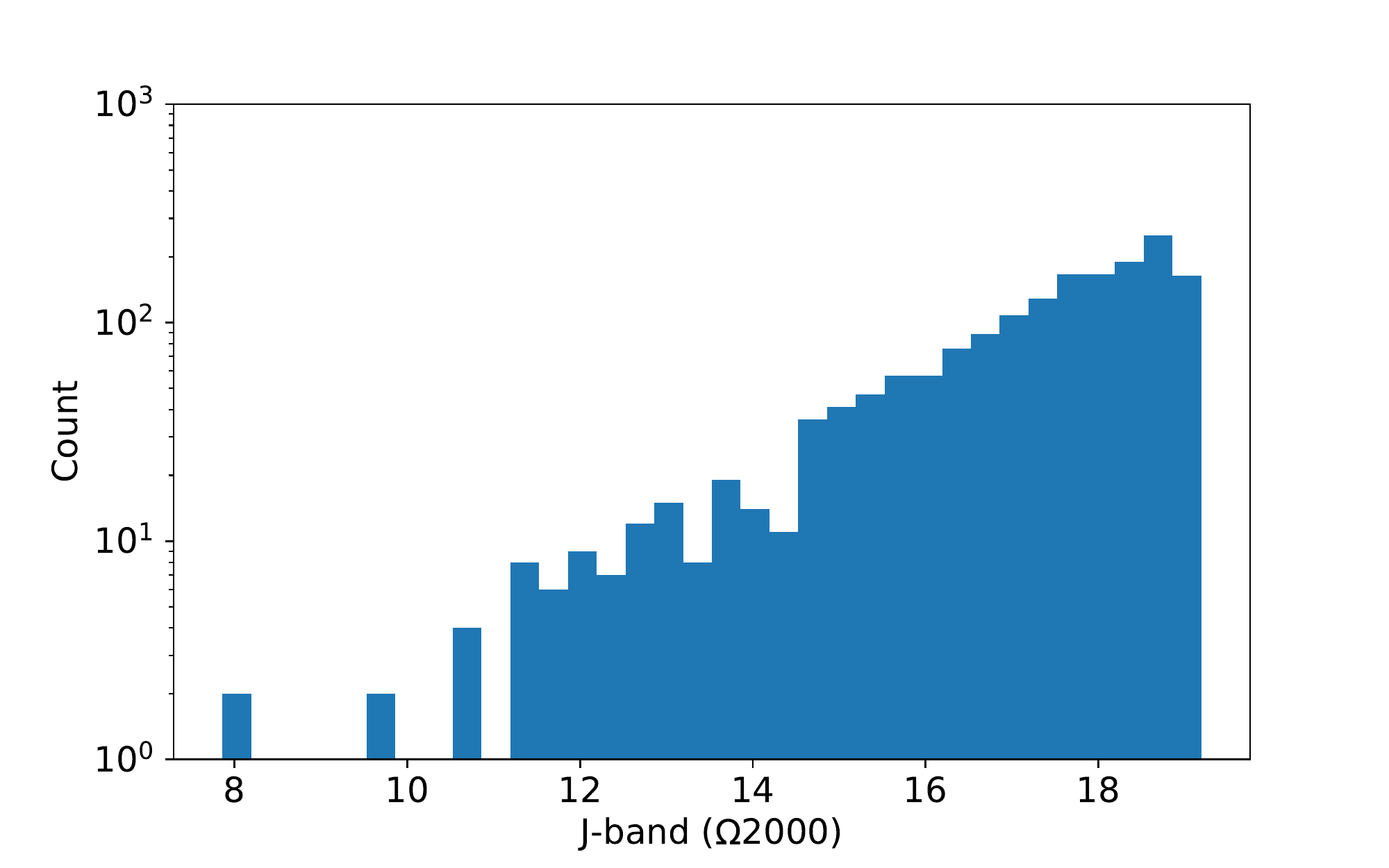} 
\caption{\label{fig_ComplJ_O2000} 
Number of near-IR detections with CAHA/O2000 in the APEX/LABOCA FOV. The bin size has been estimated via the Freedman Diaconis Estimator (resilient to outliers). We estimate
 that the completeness limit is at $J_{compl}$=18.75 mag
 and the detection limit at $J_{lim}$=20.5 mag. }.
\end{figure}

\newpage
\clearpage

\begin{figure*}
\center
\includegraphics[width=0.405\textwidth,scale=0.25]{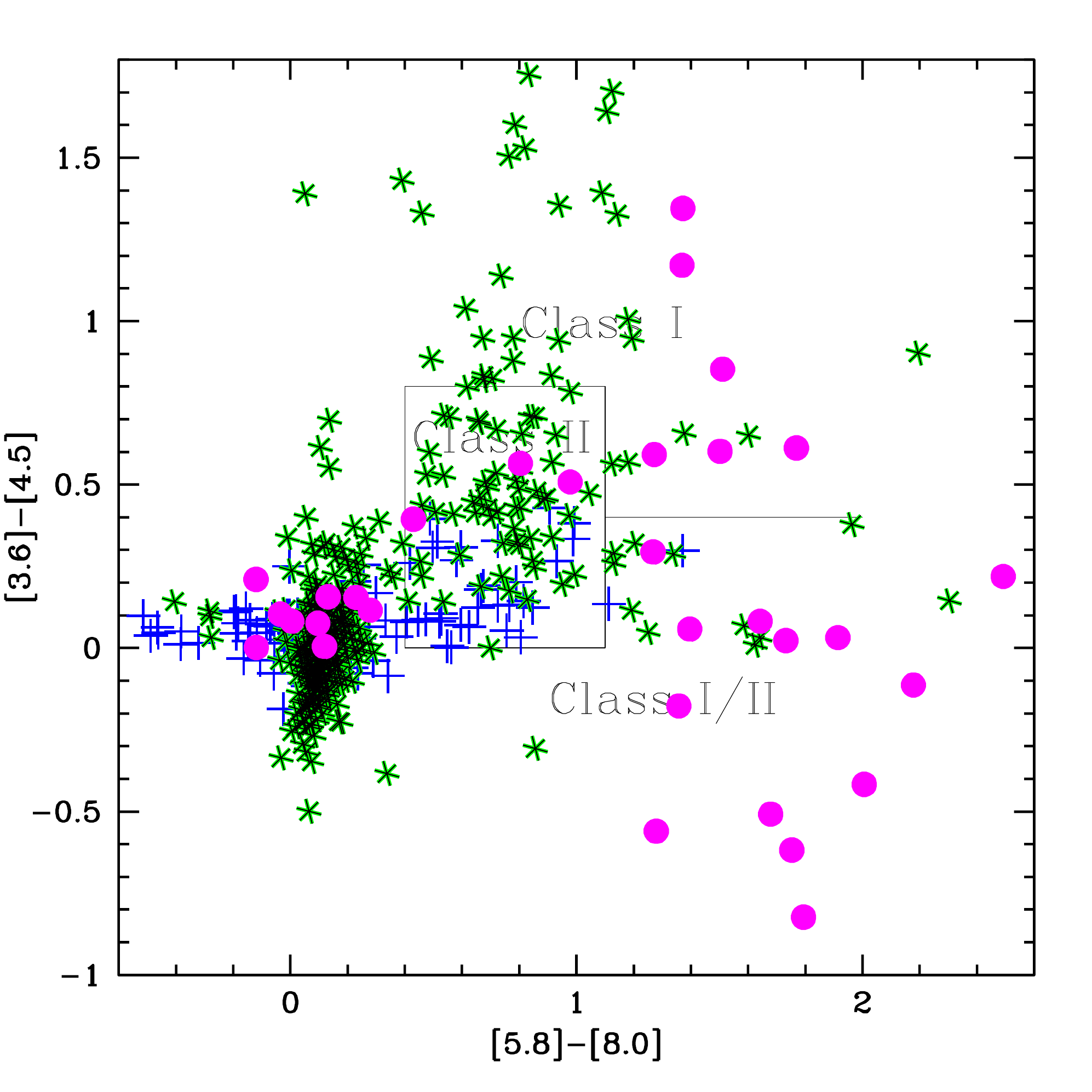} 
\includegraphics[width=0.405\textwidth,scale=0.25]{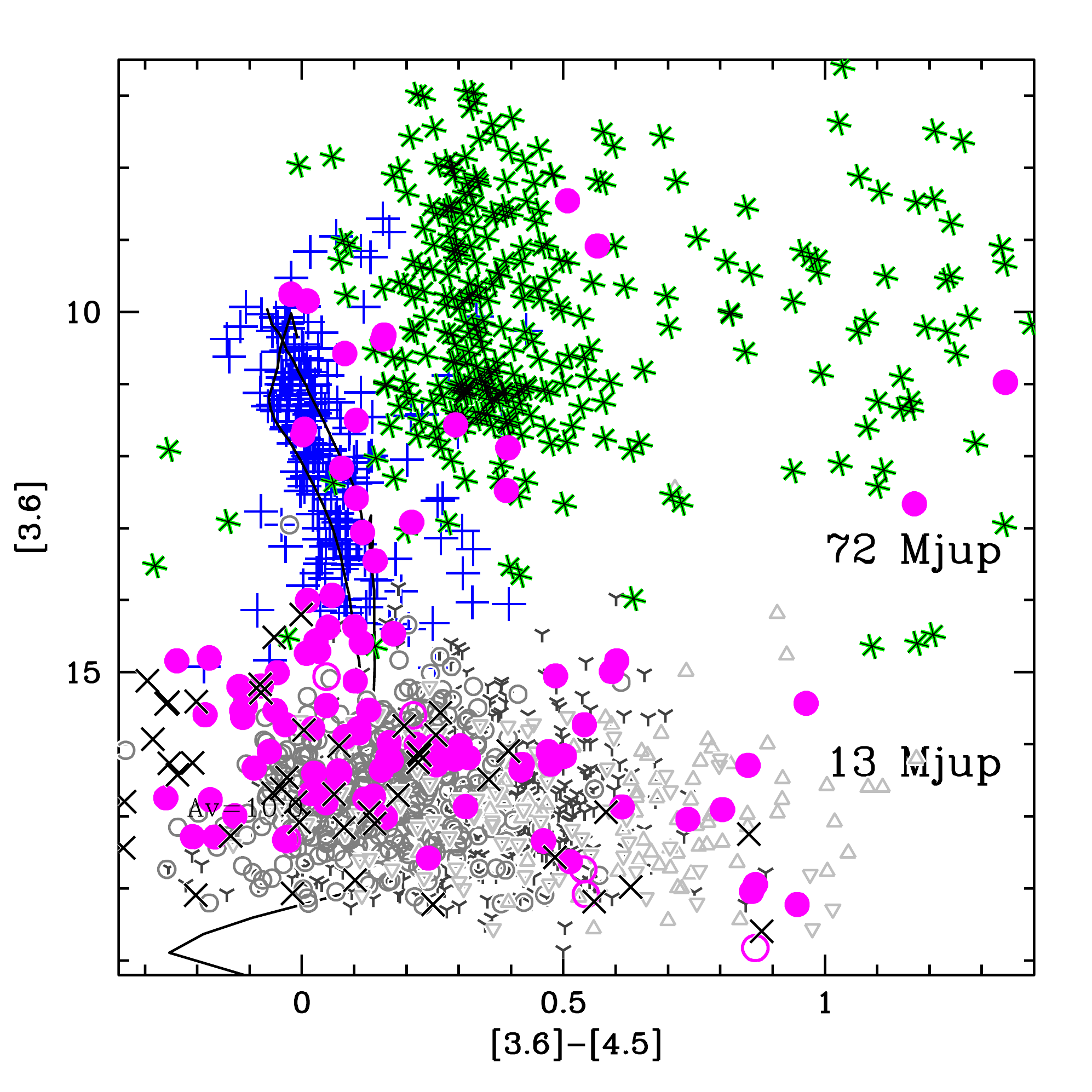} 
\includegraphics[width=0.405\textwidth,scale=0.25]{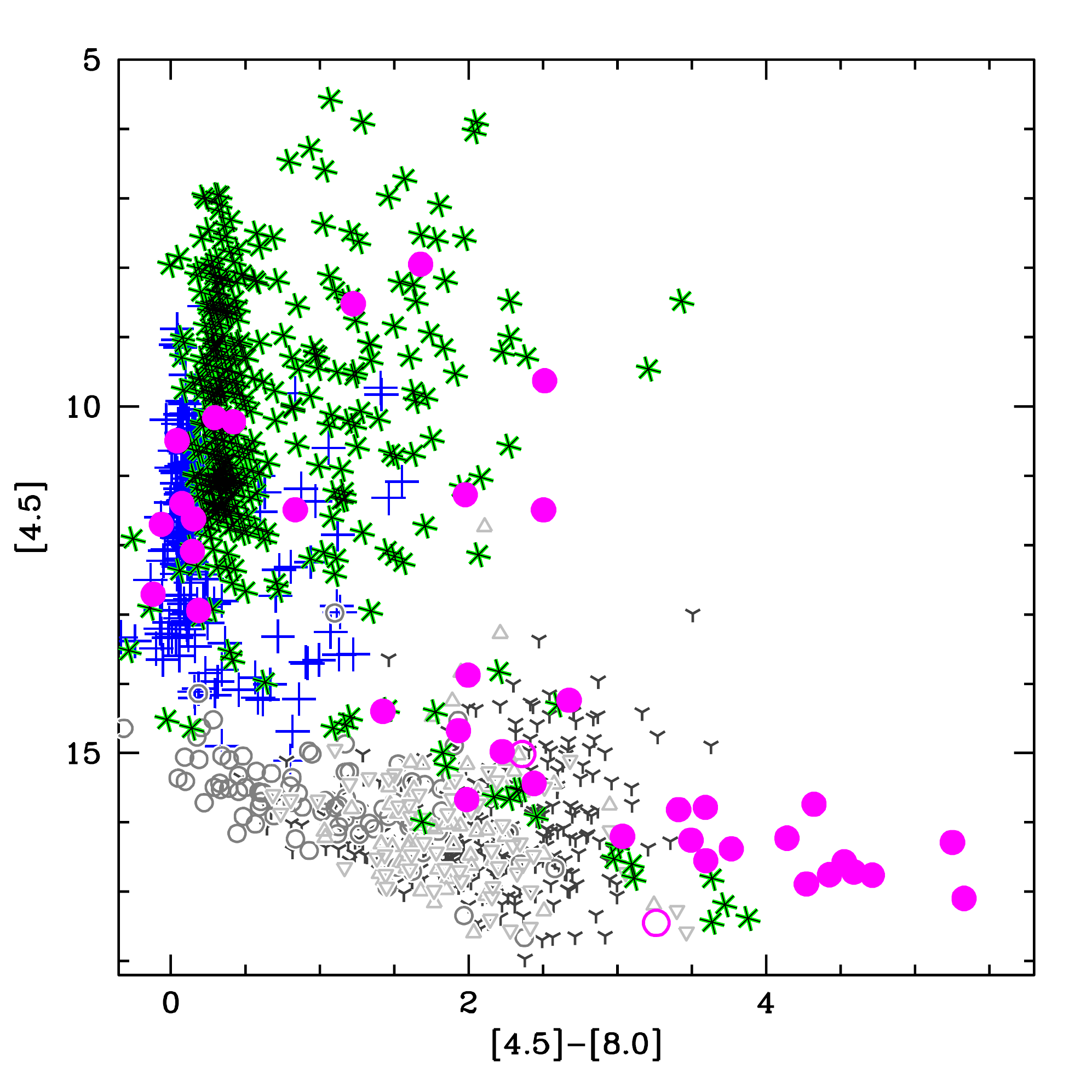} 
\includegraphics[width=0.405\textwidth,scale=0.25]{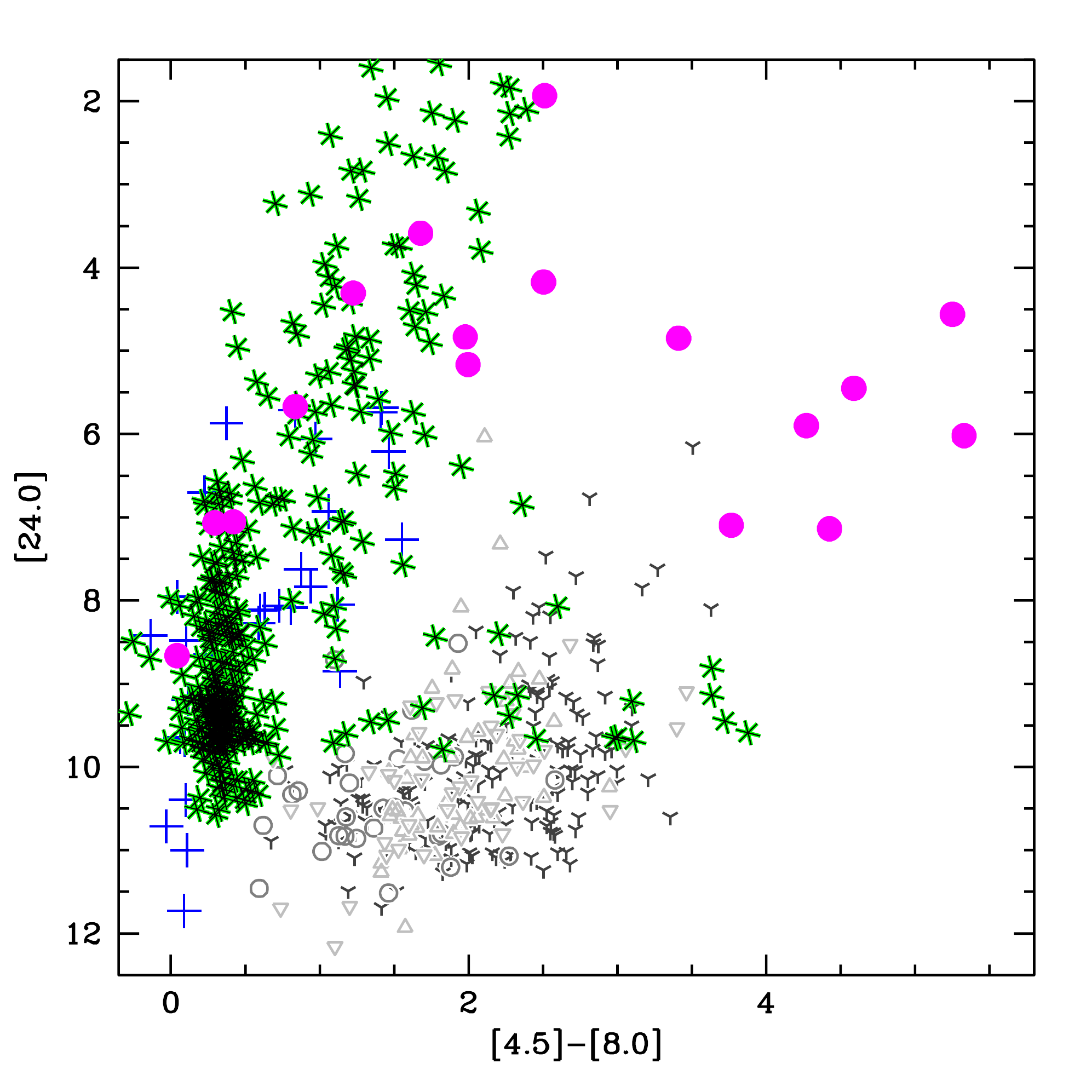} 
\includegraphics[width=0.405\textwidth,scale=0.25]{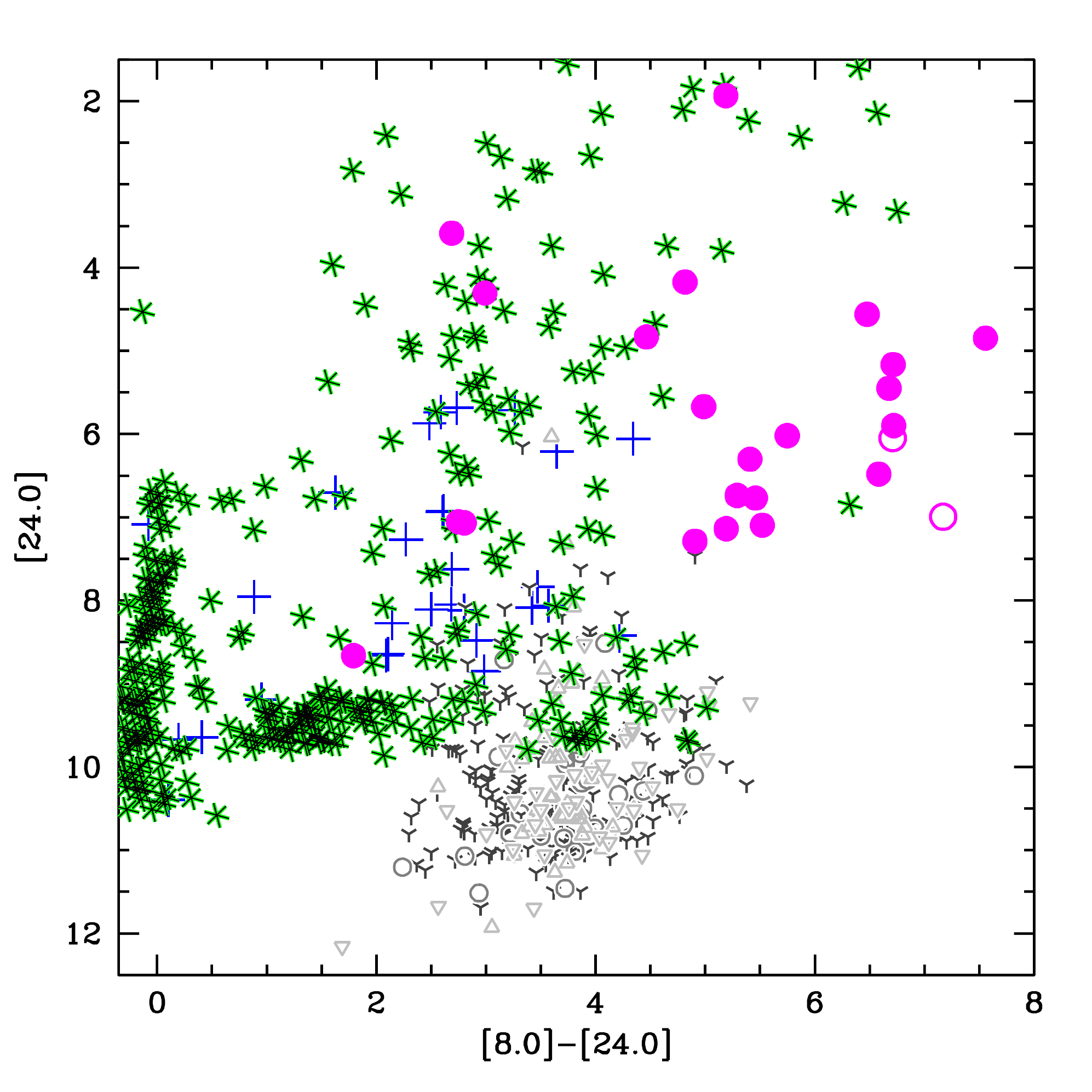} 
\caption{\label{fig_CCD_CMD_IRAC} 
   Spitzer/IRAC CCD and CMDs.
   The first panel (top, left) displays the areas where Class I, II and III objects are located (assuming a stellar or substellar nature). 
   The optical/IR counterparts of the B30 APEX/LABOCA sources are represented as magenta circles
   (solid for probable and possible members, empty circles for unknown status and a black cross for probable and possible non-members).   
   For comparison, we have added  known members of the $\sim$5-8\,Myr cluster C69  as plus, blue symbols, which is
   located in the same star forming region complex 
   (\citep{Dolan1999.1, Barrado2004.3, Barrado2007.1, Morales2008-PhD, Barrado2011.1, Bayo2011.1}).
   We have also added confirmed members of Serpens from \cite{Harvey2007-Serpens_YSO-IRAC-MIPS},
   about $\sim$2\,Myr, as green-black asterisks.
   Most of our B30 counterparts can be classified as Class~I or as transitional objects between Class~I and Class~II.
   Other panels include extragalactic sources as light gray symbols and have been extracted from \cite{Sacchi09.1}:
up- and down-ward open triangles correspond to the AGN1 and AGN2, whereas  three-point stars represent emission line galaxies and open circles other galaxies.
   We also have include BT-settl isochrones \citep{Allard2012.1} with 3 and 20 Myr. 
}
\end{figure*}

\newpage
\clearpage

\begin{figure*}
\center
\includegraphics[width=0.45\textwidth,scale=0.32]{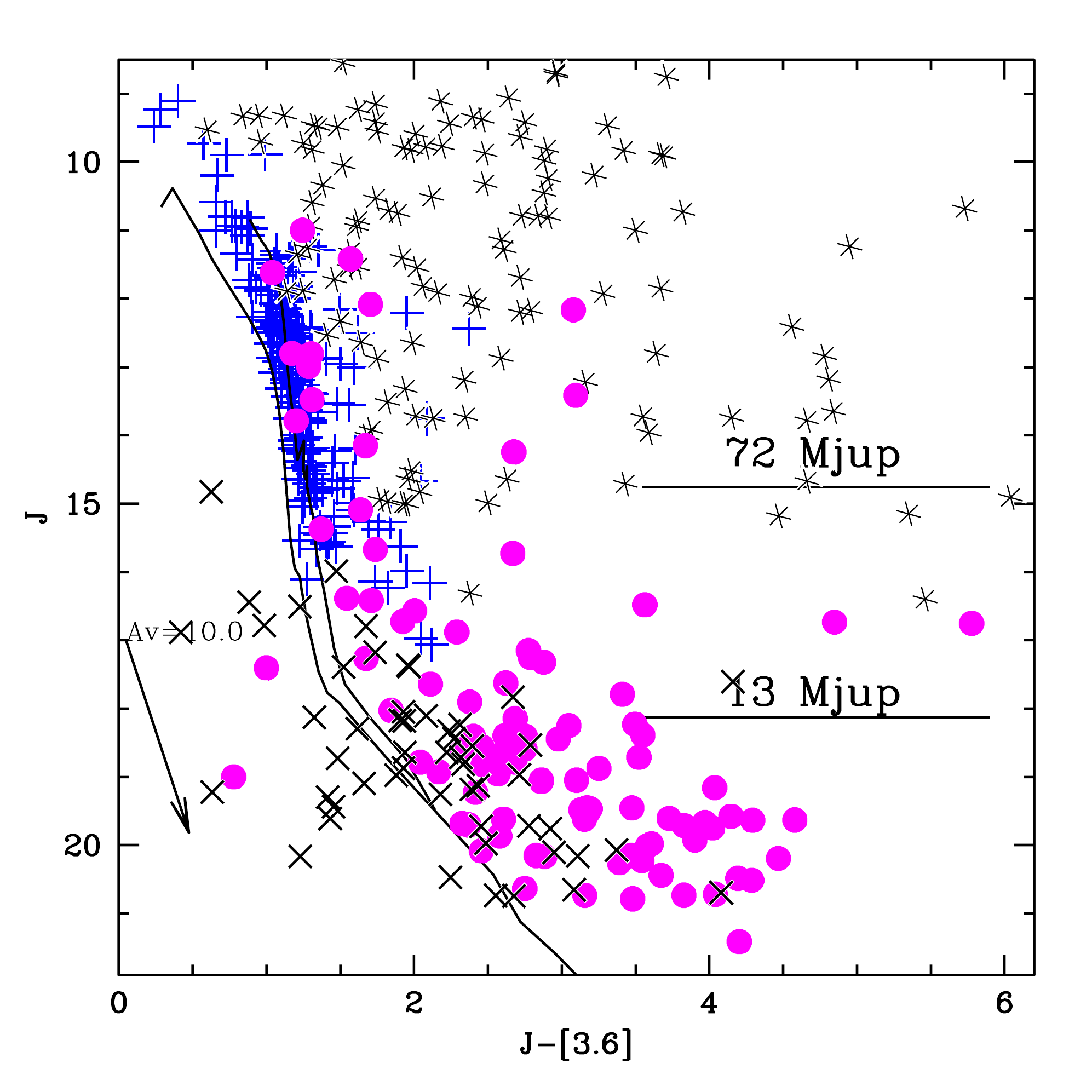} 
\includegraphics[width=0.45\textwidth,scale=0.32]{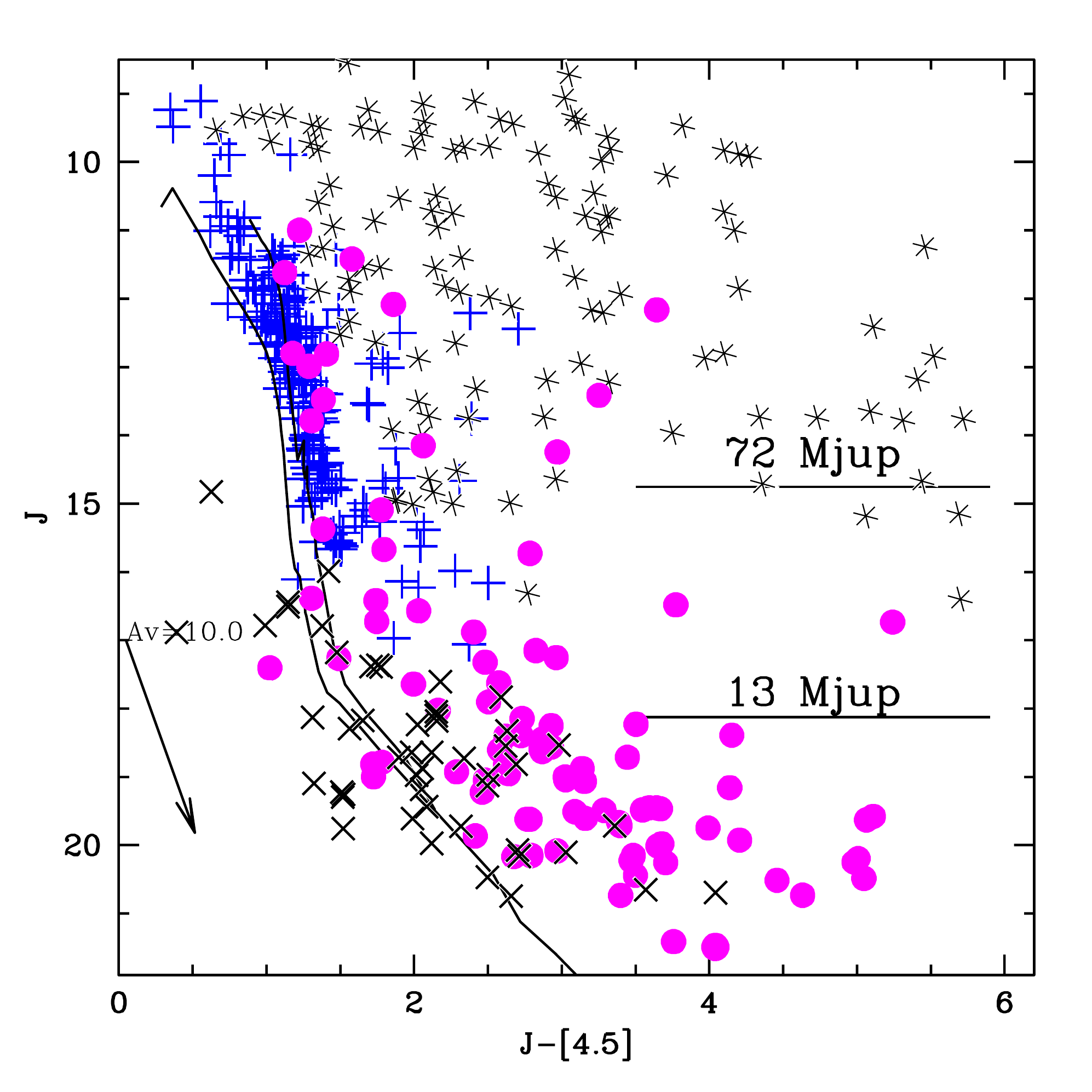} 
\includegraphics[width=0.45\textwidth,scale=0.32]{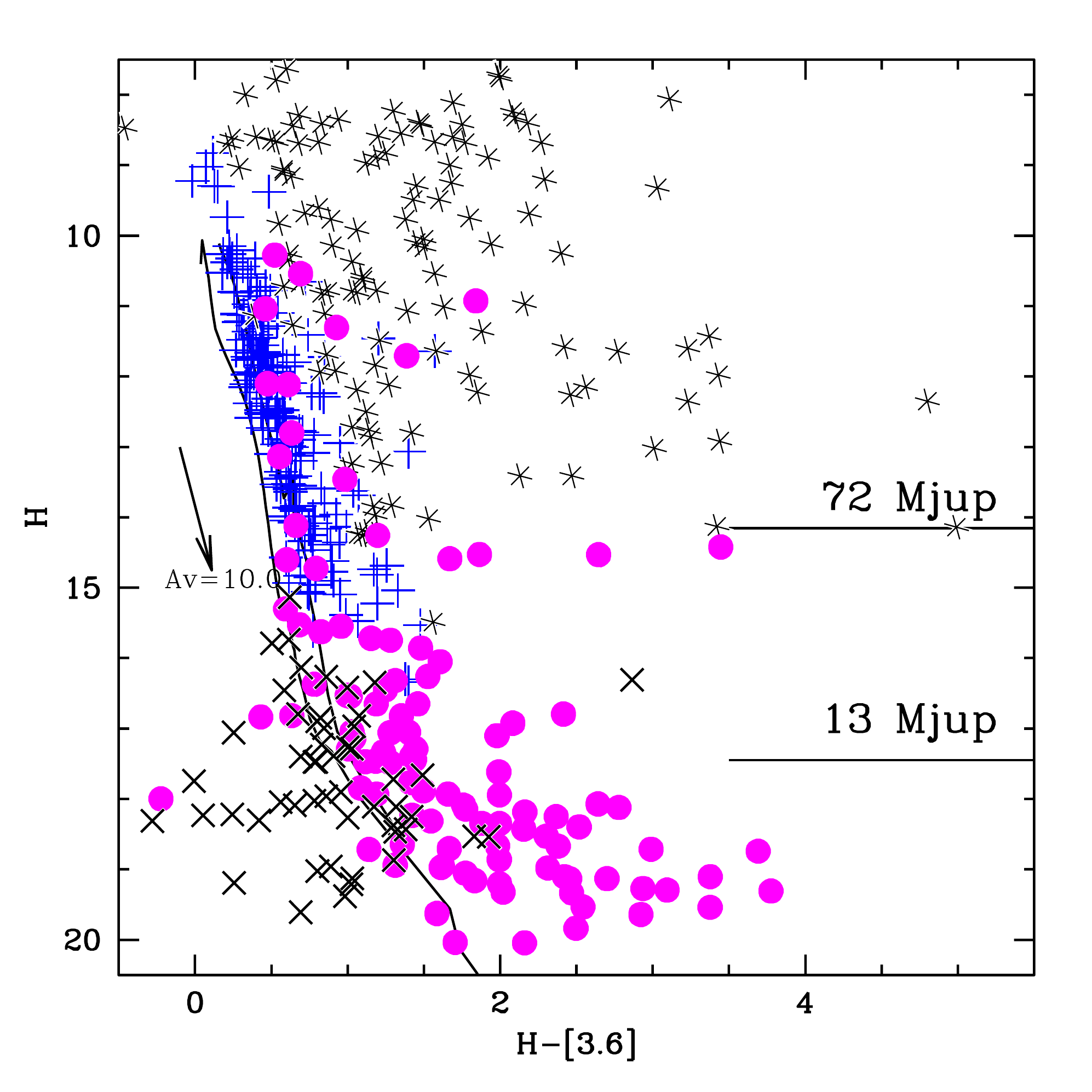} 
\includegraphics[width=0.45\textwidth,scale=0.32]{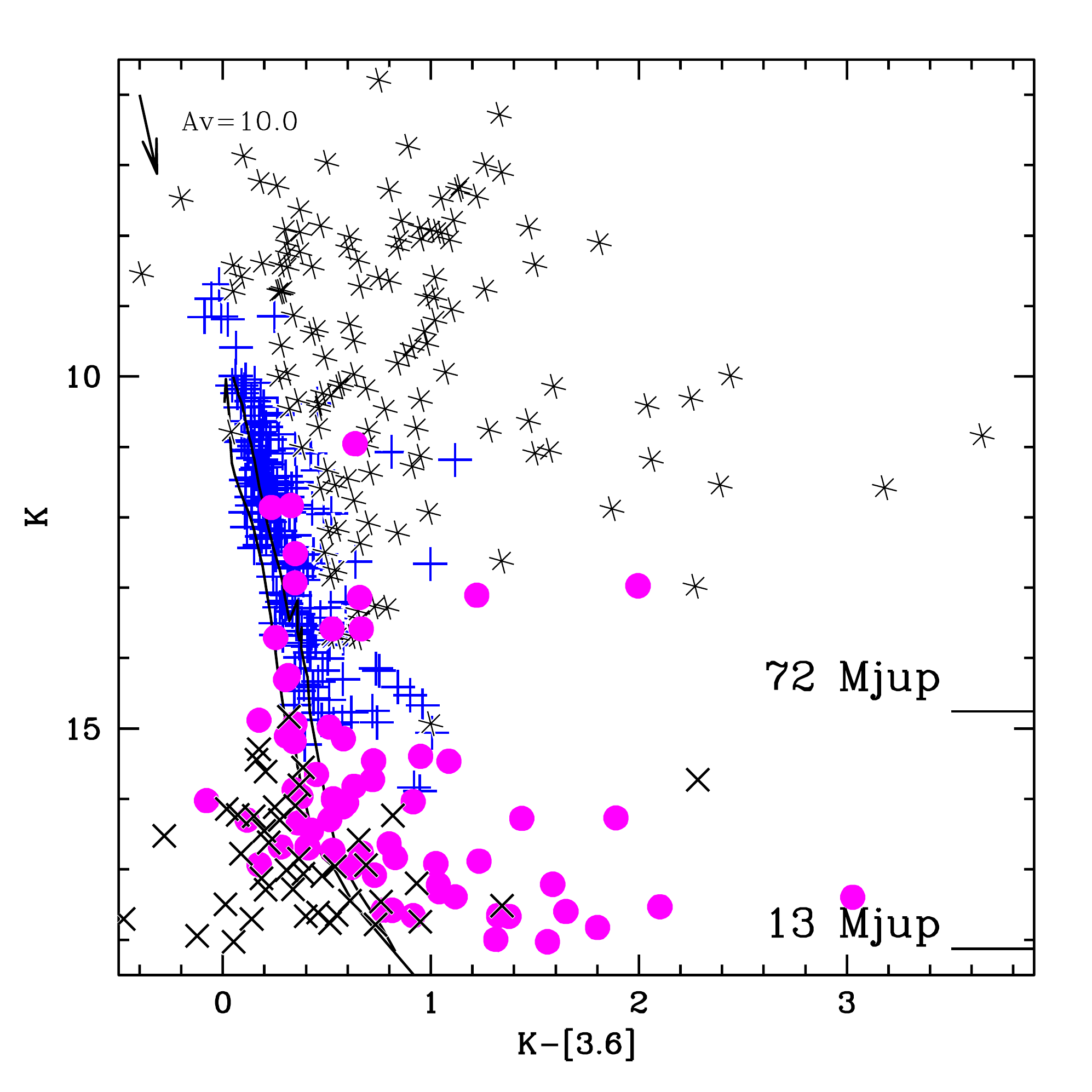} 
\caption{\label{fig_CCD_other} 
Color-magnitude diagrams for  the B30 APEX/LABOCA  counterparts.
Symbols as in Fig.~\ref{fig_CCD_CMD_IRAC}, but in these panels
we have included confirmed members of Taurus from \cite{Luhman2006.2}, about 1-2\,Myr, as asterisks.
As a reference, the locations for 72 and 13 $M_{Jupiter}$ very low-mass objects at 3 Myr are also included,
based on 3 Myr isochrone by  \citep{Allard2012.1}.}
\end{figure*}

\newpage
\clearpage

\begin{figure*}
\center
\includegraphics[width=0.55\textwidth,scale=0.35]{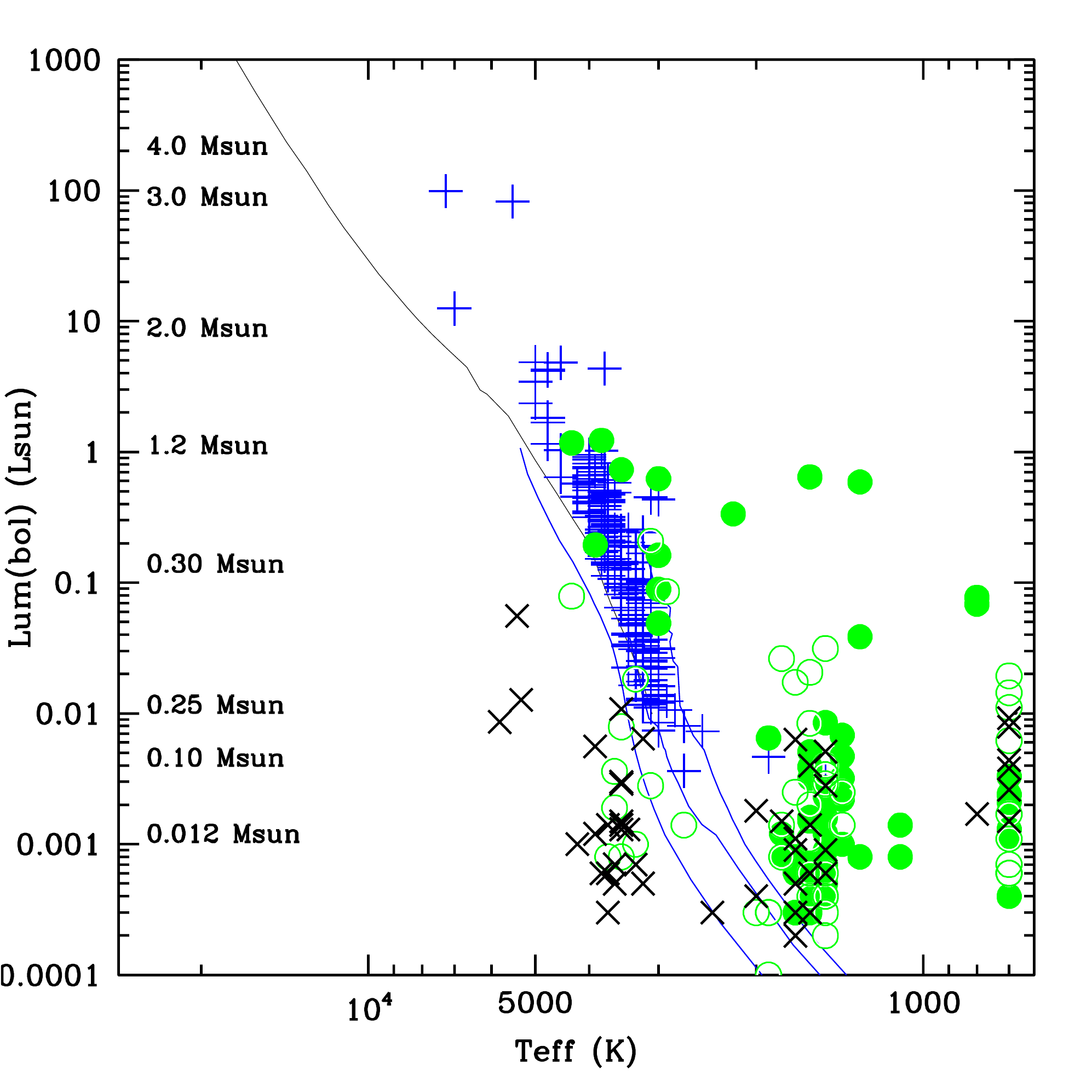} 
\includegraphics[width=0.55\textwidth,scale=0.35]{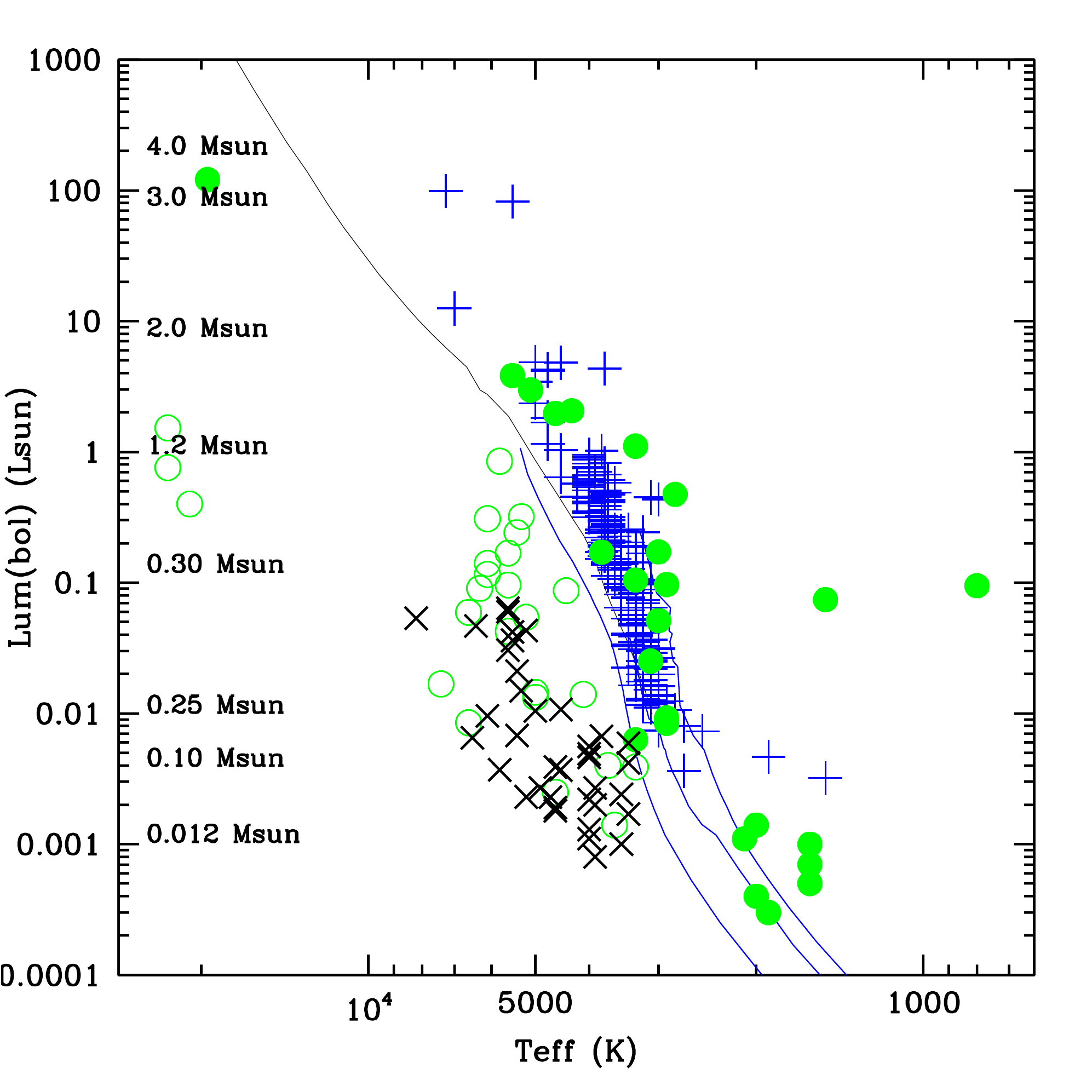} 
\caption{\label{HRD} 
Rejection of possible counterparts   based on their location on the HRD.
Big, black crosses  correspond to possible and probable  non-members.
Candidate members  are shown as open  green circles (possible members) and solid green circles
 (probable members). The bolometric luminosity and the effective temperature were derived using a fixed 
value of the reddening (Av=0.322)
 or it was derived during the fitting process (left and right hand-side panels, respectively).
Collinder 69 known members, located at similar distance, are displayed  as blue plus symbols, 
 and  were selected from \cite{Dolan1999.1}, \cite{Barrado2007.1}
 and \cite{Morales2008-PhD} and \cite{Barrado2011.1}.
Isochrones correspond to a 20 Myr isochrone from \cite{Siess2000.1}  --upper part of the main sequence, in black--
  and Settl models --blue lines with 1, 20,  and 10,000 Myr-- from the Lyon group
\citep{Allard2012.1}.
We note that some possible members can be found well below the Collinder 69 cluster sequence.
}
\end{figure*}

\newpage
\clearpage

\begin{figure}   
\center
\includegraphics[width=\textwidth,scale=0.32]{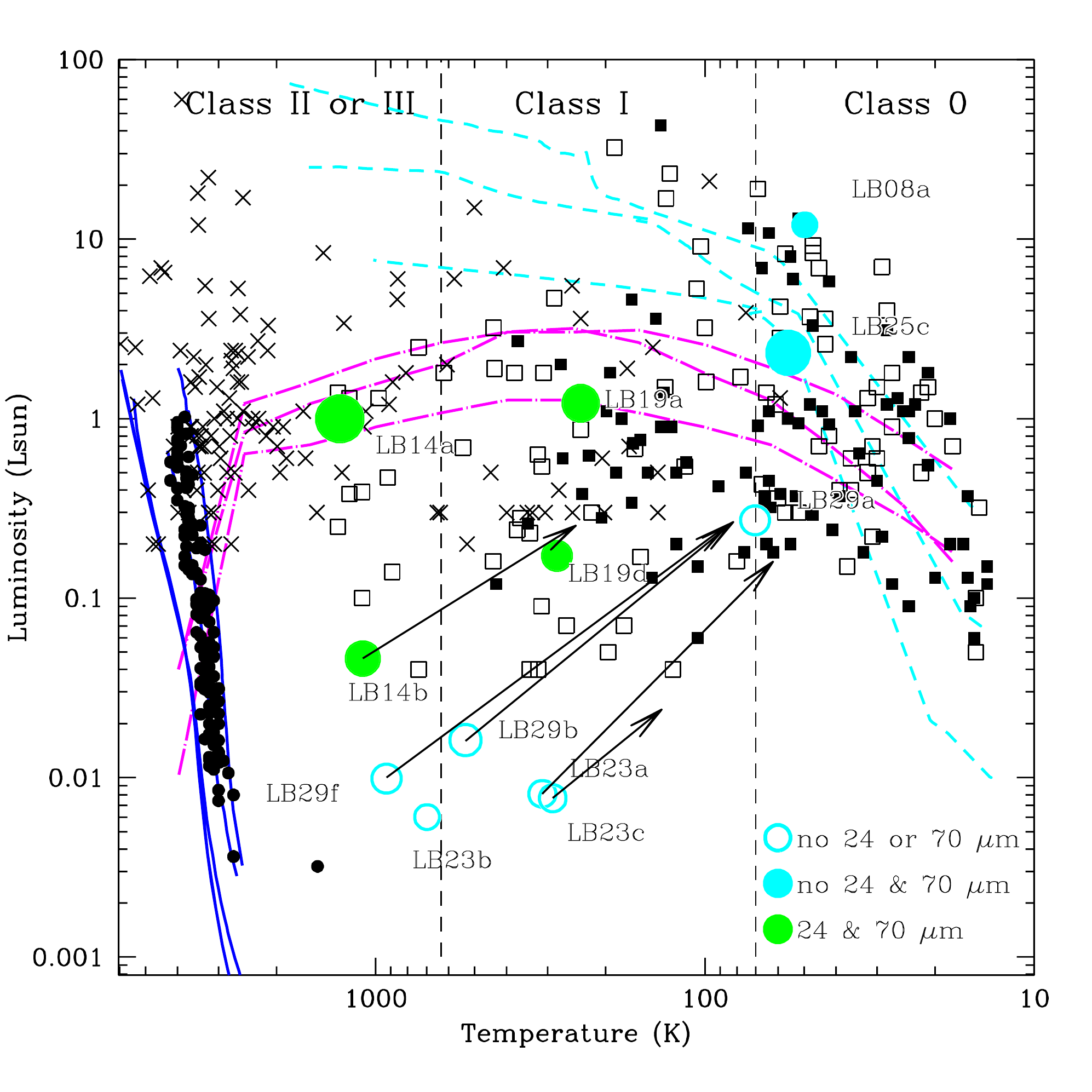} 
\caption{\label{HR_TbolLbol_Evolution_A} 
$L_{bol}$ vs. $T_{bol}$ after \cite{Young2005-SignaturesYSO} and
\cite{Dunham2008-Protostars}.
Sources compiled in these two works, with  evidences for
being embedded low luminosity sources, are shown as solid black squares. 
Young members from  \cite{Tobin2016_Multiplicity_Protostars_Perseus} in Perseus are displayed as open squares.
The black crosses correspond to Taurus members in 
\cite{Chen1995-Tbol-TaurusOph}, whereas solid, black circles come from
\citet{Bayo2011.1} and the Collinder 69 cluster.
The cyan, short-dashed lines show the evolutionary tracks for the
three models with different masses (in solar masses)
considered by \cite{Young2005-SignaturesYSO}. 
The magenta, dotted and long-dashed lines show
the evolutionary tracks for three models considered by 
\cite{Myers1998-TbolLbol-YSO}. 
The blue solid lines correspond to isochrones by \cite{Baraffe1998.1}.
The vertical dashed lines show the
Class 0--I and Class I--II $T_{bol}$ boundaries from \cite{Chen1995-Tbol-TaurusOph}.
The big, labeled circles correspond to several counterparts for our B30 sources detected at 70 $\mu$m (our group A).
Green with detections at 24 $\mu$m and cyan undetected at that band.
Solid (green or cyan) circles denote counterparts
whose properties have been derived with the 870 $\mu$m flux (for counterparts assumed to the primary origin of the submm emission),
whereas open symbols represents values estimated without this flux, 
and the arrows move their location to the values derived when including this value.
}
\end{figure}

\newpage
\clearpage

\newpage
\clearpage

\begin{figure}   
\center
\includegraphics[width=\textwidth,scale=0.32]{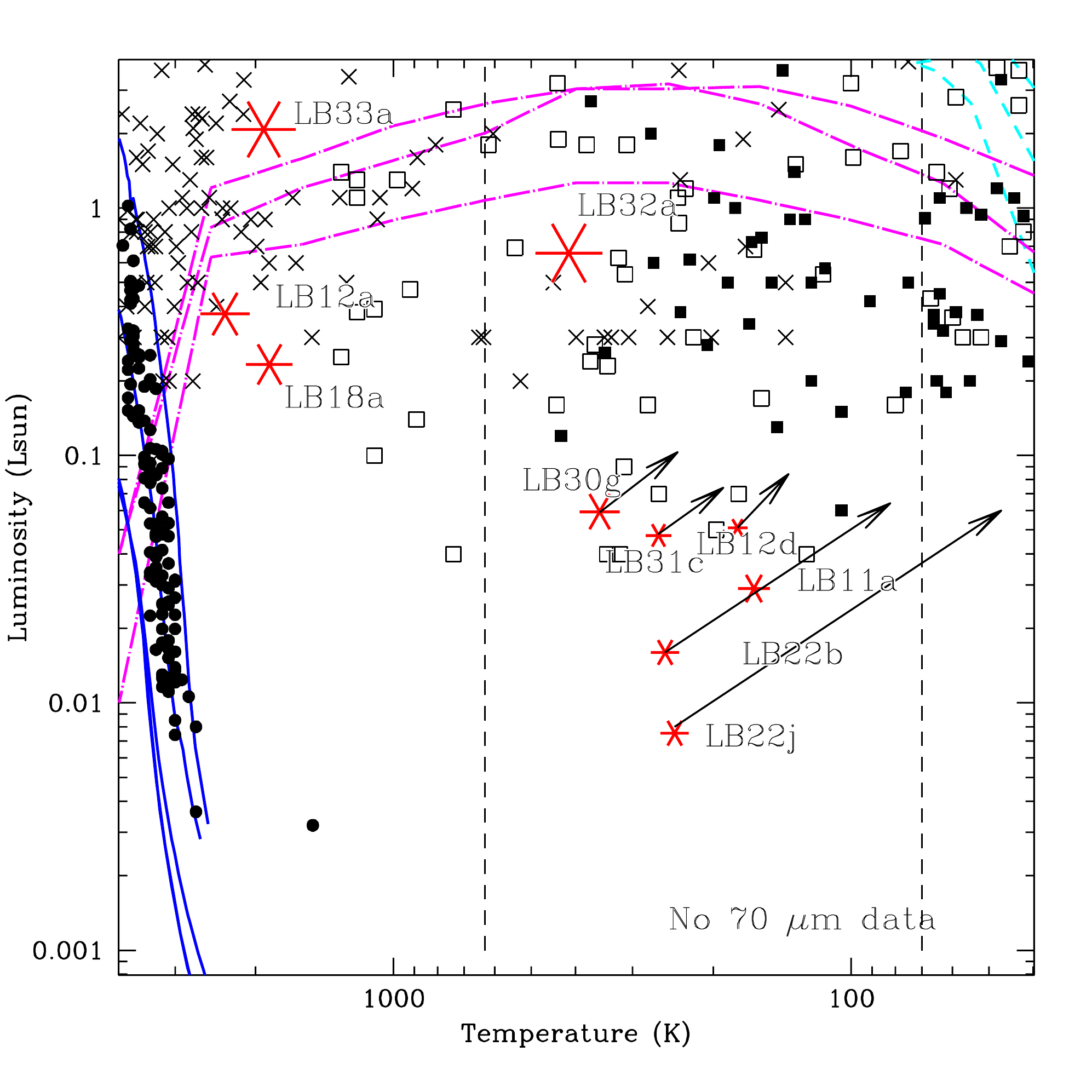} 
\caption{\label{HR_TbolLbol_Evolution_C} 
  Symbols as in Fig. \ref{HR_TbolLbol_Evolution_A} except for our B30 sources.
  Sample in group C1 are located outside the area surveyed by MIPS M2 but they have been detected at 24 $\mu$m.
  The red asterisks represent values estimated when  the flux at 870 $\mu$m has not been included,
  whereas the end of the arrows
  signals the values derived with the submm flux.
}
\end{figure}

\newpage
\clearpage

%
%

\setcounter{table}{0}
%
%
\begin{table*}[ht] 
\caption{Summary of the instruments and  photometry: beam size and pointing errors} 
\tiny
\tiny
\begin{tabular}{lcccccc} 
\hline
Instrument   & Bands              &Beam FWHM /seeing  & ErrorPos & Mag(lim)                  & Flux(lim)          & ZP$^3$ \\ %
             &                    &(arcsec)           & (arcsec) &                           &  (Jy)              &        \\ %
\hline                                                                                                               
DM           & V,Rc,Ic            & --                &  --      &21.0,21.0,19.0             &  --                & 3952.79, 3002.88, 2520.19\\ %
INT/WFC      & r,i                & --                &  --      &22.88, 21.50               &  --                & \\ %
2MASS        & JHKs               & --                &  --      &16.8, 16.1, 15.3           &  --                & 1636.77, 1115.71, 671.53\\ %
Omega2000$^1$& J                  & --                &  0.5     & 20.00                     &  --                & 1636.77\\ %
Omega2000$^2$& JHKs               & --                &  0.5     & 21.30, 20.40, 18.25       &  --                & 1636.77, 1115.71, 671.53\\ %
Spitzer/IRAC & 3.6, 4.5, 5.8, 8.0 &1.66, 1.72, 1.88, 1.98&$\sim$1&18.25, 17.75, 15.625, 15.25&  --                & 277.5, 179.5, 116.6,63.1\\ %
Spitzer/MIPS & 24, 70             & 5.8, 18.6         &  $\sim$1 &    9.0, --                &  --                & 7.18, 0.778\\ %
WISE         & 3.4, 4.6, 12, 22   &6.1, 6.4, 6.5, 12.0& $<$1     &17.40, 16.05, 11.85, 8.30  &  --                & 309.540, 171.787, 31.674, 8.363 \\ %
Akari/IRC    & 9, 18              & $\sim$2.4$^4$     & 2        &    --                     & 0.050, 0,120       & --    \\ %
Akari/FIS    & 65, 90, 140, 160   & 37, 39, 58, 61    & 8        &    --                     &2.4, 0.55, 1.4, 6.3 & --    \\ %
APEX/LABOCA  & 870                & 27.6              & 3$^5$    &    --                     &  0.03              & --   \\ %
\hline
\end{tabular} 
\\ %
$^1$ December 2007.\\
$^2$ August 2011.\\
$^3$ Flux (Jy) = ZP$\times$10$^{-0.4 \times Mag}$.  \\
$^4$ A 4 pixel binning was applied, so the actual beam size is close to 9.5 arcsec.\\ %
$^5$ A nominal value of 3 arcsec as quoted in \cite{Weiss2009-APEX}, but our own calibrations have an pointing error of 1 arcsec. In any case, for our counterpart search we have used 5 arcsec.\\ %
\label{tabInstruments} 
\end{table*}

\newpage
\clearpage

\clearpage

\setcounter{table}{1}
%
%

\tiny
\begin{table*}[ht] 
\caption{APEX/LABOCA sources in Barnard 30.} 
\tiny
\begin{tabular}{lccrrccccccrllll} 
\hline

B30-         & RA$^1$      & DEC$^1$   &S$_\nu$$^2$ & I$_{peak}$$^3$
                                                        &$A_V^4$&\multicolumn{2}{c}{Mass$^5$}&  L(int)$^6$& Flux    &Ang.& rms& Gr$^8$ & Type$^{9}$ & \multicolumn{2}{c}{BD?$^{10}$} \\ 
                                             \cline{4-5}             \cline{7-8}                                                                                    \cline{15-16}            
ID           &              &          &\multicolumn{2}{c}{(870 $\mu$m)} &       &  All    & Peak &      & 70$\mu$m        &size$^7$&870  &    &             & in & in        \\  
             &(2000.0)      & (2000.0) &\multicolumn{2}{c}{(mJy) (mJy/beam)} & (mag) &($M_\odot$)&(M$_\odot$)  & ($L_\odot$) &  (Jy)   & (")&(mJy)&    &             &beam& 5'' \\ 
\hline                                                              
LB01      & 05 31 13.64 & 12 22 22.7 &  182  &   114    & 4.109 & 0.266    & 0.166      &   $<$0.382 & $<$0.89   & 22 & 16  & B2    &  starless   & --   & --         \\ 
LB02      & 05 31 32.30 & 12 19 14.7 &   68  &    68    & 4.141 & 0.099    & 0.099      &   $<$0.230 & $<$0.52   & PL & 11  & B2    &  excess     & Y    & --         \\ 
LB03$^a$  & 05 31 29.20 & 12 17 22.7 &   39  &    39    & 2.471 & 0.057    & 0.057      &      --    &    --     & PL &  9  & C2    &  YSO        & Y    & a$^{BD}$   \\ 
LB04$^a$  & 05 31 30.56 & 12 17 18.7 &   40  &    40    & 2.841 & 0.059    & 0.059      &      --    &    --     & PL &  9  & C2    &  excess     & Y    & --        \\ 
LB05      & 05 31 29.47 & 12 14 30.7 &   56  &    35    & 3.064 & 0.081    & 0.051      &      --    &    --     & 36 & 77  & C2    &  YSO        & Y    & a         \\ 
LB06      & 05 31 30.83 & 12 12 46.7 &  144  &   527    & 4.670 & 0.210    & 0.077      &      --    &    --     & 49 &  8  & C2    &  starless   & Y    & --        \\ 
LB07      & 05 31 32.19 & 12 11 46.7 &   36  &    36    & 4.415 & 0.052    & 0.052      &      --    &    --     & PL &  7  & C2    &  YSO        & Y    & a$^{BD}$   \\ 
LB08      & 05 31 23.19 & 12 11 26.7 &   73  &    46    & 2.486 & 0.106    & 0.066      &      1.223 &    3.07   & 27 &  7  & A2    &  YSO        & --   & a         \\ 
LB09      & 05 31 18.55 & 12 11 26.7 &   30  &    30    & 3.134 & 0.043    & 0.043      &   $<$0.422 & $<$0.99   & PL &  7  & B2    &  YSO        & Y    & a$^{BD}$    \\ 
LB10      & 05 31 09.55 & 12 11 02.7 &   32  &    33    & 1.146 & 0.046    & 0.046      &      --    &    --     & PL &  7  & C2    &  YSO        & Y    & a$^{BD}$   \\ 
LB11      & 05 31 10.37 & 12 10 38.7 &   32  &    33    & 1.844 & 0.046    & 0.046      &      --    &    --     & PL &  7  & C1    &  excess     & Y    & --        \\ 
LB12      & 05 31 27.01 & 12 10 14.7 &   59  &    59    & 2.474 & 0.087    & 0.087      &      --    &    --     & PL &  7  & C1    &  excess     & Y    & --        \\ 
LB13      & 05 31 18.55 & 12 09 58.7 &   74  &    59    & 3.873 & 0.108    & 0.086      &   $<$0.382 & $<$0.89   & 23 &  7  & B2    &  starless   & Y    & --         \\ 
LB14      & 05 31 19.37 & 12 09 10.7 &   35  &    35    & 1.984 & 0.051    & 0.051      &      0.276 &    0.63   & PL &  7  & A1    &  YSO        & Y    & a         \\ 
LB15      & 05 31 27.28 & 12 07 26.7 &   38  &    38    & 2.736 & 0.056    & 0.056      &   $<$0.406 & $<$0.95   & PL &  7  & B2    &  starless   & --   & --         \\ 
LB16      & 05 31 30.01 & 12 07 06.7 &   36  &    36    & 3.115 & 0.053    & 0.053      &      --    &    --     & PL &  7  & C2    &  starless   & Y    & --        \\ 
LB17      & 05 31 09.82 & 12 06 42.7 &   40  &    40    & 3.236 & 0.059    & 0.059      &      --    &    --     & PL &  7  & C2    &  YSO        & Y    & a$^{BD}$   \\ 
LB18      & 05 31 34.92 & 12 06 22.7 &   41  &    41    & 2.917 & 0.059    & 0.059      &      --    &    --     & PL &  8  & C1    &  excess     & Y    & --        \\ 
LB19$^b$  & 05 31 28.10 & 12 05 30.7 &  125  &   793    & 2.759 & 0.182    & 0.116      &      0.402 &    0.94   & 24 &  7  & A1    &  YSO        & Y    & a, b      \\ 
LB20$^c$  & 05 31 22.91 & 12 05 30.7 &   33  &    33    & 1.775 & 0.048    & 0.048      &   $<$0.592 & $<$1.42   & PL &  7  & B2$^*$&  excess     & Y    & --         \\ 
LB21      & 05 31 36.00 & 12 05 02.7 &   74  &    42    & 3.897 & 0.108    & 0.061      &      --    &    --     & 62 &  7  & C2$^*$&  YSO        & Y    & a$^{BD}$   \\ 
LB22$^d$  & 05 31 31.64 & 12 04 14.7 &   60  &    60    & 2.710 & 0.088    & 0.088      &      --    &    --     & PL &  7  & C1    &  YSO        & Y    & a$^{BD}$   \\ 
LB23$^e$  & 05 31 29.19 & 12 04 38.7 &   32  &    32    & 3.100 & 0.046    & 0.046      &      0.189 &    0.42   & PL &  7  & A2    &  excess     & Y    & --        \\ 
LB24      & 05 31 23.46 & 12 04 30.7 &   30  &    30    & 2.376 & 0.044    & 0.044      &   $<$0.604 & $<$1.45   & PL &  7  & B1    &  excess     & Y    & --        \\ 
LB25      & 05 31 20.46 & 12 05 06.7 &   87  &    52    & 1.898 & 0.126    & 0.076      & 0.557$^{j}$&1.33$^{j}$  & 34 &  8  & A2    &  YSO        & Y    & a$^{BD}$, e\\ 
LB26      & 05 31 17.73 & 12 05 06.7 &   40  &    40    & 2.960 & 0.058    & 0.058      &      --    &    --     & PL &  7  & C2    &  excess     & Y    & --         \\ 
LB27$^f$  & 05 31 13.37 & 12 05 30.7 &  191  &   103    & 4.232 & 0.279    & 0.150      &      --    &    --     & 29 &  7  & C2$^*$&  excess     & Y    & --         \\ 
LB28$^g$  & 05 31 07.64 & 12 05 06.7 &   48  &    48    & 3.477 & 0.071    & 0.071      &   $<$0.573 & $<$1.37   & PL &  7  & B2$^*$&  excess     & Y    & --         \\ 
LB29      & 05 31 08.73 & 12 03 46.7 &   44  &    44    & 3.430 & 0.064    & 0.064      &      0.337 &    0.78   & PL &  7  & A2$^*$&  YSO        & Y    & a$^{BD}$   \\ 
LB30      & 05 31 13.37 & 12 03 34.7 &   84  &    69    & 2.840 & 0.123    & 0.101      &      --    &    --     & 29 &  7  & C1    &  excess     & Y    & --        \\ 
LB31      & 05 31 15.28 & 12 03 38.7 &   56  &    56    & 2.378 & 0.082    & 0.082      &      --    &    --     & PL &  7  & C1    &  excess     & Y    & --        \\ 
LB32$^h$  & 05 32 12.56 & 12 09 54.7 &   85  &    85    & --    & 0.123    & 0.123      &      --    &    --     & PL & 16  & C1    &  YSO        & --   & a         \\ 
LB33$^i$  & 05 30 13.36 & 12 08 46.7 &   60  &    60    & 0.714 & 0.087    & 0.087      &      --    &    --     & PL & 16  & C1    &  YSO        & --   & a         \\ 
LB34$^i$  & 05 30 13.36 & 12 08 58.7 &   44  &    44    & 0.123 & 0.064    & 0.064      &      --    &    --     & PL & 16  & C2    &  starless   & --   & --         \\ 
\hline                                                                                                                                                                                                
\end{tabular} 
\\ %
\\ %
$^1$ Position of the maximum emission of the cores. The coordinates have an error of approximately 1 arcsec (pointing rms, but each APEX/LABOCA cell has a 4 arcsec size).\\
$^2$ APEX/LABOCA flux density at 870 $\mu$m computed above 3$\sigma$ emission. The absolute uncertainty in the flux scale is 8\%.\\
$^3$ APEX/LABOCA peak intensity at 870 $\mu$m (maximum of the emission). For point-like sources, the value is equal to the previous column.\\
$^4$ Values derived from the extinction map, with a resolution of 1.5 arcmin, three times larger the LABOCA beam-size.\\
$^5$ Envelope masses derived using the 870~$\mu$m flux densities and assuming a dust temperature of 15~K (\citealt{SanchezMonge2013-DenseCores}), and a dust (and gas) mass opacity coefficient of 0.0175~cm$^2$\,g$^{-1}$ (obtained by interpolating the tabulated values of \citealt{Ossenkopf94.2}, see subsec. \ref{sub:EnvelopeMass}). The uncertainty in the masses due to the dust temperature and opacity law is estimated to be a factor of 4. \\
$^6$ Internal luminosity of the core, based on MIPS flux at 70 $\mu$m, after \cite{Dunham2008-Protostars}.\\
$^7$  Deconvolved size after fitting a Gaussian to the cores. The averaged of the major and minor axis is represented. "PL" stands for point-like source.\\
$^8$  Gr = Groups based on the detection at 70 and 24 $\mu$m  and presence of counterparts, as discussed in subsection \ref{sub:ClassificationCounterparts}:
      A1 = Detection at 70 and 24 $\mu$m;
      A2 = detection at 70 and upper limit at 24 $\mu$m;
      B1 = upper limit at 70 $\mu$m and detection at 24 $\mu$m;
      B2 = upper limit both at 70 and 24 $\mu$m;
      C1 = no data at 70 $\mu$m and detection at 24 $\mu$m;
      C2 = no data at 70 $\mu$m and upper limit at 24 $\mu$m.\\
$^{9}$ Tentative classification: \\
      YSO = Young Stellar Object, with a optical and/or near-IR sources within 5 arcsec of APEX/LABOCA central coordinate. Our tentative interpretation is that most of them are proto-stars or proto-BDs.
      Excess = Optical and/or near-IR sources with excesses within the APEX/LABOCA beam but farther than 5 arcsec. The submm source cannot be assigned unambiguously to any counterpart so its nature remains unknown. 
      Starless = Possible starless core, since there is neither counterparts closer than 5 arcsec nor an optical/IR source farther away with excess.\\
      $^{10}$ Presence of brown dwarf candidates: the first column indicates whether there is a BD candidate within APEX/LABOCA beam, whereas the second column indicates whether there is a
      optical/IR counterpart within 5 arcsec from APEX/LABOCA peak (the letter for the identification and the ``BD'' super-index for the substellar candidates). \\
$^{a}$ B30-LB03 and B30-LB04 very close to each other, see subsection \ref{sub:groupA}.\\
$^{b}$ IRAS05286$+$1203, RA=82.8671, DEC=+12.0899, $[12]$=7.79e-01$\pm$0.0779 Jy, $[25]$=1.91e+00$\pm$0.2101 Jy, $[60]$=9.35e+00$\pm$1.7765 Jy, $[100]$=7.04e+01$\pm$14.784 Jy.
      LB30-LB19d has a MIPS 2 flux at 70 $\mu$m of 0.200 Jy, which translate to $L(int)$=0.094 $L_\odot$ (i.e., a VeLLO). Note that within the LABOCA beam there is a
      APEX/SABOCA source detected at 350 $\mu$m (B30-SB08, \citealt{Huelamo2017-ALMA-B30})).\\
$^{c}$ B30-LB20 contains a APEX/SABOCA at 350 $\mu$m (B30-SB09, \citealt{Huelamo2017-ALMA-B30})).\\
$^{d}$ There are two APEX/SABOCA sources within the B30-LB22 beam (B30-SB03 and B30-SB04, \citealt{Huelamo2017-ALMA-B30}).\\
$^{e}$ There are two APEX/SABOCA sources within the B30-LB23 beam (B30-SB05 and B30-SB06, \citealt{Huelamo2017-ALMA-B30}).\\
$^{f}$ B30-LB27 contains a APEX/SABOCA at 350 $\mu$m (B30-SB12, \citealt{Huelamo2017-ALMA-B30})).\\
$^{g}$ B30-LB28 contains a APEX/SABOCA at 350 $\mu$m (B30-SB16, \citealt{Huelamo2017-ALMA-B30})).\\
$^{h}$ IRAS05293$+$1207, RA=83.0450, DEC=+12.1629, $[12]<$2.50e-01 Jy, $[25]<$2.71e-01 Jy, $[60]$=1.15e+00$\pm$0.1725 Jy, $[100]<$2.66e+01 Jy. \\
$^{i}$ B30-LB33 and B30-LB34 very close to each other, see subsection \ref{sub:groupA}. \\
$^{j}$ The MIPS M2 flux probably corresponds to the counterpart identified as LB25c. \\
$^{*}$ Extended emission at 24 $\mu$m. \\
\label{tabLABOCADetections} 
\end{table*}

\clearpage

\clearpage

\setcounter{table}{2}
%
%
\tiny
\begin{landscape}
\begin{table} 

\end{landscape}

\clearpage

\setcounter{table}{3}
%
%
\tiny
\begin{landscape}
\begin{table} 
\tiny
\caption{Photometric data for all counterparts, WISE data, release 2013.
}
\begin{tabular}{lrrrrrrrrrrrrrrr}
\hline
  \multicolumn{1}{c}{B30-ID} &
  \multicolumn{1}{c}{RA(ind)} &
  \multicolumn{1}{c}{DEC(ind)} &
  \multicolumn{1}{c}{Dist} &
  \multicolumn{1}{c}{WISE ID} &
  \multicolumn{1}{c}{W1$^1$} &
  \multicolumn{1}{c}{eW1$^1$} &
  \multicolumn{1}{c}{W2$^1$} &
  \multicolumn{1}{c}{eW2$^1$} &
  \multicolumn{1}{c}{W3$^1$} &
  \multicolumn{1}{c}{eW3$^1$} &
  \multicolumn{1}{c}{W4$^1$} &
  \multicolumn{1}{c}{eW4$^1$}  &
  \multicolumn{1}{c}{Cont$^2$}  & 
  \multicolumn{1}{c}{Ext$^3$}  &   
  \multicolumn{1}{c}{Qflag$^4$}\\
  \multicolumn{1}{c}{} &
  \multicolumn{1}{c}{(deg)} &
  \multicolumn{1}{c}{(deg)} &
  \multicolumn{1}{c}{(arcsec)} &
  \multicolumn{1}{c}{} &
  \multicolumn{1}{c}{} &
  \multicolumn{1}{c}{} &
  \multicolumn{1}{c}{} &
  \multicolumn{1}{c}{} &
  \multicolumn{1}{c}{} &
  \multicolumn{1}{c}{} &
  \multicolumn{1}{c}{} &
  \multicolumn{1}{c}{} &
  \multicolumn{1}{c}{} &
  \multicolumn{1}{c}{} &
  \multicolumn{1}{c}{} \\
\hline
 LB01\_a & 82.8103943 & 12.3743267 & 0.532 & J053114.45+122227.6  & 11.725 & 0.024 &   11.698 & 0.023  &$>$11.55  & --     &$>$8.299 & --        & 0000  & 0     & AAUU  \\ 
 LB02\_d & 82.8833542 & 12.3129959 & 0.355 & J053132.02+121846.6  & 15.092 & 0.045 &   14.208 & 0.056  &$>$11.47  & --     &$>$8.62  & --        & 0000  & 0     & AAUU  \\ 
 LB03\_c & 82.8828009 & 12.2905016 & 0.252 & J053131.87+121725.5  & 14.769 & 0.042 &   14.743 & 0.071  &$>$11.407 & --     &$>$8.283 & --        & 0000  & 0     & AAUU  \\ 
 LB03\_g & 82.8685852 & 12.2881695 & 0.113 & J053128.45+121717.3  & 15.736 & 0.073 &   15.756 & 0.168  &$>$12.019 & --     &$>$8.501 & --        & 0h00  & 0     & ABUU  \\ 
 LB03\_i & 82.877498  & 12.2929915 & 0.617 & J053130.63+121735.1  & 15.06  & 0.041 &   15.037 & 0.091  &$>$11.864 & --     &$>$8.87  & --        & 0000  & 0     & AAUU  \\
\hline
\end{tabular}
\,\\
$^1$  Values in parenthesis have not been used.\\
$^2$  Contamination and confusion flag, one per band. 0 for OK, any other value indicate a problem. See source catalog for details (\citealt{Cutri2013-WISE}).\\
$^3$  Extended source flag. 0 for OK, any other value indicate a problem. See source catalog for details.\\
$^4$  Photometric quality flag. A, B, C for S/N $>$10, in the range 3-10 and $<$3. U stand for upper limit.\\
\end{table}
\end{landscape}

\tiny
\addtocounter{table}{-1}
\begin{landscape}
\begin{table} 
\tiny
\caption{Photometric data for all counterparts, WISE data, release 2013.
}
\begin{tabular}{lrrrrrrrrrrrrrrr}
\hline
  \multicolumn{1}{c}{B30-ID} &
  \multicolumn{1}{c}{RA(ind)} &
  \multicolumn{1}{c}{DEC(ind)} &
  \multicolumn{1}{c}{Dist} &
  \multicolumn{1}{c}{WISE ID} &
  \multicolumn{1}{c}{W1$^1$} &
  \multicolumn{1}{c}{eW1$^1$} &
  \multicolumn{1}{c}{W2$^1$} &
  \multicolumn{1}{c}{eW2$^1$} &
  \multicolumn{1}{c}{W3$^1$} &
  \multicolumn{1}{c}{eW3$^1$} &
  \multicolumn{1}{c}{W4$^1$} &
  \multicolumn{1}{c}{eW4$^1$}  &
  \multicolumn{1}{c}{Cont$^2$}  & 
  \multicolumn{1}{c}{Ext$^3$}  &   
  \multicolumn{1}{c}{Qflag$^4$}\\
  \multicolumn{1}{c}{} &
  \multicolumn{1}{c}{(deg)} &
  \multicolumn{1}{c}{(deg)} &
  \multicolumn{1}{c}{(arcsec)} &
  \multicolumn{1}{c}{} &
  \multicolumn{1}{c}{} &
  \multicolumn{1}{c}{} &
  \multicolumn{1}{c}{} &
  \multicolumn{1}{c}{} &
  \multicolumn{1}{c}{} &
  \multicolumn{1}{c}{} &
  \multicolumn{1}{c}{} &
  \multicolumn{1}{c}{} &
  \multicolumn{1}{c}{} &
  \multicolumn{1}{c}{} &
  \multicolumn{1}{c}{} \\
\hline
 LB20\_a & 82.8508555 & 12.0902626 & 0.147 & J053124.19+120525.0  & 13.14  & 0.063 &   13.0   & 0.055 &   8.34   & 0.233  &   6.103 & 0.348     & 00dh  & 0     & AABB  \\ 
 LB20\_c & 82.8402513 & 12.0966039 & 0.251 & J053121.66+120547.5  & 9.854  & 0.037 &   9.866  & 0.038 &   9.261  & 0.253  &$>$8.081 & --        & hh0h  & 0     & AABU  \\ 
 LB20\_d & 82.8394462 & 12.0955432 & 0.23  & J053121.47+120543.7  & 10.136 & 0.044 &   10.085 & 0.044 &   8.984  & 0.197  &   6.766 & 0.168     & dd0h  & 0     & AABB  \\ 
 LB21\_a & 82.8988495 & 12.0846224 & 1.509 & J053135.62+120504.0  & 15.862 & 0.098 &   16.098 & 0.244 &   9.623  & 0.146  &   6.693 & 0.215     & 0000  & 0     & ABBB  \\ 
 LB21\_b & 82.899437  & 12.0812626 & 2.197 & J053135.81+120450.4  & 15.428 & 0.072 &   15.144 & 0.106 &$>$11.119 & --     &$>$7.632 & --        & 0000  & 0     & AAUU  \\ 
 LB21\_c & 82.90199   & 12.081975  & 0.278 & J053136.47+120455.3  & 15.951 & 0.129 &   15.603 & 0.157 &$>$10.397 & --     &  (7.515)& 0.493     & d000  & 0     & BBUC  \\ 
 LB21\_i & 82.8977103 & 12.0892541 & 0.862 & J053135.40+120521.7  & 16.246 & 0.107 &   16.523 & 0.347 &   9.528  & 0.058  &   6.924 & 0.134     & 0000  & 0     & ABAB  \\ 
 LB22\_b & 82.879810  & 12.070272  & 2.155 & J053130.87+120413.6  & 15.538 & 0.053 &   15.573 & 0.131 &   9.225  & 0.202  &   6.249 & 0.193     & hh0h  & 0     & ABBB  \\ 
 LB22\_c & 82.8842633 & 12.0692825 & 3.469 & J053132.00+120410.7  & 15.068 & 0.042 &   15.006 & 0.079 &   9.892  & 0.415  &   6.434 & 0.287     & hh0d  & 0     & AACB  \\ 
 LB22\_d & 82.8865316 & 12.0684093 & 0.983 & J053132.70+120406.3  & 15.118 & 0.044 &   15.053 & 0.082 &$>$9.323  & --     &   7.129 & 0.476     & hh0d  & 0     & AAUC  \\ 
 LB23\_a & 82.8733007 & 12.0754017 & 1.938 & J053129.66+120429.8  & 16.477 & 0.123 &$>$17.025 & --    &$>$10.929 & --     &   7.616 & 0.357     & 000H  & 0     & BUUC  \\ 
 LB23\_b & 82.8740083 & 12.0797718 & 0.372 & J053129.76+120447.5  & 15.412 & 0.057 &   15.807 & 0.176 &   10.851 & 0.48   &   7.43  & 0.271     & 00hH  & 0     & ABCB  \\ 
 LB23\_c & 82.8761053 & 12.0778312 & 2.299 & J053130.31+120438.0  & 16.259 & 0.114 &$>$17.089 & --    &   10.073 & 0.232  &   6.774 & 0.149     & 000d  & 0     & BUBB  \\ 
 LB23\_d & 82.8687196 & 12.0718283 & 0.696 & J053128.52+120418.0  & 15.497 & 0.057 &$>$17.053 & --    &$>$11.429 & --     &$>$8.104 & --        & 0hdH  & 0     & AUUU  \\ 
 LB23\_h & 82.8712059 & 12.0724084 & 0.669 & J053129.13+120420.3  & 15.764 & 0.059 &$>$16.636 & --    &$>$10.208 & --     &$>$7.049 & --        & 0h0h  & 0     & AUUU  \\ 
 LB24\_f & 82.844633  & 12.075634  & 0.659 & J053122.40+120411.3  & 14.561 & 0.04  &   15.886 & 0.245 &   8.705  & 0.04   &   5.64  & 0.09      & 000d  & 1     & ABAA  \\ 
 LB25\_b & 82.8303947 & 12.080038  & 0.285 & J053119.30+120447.8  & 14.362 & 0.045 &   14.719 & 0.077 &   10.049 & 0.312  &   6.299 & 0.107     & 0000  & 0     & AABA  \\ 
 LB25\_c & 82.8416812 & 12.0856044 & 0.258 & J053121.99+120507.9  & 13.222 & 0.052 &   13.078 & 0.05  &   7.605  & 0.087  &   5.207 & 0.099     & 000h  & 1     & AAAA  \\ 
 LB26\_a & 82.8254809 & 12.0871155 & 0.754 & J053118.15+120513.2  & 15.576 & 0.08  &   15.729 & 0.185 &$>$10.754 & --     &$>$7.74  & --        & 0000  & 0     & ABUU  \\ 
 LB27\_a & 82.8100128 & 12.0964232 & 0.396 & J053114.42+120546.8  & 14.307 & 0.04  &   14.204 & 0.058 &   9.94   & 0.205  &   6.935 & 0.151     & 0000  & 0     & AABB  \\ 
 LB27\_b & 82.8064117 & 12.0876884 & 0.865 & J053113.53+120514.8  & 14.952 & 0.054 &   14.959 & 0.087 &   10.135 & 0.425  &$>$6.759 & --        & 0000  & 0     & AACU  \\ 
 LB27\_c & 82.8052597 & 12.0901575 & 2.263 & J053113.27+120522.3  & 15.15  & 0.061 &   15.479 & 0.139 &   10.078 & 0.415  &   7.363 & 0.348     & 0000  & 0     & ABCB  \\ 
 LB27\_d & 82.8040515 & 12.0845908 & 1.893 & J053112.99+120506.3  & 15.074 & 0.055 &   15.05  & 0.094 &$>$10.436 & --     &$>$8.118 & --        & 0000  & 0     & AAUU  \\ 
 LB27\_f & 82.8020477 & 12.0958843 & 0.956 & J053112.49+120544.2  & 15.127 & 0.061 &   15.609 & 0.16  &   9.104  & 0.097  &   6.487 & 0.112     & 0000  & 0     & ABAB  \\ 
 LB28\_b & 82.784874  & 12.0824242 & 1.02  & J053108.43+120456.6  & 15.565 & 0.08  &   15.295 & 0.115 &   9.446  & 0.103  &   7.553 & 0.2       & 0000  & 0     & ABAB  \\ 
 LB28\_c & 82.7759399 & 12.0832968 & 1.987 & J053106.16+120501.6  & 15.726 & 0.104 &   15.422 & 0.13  &   9.528  & 0.148  &   7.648 & 0.245     & 0000  & 0     & ABBB  \\ 
 LB29\_b & 82.7845362 & 12.0612638 & 0.397 & J053108.31+120340.5  & 14.17  & 0.034 &   14.606 & 0.098 &   9.186  & 0.043  &   6.669 & 0.299     & 0000  & 1     & AAAB  \\ 
 LB29\_c & 82.784803  & 12.067079  & 0.546 & J053108.35+120400.9  & 14.388 & 0.177 &   14.483 & 0.171 &$>$10.029 & --     &$>$5.85  & --        & 0000  & 0     & BBUU  \\ 
 LB29\_d & 82.7912491 & 12.0631026 & 1.46  & J053109.99+120347.6  & 14.194 & 0.035 &   14.819 & 0.201 &   8.891  & 0.032  &   6.005 & 0.116     & 0000  & 1     & ABAB  \\ 
 LB30\_a & 82.8030423 & 12.0608531 & 0.641 & J053112.74+120339.6  & 13.913 & 0.034 &   14.364 & 0.136 &   9.68   & 0.078  &   6.239 & 0.089     & 0000  & 1     & ABAA  \\ 
 LB30\_g & 82.8103715 & 12.0592594 & 1.782 & J053114.48+120335.1  & 13.491 & 0.031 &   13.617 & 0.056 &   7.786  & 0.024  &   4.973 & 0.038     & 0000  & 1     & AAAA  \\ 
 LB31\_b & 82.8175493 & 12.0635955 & 1.118 & J053116.20+120347.8  & 14.086 & 0.083 &   14.231 & 0.094 &   8.228  & 0.04   &   5.146 & 0.071     & 0000  & 1     & AAAA  \\ 
 LB31\_c & 82.8158798 & 12.0589838 & 0.857 & J053115.84+120333.0  & 13.806 & 0.065 &   14.235 & 0.089 &   8.007  & 0.033  &   5.34  & 0.097     & 0000  & 1     & AAAA  \\ 
 LB32\_a & 83.0513535 & 12.1654158 & 0.264 & J053212.30+120955.6  & 12.049 & 0.022 &   11.299 & 0.021 &   8.437  & 0.025  &   4.976 & 0.03      & 0000  & 0     & AAAA  \\ 
 LB32\_d & 83.0553741 & 12.1640244 & 1.641 & J053213.37+120951.5  & 16.664 & 0.166 &   16.737 & 0.516 &$>$11.366 & --     &   7.272 & 0.12      & 0000  & 0     & BCUB  \\ 
 LB32\_f & 83.0588531 & 12.1689034 & 0.594 & J053214.08+121007.9  & 17.336 & 0.197 &$>$16.426 & --    &$>$11.993 & --     &   8.852 & 0.464     & 0000  & 0     & BUUC  \\ 
 LB33\_a & 82.5546799 & 12.1460333 & 0.183 & J053013.12+120845.8  & 8.578  & 0.023 &   7.594  & 0.019 &   4.929  & 0.014  &   2.93  & 0.023     & 0000  & 0     & AAAA  \\ 
\hline
\end{tabular}
\,\\
$^1$  Values in parenthesis have not been used.\\
$^2$  Contamination and confusion flag, one per band. 0 for OK, any other value indicate a problem. See source catalog for details (\citealt{Cutri2013-WISE}).\\
$^3$  Extended source flag. 0 for OK, any other value indicate a problem. See source catalog for details.\\
$^4$  Photometric quality flag. A, B, C for S/N $>$10, in the range 3-10 and $<$3. U stand for upper limit.\\
\label{tabPhotometryWISE} 
\end{table}
\end{landscape}


\clearpage

\clearpage

\setcounter{table}{4}
%
%
\tiny
\begin{landscape}
\begin{table} 
\tiny
\caption{Optical and near-IR photometry from 2MASS.
}
\begin{tabular}{ l l l r r r r l r l l l l l l l }
\hline
  \multicolumn{1}{ c }{B30-ID} &
  \multicolumn{1}{c }{RA} &
  \multicolumn{1}{c }{DEC} &
  \multicolumn{1}{c }{V$^1$} &
  \multicolumn{1}{c }{R$^1$} &
  \multicolumn{1}{c }{I$^1$} &
  \multicolumn{1}{c }{r$^2$} &
  \multicolumn{1}{c }{err\_r$^2$} &
  \multicolumn{1}{c }{i$^2$} &
  \multicolumn{1}{c }{err\_i$^2$} &
  \multicolumn{1}{c }{J2M} &
  \multicolumn{1}{c }{e\_J} &
  \multicolumn{1}{c }{H2M} &
  \multicolumn{1}{c }{e\_H} &
  \multicolumn{1}{c }{K2M} &
  \multicolumn{1}{c }{e\_K} \\
  \multicolumn{1}{c}{} &
  \multicolumn{1}{c}{(deg)} &
  \multicolumn{1}{c}{(deg)} &
  \multicolumn{1}{c}{} &
  \multicolumn{1}{c}{} &
  \multicolumn{1}{c}{} &
  \multicolumn{1}{c}{} &
  \multicolumn{1}{c}{} &
  \multicolumn{1}{c}{} &
  \multicolumn{1}{c}{} &
  \multicolumn{1}{c}{} &
  \multicolumn{1}{c}{} &
  \multicolumn{1}{c}{} &
  \multicolumn{1}{c}{} &
  \multicolumn{1}{c}{} &
  \multicolumn{1}{c}{} \\
\hline
  LB01\_a & 82.8103943 & 12.3743267 & 15.968 & 14.963 & 13.917 & 15.178 & 0.035 & 14.073 & 0.033 & 12.891 & 0.029 & 12.257 & 0.035 & 12.047 & 0.031\\
  LB03\_c & 82.8828009 & 12.2905016 & --     & --     & --     & 21.672 & 0.367 & 19.768 & 0.167 & 16.56  & 0.127 & 15.423 & 0.091 & 14.784 & 0.105\\
  LB03\_g & 82.8685852 & 12.2881695 & --     & --     & --     & 21.468 & 0.305 & 19.761 & 0.166 & 16.871 & 0.167 & 16.458 & 0.225 & 15.819 & 0.261\\
  LB05\_a & 82.872252  & 12.2411915 & 15.114 & 14.059 & 12.835 & 14.311 & 0.031 & 12.884 & 0.03  & 11.596 & 0.02  & 11.015 & 0.021 & 10.775 & 0.019\\
  LB05\_b & 82.8752217 & 12.2423194 & --     & --     & --     & 20.574 & 0.14  & 18.727 & 0.068 & 15.685 & 0.067 & 14.262 & 0.042 & 13.542 & 0.046\\
  LB06\_a & 82.8729098 & 12.2138781 & --     & --     & --     & 22.248 & 0.705 & 20.155 & 0.293 & 16.51  & 0.124 & 14.522 & 0.042 & 13.522 & 0.037\\
  LB08\_b & 82.8490389 & 12.195165  & --     & --     & --     & 22.667 & 1.04  & 20.926 & 0.613 & --     & --    & --     & --    & --     & \\
  LB10\_a & 82.7909414 & 12.1845672 & --     & --     & --     & 22.091 & 0.61  & 20.213 & 0.31  & --     & --    & --     & --    & --     & \\
  LB10\_b & 82.7897077 & 12.1863344 & --     & --     & --     & 22.64  & 1.014 & 20.864 & 0.578 & --     & --    & --     & --    & --     & \\
  LB10\_d & 82.7906454 & 12.1805146 & --     & --     & --     & 20.544 & 0.153 & 19.137 & 0.113 & --     & --    & --     & --    & --     & \\
  LB10\_f & 82.7940554 & 12.1827567 & --     & --     & --     & 22.595 & 0.973 & 20.52  & 0.416 & --     & --    & --     & --    & --     & \\
  LB10\_h & 82.7929218 & 12.1881804 & 19.43  & 17.961 & 16.956 & 18.137 & 0.311 & 17.125 & 0.241 & 15.239 & 0.045 & 14.585 & 0.051 & 14.206 & 0.07 \\
  LB10\_i & 82.795761  & 12.1864233 & --     & --     & --     & 23.063 & 1.503 & 21.497 & 1.062 & --     & --    & --     & --    & --     & \\
  LB10\_j & 82.792862  & 12.1905084 & --     & --     & --     & 22.29  & 0.733 & 20.845 & 0.567 & --     & --    & --     & --    & --     & \\
  LB10\_k & 82.7826309 & 12.1854864 & --     & --     & --     & 19.316 & 0.068 & 18.083 & 0.05  & 16.13  & 0.086 & 15.375 & 0.096 & 15.154 & 0.154\\
  LB11\_d & 82.7977769 & 12.1732308 & --     & --     & --     & 22.178 & 0.661 & 20.745 & 0.515 & --     & --    & --     & --    & --     & \\
  LB12\_a & 82.8622399 & 12.1723873 & --     & --     & --     & 17.023 & 0.113 & 14.61  & 0.039 & 11.956 & 0.022 & 11.249 & 0.02  & 10.887 & 0.018\\
  LB13\_a & 82.8314832 & 12.1614693 & 18.602 & 17.054 & 15.227 & 17.699 & 0.051 & 15.558 & 0.03  & 13.378 & 0.023 & 12.766 & 0.026 & 12.451 & 0.023\\
  LB14\_a & 82.8310571 & 12.1542772 & 17.561 & 16.016 & 14.486 & 16.116 & 0.055 & 14.584 & 0.038 & 11.987 & 0.022 & 10.776 & 0.023 & 10.091 & 0.018\\
  LB14\_b & 82.8313354 & 12.1495811 & --     & --     & --     & 21.548 & 0.37  & 19.536 & 0.163 & 16.645 & 0.132 & 14.735 & 0.059 & 13.135 & 0.036\\
  LB14\_c & 82.8355319 & 12.1547291 & 17.782 & 16.455 & 14.653 & 17.154 & 0.05  & 15.003 & 0.03  & 12.72  & 0.023 & 12.096 & 0.024 & 11.797 & 0.02 \\
  LB14\_e & 82.8256866 & 12.1508073 & --     & --     & --     & 20.745 & 0.181 & 19.911 & 0.233 & --     & --    & --     & --    & --     & \\
  LB14\_j & 82.8302892 & 12.1454723 & --     & --     & --     & 20.928 & 0.212 & 19.734 & 0.197 & --     & --    & --     & --    & --     & \\
  LB15\_a & 82.8675875 & 12.1263543 & 16.162 & 15.093 & 13.946 & 15.093 & 0.0   & 13.946 & 0.0   & 12.718 & 0.02  & 12.006 & 0.021 & 11.818 & 0.021\\
  LB15\_b & 82.8619391 & 12.1178684 & --     & --     & --     & 21.266 & 0.286 & 20.188 & 0.302 & --     & --    & --     & --    & --     & \\
  LB16\_c & 82.881731  & 12.1191153 & --     & --     & --     & 22.927 & 1.324 & 21.079 & 0.711 & 17.39  & 0.255 & 16.046 & 0.153 & 15.547 & 0.222\\
  LB17\_g & 82.7840129 & 12.110386  & --     & --     & --     & 22.869 & 1.255 & 20.835 & 0.562 & --     & --    & --     & --    & --     & \\
  LB18\_a & 82.893931  & 12.1086094 & --     & --     & --     & 20.148 & 0.112 & 17.533 & 0.038 & 13.45  & 0.026 & 11.792 & 0.024 & 11.027 & 0.023\\
  LB19\_a & 82.8658453 & 12.0919505 & --     & --     & --     & 95.001 & 0.0   & 91.001 & 0.0   & 16.319 & 0.0   & 14.596 & 0.08  & 12.763 & 0.04 \\
  LB19\_b & 82.8676553 & 12.0917727 & --     & --     & --     & 19.325 & 0.949 & 16.971 & 0.209 & 14.139 & 0.038 & 13.407 & 0.078 & 12.996 & 0.075\\
  LB19\_c & 82.8686592 & 12.0905224 & 17.473 & 16.266 & 14.906 & 16.567 & 0.077 & 15.236 & 0.052 & 13.764 & 0.04  & 13.159 & 0.035 & 12.921 & 0.038\\
  LB19\_d & 82.8733996 & 12.0925052 & --     & --     & --     & 95.001 & 0.0   & 91.001 & 0.155 & 15.903 & 0.079 & 14.548 & 0.054 & 13.838 & 0.055\\
  LB19\_e & 82.8614809 & 12.0885397 & --     & --     & --     & 22.447 & 0.848 & 21.522 & 1.089 & --     & --    & --     & --    & --     & \\
  LB20\_a & 82.8508555 & 12.0902626 & --     & --     & --     & 18.867 & 0.617 & 17.491 & 0.335 & 15.144 & 0.057 & 14.122 & 0.049 & 13.682 & 0.058\\
  LB20\_c & 82.8402513 & 12.0966039 & 13.76  & 12.809 & 11.842 & 13.132 & 0.03  & 12.273 & 0.03  & 10.946 & 0.045 & 10.24  & 0.049 & 10.037 & 0.049\\
  LB20\_d & 82.8394462 & 12.0955432 & --     & --     & --     & 14.485 & 0.032 & 13.146 & 0.031 & 11.386 & 0.029 & 10.519 & 0.025 & 10.239 & 0.027\\
  LB20\_e & 82.8394127 & 12.0930814 & --     & --     & --     & 20.823 & 0.193 & 19.423 & 0.147 & 16.629 & 0.217 & 14.801 & 0.0   & 14.81  & 0.164\\
\hline
\end{tabular}
\label{tabPhotometryOpt2M} 
\,\\
$^1$ \cite{Dolan2001.1}.\\
$^2$ INT/WFC, this work.\\
\end{table}
\end{landscape}

\tiny
\addtocounter{table}{-1}
\begin{landscape}
\begin{table} 
\tiny
\caption{Optical and near-IR photometryfrom 2MASS.
}
\begin{tabular}{ l l l r r r r l r l l l l l l l }
\hline
  \multicolumn{1}{ c }{B30-ID} &
  \multicolumn{1}{c }{RA} &
  \multicolumn{1}{c }{DEC} &
  \multicolumn{1}{c }{V$^1$} &
  \multicolumn{1}{c }{R$^1$} &
  \multicolumn{1}{c }{I$^1$} &
  \multicolumn{1}{c }{r$^2$} &
  \multicolumn{1}{c }{err\_r$^2$} &
  \multicolumn{1}{c }{i$^2$} &
  \multicolumn{1}{c }{err\_i$^2$} &
  \multicolumn{1}{c }{J2M} &
  \multicolumn{1}{c }{e\_J} &
  \multicolumn{1}{c }{H2M} &
  \multicolumn{1}{c }{e\_H} &
  \multicolumn{1}{c }{K2M} &
  \multicolumn{1}{c }{e\_K} \\
  \multicolumn{1}{c}{} &
  \multicolumn{1}{c}{(deg)} &
  \multicolumn{1}{c}{(deg)} &
  \multicolumn{1}{c}{} &
  \multicolumn{1}{c}{} &
  \multicolumn{1}{c}{} &
  \multicolumn{1}{c}{} &
  \multicolumn{1}{c}{} &
  \multicolumn{1}{c}{} &
  \multicolumn{1}{c}{} &
  \multicolumn{1}{c}{} &
  \multicolumn{1}{c}{} &
  \multicolumn{1}{c}{} &
  \multicolumn{1}{c}{} &
  \multicolumn{1}{c}{} &
  \multicolumn{1}{c}{} \\
\hline
  LB21\_e & 82.906055  & 12.0828322 & --     & --     & --     & 22.648 & 1.022 & 21.406 & 0.973 & --     & --    & --     & --    & --     & \\
  LB22\_d & 82.8865316 & 12.0684093 & --     & --     & --     & 21.039 & 0.234 & 19.358 & 0.139 & --     & --    & --     & --    & --     & \\
  LB22\_f & 82.8850515 & 12.0706474 & --     & --     & --     & 23.115 & 1.577 & 20.607 & 0.452 & --     & --    & --     & --    & --     & \\
  LB22\_i & 82.879455  & 12.0659626 & --     & --     & --     & 21.672 & 0.415 & 20.296 & 0.335 & --     & --    & --     & --    & --     & \\
  LB23\_d & 82.8687196 & 12.0718283 & --     & --     & --     & 19.246 & 0.066 & 18.105 & 0.05  & 16.587 & 0.214 & 15.539 & 0.149 & 16.019 & 0.0  \\
  LB23\_h & 82.8712059 & 12.0724084 & --     & --     & --     & 21.971 & 0.546 & 20.002 & 0.253 & --     & --    & --     & --    & --     & \\
  LB24\_e & 82.8468915 & 12.072453  & --     & --     & --     & 21.335 & 0.305 & 20.023 & 0.258 & --     & --    & --     & --    & --     & \\
  LB25\_b & 82.8303947 & 12.080038  & --     & --     & --     & 20.556 & 0.154 & 18.78  & 0.083 & 16.386 & 0.175 & 15.095 & 0.113 & 14.672 & 0.121\\
  LB25\_c & 82.8416812 & 12.0856044 & --     & --     & --     & 19.271 & 0.902 & 18.147 & 0.609 & 15.655 & 0.089 & 14.688 & 0.07  & 14.316 & 0.091\\
  LB26\_c & 82.8216294 & 12.0792754 & --     & --     & --     & 21.311 & 0.298 & 20.01  & 0.255 & --     & --    & --     & --    & --     & \\
  LB27\_a & 82.8100128 & 12.0964232 & --     & --     & --     & 22.359 & 0.782 & 20.624 & 0.459 & 16.856 & 0.0   & 15.641 & 0.181 & 14.63  & 0.117\\
  LB27\_d & 82.8040515 & 12.0845908 & --     & --     & --     & 22.676 & 1.049 & 20.396 & 0.369 & --     & --    & --     & --    & --     & \\
  LB29\_a & 82.7869358 & 12.0624632 & --     & --     & --     & 22.46  & 0.858 & 20.799 & 0.543 & --     & --    & --     & --    & --     & \\
  LB29\_c & 82.784803  & 12.067079  & --     & --     & --     & 18.783 & 0.57  & 17.883 & 0.479 & 15.99  & 0.084 & 15.147 & 0.071 & 14.794 & 0.112\\
  LB29\_d & 82.7912491 & 12.0631026 & --     & --     & --     & 22.406 & 0.816 & 21.192 & 0.792 & --     & --    & --     & --    & --     & \\
  LB29\_e & 82.7846915 & 12.0644042 & --     & --     & --     & 21.431 & 0.333 & 19.985 & 0.249 & --     & --    & --     & --    & --     & \\
  LB29\_f & 82.7910175 & 12.0668983 & --     & --     & --     & 19.744 & 0.085 & 18.725 & 0.079 & --     & --    & --     & --    & --     & \\
  LB29\_j & 82.787634  & 12.0558409 & --     & --     & --     & 18.821 & 0.058 & 18.055 & 0.049 & 17.078 & 0.195 & 16.429 & 0.223 & 15.654 & 0.236\\
  LB30\_a & 82.8030423 & 12.0608531 & --     & --     & --     & 21.18  & 0.265 & 19.31  & 0.133 & 16.703 & 0.143 & 15.399 & 0.094 & 15.285 & 0.176\\  
  LB30\_c & 82.8016933 & 12.0635964 & --     & --     & --     & 22.607 & 0.984 & 21.543 & 1.111 & --     & --    & --     & --    & --     & \\       
  LB31\_b & 82.8175493 & 12.0635955 & --     & --     & --     & 20.855 & 0.199 & 19.373 & 0.141 & --     & --    & --     & --    & --     & \\       
  LB30\_f & 82.80728   & 12.061461  & --     & --     & --     & 22.812 & 1.191 & 21.714 & 1.309 & --     & --    & --     & --    & --     & \\       
  LB31\_g & 82.8131356 & 12.0558984 & --     & --     & --     & 20.609 & 0.161 & 19.129 & 0.113 & 16.827 & 0.171 & 15.88  & 0.157 & 15.415 & 0.196\\  
  LB31\_h & 82.8147323 & 12.0538983 & --     & --     & --     & 21.231 & 0.278 & 18.916 & 0.093 & 16.295 & 0.182 & 15.855 & 0.202 & 15.163 & 0.195\\  
  LB31\_d & 82.8149427 & 12.0656361 & --     & --     & --     & 21.664 & 0.412 & 20.103 & 0.279 & --     & --    & --     & --    & --     & \\       
  LB31\_f & 82.8188762 & 12.0560849 & --     & --     & --     & 20.661 & 0.168 & 19.817 & 0.213 & --     & --    & --     & --    & --     & \\       
  LB32\_a & 83.0513535 & 12.1654158 & 16.228 & 15.63  & 15.109 & 15.679 & 0.043 & 15.179 & 0.05  & 14.217 & 0.03  & 13.427 & 0.031 & 12.653 & 0.028\\
  LB32\_b & 83.0526962 & 12.1675291 & 16.294 & 15.794 & 15.307 & 15.816 & 0.046 & 15.408 & 0.058 & 14.817 & 0.041 & 14.374 & 0.054 & 14.257 & 0.082\\
  LB32\_c & 83.0493698 & 12.1644573 & --     & --     & --     & 17.362 & 0.152 & 17.022 & 0.219 & 16.514 & 0.128 & 15.764 & 0.143 & 15.776 & 0.251\\
  LB32\_d & 83.0553741 & 12.1640244 & --     & --     & --     & 17.859 & 0.241 & 17.509 & 0.34  & --     & --    & --     & --    & --     & \\
  LB33\_a & 82.5546799 & 12.1460333 & 13.836 & 13.055 & 12.282 & 13.055 & 0.0   & 12.282 & 0.0   & 10.954 & 0.022 & 10.12  & 0.023 & 9.557  & 0.018\\
\hline
\end{tabular}
\label{tabPhotometryOpt2M} 
\,\\
$^1$ \cite{Dolan2001.1}.\\
$^2$ INT/WFC, this work.\\
\end{table}
\end{landscape}


\clearpage

\setcounter{table}{5}
%
%
\tiny
\begin{landscape}
\begin{table} 
\tiny
\caption{Akari photometry for our APEX/LABOCA sources. 
}
\begin{tabular}{ l l l r r r r r r r r r r r r l }
\hline
  \multicolumn{1}{ c }{B30-ID} &
  \multicolumn{1}{c }{RA} &
  \multicolumn{1}{c }{DEC} &
  \multicolumn{1}{c }{S09} &
  \multicolumn{1}{c }{e\_S09} &
  \multicolumn{1}{c }{S18} &
  \multicolumn{1}{c }{e\_S18} &
  \multicolumn{1}{c }{S65} &
  \multicolumn{1}{c }{e\_S65} &
  \multicolumn{1}{c }{S90} &
  \multicolumn{1}{c }{e\_S90} &
  \multicolumn{1}{c }{S140} &
  \multicolumn{1}{c }{e\_S140} &
  \multicolumn{1}{c }{S160} &
  \multicolumn{1}{c }{e\_S160} &
  \multicolumn{1}{c }{NameAkari} \\
  \multicolumn{1}{ c }{} &
  \multicolumn{1}{ c }{(deg)} &
  \multicolumn{1}{ c }{(deg)} &
  \multicolumn{1}{ c }{} &
  \multicolumn{1}{ c }{} &
  \multicolumn{1}{ c }{} &
  \multicolumn{1}{ c }{} &
  \multicolumn{1}{ c }{} &
  \multicolumn{1}{ c }{} &
  \multicolumn{1}{ c }{} &
  \multicolumn{1}{ c }{} &
  \multicolumn{1}{ c }{} &
  \multicolumn{1}{ c }{} &
  \multicolumn{1}{ c }{} &
  \multicolumn{1}{ c }{} &
  \multicolumn{1}{ c }{} \\

\hline
  08\_a      & 82.8479608 & 12.1911092 & --     & --     & --     & --      &  5.865  & 0.923 & 11.2   & 0.245 & 25.0   & 3.92 & 29.19  & 9.29 & 0531239+121115\\
  14\_a      & 82.8310571 & 12.1542772 &  0.095 & 0.0111 & --     & --      & --      & --    & --     & --    & --     & --   & --     & --   & 0531194+120915\\
  19\_a$^{a}$& 82.8658453 & 12.0919505 & --     & --     &  0.845 & 0.0319  & --      & --    & --     & --    & --     & --   & --     & --   & 0531278+120531\\
  25\_c      & 82.8416812 & 12.0856044 & --     & --     & --     & --      &  2.5    & 1.21  & 9.587  & 0.682 & 11.14  & 4.37 & 18.12  & 10.1 & 531219+120513\\
  32\_a$^{b}$& 83.0513535 & 12.1654158 & --     & --     & --     & --      &  1.201  & 0.0   & 1.124  & 0.0628& 2.112  & 0.399& --     & --   & 0532122+120959\\
  33\_a      & 82.5546799 & 12.1460333 &  0.242 & 0.0152 &  0.28  & 0.00562 & --      & --    & --     & --    & --     & --   & --     & --   & 0530131+120845\\
 %
\hline\end{tabular}
\\ %
$^{a}$ IRAS05286$+$1203, RA=82.8671, DEC=+12.0899, $[12]$=7.79e-01$\pm$0.0779 Jy, $[25]$=1.91e+00$\pm$0.2101 Jy, $[60]$=9.35e+00$\pm$1.7765 Jy, $[100]$=7.04e+01$\pm$14.784 Jy.\\
$^{b}$ IRAS05293$+$1207, RA=83.0450, DEC=+12.1629, $[12]<$2.50e-01 Jy, $[25]<$2.71e-01 Jy, $[60]$=1.15e+00$\pm$0.1725 Jy, $[100]<$2.66e+01 Jy. \\
\label{tabAkari} 
\end{table}
\end{landscape}

\clearpage

\setcounter{table}{6}
%
%
\tiny
\begin{landscape}
\begin{table} 
\caption{Properties and memberships criteria for all counterpars within the APEX/LABOCA beam.}
\label{tabmembership} 
\tiny
\begin{tabular}{llllllllllllllll}
\hline
  \multicolumn{1}{c}{B30-ID}   &
  \multicolumn{1}{c}{RA} &
  \multicolumn{1}{c}{DEC} &
  \multicolumn{1}{c}{Teff} &
  \multicolumn{1}{c}{Lbol} &
  \multicolumn{1}{c}{eLbol} &
  \multicolumn{1}{c}{$A_V$}&
  \multicolumn{1}{c}{Teff} &
  \multicolumn{1}{c}{Lbol} &
  \multicolumn{1}{c}{eLbol} &
  \multicolumn{1}{c}{Membership$^1$} &
  \multicolumn{1}{c}{Final$^2$}&
  \multicolumn{1}{c}{Npoints$^3$}&
  \multicolumn{1}{c}{Class$^4$}&
  \multicolumn{1}{c}{IRAC$^5$}   &
  \multicolumn{1}{c}{IR excess$^6$}     \\
  \multicolumn{1}{c}{}        &
  \multicolumn{1}{c}{(deg)}        &
  \multicolumn{1}{c}{(deg)}        &
  \multicolumn{1}{c}{(K)}        &
  \multicolumn{1}{c}{($L_\odot$)}        &
  \multicolumn{1}{c}{($L_\odot$)}        &
  \multicolumn{1}{c}{(mag)}        &
  \multicolumn{1}{c}{(K)}        &
  \multicolumn{1}{c}{($L_\odot$)}        &
  \multicolumn{1}{c}{($L_\odot$)}        &
  \multicolumn{1}{c}{HR \& CM }        &
  \multicolumn{1}{c}{}        &
  \multicolumn{1}{c}{}        &
  \multicolumn{1}{c}{}        &
  \multicolumn{1}{c}{slope}   &
  \multicolumn{1}{c}{}        \\
\cline{4-6}
\cline{8-10}
  \multicolumn{1}{c}{}        &
  \multicolumn{1}{c}{}        &
  \multicolumn{1}{c}{}        &
  \multicolumn{3}{c}{$A_V$=0.322}        &
  \multicolumn{1}{c}{}        &
  \multicolumn{3}{c}{$A_V$=variable}        &
  \multicolumn{1}{c}{diagrams}        &
  \multicolumn{1}{c}{}        &
  \multicolumn{1}{c}{}        &
  \multicolumn{1}{c}{}        \\
\hline
  LB01\_a & 82.8103943 & 12.3743267 & 4000 & 0.2074  & 0.0045 & 2.625 & 6000 & 2.8892  & 0.0045 & Y? Y? Y  Y  -- -- -- -- & Y  & 7 13 & III  & N  N  N & Y? -  -  N  N  ?     \\   
  LB01\_b & 82.8095387 & 12.3746628 & --   & --      & --     & --    & --   & --      & --     & -- -- -- -- -- -- -- -- & -- & -  - & -    & -  -  - & -  -  -  -  -  -     \\   
  LB01\_c & 82.8108063 & 12.376833  & --   & --      & --     & --    & --   & --      & --     & -- -- -- -- -- -- -- -- & -- & -  - & -    & -  -  - & -  -  -  -  -  -     \\   
  LB02\_a & 82.8820496 & 12.3225374 & --   & --      & --     & --    & --   & --      & --     & -- -- -- -- -- -- -- -- & -- & -  - & -    & -  -  - & -  -  -  -  -  -     \\   
  LB02\_b & 82.8837585 & 12.3257999 & --   & --      & --     & --    & --   & --      & --     & -- -- -- -- -- -- -- -- & -- & -  - & -    & -  -  - & -  -  -  -  -  -     \\   
  LB02\_c & 82.888237  & 12.3213615 & --   & --      & --     & --    & --   & --      & --     & -- -- -- -- -- -- -- -- & -- & -  - & -    & -  -  - & -  -  -  -  -  -     \\   
  LB02\_d & 82.8833542 & 12.3129959 & 1300 & 0.0018  & 1.0E-4 & --    & --   & --      & --     & Y  -- -- -- -- -- -- -- & Y? & 3  6 & I    & T  T  T & N? -  -  G  N  G     \\   
  LB03\_a & 82.8715635 & 12.2891899 & 1300 & 0.0030  & 1.0E-4 & --    & --   & --      & --     & Y  -- Y  Y  Y  Y  N? Y  & Y? & 3  5 & -    & -  -  - & -  -  -  -  -  -     \\   
  LB03\_b & 82.878997  & 12.288692  & 1500 & 7.0E-4  & 1.0E-4 & --    & --   & --      & --     & Y  -- Y  Y  Y  Y  Y  Y? & Y  & 3  5 & -    & -  -  - & -  -  -  -  -  -     \\   
  LB03\_c & 82.8828009 & 12.2905016 & 700  & 0.0151  & 2.0E-4 & 5.25  & 2900 & 0.0316  & 2.0E-4 & Y  Y  Y  Y  Y  Y  Y  Y  & Y? & 5 11 & -    & -  A  - & -  -  -  -  -  -     \\   
  LB03\_d & 82.8781682 & 12.2934734 & 700  & 0.0020  & 1.0E-4 & --    & --   & --      & --     & Y  -- -- -- -- -- -- -- & Y? & 3  3 & -    & -  -  - & -  -  -  -  -  -     \\   
  LB03\_e & 82.8704761 & 12.2850072 & 1700 & 3.0E-4  & 0.5E-4 & 2.625 & 2000 & 5.0E-4  & 0.5E-4 & Y  Y  Y  Y  Y  Y? -- -- & Y  & 4  4 & -    & -  -  - & -  -  -  -  -  -     \\   
  LB03\_f & 82.8752735 & 12.2819404 & 1500 & 6.0E-4  & 0.5E-4 & --    & --   & --      & --     & Y  -- Y  Y  Y  Y  Y  Y  & Y  & 3  7 & I    & T  T  T & N? -  -  N  N  -     \\   
  LB03\_g & 82.8685852 & 12.2881695 & 1700 & 0.0040  & 2.0E-4 & 1.575 &10000 & 0.268   & 0.0010 & Y  N  Y  Y  Y  N  N  N  & N  & 6  9 & -    & -  N  - & -  -  -  -  -  -     \\   
  LB03\_h & 82.8649428 & 12.2894562 & 3500 & 0.0013  & 1.0E-4 & --    & --   & --      & --     & N  -- -- -- -- -- N  N  & N  & 3  3 & -    & -  -  - & -  -  -  -  -  -     \\   
  LB03\_i & 82.877498  & 12.2929915 & 1500 & 0.0048  & 1.0E-4 & --    & --   & --      & --     & Y  -- Y  Y  Y  Y  Y  Y  & Y  & 3  8 & -    & -  N  - & -  -  -  -  -  -     \\   
  LB04\_a & 82.877498  & 12.2929915 & 1500 & 0.0048  & 1.0E-4 & --    & --   & --      & --     & Y  -- Y  Y  Y  Y  Y  Y  & Y  & 3  8 & -    & -  N  - & -  -  -  -  -  -     \\   
  LB05\_a & 82.872252  & 12.2411915 & 2900 & 0.5969  & 0.0064 & 5.775 &31000 &398.306  & 0.0064 & Y  ?  Y  Y  Y  Y  -- -- & Y  & 8 15 & III  & N  N  N & Y? -  -  N  -  -     \\   
  LB05\_b & 82.8752217 & 12.2423194 & 700  & 0.0468  & 2.0E-4 & 6.3   & 2900 & 0.1286  & 2.0E-4 & Y  Y  Y  Y  Y  Y  Y  Y  & Y? & 5 12 & III  & A  N  A & Y? -  -  N  N  ?     \\   
  LB05\_c & 82.8747836 & 12.247714  & 700  & 0.0025  & 1.0E-4 & --    & --   & --      & --     & Y  -- Y  Y  Y  Y  Y  Y  & Y  & 3  5 & -    & -  -  - & -  -  -  -  -  -     \\   
  LB05\_d & 82.8662881 & 12.2437099 & 3700 & 6.0E-4  & 0.5E-4 & 3.15  & 4000 & 0.0012  & 1.0E-4 & N  N  N  Y? N  N  N  N  & N  & 5  5 & -    & -  -  - & -  -  -  -  -  -     \\   
  LB05\_e & 82.8728663 & 12.2481522 & 1700 & 2.0E-4  & 0.5E-4 & --    & --   & --      & --     & Y  -- Y? -- N  -- -- -- & N? & 3  3 & -    & -  -  - & -  -  -  -  -  -     \\   
  LB06\_a & 82.8729098 & 12.2138781 & 1500 & 0.0237  & 2.0E-4 & 7.875 & 2000 & 0.0819  & 2.0E-4 & Y  Y  Y  Y  Y  Y  Y  Y  & Y  & 5 12 & III  & N  A  N & Y? -  -  N  -  -     \\   
  LB06\_b & 82.8817291 & 12.2117243 & --   & --      & --     & --    & --   & --      & --     & -- -- -- -- -- -- -- -- & -- & -  - & -    & -  -  - & -  -  -  -  -  -     \\   
  LB06\_c & 82.8716738 & 12.216362  & 1500 & 0.0016  & 1.0E-4 & --    & --   & --      & --     & Y  -- Y  Y  Y  Y  Y? Y  & Y  & 3  5 & -    & -  -  - & -  -  -  -  -  -     \\   
  LB06\_d & 82.8820695 & 12.2186419 & 1100 & 0.0013  & 1.0E-4 & --    & --   & --      & --     & Y  -- Y  Y  Y  Y  Y? N  & N? & 3  5 & -    & -  -  - & -  -  -  -  -  -     \\   
  LB07\_a & 82.8845502 & 12.1965247 & --   & --      & --     & --    & --   & --      & --     & -- -- -- -- Y  Y  -- -- & Y? & -  - & -    & -  -  - & -  -  -  -  -  -     \\   
  LB07\_b & 82.8816234 & 12.1965027 & 2000 & 2.0E-4  & 0.5E-4 & 6.825 & 4000 & 0.0012  & 1.0E-4 & N? N  Y  Y? N  N  N  N  & N  & 5  5 & -    & -  -  - & -  -  -  -  -  -     \\   
  LB07\_c & 82.8828726 & 12.1992986 & --   & --      & --     & --    & --   & --      & --     & -- -- -- -- Y? Y  -- -- & Y? & -  - & -    & -  -  - & -  -  -  -  -  -     \\   
  LB07\_d & 82.8825019 & 12.1927908 & 1600 & 3.0E-4  & 0.5E-4 & 9.45  & 4700 & 0.0032  & 1.0E-4 & Y  N  Y  Y  Y? Y  N  N  & N  & 4  5 & -    & -  -  - & -  -  -  -  -  -     \\   
  LB07\_e & 82.8775824 & 12.1930475 & 1800 & 8.0E-4  & 0.5E-4 & 1.575 & 2000 & 0.0011  & 1.0E-4 & Y  Y  Y  Y  Y  Y  -- -- & Y  & 4  4 & -    & -  -  - & -  -  -  -  -  -     \\   
  LB07\_f & 82.8885421 & 12.1999506 & --   & --      & --     & --    & --   & --      & --     & -- -- -- -- -- -- -- -- & -- & -  - & -    & -  -  - & -  -  -  -  -  -     \\   
  LB07\_g & 82.8901372 & 12.1990874 & 1500 & 7.0E-4  & 0.5E-4 & --    & --   & --      & --     & Y  -- Y  Y  Y  Y? Y  N  & N? & 3  5 & -    & -  -  - & -  -  -  -  -  -     \\   
  LB07\_h & 82.8783659 & 12.2009678 & --   & --      & --     & --    & --   & --      & --     & -- -- -- -- N  -- -- -- & N? & -  - & -    & -  -  - & -  -  -  -  -  -     \\   
  LB07\_i & 82.8767882 & 12.1977526 & --   & --      & --     & --    & --   & --      & --     & -- -- -- -- Y? -- -- -- & Y? & -  - & -    & -  -  - & -  -  -  -  -  -     \\   
  LB08\_a & 82.8479608 & 12.1911092 & 1700 & 0.0010  & 1.0E-4 & --    & --   & --      & --     & Y  -- Y  Y  Y  Y  N  N  & Y? & 3  5 & -    & -  -  - & -  -  -  -  -  -     \\   
  LB08\_b & 82.8490389 & 12.195165  & 2000 & 0.0014  & 1.0E-4 & 5.25  & 3900 & 0.0057  & 1.0E-4 & Y  N  Y  Y  Y  Y  N  N  & N  & 4  5 & -    & -  -  - & -  -  -  -  -  -     \\   
  LB08\_c & 82.8478087 & 12.1849561 & 1500 & 0.0030  & 1.0E-4 & --    & --   & --      & --     & Y  -- Y  Y  Y  Y  N  Y  & N? & 3  5 & -    & -  -  - & -  -  -  -  -  -     \\   
  LB08\_d & 82.8487087 & 12.1976932 & --   & --      & --     & --    & --   & --      & --     & -- -- N  -- N  -- -- -- & N? & -  - & -    & -  -  - & -  -  -  -  -  -     \\   
  LB08\_e & 82.8450846 & 12.1841564 & 2000 & 2.0E-4  & 0.5E-4 & 7.35  & 5100 & 0.0024  & 1.0E-4 & N? N  N  Y? N  N  N  N  & N  & 5  5 & -    & -  -  - & -  -  -  -  -  -     \\   
  LB08\_f & 82.8393117 & 12.1900185 & --   & --      & --     & --    & --   & --      & --     & -- -- -- -- -- -- -- -- & -- & -  - & -    & -  -  - & -  -  -  -  -  -     \\   
  LB08\_g & 82.8410187 & 12.1867886 & --   & --      & --     & --    & --   & --      & --     & -- -- -- -- -- -- -- -- & -- & -  - & -    & -  -  - & -  -  -  -  -  -     \\   
  LB08\_h & 82.8538342 & 12.1925179 & 1500 & 7.0E-4  & 0.5E-4 & --    & --   & --      & --     & Y  -- Y  Y  Y  N  Y  N  & N? & 3  5 & -    & -  -  - & -  -  -  -  -  -     \\   
\hline
\end{tabular}
$\,$ \\
$^1$ Membership criteria based on the location on Herzprung-Russell, optical-IR CCDs and CMDs:
HRD($A_V$=0.322), HRD($A_V$ variable), $(J,J-I1)$,  $(J,J-I2$, $(H,H-I1)$, $(H,H-I2)$, $(K,K-I1)$, $(K,K-I2)$.        \\   
$^2$ Final membership assesment.         \\   
$^3$ Number of photometric points on the SED: Fit and total.         \\   
$^4$ Evolutionary class based on a Spitzer/IRAC $(I1-I2,I3-I4)$ CCD.        \\   
$^5$ IRAC slope.         \\   
$^6$ IR excess based on optical, near-IR and Spitzer Color-Color and Color-Magnitude Diagrams: 
$(I1,I1-I4)$, $(H-K,K-M1)$, $(I1-I4,I4-M1)$, $(I2-I3,I3-I4)$, $(I1-I3,I2-I4)$, $(I2,I2-I4)$.           \\   
\end{table}
\end{landscape}

\addtocounter{table}{-1}

\tiny
\begin{landscape}
\begin{table} 
\caption{Properties and memberships criteria for all counterpars within the APEX/LABOCA beam (continuation).}
\label{tabmembership} 
\tiny
\begin{tabular}{llllllllllllllll}
\hline
  \multicolumn{1}{c}{B30-ID}   &
  \multicolumn{1}{c}{RA} &
  \multicolumn{1}{c}{DEC} &
  \multicolumn{1}{c}{Teff} &
  \multicolumn{1}{c}{Lbol} &
  \multicolumn{1}{c}{eLbol} &
  \multicolumn{1}{c}{$A_V$}&
  \multicolumn{1}{c}{Teff} &
  \multicolumn{1}{c}{Lbol} &
  \multicolumn{1}{c}{eLbol} &
  \multicolumn{1}{c}{Membership$^1$} &
  \multicolumn{1}{c}{Final$^2$}&
  \multicolumn{1}{c}{Npoints$^3$}&
  \multicolumn{1}{c}{Class$^4$}&
  \multicolumn{1}{c}{IRAC$^5$}   &
  \multicolumn{1}{c}{IR excess$^6$}     \\
  \multicolumn{1}{c}{}        &
  \multicolumn{1}{c}{(deg)}        &
  \multicolumn{1}{c}{(deg)}        &
  \multicolumn{1}{c}{(K)}        &
  \multicolumn{1}{c}{($L_\odot$)}        &
  \multicolumn{1}{c}{($L_\odot$)}        &
  \multicolumn{1}{c}{(mag)}        &
  \multicolumn{1}{c}{(K)}        &
  \multicolumn{1}{c}{($L_\odot$)}        &
  \multicolumn{1}{c}{($L_\odot$)}        &
  \multicolumn{1}{c}{HR \& CM }        &
  \multicolumn{1}{c}{}        &
  \multicolumn{1}{c}{}        &
  \multicolumn{1}{c}{}        &
  \multicolumn{1}{c}{slope}   &
  \multicolumn{1}{c}{}        \\
\cline{4-6}
\cline{8-10}
  \multicolumn{1}{c}{}        &
  \multicolumn{1}{c}{}        &
  \multicolumn{1}{c}{}        &
  \multicolumn{3}{c}{$A_V$=0.322}        &
  \multicolumn{1}{c}{}        &
  \multicolumn{3}{c}{$A_V$=variable}        &
  \multicolumn{1}{c}{diagrams}        &
  \multicolumn{1}{c}{}        &
  \multicolumn{1}{c}{}        &
  \multicolumn{1}{c}{}        \\
\hline
  LB09\_a & 82.8274595 & 12.1912422 & 700  & 8.0E-4  & 0.5E-4 & --    & --   & --      & --     & Y  -- Y  Y  Y  Y  Y  Y  & Y  & 3  5 & -    & -  -  - & -  -  -  -  -  -     \\   
  LB09\_b & 82.8286146 & 12.192915  & 1500 & 3.0E-4  & 0.5E-4 & --    & --   & --      & --     & Y  -- -- -- Y  Y  -- -- & Y? & 3  3 & -    & -  -  - & -  -  -  -  -  -     \\   
  LB09\_c & 82.8233122 & 12.1910026 & 1500 & 0.0034  & 1.0E-4 & --    & --   & --      & --     & Y  -- Y  Y  Y  Y  Y  Y  & Y  & 3  6 & -    & -  A  - & -  -  -  -  -  -     \\   
  LB09\_d & 82.8244826 & 12.1877228 & 1500 & 7.0E-4  & 0.5E-4 & --    & --   & --      & --     & Y  -- Y  Y  Y  Y  N  N  & N? & 3  5 & -    & -  -  - & -  -  -  -  -  -     \\   
  LB09\_e & 82.831573  & 12.1865735 & 700  & 0.0035  & 1.0E-4 & --    & --   & --      & --     & Y  -- Y  Y  Y  Y  Y  Y  & Y  & 3  5 & -    & -  -  - & -  -  -  -  -  -     \\   
  LB09\_f & 82.8338717 & 12.1928557 & 1700 & 5.0E-4  & 0.5E-4 & 7.875 & 5100 & 0.0052  & 1.0E-4 & Y  N  Y  N  Y  N  Y? N  & N  & 5  5 & -    & -  -  - & -  -  -  -  -  -     \\   
  LB09\_g & 82.8307994 & 12.1940775 & 1500 & 2.0E-4  & 0.5E-4 & 9.975 & 3600 & 0.0015  & 1.0E-4 & Y  N  Y  Y  Y  Y? -- -- & Y? & 4  4 & -    & -  -  - & -  -  -  -  -  -     \\   
  LB09\_h & 82.8265858 & 12.1960763 & 2500 & 3.0E-4  & 0.5E-4 & --    & --   & --      & --     & N  -- -- -- N  N  -- -- & N  & 3  3 & -    & -  -  - & -  -  -  -  -  -     \\   
  LB10\_a & 82.7909414 & 12.1845672 & 3700 & 0.0023  & 1.0E-4 & 4.2   & 4600 & 0.0078  & 1.0E-4 & N  N  Y  Y  N  Y  N  N  & Y? & 4  6 & -    & -  T  - & -  -  -  -  -  -     \\   
  LB10\_b & 82.7897077 & 12.1863344 & 3700 & 8.0E-4  & 0.5E-4 & --    & --   & --      & --     & N  -- Y? Y? N  Y? Y? Y? & N  & 3  5 & -    & T  A  T & Y  -  -  -  -  -     \\   
  LB10\_c & 82.789993  & 12.1816228 & 700  & 0.0030  & 1.0E-4 & --    & --   & --      & --     & Y  -- Y  Y  Y  Y  Y  Y  & Y  & 3  5 & -    & -  -  - & -  -  -  -  -  -     \\   
  LB10\_d & 82.7906454 & 12.1805146 & 3500 & 0.0033  & 1.0E-4 & --    & --   & --      & --     & N  -- Y  Y  N  Y  N  Y? & N  & 3  7 & -    & -  -  - & -  -  -  -  -  -     \\   
  LB10\_e & 82.7929654 & 12.1829769 & 700  & 0.0014  & 1.0E-4 & --    & --   & --      & --     & Y  -- Y  Y  Y  Y  Y  Y  & Y  & 3  5 & -    & -  -  - & -  -  -  -  -  -     \\   
  LB10\_f & 82.7940554 & 12.1827567 & 700  & 0.0046  & 1.0E-4 & --    & --   & --      & --     & Y  -- Y  Y  Y  Y  N  N  & N  & 3  5 & -    & -  -  - & -  -  -  -  -  -     \\   
  LB10\_g & 82.7954744 & 12.181736  & 700  & 0.0030  & 1.0E-4 & --    & --   & --      & --     & Y  -- Y  Y  Y  Y  Y? Y  & Y  & 3  5 & -    & -  -  - & -  -  -  -  -  -     \\   
  LB10\_h & 82.7929218 & 12.1881804 & 3100 & 0.0195  & 3.0E-4 & 4.725 & 7000 & 0.4199  & 3.0E-4 & Y  N  Y  Y  Y  Y  Y  Y? & Y? & 8 14 & -    & -  N  - & -  -  -  -  -  -     \\   
  LB10\_i & 82.795761  & 12.1864233 & 1600 & 0.0021  & 1.0E-4 & --    & --   & --      & --     & Y  -- Y  Y  Y  Y  Y  Y  & Y? & 3  7 & -    & -  -  T & N? -  -  -  -  -     \\   
  LB10\_j & 82.792862  & 12.1905084 & 3700 & 6.0E-4  & 0.5E-4 & --    & --   & --      & --     & N  -- N  N  N  N  -- -- & N  & 3  4 & -    & -  -  - & -  -  -  -  -  -     \\   
  LB10\_k & 82.7826309 & 12.1854864 & 2100 & 0.0068  & 2.0E-4 & 4.725 & 6400 & 0.0597  & 2.0E-4 & Y  N  Y  Y? Y  N  Y  N  & Y? & 5 10 & -    & -  -  - & -  -  -  -  -  -     \\   
  LB10\_l & 82.7848876 & 12.1790492 & 3700 & 3.0E-4  & 0.5E-4 & --    & --   & --      & --     & N  -- N  -- N  -- -- -- & N  & 3  3 & -    & -  -  - & -  -  -  -  -  -     \\   
  LB10\_m & 82.7873969 & 12.1780491 & 1800 & 3.0E-4  & 0.5E-4 & --    & --   & --      & --     & Y  -- Y  Y  Y? Y  -- -- & Y  & 3  4 & -    & -  -  - & -  -  -  -  -  -     \\   
  LB11\_a & 82.7919693 & 12.1718931 & --   & --      & --     & --    & --   & --      & --     & -- -- -- -- -- -- -- -- & -- & -  - & -    & -  -  - & -  -  -  -  -  -     \\   
  LB11\_b & 82.7991141 & 12.1765574 & 600  & 3.0E-4  & 0.5E-4 & --    & --   & --      & --     & Y  -- Y  -- Y  -- -- -- & Y? & 3  3 & -    & -  -  - & -  -  -  -  -  -     \\   
  LB11\_c & 82.7959332 & 12.1745673 & 600  & 6.0E-4  & 0.5E-4 & --    & --   & --      & --     & Y  -- -- -- -- -- -- -- & Y? & 3  3 & -    & -  -  - & -  -  -  -  -  -     \\   
  LB11\_d & 82.7977769 & 12.1732308 & 3600 & 7.0E-4  & 0.5E-4 & 2.625 & 4600 & 0.0017  & 1.0E-4 & N  N  N  N  N  N  -- -- & N  & 4  4 & -    & -  -  - & -  -  -  -  -  -     \\   
  LB11\_e & 82.7945144 & 12.170888  & 1600 & 3.0E-4  & 0.5E-4 & --    & --   & --      & --     & Y  -- Y  -- Y  -- -- -- & Y? & 3  3 & -    & -  -  - & -  -  -  -  -  -     \\   
  LB11\_f & 82.7871577 & 12.1760213 & 3700 & 6.0E-4  & 0.5E-4 & 2.1   & 4000 & 0.0010  & 1.0E-4 & N  N  N  Y? N  N  N  Y? & N  & 4  5 & -    & -  -  - & -  -  -  -  -  -     \\   
  LB12\_a & 82.8622399 & 12.1723873 & 2200 & 0.3283  & 0.0025 & 6.825 & 8000 & 7.911   & 0.0025 & Y  Y? Y  Y  Y  Y  -- -- & Y  & 5 11 & III  & A  A  A & Y  T  T  N  N  ?     \\   
  LB12\_b & 82.860249  & 12.1732605 & --   & --      & --     & --    & --   & --      & --     & -- -- -- -- -- -- Y  Y  & Y? & -  - & -    & -  -  - & -  -  -  -  -  -     \\   
  LB12\_c & 82.8616483 & 12.1773707 & --   & --      & --     & --    & --   & --      & --     & -- -- -- -- Y  Y  -- -- & Y  & -  - & I    & T  T  T & N? -  -  N  N  G     \\   
  LB12\_d & 82.8578644 & 12.1650391 & --   & --      & --     & --    & --   & --      & --     & -- -- -- -- -- -- -- -- & -- & -  - & -    & -  T  - & -  -  -  -  -  -     \\   
  LB12\_e & 82.8602536 & 12.1663629 & 1100 & 8.0E-4  & 0.5E-4 & --    & --   & --      & --     & Y  -- -- -- Y  Y  -- -- & Y  & 3  4 & -    & -  T  - & -  -  -  -  -  -     \\   
  LB12\_f & 82.8658218 & 12.169281  & --   & --      & --     & --    & --   & --      & --     & -- -- -- -- -- -- -- -- & -- & -  - & -    & -  -  - & -  -  -  -  -  -     \\   
  LB12\_g & 82.8675273 & 12.1702059 & 1300 & 0.0039  & 1.0E-4 & --    & --   & --      & --     & Y  -- Y  Y  Y  Y  Y  Y  & Y  & 3  6 & -    & -  T  - & -  -  -  -  -  -     \\   
  LB12\_h & 82.8669147 & 12.1677647 & 700  & 0.0012  & 1.0E-4 & --    & --   & --      & --     & Y  -- -- -- -- -- -- -- & Y? & 3  4 & -    & -  -  - & -  -  -  -  -  -     \\   
  LB12\_i & 82.8647003 & 12.1651821 & --   & --      & --     & --    & --   & --      & --     & -- -- -- -- -- -- -- -- & -- & -  - & -    & -  -  - & -  -  -  -  -  -     \\   
  LB12\_j & 82.8687731 & 12.1739057 & 1500 & 0.0012  & 1.0E-4 & --    & --   & --      & --     & Y  -- Y  Y  Y  Y  Y  Y  & Y  & 3  6 & -    & -  -  T & N? -  -  -  -  G     \\   
  LB12\_k & 82.8610365 & 12.1758133 & --   & --      & --     & --    & --   & --      & --     & -- -- -- -- Y  Y  -- -- & Y? & -  - & -    & -  -  - & -  -  -  -  -  -     \\   
  LB12\_l & 82.86409   & 12.1714582 & --   & --      & --     & --    & --   & --      & --     & -- -- -- -- -- -- -- -- & -- & -  - & -    & -  -  - & -  -  -  -  -  -     \\   
  LB13\_a & 82.8314832 & 12.1614693 & 2800 & 0.0997  & 0.0017 & 0.0   & 2900 & 1.0535  & 0.0017 & Y  Y  Y  Y  Y  Y  Y  Y  & Y  & 9 15 & III  & N  N  N & Y? -  -  N  N  ?     \\   
  LB13\_b & 82.8248788 & 12.1638624 & 1600 & 3.0E-4  & 0.5E-4 & --    & --   & --      & --     & Y  -- Y  Y  Y? Y? -- -- & Y  & 3  4 & -    & -  -  - & -  -  -  -  -  -     \\   
\hline
\end{tabular}
$\,$ \\
$^1$ Membership criteria based on the location on Herzprung-Russell, optical-IR Color-Color and Color-Magnitude Diagrams:
HRD($A_V$=0.322), HRD($A_V$ variable), $(J,J-I1)$,  $(J,J-I2$, $(H,H-I1)$, $(H,H-I2)$, $(K,K-I1)$, $(K,K-I2)$.        \\   
$^2$ Final membership assesment.         \\   
$^3$ Number of photometric points on the SED: Fit and total.         \\   
$^4$ Evolutionary class based on a Spitzer/IRAC $(I1-I2,I3-I4)$ CCD.        \\   
$^5$ IRAC slope.         \\   
$^6$ IR excess based on optical, near-IR and Spitzer Color-Color and Color-Magnitude Diagrams: 
$(I1,I1-I4)$, $(H-K,K-M1)$, $(I1-I4,I4-M1)$, $(I2-I3,I3-I4)$, $(I1-I3,I2-I4)$, $(I2,I2-I4)$.           \\   
\end{table}
\end{landscape}

\addtocounter{table}{-1}

\tiny
\begin{landscape}
\begin{table} 
\caption{Properties and memberships criteria for all counterpars within the APEX/LABOCA beam  (continuation).}
\label{tabmembership} 
\tiny
\begin{tabular}{llllllllllllllll}
\hline
  \multicolumn{1}{c}{B30-ID}   &
  \multicolumn{1}{c}{RA} &
  \multicolumn{1}{c}{DEC} &
  \multicolumn{1}{c}{Teff} &
  \multicolumn{1}{c}{Lbol} &
  \multicolumn{1}{c}{eLbol} &
  \multicolumn{1}{c}{$A_V$}&
  \multicolumn{1}{c}{Teff} &
  \multicolumn{1}{c}{Lbol} &
  \multicolumn{1}{c}{eLbol} &
  \multicolumn{1}{c}{Membership$^1$} &
  \multicolumn{1}{c}{Final$^2$}&
  \multicolumn{1}{c}{Npoints$^3$}&
  \multicolumn{1}{c}{Class$^4$}&
  \multicolumn{1}{c}{IRAC$^5$}   &
  \multicolumn{1}{c}{IR excess$^6$}     \\
  \multicolumn{1}{c}{}        &
  \multicolumn{1}{c}{(deg)}        &
  \multicolumn{1}{c}{(deg)}        &
  \multicolumn{1}{c}{(K)}        &
  \multicolumn{1}{c}{($L_\odot$)}        &
  \multicolumn{1}{c}{($L_\odot$)}        &
  \multicolumn{1}{c}{(mag)}        &
  \multicolumn{1}{c}{(K)}        &
  \multicolumn{1}{c}{($L_\odot$)}        &
  \multicolumn{1}{c}{($L_\odot$)}        &
  \multicolumn{1}{c}{HR \& CM }        &
  \multicolumn{1}{c}{}        &
  \multicolumn{1}{c}{}        &
  \multicolumn{1}{c}{}        &
  \multicolumn{1}{c}{slope}   &
  \multicolumn{1}{c}{}        \\
\cline{4-6}
\cline{8-10}
  \multicolumn{1}{c}{}        &
  \multicolumn{1}{c}{}        &
  \multicolumn{1}{c}{}        &
  \multicolumn{3}{c}{$A_V$=0.322}        &
  \multicolumn{1}{c}{}        &
  \multicolumn{3}{c}{$A_V$=variable}        &
  \multicolumn{1}{c}{diagrams}        &
  \multicolumn{1}{c}{}        &
  \multicolumn{1}{c}{}        &
  \multicolumn{1}{c}{}        \\
\hline
  LB14\_a & 82.8310571 & 12.1542772 & 1600 & 0.6658  & 0.0042 & 6.3   & 5500 & 6.8753  & 0.0042 & Y  Y  Y  Y  Y  Y  -- -- & Y  & 8 16 & II   & T  T  T & Y  Y  Y  N  N  Y     \\   
  LB14\_b & 82.8313354 & 12.1495811 & 1100 & 0.051   & 3.0E-4 & 2.1   & 1700 & 0.1135  & 3.0E-4 & Y  Y  Y  Y  Y  Y  Y  Y  & Y  & 6 13 & II   & T  T  T & Y  Y  Y  N  N  Y     \\   
  LB14\_c & 82.8355319 & 12.1547291 & 2900 & 0.1839  & 0.0021 & 0.0   & 2900 & 1.923   & 0.0021 & Y  Y  Y  Y  Y  Y  Y  Y  & Y  & 8 15 & III  & N  A  N & Y? -  -  N  N  ?     \\   
  LB14\_d & 82.8285637 & 12.1530748 & 1300 & 0.0013  & 1.0E-4 & --    & --   & --      & --     & Y  -- Y  Y  Y  Y  -- -- & Y  & 3  4 & -    & -  -  - & -  -  -  -  -  -     \\   
  LB14\_e & 82.8256866 & 12.1508073 & 2900 & 8.0E-4  & 0.5E-4 & --    & --   & --      & --     & N  -- N  Y  N  Y  N  Y  & N  & 3  5 & -    & -  -  - & -  -  -  -  -  -     \\   
  LB14\_f & 82.8350675 & 12.1520706 & 700  & 7.0E-4  & 0.5E-4 & --    & --   & --      & --     & Y  -- -- Y  -- Y  -- -- & Y? & 3  3 & -    & -  -  - & -  -  -  -  -  -     \\   
  LB14\_g & 82.835154  & 12.1492848 & 3700 & 0.0014  & 1.0E-4 & 5.25  & 4200 & 0.0050  & 1.0E-4 & N  N  Y  Y? Y? N  N  N  & N  & 5  5 & -    & -  -  - & -  -  -  -  -  -     \\   
  LB14\_h & 82.8313458 & 12.147645  & 1300 & 8.0E-4  & 0.5E-4 & --    & --   & --      & --     & Y  -- Y  Y  Y  Y  -- -- & Y  & 3  4 & -    & -  -  - & -  -  -  -  -  -     \\   
  LB14\_i & 82.8294198 & 12.1463334 & 1900 & 4.0E-4  & 0.5E-4 & --    & --   & --      & --     & Y  -- Y  -- Y  -- -- -- & Y? & 3  3 & -    & -  -  - & -  -  -  -  -  -     \\   
  LB14\_j & 82.8302892 & 12.1454723 & --   & --      & --     & --    & --   & --      & --     & -- -- Y  Y  -- -- Y  Y  & N  & -  - & -    & -  -  - & -  -  -  -  -  -     \\   
  LB14\_k & 82.8271388 & 12.1468297 & --   & --      & --     & --    & --   & --      & --     & -- -- -- -- -- -- -- -- & -- & -  - & -    & -  -  - & -  -  -  -  -  -     \\   
  LB15\_a & 82.8675875 & 12.1263543 & 2900 & 0.2189  & 0.0088 & 3.15  & 6200 & 1.2524  & 0.0088 & Y  N  Y  Y  Y  Y  Y  Y  & Y? & 9 15 & III  & N  N  N & Y? -  -  N  N  ?     \\   
  LB15\_b & 82.8619391 & 12.1178684 & 3600 & 0.0023  & 1.0E-4 & 5.25  & 4900 & 0.0112  & 1.0E-4 & N  N  Y  Y  Y? Y  N  N  & N  & 5  5 & -    & -  -  - & -  -  -  -  -  -     \\   
  LB15\_c & 82.8632264 & 12.1307359 & 4200 & 7.0E-4  & 0.5E-4 & --    & --   & --      & --     & N  -- N  N  N  N  -- -- & N  & 3  4 & -    & -  -  - & -  -  -  -  -  -     \\   
  LB15\_d & 82.8631076 & 12.1172072 & --   & --      & --     & --    & --   & --      & --     & -- -- -- -- -- -- -- -- & -- & -  - & -    & -  -  - & -  -  -  -  -  -     \\   
  LB16\_a & 82.8742455 & 12.1128436 & 1500 & 0.0042  & 1.0E-4 & --    & --   & --      & --     & Y  -- Y  Y  Y  Y  Y  N  & Y? & 3  9 & -    & -  N  - & -  -  -  -  -  -     \\   
  LB16\_b & 82.8797155 & 12.112797  & 700  & 0.0042  & 1.0E-4 & --    & --   & --      & --     & Y  -- Y  Y  Y  Y  N  Y  & N? & 3  6 & -    & -  -  - & -  -  -  -  -  -     \\   
  LB16\_c & 82.881731  & 12.1191153 & 700  & 0.0071  & 2.0E-4 & 4.725 & 2000 & 0.0103  & 2.0E-4 & Y  Y  Y  Y  Y  Y  Y  N  & N  & 5  9 & -    & -  N  - & -  -  -  -  -  -     \\   
  LB16\_d & 82.8767443 & 12.1161898 & --   & --      & --     & --    & --   & --      & --     & -- -- -- -- N  -- -- -- & N? & -  - & -    & -  -  - & -  -  -  -  -  -     \\   
  LB16\_e & 82.87578313& 12.1217787 & --   & --      & --     &       & --   & --      & --     & -- -- -- Y  -- Y  -- -- & -- & -  - & -    & -  -  - & -  -  -  -  -  -     \\   
  LB16\_f & 82.8718226 & 12.1202306 & --   & --      & --     &       & --   & --      & --     & -- -- -- -- -- -- -- -- & -- & -  - & -    & -  -  - & -  -  -  -  -  -     \\   
  LB17\_a & 82.7896196 & 12.1125454 & 2000 & 0.0013  & 1.0E-4 & 7.875 & 5500 & 0.0171  & 1.0E-4 & Y  N  Y  Y  Y  Y  N  N  & Y? & 4  8 & -    & -  T  - & -  -  -  -  -  -     \\   
  LB17\_b & 82.7895729 & 12.1169664 & 700  & 0.0059  & 1.0E-4 & --    & --   & --      & --     & Y  -- Y  Y  Y  Y  Y? Y? & Y  & 3  8 & -    & -  N  - & -  -  -  -  -  -     \\   
  LB17\_c & 82.7956702 & 12.1080471 & --   & --      & --     & --    & --   & --      & --     & -- -- -- -- -- -- -- -- & -- & -  - & -    & -  -  - & -  -  -  -  -  -     \\   
  LB17\_d & 82.7934952 & 12.1105013 & --   & --      & --     & --    & --   & --      & --     & -- -- -- -- -- -- -- -- & -- & -  - & -    & -  T  - & -  -  -  -  -  -     \\   
  LB17\_e & 82.7967072 & 12.1139727 & --   & --      & --     & --    & --   & --      & --     & -- -- -- -- -- -- -- -- & -- & -  - & -    & -  -  - & -  -  -  -  -  -     \\   
  LB17\_f & 82.7934146 & 12.116857  & 2000 & 3.0E-4  & 0.5E-4 & 9.975 &14000 & 0.0887  & 1.0E-4 & Y? N  Y  Y? Y  Y? -- -- & Y? & 4  4 & -    & -  -  - & -  -  -  -  -  -     \\   
  LB17\_g & 82.7840129 & 12.110386  & 700  & 0.0027  & 1.0E-4 & --    & --   & --      & --     & Y  -- Y  Y  N  Y? N  N  & N  & 3  5 & -    & -  -  - & -  -  -  -  -  -     \\   
  LB18\_a & 82.893931  & 12.1086094 & 1500 & 0.3083  & 0.0014 & 7.875 & 3000 & 0.5961  & 0.0014 & Y  Y  Y  Y  Y  Y  Y  Y  & Y  & 6 14 & III  & A  A  A & Y? T  T  N  N  ?     \\   
  LB18\_b & 82.8941585 & 12.1112154 & 700  & 0.0014  & 1.0E-4 & --    & --   & --      & --     & Y  -- Y  Y  Y  Y  Y  Y  & Y  & 3  5 & -    & -  -  - & -  -  -  -  -  -     \\   
  LB18\_c & 82.900058  & 12.1082994 & 700  & 6.0E-4  & 0.5E-4 & --    & --   & --      & --     & Y  -- Y  -- Y  -- -- -- & Y? & 3  6 & -    & -  -  - & -  -  -  -  -  -     \\   
  LB18\_d & 82.8994244 & 12.1055759 & --   & --      & --     & --    & --   & --      & --     & -- -- -- -- -- -- -- -- & -- & -  - & -    & -  -  - & -  -  -  -  -  -     \\   
  LB18\_e & 82.8907647 & 12.1071047 & 700  & 0.0011  & 1.0E-4 & --    & --   & --      & --     & Y  -- Y  Y  Y  Y  Y  Y  & Y  & 3  5 & -    & -  -  - & -  -  -  -  -  -     \\   
  LB18\_f & 82.8979923 & 12.0998372 & 700  & 0.0012  & 1.0E-4 & --    & --   & --      & --     & Y  -- Y  Y? Y? N  N  N  & N? & 3  5 & -    & -  -  - & -  -  -  -  -  -     \\   
  LB18\_g & 82.8914648 & 12.1000721 & 1500 & 0.0010  & 1.0E-4 & --    & --   & --      & --     & Y  -- Y  Y  Y  Y  N  Y  & N? & 3  5 & -    & -  -  - & -  -  -  -  -  -     \\   
  LB18\_h & 82.8959583 & 12.1037709 & --   & --      & --     & --    & --   & --      & --     & -- -- -- -- Y  Y  -- -- & Y? & -  - & -    & -  -  - & -  -  -  -  -  -     \\   
  LB19\_a & 82.8658453 & 12.0919505 & 800  & 0.0787  & 2.0E-4 & 1.05  & 1200 & 0.0612  & 2.0E-4 & Y  Y  Y  Y  Y  Y  Y  Y  & Y  & 5 15 & I    & T  T  T & N? Y  Y  N  N  Y     \\   
  LB19\_b & 82.8676553 & 12.0917727 & 2700 & 0.0505  & 7.0E-4 & 4.725 & 7600 & 0.7158  & 7.0E-4 & Y  N  Y  Y  Y  Y  Y  Y  & Y  & 5 11 & -    & -  T  - & -  -  -  -  -  -     \\   
  LB19\_c & 82.8686592 & 12.0905224 & 2800 & 0.0836  & 0.0015 & 3.15  & 6500 & 0.3575  & 0.0015 & Y  N  Y  Y  Y  Y  Y  Y  & Y? & 8 12 & -    & -  N  - & -  -  -  -  -  -     \\   
  LB19\_d & 82.8733996 & 12.0925052 & 1600 & 0.0203  & 4.0E-4 & 2.625 & 3000 & 0.0815  & 4.0E-4 & Y  Y  -- -- Y  Y  -- -- & Y? & 4 13 & I    & T  T  T & N? Y  Y  N  N  Y     \\   
  LB19\_e & 82.8614809 & 12.0885397 & 700  & 0.0040  & 1.0E-4 & --    & --   & --      & --     & Y  -- Y  Y  Y  Y  Y  Y  & Y  & 3  7 & I    & T  T  T & N? -  -  N  N  G     \\   
  LB19\_f & 82.8678788 & 12.0866633 & --   & --      & --     & --    & --   & --      & --     & -- -- -- -- -- -- Y  Y  & Y  & -  - & I/II & T  T  T & N? -  Y  N  N  G     \\   
  LB19\_g & 82.8612144 & 12.0926726 & --   & --      & --     & --    & --   & --      & --     & -- -- -- -- -- -- -- -- & -- & -  - & -    & -  -  - & -  -  -  -  -  -     \\   
  LB19\_h & 82.8668776 & 12.0875591 & --   & --      & --     & --    & --   & --      & --     & -- -- -- -- -- Y  -- -- & Y? & -  - & -    & -  -  - & -  -  -  N  -  G     \\   
\hline
\end{tabular}
$\,$ \\
$^1$ Membership criteria based on the location on Herzprung-Russell, optical-IR Color-Color and Color-Magnitude Diagrams:
HRD($A_V$=0.322), HRD($A_V$ variable), $(J,J-I1)$,  $(J,J-I2$, $(H,H-I1)$, $(H,H-I2)$, $(K,K-I1)$, $(K,K-I2)$.        \\   
$^2$ Final membership assesment.         \\   
$^3$ Number of photometric points on the SED: Fit and total.         \\   
$^4$ Evolutionary class based on a Spitzer/IRAC $(I1-I2,I3-I4)$ CCD.        \\   
$^5$ IRAC slope.         \\   
$^6$ IR excess based on optical, near-IR and Spitzer Color-Color and Color-Magnitude Diagrams: 
$(I1,I1-I4)$, $(H-K,K-M1)$, $(I1-I4,I4-M1)$, $(I2-I3,I3-I4)$, $(I1-I3,I2-I4)$, $(I2,I2-I4)$.           \\   
\end{table}
\end{landscape}

\addtocounter{table}{-1}

\tiny
\begin{landscape}
\begin{table} 
\caption{Properties and memberships criteria for all counterpars within the APEX/LABOCA beam (continuation).}
\label{tabmembership} 
\tiny
\begin{tabular}{llllllllllllllll}
\hline
  \multicolumn{1}{c}{B30-ID}   &
  \multicolumn{1}{c}{RA} &
  \multicolumn{1}{c}{DEC} &
  \multicolumn{1}{c}{Teff} &
  \multicolumn{1}{c}{Lbol} &
  \multicolumn{1}{c}{eLbol} &
  \multicolumn{1}{c}{$A_V$}&
  \multicolumn{1}{c}{Teff} &
  \multicolumn{1}{c}{Lbol} &
  \multicolumn{1}{c}{eLbol} &
  \multicolumn{1}{c}{Membership$^1$} &
  \multicolumn{1}{c}{Final$^2$}&
  \multicolumn{1}{c}{Npoints$^3$}&
  \multicolumn{1}{c}{Class$^4$}&
  \multicolumn{1}{c}{IRAC$^5$}   &
  \multicolumn{1}{c}{IR excess$^6$}     \\
  \multicolumn{1}{c}{}        &
  \multicolumn{1}{c}{(deg)}        &
  \multicolumn{1}{c}{(deg)}        &
  \multicolumn{1}{c}{(K)}        &
  \multicolumn{1}{c}{($L_\odot$)}        &
  \multicolumn{1}{c}{($L_\odot$)}        &
  \multicolumn{1}{c}{(mag)}        &
  \multicolumn{1}{c}{(K)}        &
  \multicolumn{1}{c}{($L_\odot$)}        &
  \multicolumn{1}{c}{($L_\odot$)}        &
  \multicolumn{1}{c}{HR \& CM }        &
  \multicolumn{1}{c}{}        &
  \multicolumn{1}{c}{}        &
  \multicolumn{1}{c}{}        &
  \multicolumn{1}{c}{slope}   &
  \multicolumn{1}{c}{}        \\
\cline{4-6}
\cline{8-10}
  \multicolumn{1}{c}{}        &
  \multicolumn{1}{c}{}        &
  \multicolumn{1}{c}{}        &
  \multicolumn{3}{c}{$A_V$=0.322}        &
  \multicolumn{1}{c}{}        &
  \multicolumn{3}{c}{$A_V$=variable}        &
  \multicolumn{1}{c}{diagrams}        &
  \multicolumn{1}{c}{}        &
  \multicolumn{1}{c}{}        &
  \multicolumn{1}{c}{}        \\
\hline
  LB20\_a & 82.8508555 & 12.0902626 & 2000 & 0.0231  & 3.0E-4 & 5.775 & 6500 & 0.296   & 3.0E-4 & Y  N  Y  Y  Y  Y  Y  Y  & Y? & 5 11 & -    & -  -  - & -  -  -  -  -  -     \\   
  LB20\_b & 82.8513261 & 12.0917094 & 1500 & 0.0011  & 1.0E-4 & --    & --   & --      & --     & Y  -- Y  -- Y  -- -- -- & Y? & 3  3 & -    & -  -  - & -  -  -  -  -  -     \\   
  LB20\_c & 82.8402513 & 12.0966039 & 4500 & 1.3842  & 0.0229 & 1.575 & 5100 & 2.1228  & 0.0229 & Y  Y  Y  Y  Y  Y  -- -- & Y  & 8 13 & -    & -  -  - & -  -  -  -  -  -     \\   
  LB20\_d & 82.8394462 & 12.0955432 & 3400 & 0.7616  & 0.0054 & 3.675 & 5600 & 3.9349  & 0.0054 & Y  Y  Y  Y  Y  Y  -- -- & Y  & 5 11 & -    & -  N  - & -  -  -  -  -  -     \\   
  LB20\_e & 82.8394127 & 12.0930814 & 700  & 0.0123  & 5.0E-4 & 3.15  & 3500 & 0.0126  & 5.0E-4 & Y  N  Y  Y  Y  Y  Y  Y  & Y? & 5  9 & -    & -  N  - & -  -  -  -  -  -     \\   
  LB20\_f & 82.8468218 & 12.096629  & 2000 & 8.0E-4  & 0.5E-4 & 5.775 & 4000 & 0.0040  & 1.0E-4 & Y  N  Y  Y  Y? N  N  N  & N  & 5  5 & -    & -  -  - & -  -  -  -  -  -     \\   
  LB20\_g & 82.8439374 & 12.0959187 & 1500 & 4.0E-4  & 0.5E-4 & --    & --   & --      & --     & Y  -- Y  -- Y  -- -- -- & Y? & 3  3 & -    & -  -  - & -  -  -  -  -  -     \\   
  LB20\_h & 82.844606  & 12.0966218 & --   & --      & --     & --    & --   & --      & --     & -- -- -- -- Y  Y  -- -- & Y? & -  - & -    & -  -  - & -  -  -  -  -  -     \\   
  LB20\_i & 82.8430989 & 12.0983201 & 1500 & 7.0E-4  & 0.5E-4 & 2.625 & 1900 & 0.0011  & 1.0E-4 & Y  Y  Y  Y  Y  Y  -- -- & Y  & 4  4 & -    & -  -  - & -  -  -  -  -  -     \\   
  LB20\_j & 82.8529105 & 12.0926289 & 1600 & 4.0E-4  & 0.5E-4 & --    & --   & --      & --     & Y  -- -- -- Y  Y  -- -- & Y? & 3  3 & -    & -  -  - & -  -  -  -  -  -     \\   
  LB21\_a & 82.8988495 & 12.0846224 & --   & --      & --     & --    & --   & --      & --     & -- -- -- -- -- -- -- -- & -- & -  - & -    & -  -  - & -  -  -  -  -  -     \\   
  LB21\_b & 82.899437  & 12.0812626 & --   & --      & --     & --    & --   & --      & --     & -- -- -- -- -- -- -- -- & -- & -  - & -    & T  N  T & N? -  -  -  -  G     \\   
  LB21\_c & 82.90199   & 12.081975  & --   & --      & --     & --    & --   & --      & --     & -- -- Y  Y  -- -- Y  Y  & Y? & -  - & -    & -  -  - & -  -  -  -  -  -     \\   
  LB21\_d & 82.8997765 & 12.0891346 & 2000 & 6.0E-4  & 0.5E-4 & --    & --   & --      & --     & Y  -- N  Y? N  N  N  N  & Y? & 3  8 & -    & -  -  - & -  -  -  -  -  -     \\   
  LB21\_e & 82.906055  & 12.0828322 & 1800 & 6.0E-4  & 0.5E-4 & --    & --   & --      & --     & Y  -- Y  Y? Y  N  Y  N  & N  & 3  5 & -    & -  -  - & -  -  -  -  -  -     \\   
  LB21\_f & 82.8998506 & 12.0914761 & 1700 & 3.0E-4  & 0.5E-4 & 1.05  & 1900 & 4.0E-4  & 0.5E-4 & Y  Y  Y  Y  Y  Y  -- -- & Y  & 4  4 & -    & -  -  - & -  -  -  -  -  -     \\   
  LB21\_g & 82.8943406 & 12.0873348 & 1600 & 7.0E-4  & 0.5E-4 & --    & --   & --      & --     & Y  -- Y  Y  Y  Y  -- -- & Y  & 3  4 & -    & -  -  - & -  -  -  -  -  -     \\   
  LB21\_h & 82.8972397 & 12.0861282 & --   & --      & --     & --    & --   & --      & --     & -- -- -- -- -- -- -- -- & -- & -  - & -    & -  -  - & -  -  -  -  -  -     \\   
  LB21\_i & 82.8977103 & 12.0892541 & --   & --      & --     & --    & --   & --      & --     & -- -- -- -- -- -- -- -- & -- & -  - & -    & -  -  - & -  -  -  -  -  -     \\   
  LB22\_a & 82.8822765 & 12.0712583 & 700  & 0.0035  & 1.0E-4 & --    & --   & --      & --     & Y  -- Y  -- Y  -- Y  -- & Y  & 3  6 & -    & -  -  - & -  -  -  -  -  -     \\   
  LB22\_b & 82.8780746 & 12.0702267 & --   & --      & --     & --    & --   & --      & --     & -- -- -- -- -- -- -- -- & -- & -  - & -    & -  -  - & -  -  -  -  -  -     \\   
  LB22\_c & 82.8842633 & 12.0692825 & 1600 & 7.0E-4  & 0.5E-4 & 5.25  & 2000 & 0.0017  & 1.0E-4 & Y  Y  Y  Y  Y  Y  -- -- & Y  & 4  4 & -    & -  -  - & -  -  -  -  -  -     \\   
  LB22\_d & 82.8865316 & 12.0684093 & 3700 & 0.0044  & 1.0E-4 & --    & --   & --      & --     & N  -- Y  Y  Y  N  Y? N  & N  & 3  8 & -    & -  -  - & -  -  -  -  -  -     \\   
  LB22\_e & 82.883323  & 12.069021  & 700  & 0.0040  & 1.0E-4 & --    & --   & --      & --     & Y  -- Y  Y  Y  N  Y  N  & N? & 3  5 & -    & -  -  - & -  -  -  -  -  -     \\   
  LB22\_f & 82.8850515 & 12.0706474 & 3700 & 0.0019  & 1.0E-4 & 2.625 & 4400 & 0.0044  & 1.0E-4 & N  N  N  N  N  N  N  N  & N  & 5  5 & -    & -  -  - & -  -  -  -  -  -     \\   
  LB22\_g & 82.886582  & 12.0713157 & 700  & 0.0011  & 1.0E-4 & --    & --   & --      & --     & Y  -- Y  Y  Y  Y  Y  Y  & Y  & 3  5 & -    & -  -  - & -  -  -  -  -  -     \\   
  LB22\_h & 82.876306  & 12.0709514 & 1600 & 4.0E-4  & 0.5E-4 & 1.575 & 1900 & 8.0E-4  & 1.0E-4 & Y  Y  Y  Y  Y  Y  -- -- & Y  & 4  5 & -    & -  T  - & -  -  -  -  -  -     \\   
  LB22\_i & 82.879455  & 12.0659626 & 3100 & 7.0E-4  & 0.5E-4 & --    & --   & --      & --     & N  -- -- -- -- -- -- -- & N? & 3  3 & -    & -  -  - & -  -  -  -  -  -     \\   
  LB22\_j & 82.8845747 & 12.0683098 & 1500 & 4.0E-4  & 0.5E-4 & --    & --   & --      & --     & Y  -- -- -- Y  Y  -- -- & Y  & 3  5 & I/II & T  T  T & N? -  -  N  N  -     \\   
  LB23\_a & 82.8733007 & 12.0754017 & 3300 & 8.0E-4  & 0.5E-4 & --    & --   & --      & --     & N  -- Y  Y? Y  N  Y  Y? & Y? & 3  8 & -    & -  T  - & -  -  -  -  -  -     \\   
  LB23\_b & 82.8740083 & 12.0797718 & 1500 & 0.0031  & 1.0E-4 & --    & --   & --      & --     & Y  -- Y  Y  Y  Y  Y  N  & Y? & 3  9 & -    & -  -  - & -  -  -  -  -  -     \\   
  LB23\_c & 82.8761053 & 12.0778312 & 1500 & 4.0E-4  & 0.5E-4 & 2.1   & 1900 & 7.0E-4  & 1.0E-4 & Y  Y  Y  Y  Y  Y  -- -- & Y  & 4  7 & -    & -  -  T & N? -  -  -  -  G     \\   
  LB23\_d & 82.8687196 & 12.0718283 & 2900 & 0.0060  & 3.0E-4 & 6.3   &25000 & 3.1204  & 3.0E-4 & Y  N  Y? N  N  N  N  N  & N  & 6 11 & -    & -  N  - & -  -  -  -  -  -     \\   
  LB23\_e & 82.8684924 & 12.0782548 & 2000 & 4.0E-4  & 0.5E-4 & 5.775 & 4000 & 0.0021  & 1.0E-4 & Y  N  Y  Y? N  N  N  N  & N  & 5  5 & -    & -  -  - & -  -  -  -  -  -     \\   
  LB23\_f & 82.8684686 & 12.0757307 & 2000 & 0.0010  & 1.0E-4 & 6.3   & 4000 & 0.0053  & 1.0E-4 & Y  N  Y  Y  Y  N  N  N  & N  & 5  5 & -    & -  N  - & -  -  -  -  -  -     \\   
  LB23\_g & 82.8733448 & 12.0814827 & 1600 & 6.0E-4  & 0.5E-4 & 8.925 & 3100 & 0.0032  & 1.0E-4 & Y  Y? Y  Y  Y  Y  -- -- & Y? & 4  4 & -    & -  -  - & -  -  -  -  -  -     \\   
  LB23\_h & 82.8712059 & 12.0724084 & 1800 & 0.0023  & 1.0E-4 & --    & --   & --      & --     & Y  -- Y  -- Y  -- Y  -- & Y? & 3  4 & -    & -  N  - & -  -  -  -  -  -     \\   
  LB23\_i & 82.87796   & 12.080151  & --   & --      & --     & --    & --   & --      & --     & -- -- -- -- -- -- -- -- & -- & -  - & -    & -  -  - & -  -  -  -  -  -     \\   
  LB24\_a & 82.8498907 & 12.0748369 & 1900 & 7.0E-4  & 0.5E-4 & --    & --   & --      & --     & Y  -- -- Y  -- Y  -- Y  & Y  & 3  7 & -    & -  T  - & -  -  -  -  -  -     \\   
  LB24\_b & 82.8529587 & 12.0784426 & --   & --      & --     & --    & --   & --      & --     & -- -- -- -- -- -- -- -- & -- & -  - & -    & -  -  - & -  -  -  -  -  -     \\   
  LB24\_c & 82.8439484 & 12.0800447 & --   & --      & --     & --    & --   & --      & --     & -- -- -- -- -- -- -- -- & -- & -  - & -    & -  -  - & -  -  -  -  -  -     \\   
  LB24\_d & 82.8440552 & 12.0770445 & --   & --      & --     & --    & --   & --      & --     & -- -- -- -- -- -- -- -- & -- & -  - & -    & -  -  - & -  -  -  N  -  G     \\   
  LB24\_e & 82.8468915 & 12.072453  & 1700 & 0.0031  & 1.0E-4 & --    & --   & --      & --     & Y  -- Y  Y  Y  Y  Y  Y? & Y? & 3  6 & -    & -  T  - & -  -  -  -  -  -     \\   
  LB24\_f & 82.8432544 & 12.0696761 & 1800 & 8.0E-4  & 0.5E-4 & --    & --   & --      & --     & Y  -- Y  Y  Y? Y? N  N  & N? & 3  5 & -    & -  -  - & -  -  -  -  -  -     \\   
  LB24\_g & 82.8469663 & 12.08225   & 700  & 0.0022  & 1.0E-4 & --    & --   & --      & --     & Y  -- Y  Y  Y  Y  Y  Y  & Y  & 3  7 & I/II & T  T  T & N? -  -  N  N  G     \\   
\hline
\end{tabular}
$\,$ \\
$^1$ Membership criteria based on the location on Herzprung-Russell, optical-IR Color-Color and Color-Magnitude Diagrams:
HRD($A_V$=0.322), HRD($A_V$ variable), $(J,J-I1)$,  $(J,J-I2$, $(H,H-I1)$, $(H,H-I2)$, $(K,K-I1)$, $(K,K-I2)$.        \\   
$^2$ Final membership assesment.         \\   
$^3$ Number of photometric points on the SED: Fit and total.         \\   
$^4$ Evolutionary class based on a Spitzer/IRAC $(I1-I2,I3-I4)$ CCD.        \\   
$^5$ IRAC slope.         \\   
$^6$ IR excess based on optical, near-IR and Spitzer Color-Color and Color-Magnitude Diagrams: 
$(I1,I1-I4)$, $(H-K,K-M1)$, $(I1-I4,I4-M1)$, $(I2-I3,I3-I4)$, $(I1-I3,I2-I4)$, $(I2,I2-I4)$.           \\   
\end{table}
\end{landscape}

\addtocounter{table}{-1}

\tiny
\begin{landscape}
\begin{table} 
\caption{Properties and memberships criteria for all counterpars within the APEX/LABOCA beam  (continuation).}
\label{tabmembership} 
\tiny
\begin{tabular}{llllllllllllllll}
\hline
  \multicolumn{1}{c}{B30-ID}   &
  \multicolumn{1}{c}{RA} &
  \multicolumn{1}{c}{DEC} &
  \multicolumn{1}{c}{Teff} &
  \multicolumn{1}{c}{Lbol} &
  \multicolumn{1}{c}{eLbol} &
  \multicolumn{1}{c}{$A_V$}&
  \multicolumn{1}{c}{Teff} &
  \multicolumn{1}{c}{Lbol} &
  \multicolumn{1}{c}{eLbol} &
  \multicolumn{1}{c}{Membership$^1$} &
  \multicolumn{1}{c}{Final$^2$}&
  \multicolumn{1}{c}{Npoints$^3$}&
  \multicolumn{1}{c}{Class$^4$}&
  \multicolumn{1}{c}{IRAC$^5$}   &
  \multicolumn{1}{c}{IR excess$^6$}     \\
  \multicolumn{1}{c}{}        &
  \multicolumn{1}{c}{(deg)}        &
  \multicolumn{1}{c}{(deg)}        &
  \multicolumn{1}{c}{(K)}        &
  \multicolumn{1}{c}{($L_\odot$)}        &
  \multicolumn{1}{c}{($L_\odot$)}        &
  \multicolumn{1}{c}{(mag)}        &
  \multicolumn{1}{c}{(K)}        &
  \multicolumn{1}{c}{($L_\odot$)}        &
  \multicolumn{1}{c}{($L_\odot$)}        &
  \multicolumn{1}{c}{HR \& CM }        &
  \multicolumn{1}{c}{}        &
  \multicolumn{1}{c}{}        &
  \multicolumn{1}{c}{}        &
  \multicolumn{1}{c}{slope}   &
  \multicolumn{1}{c}{}        \\
\cline{4-6}
\cline{8-10}
  \multicolumn{1}{c}{}        &
  \multicolumn{1}{c}{}        &
  \multicolumn{1}{c}{}        &
  \multicolumn{3}{c}{$A_V$=0.322}        &
  \multicolumn{1}{c}{}        &
  \multicolumn{3}{c}{$A_V$=variable}        &
  \multicolumn{1}{c}{diagrams}        &
  \multicolumn{1}{c}{}        &
  \multicolumn{1}{c}{}        &
  \multicolumn{1}{c}{}        \\
\hline
  LB25\_a & 82.8354366 & 12.0851035 & 1500 & 0.0017  & 1.0E-4 & --    & --   & --      & --     & Y  -- Y  Y  Y  Y  Y  Y  & Y  & 3  8 & -    & -  -  - & -  -  -  -  -  -     \\   
  LB25\_b & 82.8303947 & 12.080038  & 1700 & 0.0097  & 3.0E-4 & 5.25  & 5400 & 0.0558  & 3.0E-4 & Y  N  Y  Y  Y  Y? N  N  & Y? & 5 12 & I/II & T  N  T & Y  -  -  G  G  G     \\   
  LB25\_c & 82.8416812 & 12.0856044 & 2000 & 0.0134  & 3.0E-4 & 7.875 &22000 & 7.7318  & 3.0E-4 & Y  N  Y  Y  Y  Y  Y  Y  & Y? & 6 13 & I/II & T  T  T & N? -  -  N  G  G     \\   
  LB25\_d & 82.8393083 & 12.0822474 & 1900 & 0.0010  & 1.0E-4 & --    & --   & --      & --     & Y  -- Y  Y  Y  Y? Y  Y  & N? & 3  5 & -    & -  -  - & -  -  -  -  -  -     \\   
  LB25\_e & 82.8342835 & 12.0861197 & 1500 & 6.0E-4  & 0.5E-4 & --    & --   & --      & --     & Y  -- Y  Y  Y  Y  -- Y  & Y? & 3  5 & -    & -  T  - & -  -  -  -  -  -     \\   
  LB25\_f & 82.8397243 & 12.0872619 & 1500 & 4.0E-4  & 0.5E-4 & --    & --   & --      & --     & Y  -- -- -- Y  Y  -- Y?-& Y? & 3  5 & I/II & T  T  T & N? -  -  N  N  G     \\   
  LB26\_a & 82.8254809 & 12.0871155 & 700  & 0.0051  & 1.0E-4 & --    & --   & --      & --     & Y  -- Y  Y  Y  Y  Y  Y  & Y? & 3  8 & -    & -  T  - & -  -  -  -  -  -     \\   
  LB26\_b & 82.82662   & 12.0823778 & 700  & 0.0026  & 1.0E-4 & --    & --   & --      & --     & Y  -- Y  Y  Y? Y  N  Y  & N? & 3  5 & -    & -  -  - & -  -  -  -  -  -     \\   
  LB26\_c & 82.8216294 & 12.0792754 & 4100 & 0.0019  & 1.0E-4 & --    & --   & --      & --     & N  -- Y  Y? Y  N  Y  Y? & N  & 3  6 & -    & -  -  - & -  -  -  -  -  -     \\   
  LB27\_a & 82.8100128 & 12.0964232 & 700  & 0.012   & 2.0E-4 & 6.825 & 2700 & 0.0207  & 2.0E-4 & Y  Y  Y  Y  Y  Y  Y  Y  & Y  & 5 11 & -    & N  A  N & Y? -  -  -  -  -     \\   
  LB27\_b & 82.8064117 & 12.0876884 & --   & --      & --     & --    & --   & --      & --     & -- -- -- -- -- -- -- -- & -- & -  - & -    & -  -  - & -  -  -  N  -  G     \\   
  LB27\_c & 82.8052597 & 12.0901575 & --   & --      & --     & --    & --   & --      & --     & -- -- -- -- -- -- -- -- & -- & -  - & -    & -  -  - & -  -  -  -  -  -     \\   
  LB27\_d & 82.8040515 & 12.0845908 & 700  & 0.0074  & 1.0E-4 & --    & --   & --      & --     & Y  -- Y  Y  Y  Y  Y  N  & N  & 3  7 & -    & -  -  - & -  -  -  -  -  -     \\   
  LB27\_e & 82.8021317 & 12.0896969 & --   & --      & --     & --    & --   & --      & --     & -- -- -- -- -- -- -- -- & -- & -  - & -    & -  -  - & -  -  -  -  -  -     \\   
  LB27\_f & 82.8020477 & 12.0958843 & --   & --      & --     & --    & --   & --      & --     & -- -- -- -- -- -- -- -- & -- & -  - & -    & -  -  - & -  -  -  -  -  -     \\   
  LB28\_a & 82.7841035 & 12.0847157 & 700  & 0.0028  & 1.0E-4 & --    & --   & --      & --     & Y  -- Y  Y  Y  Y  Y  Y  & Y  & 3  5 & -    & -  -  - & -  -  -  -  -  -     \\   
  LB28\_b & 82.784874  & 12.0824242 & --   & --      & --     & --    & --   & --      & --     & -- -- -- -- -- -- -- -- & -- & -  - & -    & -  -  - & -  -  -  -  -  -     \\   
  LB28\_c & 82.7759399 & 12.0832968 & --   & --      & --     & --    & --   & --      & --     & -- -- -- -- -- -- -- -- & -- & -  - & -    & -  -  - & -  -  -  -  -  -     \\   
  LB28\_d & 82.7767727 & 12.085662  & 1800 & 0.0013  & 1.0E-4 & --    & --   & --      & --     & Y  -- Y  Y  Y  Y  Y  Y  & Y  & 3  7 & I/II & T  T  T & N? -  -  N  G  -     \\   
  LB29\_a & 82.7869358 & 12.0624632 & 2000 & 7.0E-4  & 0.5E-4 & --    & --   & --      & --     & Y  -- Y  Y? Y  N  Y  N  & Y? & 3 12 & I/II & -  -  - & -  -  -  -  -  -     \\   
  LB29\_b & 82.7845362 & 12.0612638 & 2100 & 0.0030  & 1.0E-4 & --    & --   & --      & --     & Y  -- Y  Y  Y  N  Y? N  & Y? & 3  9 & -    & -  T  - & -  -  -  -  -  -     \\   
  LB29\_c & 82.784803  & 12.067079  & 3300 & 0.0109  & 2.0E-4 & 3.675 & 5700 & 0.0501  & 2.0E-4 & N? N  Y  Y  Y  N  Y  N  & N  & 5 11 & -    & -  -  - & -  -  -  -  -  -     \\   
  LB29\_d & 82.7912491 & 12.0631026 & 3200 & 6.0E-4  & 0.5E-4 & 7.35  &14000 & 0.0708  & 1.0E-4 & N  N  Y  N  Y  N  -- -- & N  & 4  9 & -    & -  -  - & -  -  -  -  -  -     \\   
  LB29\_e & 82.7846915 & 12.0644042 & 2200 & 0.0011  & 1.0E-4 & --    & --   & --      & --     & Y  -- Y? Y? Y? N  N  N  & N  & 3  6 & -    & -  T  - & -  -  -  -  -  -     \\   
  LB29\_f & 82.7910175 & 12.0668983 & 5100 & 0.0048  & 1.0E-4 & --    & --   & --      & --     & N  -- N  N  N  N  N  N  & Y? & 3  8 & I/II & T  T  T & N? -  Y  N  N  G     \\   
  LB29\_g & 82.7809669 & 12.0645025 & 700  & 4.0E-4  & 0.5E-4 & --    & --   & --      & --     & Y  -- Y  Y  -- -- -- -- & Y  & 3  7 & -    & -  -  - & -  -  -  -  -  -     \\   
  LB29\_h & 82.7803662 & 12.0659296 & 2200 & 3.0E-4  & 0.5E-4 & --    & --   & --      & --     & N? -- Y  -- Y  -- -- -- & Y? & 3  3 & -    & -  -  - & -  -  -  -  -  -     \\   
  LB29\_i & 82.7808674 & 12.0589337 & 1700 & 6.0E-4  & 0.5E-4 & --    & --   & --      & --     & Y  -- Y  Y  Y  Y  Y  Y  & Y  & 3  5 & -    & -  -  - & -  -  -  -  -  -     \\   
  LB29\_j & 82.787634  & 12.0558409 & 4400 & 0.0069  & 2.0E-4 & 4.2   &13500 & 0.2458  & 2.0E-4 & N  N  N  N  -- -- -- -- & N  & 6  6 & -    & -  -  T & N? -  -  -  -  G     \\   
  LB30\_a & 82.8030423 & 12.0608531 & 1700 & 0.0071  & 2.0E-4 & 7.35  & 7000 & 0.1219  & 2.0E-4 & Y  N  Y  Y  Y  Y? Y? N  & Y? & 6 12 & I/II & T  T  T & N? -  -  N  G  G     \\   
  LB30\_b & 82.801508  & 12.0605056 & 1800 & 6.0E-4  & 0.5E-4 & --    & --   & --      & --     & Y  -- Y  Y  Y  Y  Y  Y  & Y  & 3  5 & -    & -  -  - & -  -  -  -  -  -     \\   
  LB30\_c & 82.8016933 & 12.0635964 & 1800 & 7.0E-4  & 0.5E-4 & --    & --   & --      & --     & Y  -- Y  Y  Y  Y  Y  Y  & Y? & 3  7 & I/II & T  T  T & N? -  -  N  N  G     \\   
  LB30\_d & 82.8019218 & 12.064775  & --   & --      & --     & --    & --   & --      & --     & -- -- -- -- -- -- -- -- & -- & -  - & -    & -  -  - & -  -  -  -  -  -     \\   
  LB30\_g & 82.8103715 & 12.0592594 & 700  & 0.0096  & 1.0E-4 & --    & --   & --      & --     & Y  -- Y  Y  Y  Y  Y  Y  & Y  & 3  9 & -    & -  T  - & -  -  -  -  -  -     \\   
  LB31\_b & 82.8175493 & 12.0635955 & 3700 & 0.0035  & 1.0E-4 & --    & --   & --      & --     & N  -- Y  Y  Y  Y? Y  Y  & N  & 3  8 & -    & -  T  - & -  -  -  -  -  -     \\   
  LB31\_c & 82.8158798 & 12.0589838 & --   & --      & --     & --    & --   & --      & --     & -- -- -- -- -- -- -- -- & -- & -  - & -    & -  -  - & -  -  -  -  -  -     \\   
  LB31\_a & 82.8140335 & 12.0624819 & --   & --      & --     & --    & --   & --      & --     & -- -- -- -- -- -- -- -- & -- & -  - & I/II & T  T  T & N? -  -  N  N  G     \\   
  LB30\_e & 82.808399  & 12.0577028 & 700  & 0.0074  & 1.0E-4 & --    & --   & --      & --     & Y  -- Y  Y  Y  Y  Y  Y  & Y  & 3  6 & -    & -  T  - & -  -  -  -  -  -     \\   
  LB30\_f & 82.80728   & 12.061461  & 1500 & 6.0E-4  & 0.5E-4 & --    & --   & --      & --     & Y  -- -- Y  -- Y  -- Y  & Y  & 3  4 & -    & -  -  - & -  -  -  -  -  -     \\   
  LB31\_g & 82.8131356 & 12.0558984 & 1800 & 0.0055  & 3.0E-4 & 6.3   & 6400 & 0.0709  & 3.0E-4 & Y  N  Y  Y? Y? N  N  N  & N  & 6  8 & -    & -  -  - & -  -  -  -  -  -     \\   
  LB31\_h & 82.8147323 & 12.0538983 & 2100 & 0.0056  & 4.0E-4 & 6.825 & 9200 & 0.2404  & 4.0E-4 & Y  N  Y  Y  Y  Y  Y  Y  & Y  & 6  9 & -    & -  T  - & -  -  -  -  -  -     \\   
  LB31\_d & 82.8149427 & 12.0656361 & 2200 & 9.0E-4  & 0.5E-4 & --    & --   & --      & --     & Y  -- -- -- -- -- -- -- & Y? & 3  3 & -    & -  -  - & -  -  -  -  -  -     \\   
  LB31\_e & 82.8187075 & 12.0627859 & 1600 & 7.0E-4  & 0.5E-4 & 0.0   & 1900 & 0.0010  & 1.0E-4 & Y  Y  Y  Y  Y  Y  -- -- & Y  & 4  4 & -    & -  -  - & -  -  -  -  -  -     \\   
  LB31\_f & 82.8188762 & 12.0560849 & 4400 & 0.0016  & 1.0E-4 & --    & --   & --      & --     & N  -- Y  -- Y  -- Y  -- & N  & 3  4 & -    & -  -  - & -  -  -  -  -  -     \\   
  LB32\_a & 83.0513535 & 12.1654158 & 4300 & 0.082   & 0.0021 & 0.525 & 4100 & 0.0848  & 0.0021 & N  N  Y  Y  -- -- -- -- & Y? & 7 16 & I/II & T  T  T & N? Y  Y  N  N  -     \\   
  LB32\_b & 83.0526962 & 12.1675291 & 5400 & 0.0574  & 0.0017 & 0.525 & 5700 & 0.0668  & 0.0017 & N  N  N  N  -- -- -- -- & N  &10 10 & -    & -  N  - & -  -  -  -  -  -     \\   
  LB32\_c & 83.0493698 & 12.1644573 & 4800 & 0.0104  & 3.0E-4 & 0.0   & 4600 & 0.0090  & 3.0E-4 & N  N  N  N  -- -- -- -- & N  & 4  6 & -    & -  -  - & -  -  -  -  -  -     \\   
  LB32\_d & 83.0553741 & 12.1640244 & 6200 & 0.0106  & 1.0E-4 & --    & --   & --      & --     & N  -- N  N  -- -- -- -- & N  & 3  3 & -    & -  -  - & -  -  -  -  -  -     \\   
  LB32\_e & 83.0466461 & 12.1649246 & 1600 & 9.0E-4  & 0.5E-4 & --    & --   & --      & --     & Y  -- Y  Y  -- -- -- -- & Y? & 3  3 & -    & -  -  - & -  -  -  -  -  -     \\   
  LB32\_f & 83.0588531 & 12.1689034 & --   & --      & --     & --    & --   & --      & --     & -- -- Y  Y  -- -- -- -- & Y? & -  - & -    & -  -  - & -  -  -  -  -  -     \\   
  LB33\_a & 82.5546799 & 12.1460333 & 3900 & 1.3239  & 0.0121 & 1.575 & 4500 & 2.7174  & 0.0121 & Y  Y  Y  Y  -- -- -- -- & Y  & 6 16 & II   & T  T  T & N? Y  Y  N  N  Y     \\   
\hline
\end{tabular}
$\,$ \\
$^1$ Membership criteria based on the location on Herzprung-Russell, optical-IR Color-Color and Color-Magnitude Diagrams:
HRD($A_V$=0.322), HRD($A_V$ variable), $(J,J-I1)$,  $(J,J-I2$, $(H,H-I1)$, $(H,H-I2)$, $(K,K-I1)$, $(K,K-I2)$.        \\   
$^2$ Final membership assesment.         \\   
$^3$ Number of photometric points on the SED: Fit and total.         \\   
$^4$ Evolutionary class based on a Spitzer/IRAC $(I1-I2,I3-I4)$ CCD.        \\   
$^5$ IRAC slope.         \\   
$^6$ IR excess based on optical, near-IR and Spitzer Color-Color and Color-Magnitude Diagrams: 
$(I1,I1-I4)$, $(H-K,K-M1)$, $(I1-I4,I4-M1)$, $(I2-I3,I3-I4)$, $(I1-I3,I2-I4)$, $(I2,I2-I4)$.           \\   
\end{table}
\end{landscape}


\clearpage

\setcounter{table}{7}
%
%

\tiny
\begin{landscape}
\begin{table} 
\caption{Temperatures and luminosities for Barnard 30 objects with significant infrared excesses.}
\label{TABbolometric} 
\begin{tabular}{lrrrrlrrlrrrrrllr}
\hline
  \multicolumn{1}{c}{B30-ID} &
  \multicolumn{1}{c}{RA2000} &
  \multicolumn{1}{c}{DEC200} &
  \multicolumn{1}{c}{Teff} &
  \multicolumn{1}{c}{Lbol} &
  \multicolumn{1}{c}{Group$^1$} &
  \multicolumn{1}{c}{Teff} &
  \multicolumn{1}{c}{Lbol} &
  \multicolumn{1}{c}{MemF} &
  \multicolumn{1}{c}{Tbol} &
  \multicolumn{1}{c}{Tbol$^2$} &
  \multicolumn{1}{c}{Lbol} &
  \multicolumn{1}{c}{Lbol$^2$} &
  \multicolumn{1}{c}{L(int)$^3$} &
  \multicolumn{1}{c}{Class$^4$} &
  \multicolumn{1}{c}{Type} &
  \multicolumn{1}{c}{Npoint} \\
\cline{4-5}
\cline{7-8}
\cline{10-13}
  \multicolumn{1}{c}{} &
  \multicolumn{1}{c}{(deg)} &
  \multicolumn{1}{c}{(deg)} &
  \multicolumn{2}{c}{$A_V$=0.322} &
  \multicolumn{1}{c}{APEX} &
  \multicolumn{2}{c}{$A_V$=variable} &
  \multicolumn{1}{c}{} &
  \multicolumn{4}{c}{Integration} &
  \multicolumn{1}{c}{} &
  \multicolumn{1}{c}{} &
  \multicolumn{1}{c}{} &
  \multicolumn{1}{c}{} \\
\hline

  LB14a        & 82.8310571  & 12.1542772  &   1600 & 0.6658   & A1, YSO    &      5500    &     6.8753    &   Y   &      --      &1284     &  --      &     0.998    &  0.276    & II                & CTT        &  20    \\ 
  LB14b        & 82.8313354  & 12.1495811  &   1100 & 0.0510   & A1, YSO    &      1700    &     0.1135    &   Y   &    1093      & 246     &  0.046   &     0.253    &  --       & II$\rightarrow$I  & CTT/YSO    &  14    \\ 
  LB19a$^{5}$   & 82.8658453  & 12.0919505  &   800  & 0.0787   & A1, YSO   &      1200    &     0.0612    &   Y   &      --      & 238     &  --      &     1.218     &  0.402    & I                 & YSO         &  15    \\ 
  LB19d        & 82.8733996  & 12.0925052  &   1600 & 0.0203   & A1, YSO    &      3000    &     0.0815    &   Y?  &     281      & 249     &  0.172   &     0.199    &  0.094    & I                 & YSO/VeLLO  &  12    \\ 
  LB24d        & 82.8440552  & 12.0770445  &   --   & --       & B1, excess  &      --      &     --        &Y?$^\dagger$& 353      &   88    &  0.056      &$\le$0.524   &$\le$0.604 & I                & YSO         &  12    \\ 
  LB11a        & 82.7921273  & 12.1737653  &   --    &   --       &  C1, excess &  --    &  --     &   Y?      &   --     &  163     & --        &  0.029     &   --      & I                & YSO/protoBD     &  7     \\ 
  LB12a        & 82.8622399  & 12.1723873  &   2200  &   0.3283   &  C1, excess &  8000  &  7.911  &   Y       &   --     & 2234     & --        &  0.392     &   --      & II               & WTT             &  15    \\ 
  LB12d        & 82.858020   & 12.166592   &   --    &   --       &  C1, excess &  --    &  --     &   Y?      &  177     &  136     & 0.051     &  0.084     &   --      & I                & YSO/protoBD     &  4     \\ 
  LB18a        & 82.893931   & 12.1086094  &   1500  &   0.3083   &  C1, excess &  3000  &  0.5961 &   Y       &   --     & 1753     & --        &  0.249     &   --      & II               & WTT             &  14    \\ 
  LB22b$^7$    & 82.8780746  & 12.0702267  &   --    &   --       &  C1, YSO    &  --    &  --     &Y?$^\dagger$&  255     &   82     & 0.016     &  0.064     &   --      & I                & YSO/protoBD       &  8     \\ 
  LB22j$^8$    & 82.8845747  & 12.0683098  &   1500  &   0.0004   &  C1, YSO    &  --    &  --     &   Y       &  243     &   47     & 0.008     &  0.060     &   --      & I$\rightarrow$0  & YSO/protoBD     &  6     \\    
  LB30g        & 82.8103715  & 12.0592594  &   700   &   0.0096   &  C1, excess &  --    &  --     &   Y       &  354     &  239     & 0.059     &  0.103     &   --      & I                & YSO/protoBD     &  11    \\ 
  LB31c        & 82.8158798  & 12.0589838  &   --    &   --       &  C1, excess &  --    &  --     &   --      &  264     &  190     & 0.048     &  0.074     &   --      & I                & YSO/protoBD       &  6     \\ 
  LB32a        & 83.0513535  & 12.1654158  &   4300  &   0.0820   &  C1, YSO    &  4100  &  0.0848 &   Y?      &   --     &  414     & --        &  0.657     &   --      & I                & YSO             &  21    \\ 
  LB33a        & 82.5546799  & 12.1460333  &   3900  &   1.3239   &  C1, YSO    &  4500  &  2.7174 &   Y       &   --     & 1920     & --        &  2.079     &   --      & II               & CTT             &  20    \\ 
  LB08a        & 82.8479608  & 12.1911092  &   1700 & 0.0010   & A2, YSO    &      --      &     --        &   Y?  &      --      &   50    &  --      &     12.01    &  1.223    & 0                 & YSO        &  11    \\ 
  LB23a$^{6}$  & 82.8733007  & 12.0754017  &   3300 & 0.0008   & A2, excess  &      --      &     --        &   Y?  &     312     &   62    &  0.008   &     0.160    &  0.189    & I$\rightarrow$0   & YSO/protoBD&  11    \\ 
  LB23b        & 82.8740083  & 12.0797718  &   1500 & 0.0031   & A2, excess  &      --      &     --        &   Y?  &     697     &   --    &  0.006   &    --       &  --       & II                & CTT/BD     &  10    \\   
  LB23c        & 82.8761053  & 12.0778312  &   1500 & 0.0004   & A2, excess  &      1900    &     7.0E-4    &   Y   &     290     &  135    &  0.008   &     0.024   &  --       & I                 & YSO/protoBD&   11    \\    
  LB25c        & 82.8416812  & 12.0856044  &   2000 & 0.0134   & A2, YSO    &      22000   &     7.7318    &   Y?  &      69      &   56     &  1.709   &     2.325   &  0.557    & 0                 & YSO        &  20    \\ 
  LB29a        & 82.7869358  & 12.0624632  &   2000 & 0.0007   & A2$^*$, YSO&      --      &     --        &   Y?  &      --      & (70)$^*$&  --      &    (0.271)$^*$&  0.337    & I/0               & YSO/protoBD$^*$&  12    \\ 
  LB29b        & 82.7845362  & 12.0612638  &   2100 & 0.0030   & A2$^*$, YSO&      --      &     --        &   Y?  &    (533)$^*$ & (82)$^*$&(0.016)$^*$&   (0.266)$^*$&  --       & I                 & YSO/protoBD$^*$&  13    \\ 
  LB29f        & 82.7910175  & 12.0668983  &   5100 & 0.0048   & A2$^*$, YSO&      --      &     --        &   Y?  &    (926)$^*$ & (85)$^*$&(0.010)$^*$&   (0.257)$^*$&  --       & II$\rightarrow$I  & YSO/protoBD$^*$&  12    \\ 
\hline
\end{tabular}
$\,$ \\
$^1$  Grouping based on the detection at 70 and 24 $\mu$m  and presence of counterparts, as discussed in subsection \ref{sub:ClassificationCounterparts}:
      A1 = Detection at 70 and 24 $\mu$m;
      A2 = detection at 70 and upper limit at 24 $\mu$m;
      B1 = upper limit at 70 $\mu$m and detection at 24 $\mu$m;
      C1 = no data at 70 $\mu$m and detection at 24 $\mu$m.
      Objects without detected emision at 24 and/or 70  $\mu$m are not listed here (B2 and C2).
Tentative classification: \\
      YSO = Young Stelalr Object, with a optical and/or nearIR sources within 5 arcsec of APEX/LABOCA central coordinate. Our tentative interpretation is that most of them are proto-stars or proto-BDs.
      Excess = Optical and/or nearIR sources with excesses within the APEX/LABOCA beam but farther than 5 arcsec. The submm source cannot be asssigned unambiguously to any counterpart so its nature remains unknown. 
      Starless = Possible starless core, since there is neither counterparts closer than 5 arcsec nor a optical/IR source farther away with excess.\\
$^2$ Including the flux at 870 $\mu$m from APEX/LaBoca and, if present, MIPS M2.\\
$^3$ Internal luminosity, based on MIPS flux at 70 $\mu$m, after \cite{Dunham2008-Protostars}.\\%
$^4$ SED classification based on the T$_{bol}$ vs. $L_{bol}$ diagram.\\
$^5$ There is a submm source nearby, namely SB08, detected with APEX/SABOCA at 350 $\mu$m (\citealt{Huelamo2017-ALMA-B30}).\\ 
$^6$ There is a submm source nearby, namely SB05, detected with APEX/SABOCA at 350 $\mu$m  (\citealt{Huelamo2017-ALMA-B30}).\\\
$^7$ There is a submm source nearby, namely SB04, detected with APEX/SABOCA at 350 $\mu$m (\citealt{Huelamo2017-ALMA-B30}).\\ 
$^8$ There is a submm source nearby, namely SB03, detected with APEX/SABOCA at 350 $\mu$m  (\citealt{Huelamo2017-ALMA-B30}).\\
$^\dagger$ Membership status updated from -- (Tab. \ref{tabmembership}) to Y? based on all the available data and our analysis.\\
$^{*}$ Possible detection at MIPS 24 $mu$m, but it real, it might be affected by the inhomogeneous extended emission by the nebulosity.\\
\end{table}
\end{landscape}

\clearpage



\appendix

%
\section{Classification of the  LABOCA submm sources and the counterparts\label{appendix}}
%


\subsection{Group A: LABOCA sources detected at 70 $\mu$m\label{sub:groupA}}

Maps obtained with Spitzer/MIPS M2 might cover only half of the surveyed area due to the loss of one detector.
This is our case and we do not have measured fluxes at 70 $\mu$m for 20 APEX/LABOCA sources. Of the other 14, only six have been detected
at this wavelength and they are discussed in this section.
The size of the MIPS M2 beam is about  18.6 arcsec (\citealt{Frayer2009-COSMOS-MIPSinfo}), about half of the value of APEX/LABOCA,
whereas the MIPS M1 beam is much small, 5.8 arcsec.
Thus, a visual inspection has been carried out in order to confirm our identifications (Fig. \ref{FCh_groupA1} and Fig. \ref{FCh_groupA2}). 
We note that APEX/LABOCA has a nominal pointing error below $\sim$2 arcsec. 
As stated before, the actual rms of the pointing is about 1 arcsec.
Although these optical/IR sources are the most probable  counterparts to the submm detections, 
there are additional young sources  within the LABOCA beam that might contribute to (or less probably be responsible of) the measured flux.
Thus, we discuss here all sources within the LABOCA beam which have been detected and present mid-IR emission.



\subsubsection{Group A1: LABOCA sources detected at 24 $\mu$m\label{sub:groupA1notes}}


{\bf B30-LB14.-}
B30-LB14a is a  Class II object
(based on the IRAC data  displayed  Fig. \ref{fig_CCD_CMD_IRAC},
and in the SED, shown in Fig. \ref{SED_groupA1}), as classified based on IRAC photometry, located
within the central 5 arcsec of the LABOCA beam. 
 It is one of the brightest objects in our sample, fully within the stellar domain.
The SED has been represented in Fig. \ref{SED_groupA1}.
For comparison, we have overplotted in the SED panel the expected emission of a 3 Myr object with 0.072 $M_\odot$
 without interstellar absorption and with a value of A$_V=10$ magnitudes (black solid lines).
We note, however, that the interstellar extinction as estimated with the A$_v$ map from 2MASS data is 2.471 mag (subsect.~\ref{subsub:Av}).
 The SED allows to derived bolometric temperature and luminosity of 
 1284 K and 0.998 $L_\odot$. The internal luminosity, as derived from the MIPS M2 flux,  is 0.276  $L_\odot$.
 This source has been detected with ALMA, see details in \cite{Huelamo2017-ALMA-B30} and it can be classified as a CTT.
The  envelope mass can be estimated as 0.051 $M_\odot$, based on the flux at 870 $\mu$m.

\begin{figure*}  
\center
%
\includegraphics[width=0.450\textwidth,scale=0.330]{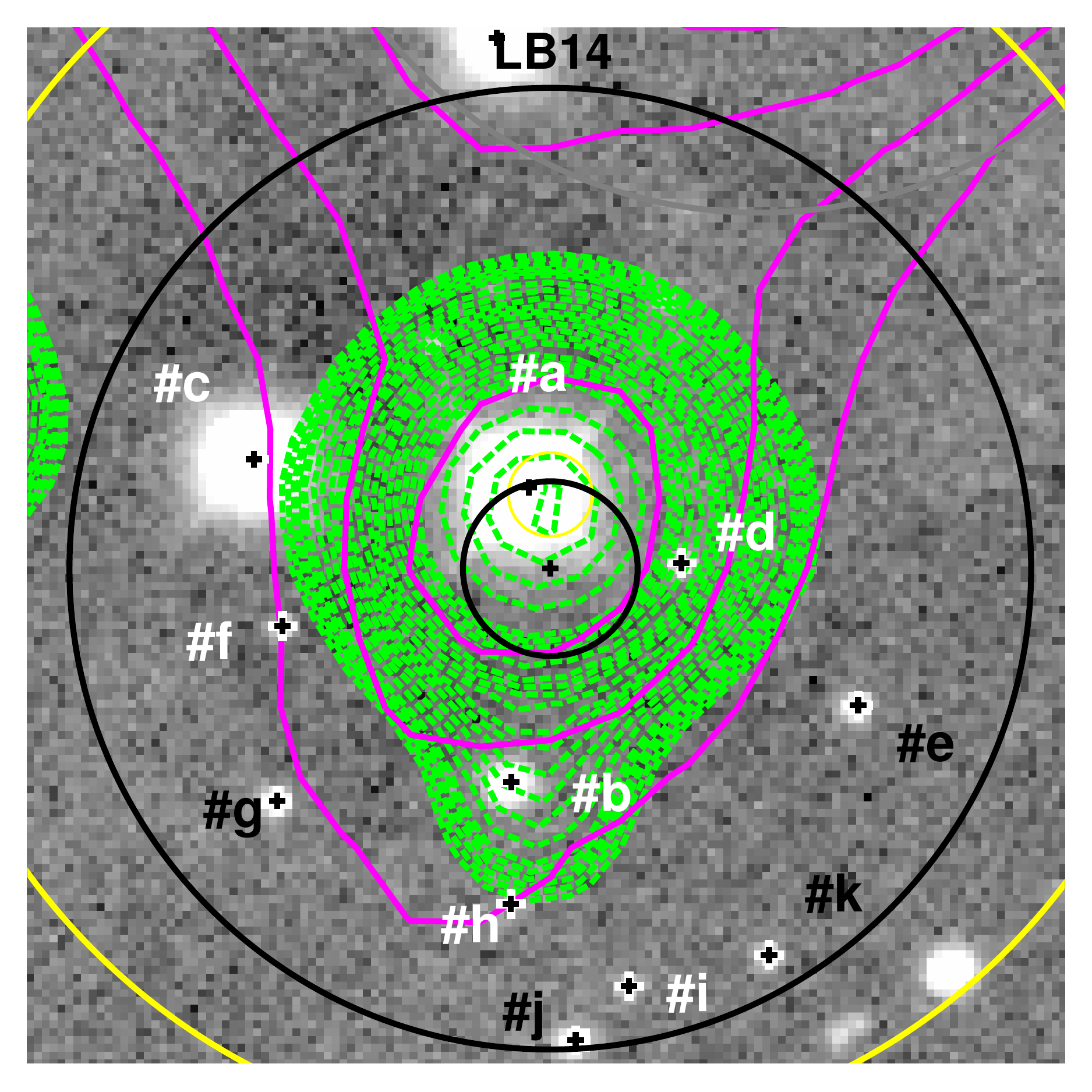}
\includegraphics[width=0.450\textwidth,scale=0.330]{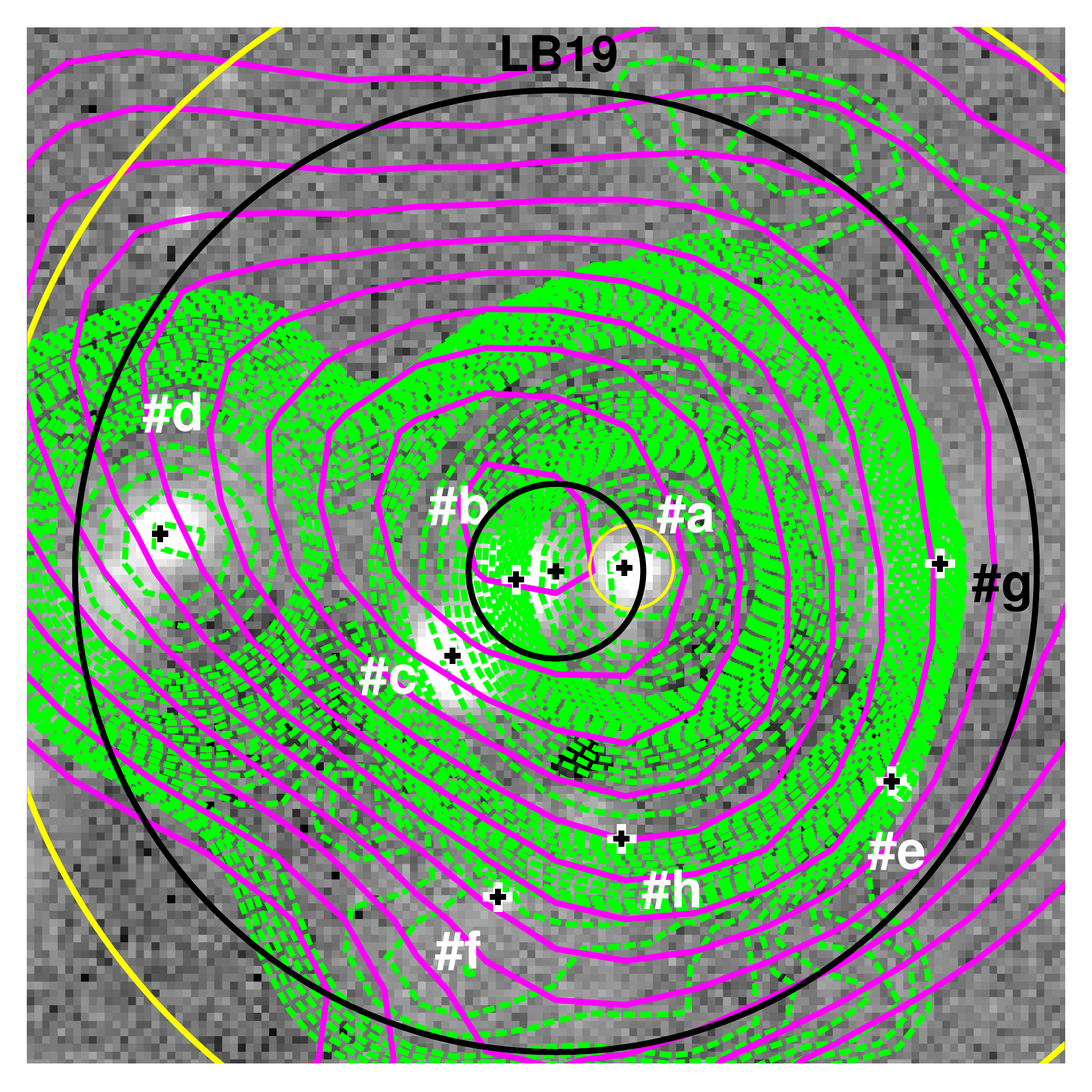}
\caption{\label{FCh_groupA1} 
Finding charts for the relevant sources detected also both at 70 and 24 $\mu$m with MIPS (A1 group). 
The black large and small circles correspond to APEX/LABOCA beam size and the search radius, 27.6 and 5 arcsec respectively,
whereas the yellow big and small circles represent the Akari/FIS and Akari/IRC beams. 
White labels denote possible and probable members of
B30, whereas black labels are used for non-members or objects with unknown status.
The background images correspond to CAHA/Omega2000 in the $J$ band. 
The  green, dashed contours come from the Spitzer/MIPS image at 24 micron after removing the difuse emission  by the nebulosity
(see subsection \ref{sub:Spitzer}), whereas the magenta 
levels comes from the APEX/LABOCA and correspond to S/N starting at 2 with increments of 1. 
}
\end{figure*}

\begin{figure*}   
\center
%
\includegraphics[width=0.450\textwidth,scale=0.33]{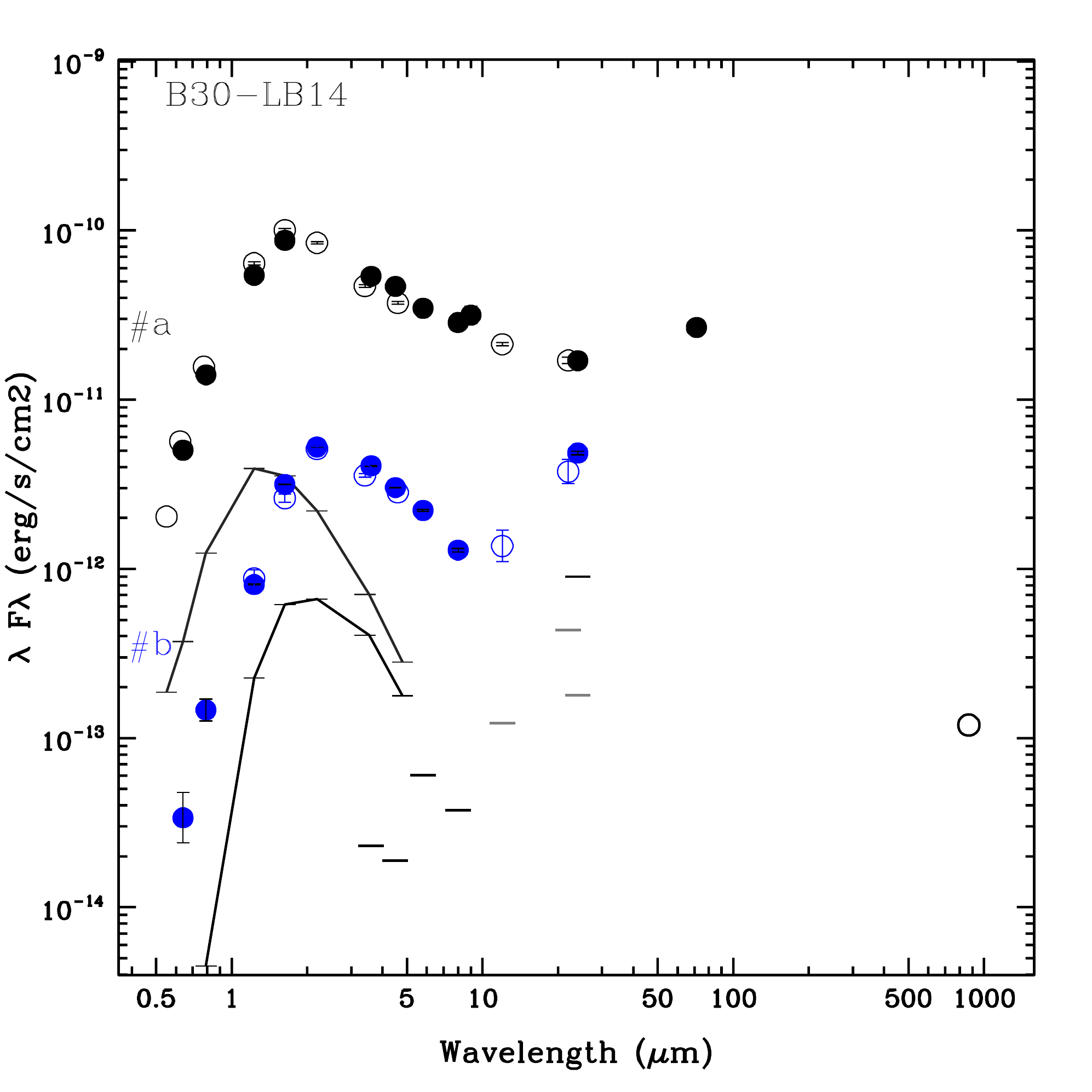} 
\includegraphics[width=0.450\textwidth,scale=0.33]{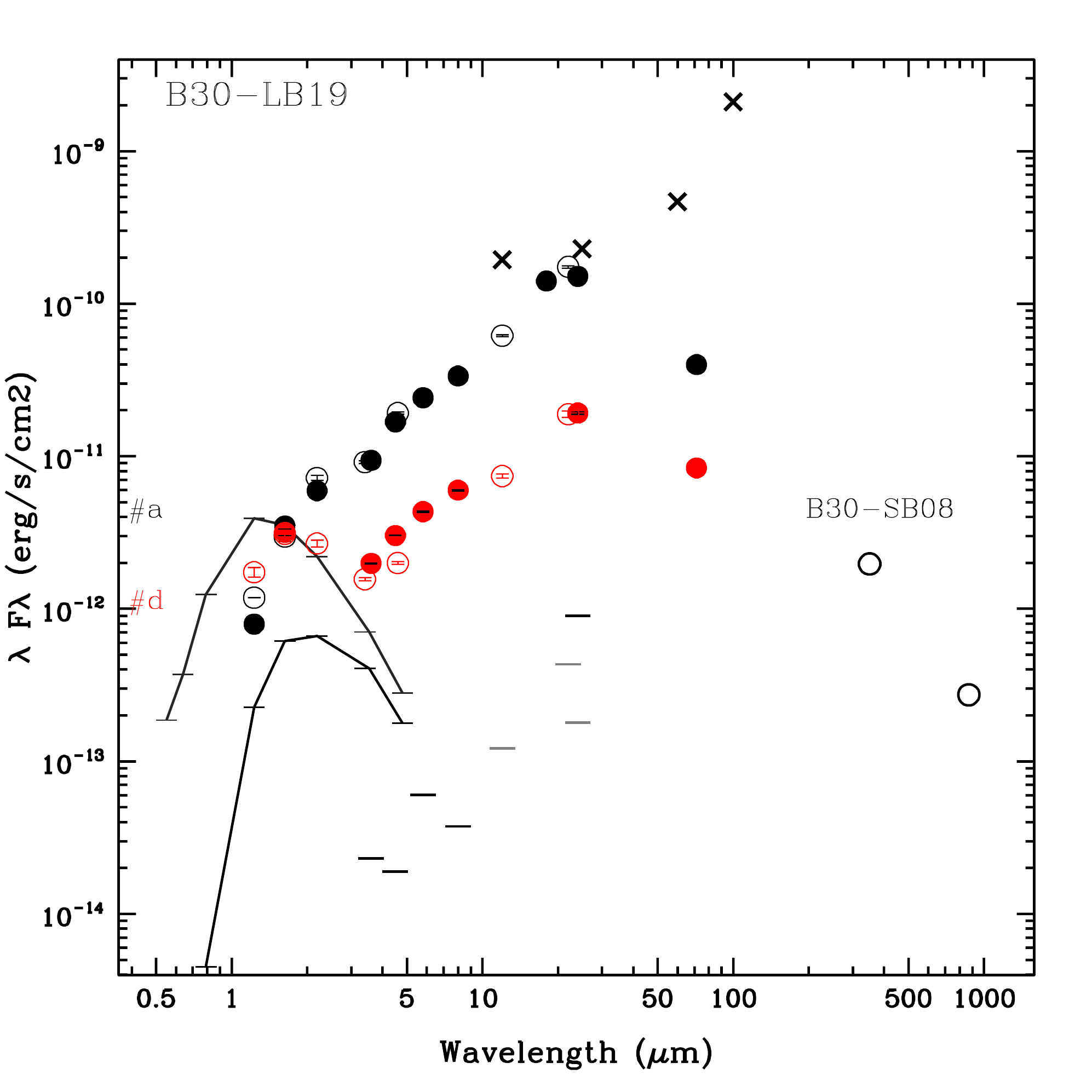} 
\caption{\label{SED_groupA1} 
Spectral energy distribution for the relevant sources detected also both at 70 and 24 $\mu$m with MIPS (A1 group).
 Completeness and detection limits have been included for Spitzer data, as gray and black line  segments, 
respectively.
For comparison, we have overplotted in the SED panel the expected emission of a 3 Myr object
 with 0.072 $M_\odot$ at the distance of B30 --400pc,  
without interstellar absorption and with a value of A$_V=10$ magnitudes (black solid lines).
}
\end{figure*}

South of the central counterpart is located B30-LB14b (Fig. \ref{FCh_groupA1}), also detected at 24 $\mu$m (both objects have been detected
with WISE W3 and W4 filters at 12 and 22 $\mu$m).
The bolometric temperature and luminosity  of \#b, taken into account the fluxes up to 24 $\mu$m (i.e., including MIPS M1), are
1093 K and 0.046  $L_\odot$ (i.e., within the substellar domain, in full agreement with the SED analysis with VOSA).
Thus, the evolutionary status would correspond to a Class II brown dwarf.
However, due to the uncertainty in the assignment of the 70 and 870 fluxes,
we have repeated the calculation assuming that these
fluxes correspond to source \#b. Thus, we have derived T$_{bol}^{70+870}$=246 K and L$_{bol}^{70+870}$=0.253 $L_\odot$.
Figure \ref{HR_TbolLbol_Evolution_A} displays L$_{bol}$ versus T$_{bol}$. The arrow links both sets of values for B30-LB14b and suggests this
counterpart might be a Class I object at the border between stars and brown dwarfs
(a very low mass YSO or massive  substellar analog).

Other interesting objects within the B30-LB14 beam are \#d and \#h. Both are very faint and have been
detected only at $J$, $H$, I1 and I2 and with increasing fluxes. The data for \#f and \#i are even more scarce.
All four are worth further follow-up, since they might be in the substellar domain.

{\bf B30-LB19.-}
B30-LB19 can be identified with the source IRAS05286$+$1203 (RA=82.8671, DEC=+12.0899). 
This object was included in the work by \cite{Connelley2008-Protostars}, which focused on Class I stellar sources.
However, due to the lack of spatial resolution and the large number of objects, we have not included the IRAS data
 (indicated in Fig.~\ref{SED_groupA1}) in any calculation.
In any case, the APEX/LABOCA source is also extended with a size of 24 arcsec.
The central source,  B30-LB19a,  is within the 5 arcsec peak and is a probable member.
It can also be classified as a very low-mass,  Class I  member based on the IRAC data and the near-IR photometry
(see also Fig. \ref{HR_TbolLbol_Evolution_A}).
It does  not have an optical counterpart and it is at the border line between brown dwarfs and stars. 
We estimate the  mass of the envelope as 0.116 or 0.182 $M_\odot$, 
depending whether we take the peak intensity or all the submm emission -- the total flux density, respectively.
The bolometric luminosity is 1.218 $L_\odot$ with  T$_{bol}$=238 K,  
whereas the internal luminosity can be estimated as 0.402 $L_\odot$ from the flux at 70 $\mu$m.
This source has been detected with ALMA and by APEX/SABOCA, see details in \cite{Huelamo2017-ALMA-B30}.
It can be classified as a YSO.
B30-LB19b is also within 5 arcsec of the nominal center of the LABOCA emission. 
Its photometric data indicate it is a very low-mass stellar object with IR excess. 

The MIPS M2 image also contains another source which nicely overlaps in near and mid-IR with B30-LB19d.
The SED of this object is consistent with a Class I object with T$_{bol}$=281 K and L$_{bol}$=0.172 $L_\odot$, in agreement with the IRAC data.
Adding the 870 $\mu$m flux (i.e., estimating a upper limit for the luminosity) would modify these values by about a 10\%.
We note however that the near-IR images reveal/suggest a high-inclined system with a bipolar nebula (\citealt{Huelamo2017-ALMA-B30}).
The L$_{int}$, as derived by the 70 $\mu$m flux, is 0.094 $L_\odot$.
Thus, it seems we are dealing with a bona fide VeLLO in what seems to be a wide multiple system.

Other interesting objects, based on their membership and SED, are \#e and \#f. Both have been classified as Class
I  and I/II based on the IRAC data and seem to be substellar. 
This fact reinforces the idea that we are dealing here with a probable young, multiple system
with very low individual masses.

\subsubsection{Group A2: LABOCA sources undetected at 24 $\mu$m\label{sub:groupA2notes}}

{\bf B30-LB08.-}
This source is extended, with an estimated size of 27 arcsec, with elongated  contours
(Fig. \ref{FCh_groupA2}).  
It has been detected by ALMA (\citealt{Huelamo2017-ALMA-B30}).
The SED corresponding to \#a, as displayed in Fig.~\ref{SED_groupA2}, is quite remarkable.
 A possible member of the association, 
 this object is barely within the central 5 arcsec and seems to be, in principle, substellar,
 has no excess in IRAC or MIPS/M1 (undetected)  and has a very strong excess in the 
far-IR, as measured with Akari/FIS and MIPS/M2.
However, there is an offset with the near-IR source (the M2 center is closer to the \#g counterpart),
although in part it might be due to the beam size and the positional uncertainty.
The alternative is that this contribution comes from another source.
 The envelope mass is 0.106 $M_\odot$ or 0.066 $M_\odot$ (total submm flux or the peak intensity, 
respectively).
 Based on these data, it would be a substellar  transitional object between embedded and 
Classical T-Tauri stages (a YSO).
However, the integrated bolometric luminosity for \#a, as derived from the SED and taking into account the far-IR fluxes,
is 12.007 $L_\odot$, and T$_{bol}$ reaches 50 K, although these values should be used with caution, based on the above discussion.
The internal luminosity, following the prescription by \cite{Dunham2008-Protostars}
and using the emission at 70 $\mu$m, is much smaller, L$_{int}$=1.223 $L_\odot$, but still much higher than typical values for VeLLOs.
 Other counterparts have been classified as non-members and have not been plotted on the figure.

 {\bf B30-LB23.-}
 The dust mass from the submm emission can be estimated as 0.046 $M_\odot$
 whereas the internal luminosity is 0.189  $L_\odot$.
 There is no optical or near-IR counterpart within 5 arcsec of the nominal center of the APEX/LABOCA beam.
 The closest is  B30-LB23a and neither it or any other possible counterpart  have a detected emission at 24 $\mu$m with MIPS.
 However,  another two sources, namely \#b and \#c, have been detected with WISE W4.
 In any event, the M2 detection is located between \#a and \#c and we have computed their bolometric luminosity and
 temperature with and without including the fluxes at 72 and 870 $\mu$m, yielding
 T$_{bol}$=312 K and L$_{bol}$=0.008 $L_\odot$, or
 T$_{bol}^{M2+870}$=62 K and L$_{bol}^{M2+870}$=0.160 $L_\odot$ for \#a;
 T$_{bol}$=290 K and L$_{bol}$=0.008 $L_\odot$, or
 T$_{bol}^{M2+870}$=62 K and L$_{bol}^{M2+870}$=0.161 $L_\odot$ for \#c.
 With these properties, they would be YSO of possible substellar nature (proto-BDs).

 In addition, counterpart \#b is characterized by 697 K and 0.006 $L_\odot$, assuming that the W4 measurement is real.
 These values are represented in Fig. \ref{HR_TbolLbol_Evolution_A}. 
As in the other counterparts, we may be dealing with a proto-BD, in any case a YSO.
 
  Other interesting counterparts, red and faint but not detected at mid-IR, are
  \#g and \#h.
 We note that the MIPS M1 image taken in 2005 has an artifact centered at counterpart \#f,
 as can be easily detected by comparison with the image obtained at the beginning of the mission one year earlier.

{\bf B30-LB25.-}
B30-LB25 is extended (34 arcsec)
with a total mass of the envelope of 0.0126 $M_\odot$ (0.076 $M_\odot$ for the peak emission).
It is  close to B30-LB20 and B30-LB26 (Fig.\ref{FOV_LABOCA}). 
To avoid ambiguity, the common sources have been assigned to B30-LB25.
Another nearby submm source is B30-LB24. 
The maximum emission in Spitzer/MIPS at 24 $\mu$m,  but is diffuse, and happens to be in between both APEX/LABOCA sources.

We have assigned the M2 flux to the component B30-LB25c,
which is located close to a source detected with the Akari satellite with FIS (Fig. \ref{FCh_groupA2}).
The integrated fluxes for \#c, including the 70 $\mu$m value, produce T$_{bol}$=69 K and L$_{bol}$=1.709 $L_\odot$.
Adding the flux at 870 $\mu$m  modifies  these values resulting into T$_{bol}^{70+870}$=56 K and L$_{bol}^{70+870}$=2.325 $L_\odot$.
The internal luminosity is 0.557 $L_\odot$.
In any event, B30-LB25c would be a YSO o Class 0 proto-star (Class I/II from the IRAC CCD, Fig. \ref{HR_TbolLbol_Evolution_A}). 

At the center of the B30-LB25 submm emission is \#a, likely substellar.
Moreover,  the counterpart   \#e is within the  central 5 arcsec, 
fainter than \#a and seems to be also a Class I object.
Thus, it is quite possible we have 
identified a visual very young brown dwarf binary with an angular
separation of 5.5 arcsec which might be (or not) bounded to a more massive component. 
Although it might be a projection effect, in principle the likelihood
is very small and more probable they are coming from the same clump. 

In addition, \#b resembles a massive BD with an excess at 8 $\mu$m. The counterpart \#f is interesting, but there is not enough
data to characterize it properly, except that the IRAC data indicates it is a class I/II object.

{\bf B30-LB29.-}
Counterpart \#a is a possible member and has an extended emission with MIPS at  24 $\mu$m. It also has been detected at 70 $\mu$m. 
It is within the central 5 arcsec of the submm peak.
We note, however, that  we might be dealing with a multiple system or at least with
several unevolved BDs which have been born inside the same clump.
Its bolometric temperature is 70 K, at the border between Class 0 and Class I (\citealt{Chen1995-Tbol-TaurusOph}),
and the integrated luminosity reaches 0.271 $L_\odot$, whereas
the internal luminosity, as derived from 
\cite{Dunham2008-Protostars}, is L$_{int}$=0.337 $L_\odot$.
This value is larger than L$_{bol}$ derived from the integration along the SED. This fact might be due to the way 
L$_{int}$ has been estimated. Firstable, L$_{int}$  is  affected by a factor two uncertainty (as derived by the
 spread in their Figure 4). On the other hand, our fluxes in B30 are below the lower limit in the
 case of the Taurus members used by \cite{Dunham2008-Protostars}
and this extrapolation might have a consequence on the derived L$_{int}$.
In any case, as in previous candidates, the mass from the envelope, 0.064 $M_\odot$, is inside the substellar 
domain. Our tentative classification corresponds to a YSO, perhaps in the substellar domain (a  massive proto-BD).
On the other hand, B30-LB29  \#b and \#f have near-IR excesses and can be classified as YSO.
Their bolometric temperature and luminosity are
533 K and 0.016  $L_\odot$ and
926 K and 0.010 $L_\odot$, respectively.
Counterpart \#g,  faint and detected only at $J$, I1, and I2, present increasing fluxes at longer wavelengths.
Another interesting object is \#i, substellar if member.
Again, we may be dealing with a multiple system or at least with
several unevolved BD, perhaps physically associated.

%
\subsection{Group B: LABOCA sources undetected at 70 $\mu$m\label{sub:groupB}}
%


\subsubsection{Group B1: LABOCA sources with detection at 24 $\mu$m\label{sub:groupB1notes}}


{\bf B30-LB24.-}
The MIPS M1 image (or the contours in the $J$ image, Fig. \ref{FCh_groupB1})
indicates that there is an extended emission at 24 $\mu$m.
Two sources are within it, B30-LB24 \#c and \#d. 
None of them is within the central 5 arcsec.
The first one is characterized by L$_{bol}$ $\le$ 0.555 $L_\odot$ and T$_{bol}$=71 K, whereas the other has
L$_{bol}$=0.099 $L_\odot$ and T$_{bol}$=254 K ($\le$0.531 $L_\odot$ and 88 K if the upper limit at 70 $\mu$m
is included). Both look like Class I stellar members (YSO).
The mass of the envelope is 0.044 $M_\odot$
with L$_{int}$ $\le$ 0.604 L$_{bol}$.
Other relevant possible members within the LABOCA beam are  \#a, \#e and \#g, especially this last one, whose SED
corresponds to a class I or II object, possibly of substellar nature.


\subsubsection{Group B2: LABOCA sources undetected at 24 $\mu$m\label{sub:groupB2notes}}


There are seven LABOCA sources without emission at 24 $\mu$m and detection limits at 70 $\mu$m, namely
  B30-LB01,
  B30-LB02,
  B30-LB09,
  B30-LB13,
  B30-LB15,
  B30-LB20 and
  B30-LB28.
  B30-LB01, B30-LB13 and B30-LB15 do not display any special behavior (apart of the substellar candidate B30-LB13b)
  and they will not be discussed in this subsection.
  However, in all these three cases the optical
  or near-IR counterpart are farther than 5 arcsec from the center of the LABOCA emission and this could indicate 
  they are, in fact, starless cores (subsection \ref{sub:StarlessCores}). 

{\bf B30-LB02.-}
The counterpart \#a only has been detected with IRAC I1 and I2, being the flux at 4.5 $\mu$m stronger than at 3.6 $\mu$m, although close to the detection limits.
It is outside the central 5 arcsec.
The mass of the submm envelope can be estimated as 0.099 $M_\odot$
and the internal luminosity as  L$_{int}$ $<$ 0.230 L$_{bol}$. Thus, this source is an interesting candidate since, if member,
it would be substellar.
Counterpart \#d has only been detected with IRAC and its colors correspond to a Class I object (Fig. \ref{fig_CCD_CMD_IRAC}),
but the SED suggests it is a Class II BD, if membership is confirmed.

{\bf B30-LB09.-}
There are several probable and possible members within the B30-LB09 beam (Fig.~\ref{FCh_groupB2}). 
All three seem to be substellar (Fig.~\ref{SED_groupB2}) and,
in fact, \#a is located within the submm peak and has not  been detected with MIPS either at 24 $\mu$m or at 70 $\mu$m. 
The upper limit at 70 $\mu$m an internal luminosity of L$_{int}$ $\le$ 0.422 L$_{bol}$.
Therefore, we have identified it as
the origin of the LABOCA emission and it is another excellent
brown dwarf candidate with a significant submm emission and an envelope mass of 0.043 $M_\odot$ (subsection \ref{sub:EnvelopeMass}).
We note, however, that the near-IR slope in the SED is not rising.
Apart from the central source, there are two probable members within the B30-LB09 beam, namely \#c and \#e
 (Fig.~\ref{FCh_groupB2}).  Both  seem to be substellar.

{\bf B30-LB20.-}
This LABOCA source has two interesting counterparts, identified as \#a and \#d.
Both are far from the submm peak  (Fig. \ref{FCh_groupB2}).
Moreover, they have been detected  with WISE W3 and W4, but only \#d has a possible, extended flux with MIPS M1.
Most likely, both are class II stars.
The total mass of the envelope is 0.048 $M_\odot$ and the 
internal luminosity is L$_{int}$ $\le$ 0.592 $L_\odot$.

In addition, there are another two interesting objects, namely \#b and \#g, which could be substellar. They only have
three data-points in the SED, but the shape suggests  they might be Class II or even Class I.
We note that B30-LB20c is DM142, but it does no show any IR excess.
Moreover, \cite{Huelamo2017-ALMA-B30} have detected a 350 $\mu$m source with APEX/SABOCA within the LABOCA beam,
but it does not have any optical or IR counterpart.

{\bf B30-LB28.-}
The envelope mass for this submm source is 0.071 $M_\odot$.
Although the WISE fluxes measured for components \#b and \#c are uncertain, both have dubious emission at 24 $\mu$m
and their SED indicates they are Class I or even Class 0 (the shortest wavelength they display correspond
to WISE W1).
Thus, they are potential proto-BD candidates but additional data are required.
In any case, there is no optical or IR source within the central 5 arcsec and the LABOCA source is elongated in the north-east direction, toward the very intense submm source B30-LB27.

Counterpart \#d has been classified as I/II from the IRAC CCD (Fig.  \ref{fig_CCD_CMD_IRAC})
and the complete SED agrees with this classification. If member, it would be substellar.
This is the case of \#a, but contrary to \#d, there is no IR excesses. We note, however, that there is submm emission
as seen by APEX/SABOCA at 350 $\mu$m (\citealt{Huelamo2017-ALMA-B30}), However, because the shape of the SED
(Fig.\ref{SED_groupB2}), we believe that they are unrelated
although we cannot discard completely that B30-LB28a has a submm excess.

%
\subsection{Group C: LABOCA sources with no data at 70 $\mu$m\label{sub:groupC}}
%


\subsubsection{Group C1: LABOCA sources detected at 24 $\mu$m\label{sub:groupC1notes}}


{\bf B30-LB11.-}
The counterpart \#a, detected with MIPS M1, is quite distant from the peak of the submm emission
(Fig. \ref{FCh_groupC1}).
The integrated properties are T$_{bol}$=145 K,  L$_{bol}$=0.028 L$_{bol}$ and an envelope mass of 0.046 $M_\odot$.
Thus, is seems it could be a Class I object in the substellar domain (a proto-BD candidate).

Moreover, counterpart \#c displays an interesting SED with a significant flux at 8 $\mu$m.
If member, it  would also be substellar. Other interesting sources are \#b and \#e, since there are also red and faint.

{\bf B30-LB12.-}
From the flux at 870 $\mu$m we derive an envelope mass of 0.087 $M_\odot$.
Counterparts \#a and \#d have been detected at 24 $\mu$m. The first one is the probable origin of the submm emission,
since it is almost within the central 5 arcsec, and
has L$_{bol}$=0.418 $L_\odot$ and T$_{bol}$=2108 K. The IRAC data indicate it is a class III star and the SED shows that
the excess is only at 24 $\mu$m (i.e., a WTT).

More interesting is \#d, with
L$_{bol}$=0.058 $L_\odot$ and T$_{bol}$=174 K.
If the 870 $\mu$m  flux is considered, the values change to
L$_{bol}^{870}$=0.094 $L_\odot$ and T$_{bol}^{870}$=138 K, although it is at the edge of the submm beam.
In any case, its properties make it a substellar Class I object (proto-BD candidate).

Other interesting possible or probable members are \#c, \#g and \#j, specially the first one,
very faint and classified as Class I based on the IRAC colors. In other words, another proto-BD candidate.

{\bf B30-LB18.-}
Counterpart \#a is not within the submm peak. It is a class III star (possibly a WTT) with 
L$_{bol}$=0.249 L$_{bol}$, T$_{bol}$=1748 K and an excess only at 24 $\mu$m, being the mass for the dust envelope, as
estimated with the flux at 870 $\mu$m,
0.059 $M_\odot$.

The counterparts \#b and \#c might be of substellar nature.
The first one, close to \#a,  presents increasing fluxes up to I2, suggesting
it might be a class I BD associated to \#a (i.e., a proto-BD candidate y physical association to a more evolved low-mass star).

{\bf B30-LB22.-}
Counterpart \#a is located very close to the nominal center of the submm, it is a probable member and has a
 peculiar SED (although  with only four points, Fig.~\ref{SED_groupC1}), indicating it is a
 Class I source of the substellar nature, with an envelope  mass  of 0.088 $M_\odot$. 
 However, although \#a is not detected at 24 $\mu$m, there  is a SABOCA source at 350 $\mu$m (\citealt{Huelamo2017-ALMA-B30}).
 In addition, there is a source at this wavelength west of it, 
halfway toward component \#b, which seems to the  origin of it.

In the case of the counterpart B30-LB22b, Without taking into account the 870 $\mu$m flux,
 we derive L$_{bol}$=0.016 $L_\odot$ and 
T$_{bol}$=255 K. 
When including the flux at 870 $\mu$m the estimates
are  L$_{bol}^{870}$=0.064 $L_\odot$ and  T$_{bol}^{870}$=82 K. 
In either case, a Class I YSO of possible substellar nature.
Counterpart \#j is also detected at 24 $\mu$m and has 243 K and 0.008  $L_\odot$, possibly a Class I BD (proto-BD candidate).

The LABOCA source  includes  another possible embedded object, namely \#h. 
Finally, there are other two probable members: \#c and \#g.
In addition, there is some overlap with B30-LB23 (Fig. \ref{FOV_LABOCA}), but there are not common optical or IR sources.
B30-LB22f is also a submm source at 350 $\mu$m (\citealt{Huelamo2017-ALMA-B30}), but it does not seem to belong to the
association and might be extra-galactic.

{\bf B30-LB30.-}
The submm emission is extended and the derived mass for the envelop is
0.123 $M_\odot$ (0.101 $M_\odot$ at peak intensity).
There is no counterpart within the submm peak, but nearby there are two objects detected either with
MIPS M1 (\#g) or W4 (\#a).
We have derived
0.059 $L_\odot$ and 354 K for B30-LB30g without taking into account the 870 $\mu$m flux.
With it, the bolometric luminosity and temperature for each of them would be
L$_{bol}^{870}$=0.103 $L_\odot$ and T$_{bol}^{870}$=239 K.
In any case, it seems we are dealing with very low-mass Class I object (YSO), perhaps of the substellar nature.
Counterpart \#b lacks excess and might be substellar. Finally, \#c has been classified, based on the IR data,
as class I/II BD (Fig. \ref{fig_CCD_CMD_IRAC}).

{\bf B30-LB31.-}
This LABOCA source  has also been detected by ALMA (\citealt{Huelamo2017-ALMA-B30}).
The mass of the envelope is 0.082 $M_\odot$.
Counterpart \#a is almost within the 5 arcsec beam of the submm source and presents a possible extended emission at 24 $\mu$m.

In addition, LB30-LB31c has been detected at 24 $\mu$m and
its properties are L$_{bol}$=0.048 $L_\odot$ and T$_{bol}$=264 K.
Adding up the 870 $\mu$m would change these values to
L$_{bol}^{870}$=0.074 $L_\odot$ and T$_{bol}^{870}$=190 K.
Then, it is  possible Class I proto-BD candidate.
Other relevant objects, because their faintest and red SEDs, are \#h, \#e and, especially because of the
increased fluxes toward longer wavelength, \#d.

{\bf B30-LB32.-}
LB32a can be identified with the source IRAS\,05293$+$1207 (RA=83.0450, DEC=+12.1629).
It was only detected at 60 $\mu$m and the flux, $[60]$=1.15e+00$\pm$0.1725 Jy,
agrees very well with the current SED (Fig.~\ref{SED_groupC1}) and the fluxes detected
with Akari/FIS.
It is outside the M2 FOV but it is detected with MIPS/M1.
The integrated bolometric luminosity is 0.657 $L_\odot$ with T$_{bol}$=414 K, and the mass of the envelope is
0.123 $M_\odot$.
 In any event, we are dealing with a Class I stellar member of the association (a YSO), although the
small dip observed at 4.5 $\mu$m might indicate it has already started to lose part of the envelope.

{\bf B30-LB33 and LB34.-}
This double peak submm source  
is located at the 
Western border of our LABOCA map and outside the clustering 
of submm sources (or the dust structure seen with IRAC). 
The individual values for the masses of each envelope are 0.087 $M_\odot$ and  0.064 $M_\odot$, respectively.
There is only one optical and near-IR counterpart located closer to B30-LB33
 (counterpart \#a, Fig.~\ref{FCh_groupC1}), both identified previously by 
\cite{Dolan1999.1, Dolan2002.1} and by \cite{Duerr1982-Orion-SFR}. 
It is GY Ori, identified by those works as DM115 and DIL19, respectively.
There is no MIPS/M2 data but it has been has been detected at  24 $\mu$m.
Thus, B30-LB33a has  T$_{bol}$=1920 K and L$_{bol}$=2.079 $L_\odot$.
We have classified  B30-LB33a   as a Class~II stellar member.

\subsubsection{Group C2: LABOCA sources undetected at 24 $\mu$m\label{sub:groupC2notes}}

{\bf B30-LB03 and LB04.-}
The sources B30-LB03 and B30-LB04 are very close to each other, as shown in 
Fig.~\ref{FCh_groupC2}. While B30-LB04 does not show any counterpart within 5~arcsec, 
B30-LB03 shows one source, a YSO. 

The SED of central source of  B30-LB03, \#a for short,
together with the previous analysis based on CC and CM diagrams,
indicates its is a possible member with substellar nature, but without IR excesses up to 3.6 $\mu$m. 
Some excess might be at 4.5 $\mu$m, although it is unclear. 
The center of the LABOCA source is outside the MIPS/M2 images (due to the malfunctioning of one of the detectors), but the counterpart 
\#a is not detectable at 24 $\mu$m.
Thus, it is not completely clear whether this object is the origin of the submm emission.
In any case, the mass of the envelope can be estimated as 0.057 $M_\odot$ (subsection \ref{sub:EnvelopeMass}),
 well below  the stellar/substellar limit.
 
Source  \#b, located close to the B30-LB04 peak, is a probable substellar 
member without any detectable excess. 
The mass of the envelope is 0.059  $M_\odot$.

In addition, we have classified as cluster candidates \#d and \#c --possible, and \#e, \#i,  and \#f --probable members.
This last  one is a very red object with a step slope which can be classified as a Class I source based on the IRAC CCD.
Therefore, we are dealing with a very low luminosity object (in the optical to mid-IR range, 
lower than the characteristic
luminosity of Class II or III brown dwarfs) and a quite extended emission in the submm, 
which might not be directly associated to it, although the visual inspection indicates that B30-LB03f 
is in the middle of a filament like asymmetric structure about 60$\times$30 arcsec.

{\bf B30-LB05.-}

This is an extended submm source with an estimated size,
 after deconvolving with a Gaussian to take into account 
the core, of 36 arcsec.
The mass of the envelope can be estimated as 0.081  $M_\odot$ (full beam) or  0.051  $M_\odot$ (peak).

B30-LB05a is located within 5 arcsec of the submm peak intensity
and it is not detected at 24 $\mu$m, as measured with Spitzer/MIPS,
but it is outside the M2 FOV. 
This source seems to be the origin of the submm excess.
We note, however, that \#a is a Class III star based on its IRAC data.

In addition,  B30-LB05c, at $\sim$25 arcsec from the main source (Fig.~\ref{FCh_groupC2}),
 seems to be a bona-fide, substellar member.
We note that both optical/IR counterparts (\#a and \#c)  are within the APEX/LABOCA beam.

{\bf B30-LB07.- }
Counterpart \#a, located close to the center, might be the origin of the submm emission (Fig.~\ref{FCh_groupC2}). 
The SED suggests it is a substellar object, if it is indeed a member of B30 (its membership status is Y?,
 although it is based on only three data-points).
The total mass of the submm envelope  (subsection \ref{sub:EnvelopeMass}) is 0.052 $M_\odot$, which provides a 
stronger hint regarding its proto-BD status. However, there is no information regarding emission at 70 $\mu$m 
and it lacks emission at 24 $\mu$m, which casts some doubts about the actual connection between the near-IR and the submm sources.

 B30-LB07e is also an interesting object, since it is a probable member whose luminosity indicates it is substellar. 
Other interesting sources are \#c and \#i, which have a relatively flat spectrum 
and have been classified as possible members. In addition, B30-LB07f has been detected in only two bands, so no membership
status has been assigned, but its SED slope is positive.

{\bf B30-LB10.-}
As can be seen in Fig.~\ref{FOV_LABOCA}, 
B30-LB11 is located southeast of B30-LB10 and overlaps with it. All sources in the common area have been assigned to
B30-LB10.
It has been detected by ALMA (\citealt{Huelamo2017-ALMA-B30}).
The source \#a  is just at the center, with a weak mid-IR excess as derived with IRAC.
There is no M2 flux (outside the FOV) and it is not detected with M1 at 24 $\mu$m.
It is a possible member and, if membership is confirmed,
it would be substellar.
The envelope mass has been estimated as 0.046 $M_\odot$. 
 
 This submm source includes eight probable  (\#c, \#e, \#g and \#m) and possible members (\#a, \#h, \#i and \#k, 
see Fig.~\ref{FCh_groupC2}). Some of them might be substellar, but 
they display no special feature in the SED.

 {\bf B30-LB17.-}
 Although the closest counterpart to the submm peak is  \#a,
 \#d shows a significant IR excess, including the emission at 12 and 22 $\mu$m (WISE W3 and W4, with extended with MIPS M1 and no data for M2), as can be seen 
 in the SED included in Fig.~\ref{SED_groupC1}.
 Both seem to be below the substellar limit.
The envelope mass is about half of the limit between stars and brown dwarfs, with a value
of  0.059 $M_\odot$ (subsection \ref{sub:EnvelopeMass}).
Thus, B30-LB17a is a BD candidate  and  B30-LB17d might be a proto-BD candidate.
Figure~\ref{FCh_groupC1}  includes another interesting substellar candidate:
\#b is a probable member without IR excess.

{\bf B30-LB21.-}
Counterpart \#a, located at the submm  peak and with a dubious extended emission at 24 $\mu$m,  is possibly a Class I object. This source is also extended and the
Gaussian fit gives an angular size of 62 arcsec, the largest among all our APEX/LABOCA sources.
We note that this component is located in the ionizing edge
 produced by the star $\lambda$ Ori (Fig.\ref{FOV_LABOCA}),  as projected on the sky, 
and is also very close to B30-LB22.
The mass of the envelope can be estimated as 0.108 $M_\odot$
(0.061 $M_\odot$ just for the peak, see detail in subsection \ref{sub:EnvelopeMass}) 
 and based on the SED it might be substellar, making it a 
 candidate as an irradiated brown dwarf embryo, although a more detail analysis is required.

LB21f and B30-LB21g are two probable members of substellar nature. However, they do not show IR excesses, 
 The SED of counterpart \#c is similar to \#f and \#g, except that it
 has been detected with WISE W4 although the measured flux seems dubious,
since might be affected by source confusion.
Other possible members are   \#d and \#h.

{\bf B30-LB27.-}
No object is within the central 5 arcsec of the LABOCA peak.
The brightest counterpart in the optical/near-IR within the B30-LB27 beam is \#a,
but it seems to be a Class II BD or very low-mass star,
if member, detected with WISE W3 and W4 but with no measurement at MIPS M1
(L$_{bol}$=0.014 $L_\odot$ T$_{bol}$=988 K).

The flux at 870 $\mu$m has been assigned to \#b, which has a possible extended emission at 24 $\mu$m.
We note that we have also assigned to this source the SABOCA flux at 350 $\mu$m, although
it is approximately NW from it (elongated in the same direction, (\citealt{Huelamo2017-ALMA-B30}).

There are two other interesting objects:
components  \#c and \#f.
Both have been detected with WISE W3 and W4 and in the case of \#f, there is also a possible detection at 24 $\mu$m.
Since the mass of the envelope is 
0.279 $M_\odot$ (0.150 $M_\odot$ at peak intensity)
and these three objects have low values of the bolometric temperature and luminosity,
regardless the assignment of the submm flux, they look like bona fide BDs candidates.

\begin{figure*}  
\center
%
\includegraphics[width=0.450\textwidth,scale=0.330]{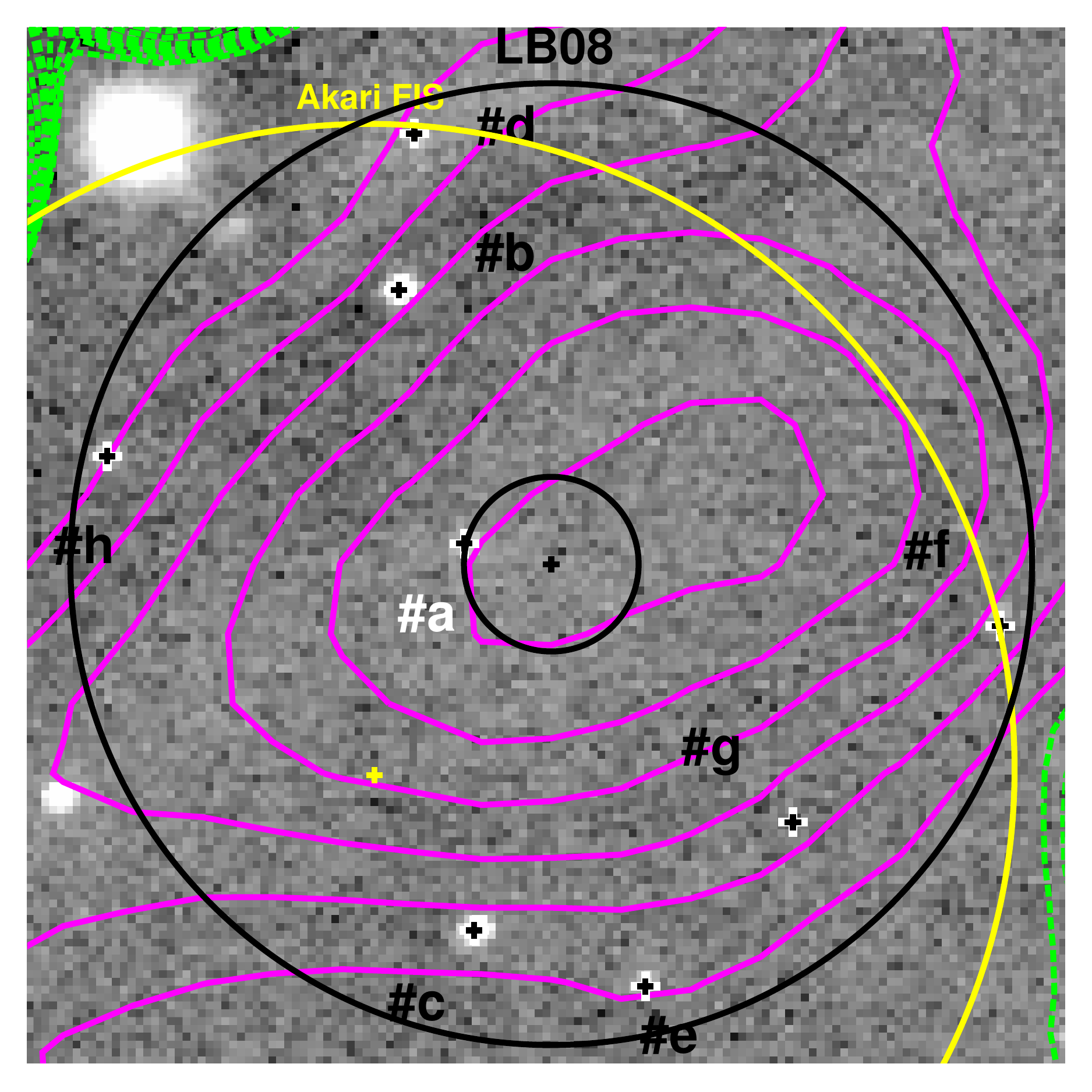}
\includegraphics[width=0.450\textwidth,scale=0.330]{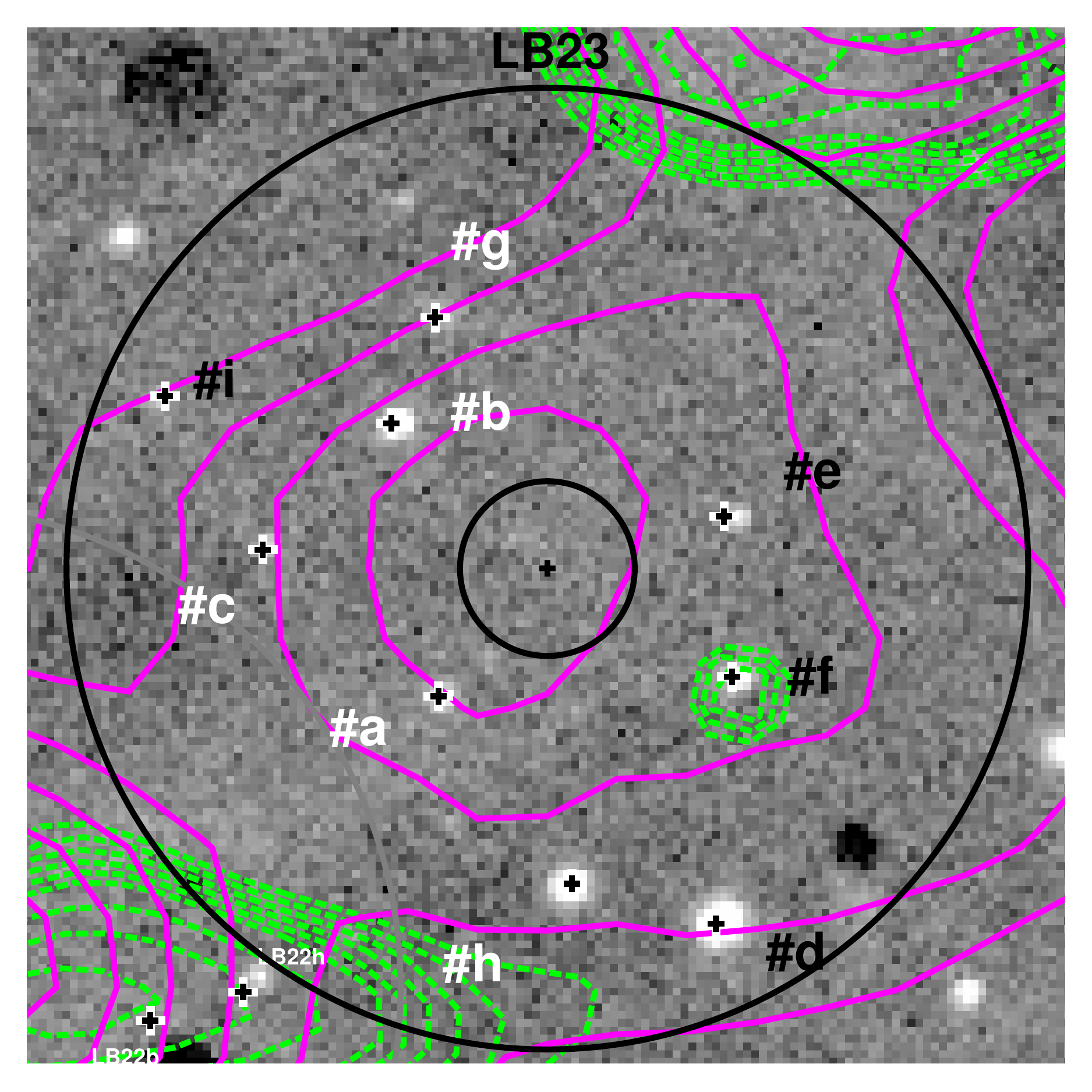}
\includegraphics[width=0.450\textwidth,scale=0.330]{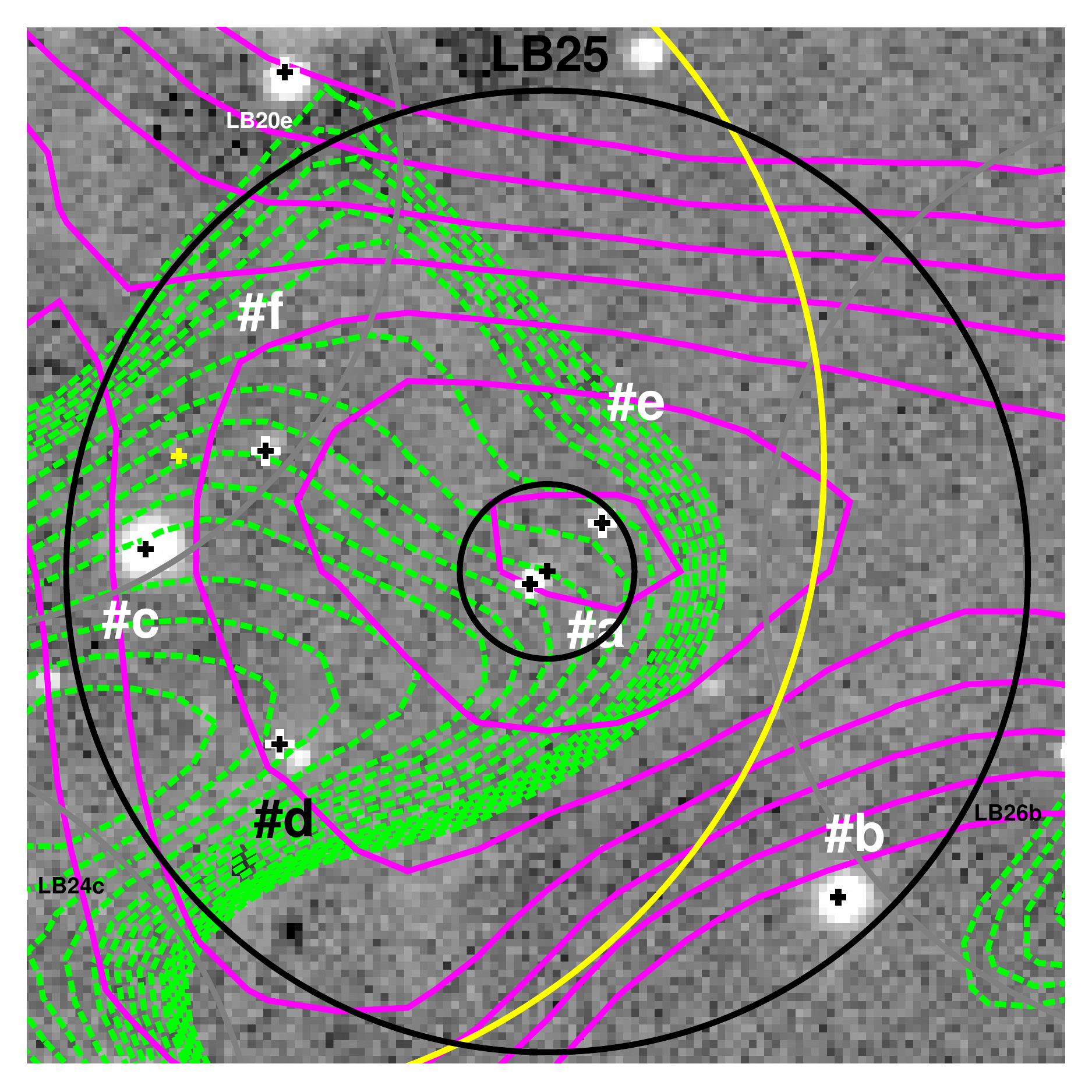} 
\includegraphics[width=0.450\textwidth,scale=0.330]{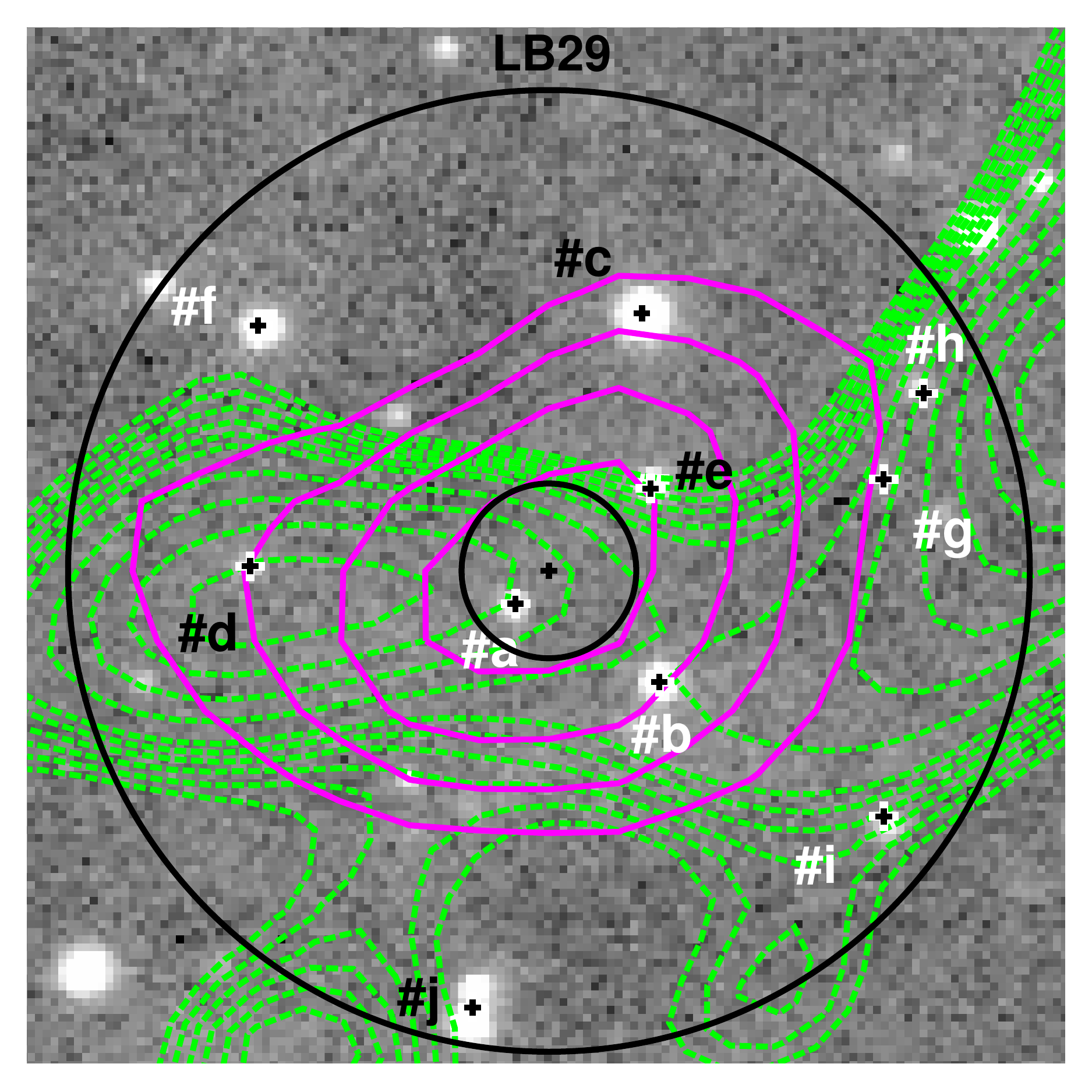}
\caption{\label{FCh_groupA2} 
Finding charts for the relevant sources detected also  at 70 and undetected at 24 $\mu$m (A2 group).
Symbols are in Fig. \ref{FCh_groupA1}.
We note that the feature displayed at 24 $\mu$m for B30-LB23f corresponds to an artifact.
}
\end{figure*}

\begin{figure*}   
\center
%
\includegraphics[width=0.450\textwidth,scale=0.33]{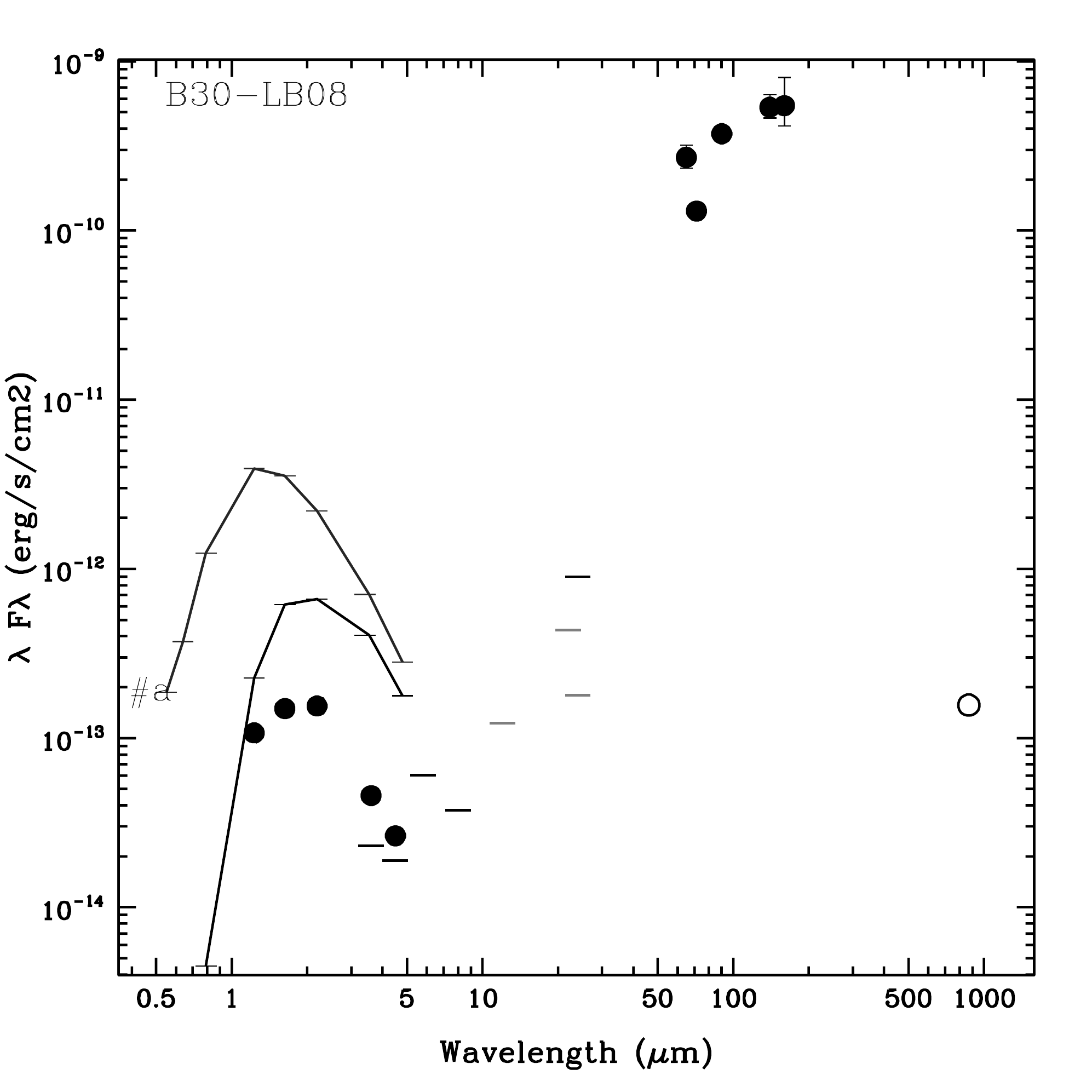} 
\includegraphics[width=0.450\textwidth,scale=0.33]{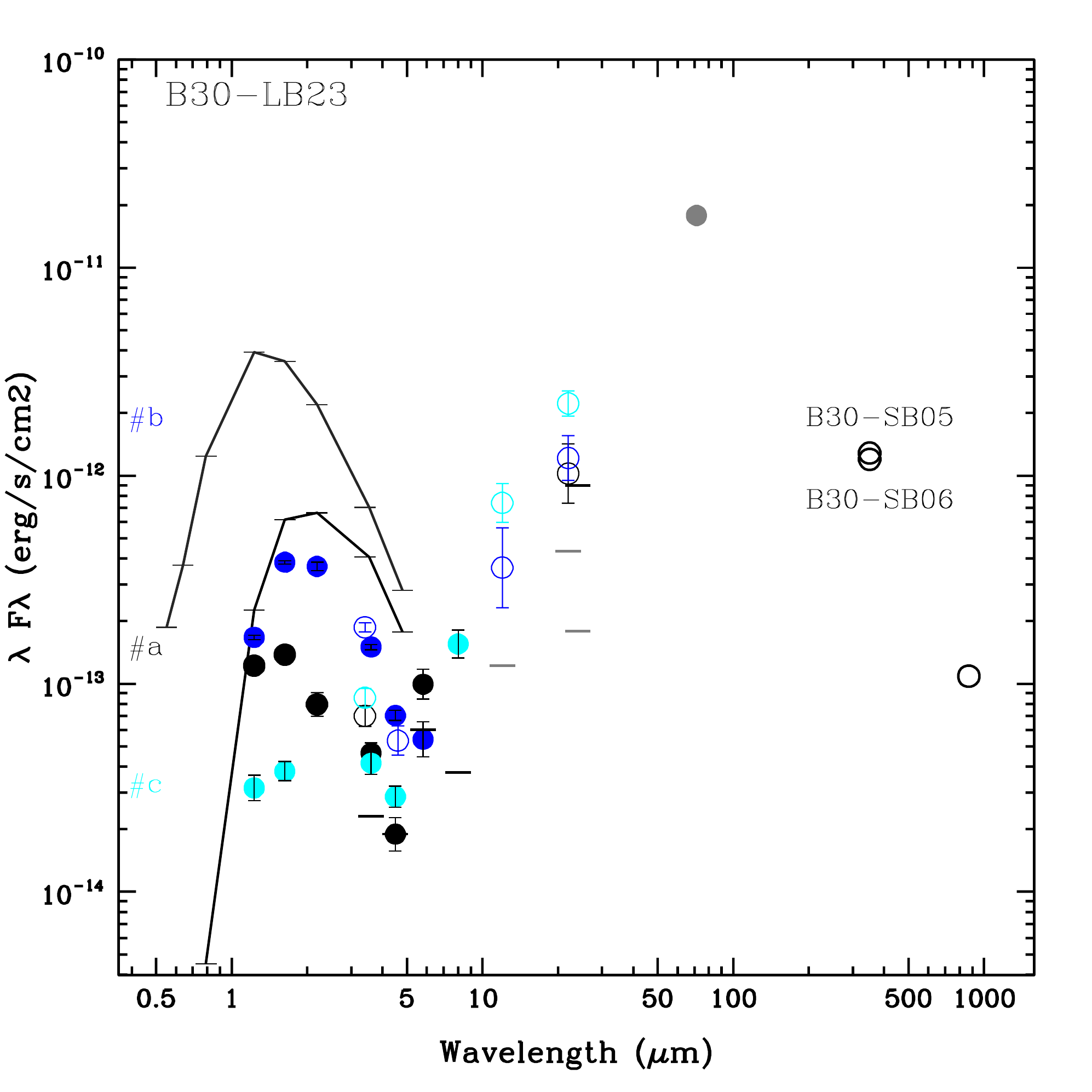} 
\includegraphics[width=0.450\textwidth,scale=0.33]{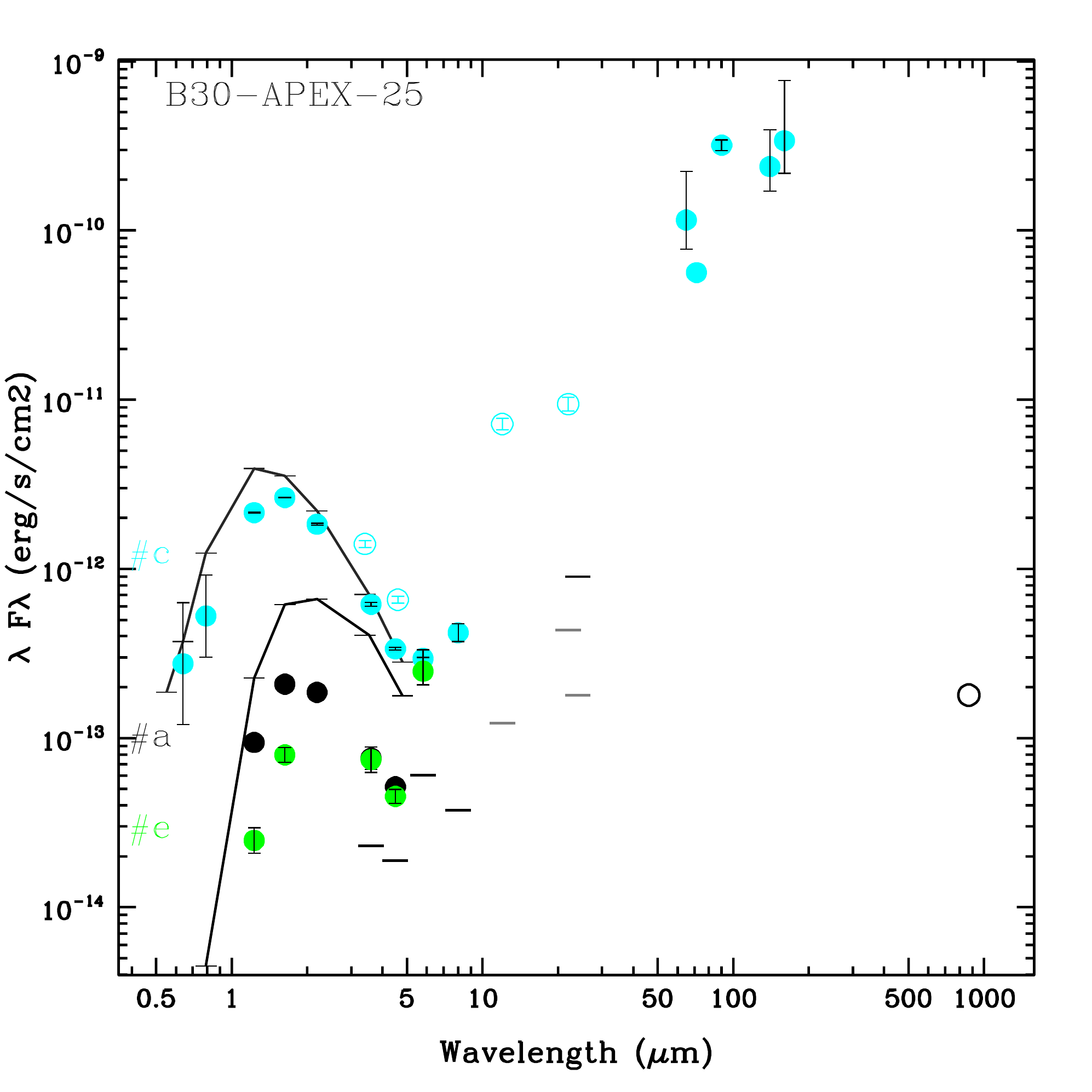} 
\includegraphics[width=0.450\textwidth,scale=0.33]{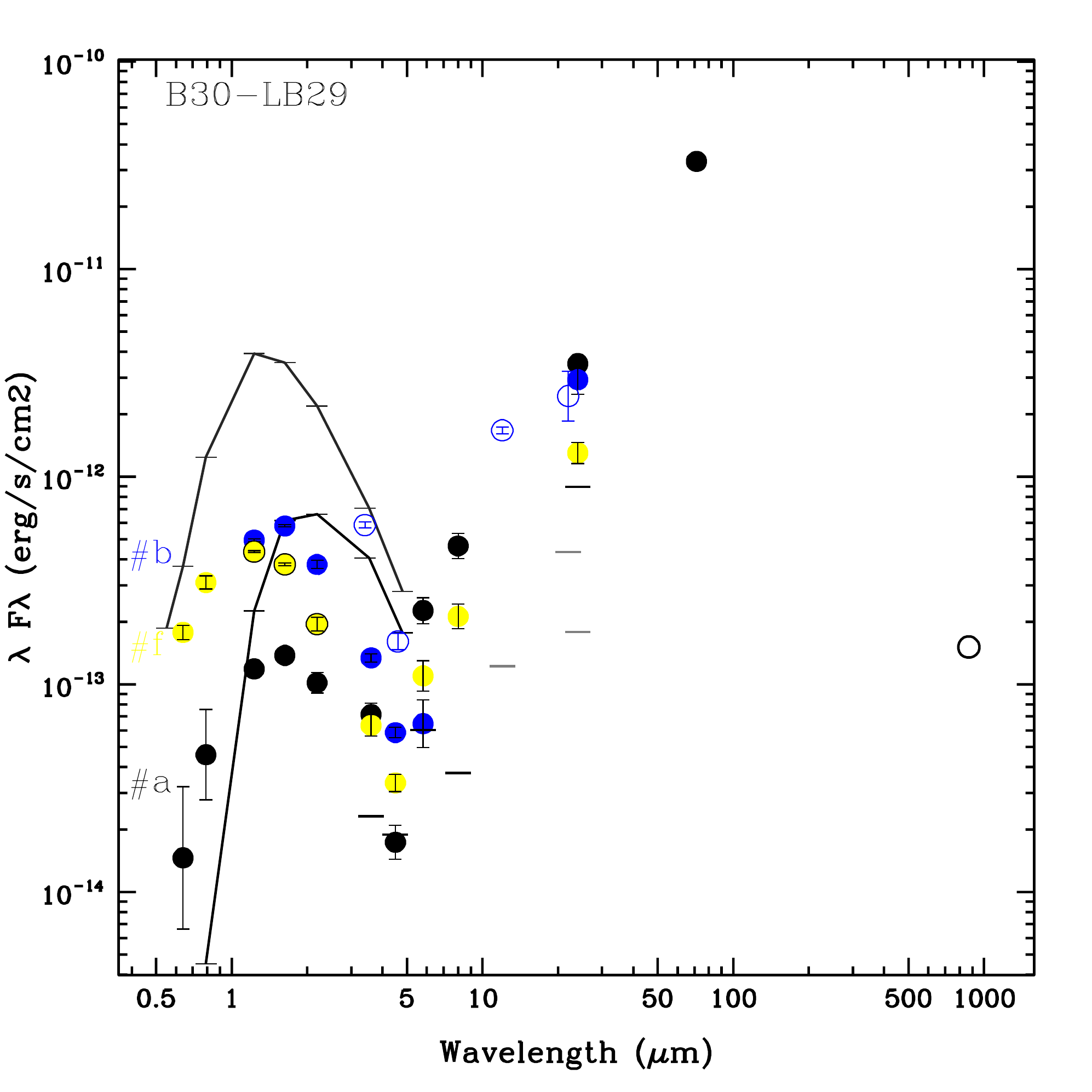} 
\caption{\label{SED_groupA2} 
  Spectral energy distribution for the relevant sources
  with emission at 70 $\mu$m but undetected at 24 $\mu$m (A2 group). We note that several counterparts for B30-LB29 have a possible extended emission at 24 $\mu$m.
Symbols as in Fig. \ref{SED_groupA1}.
}
\end{figure*}

\begin{figure*}  
\center
%
\includegraphics[width=0.45\textwidth,scale=0.33]{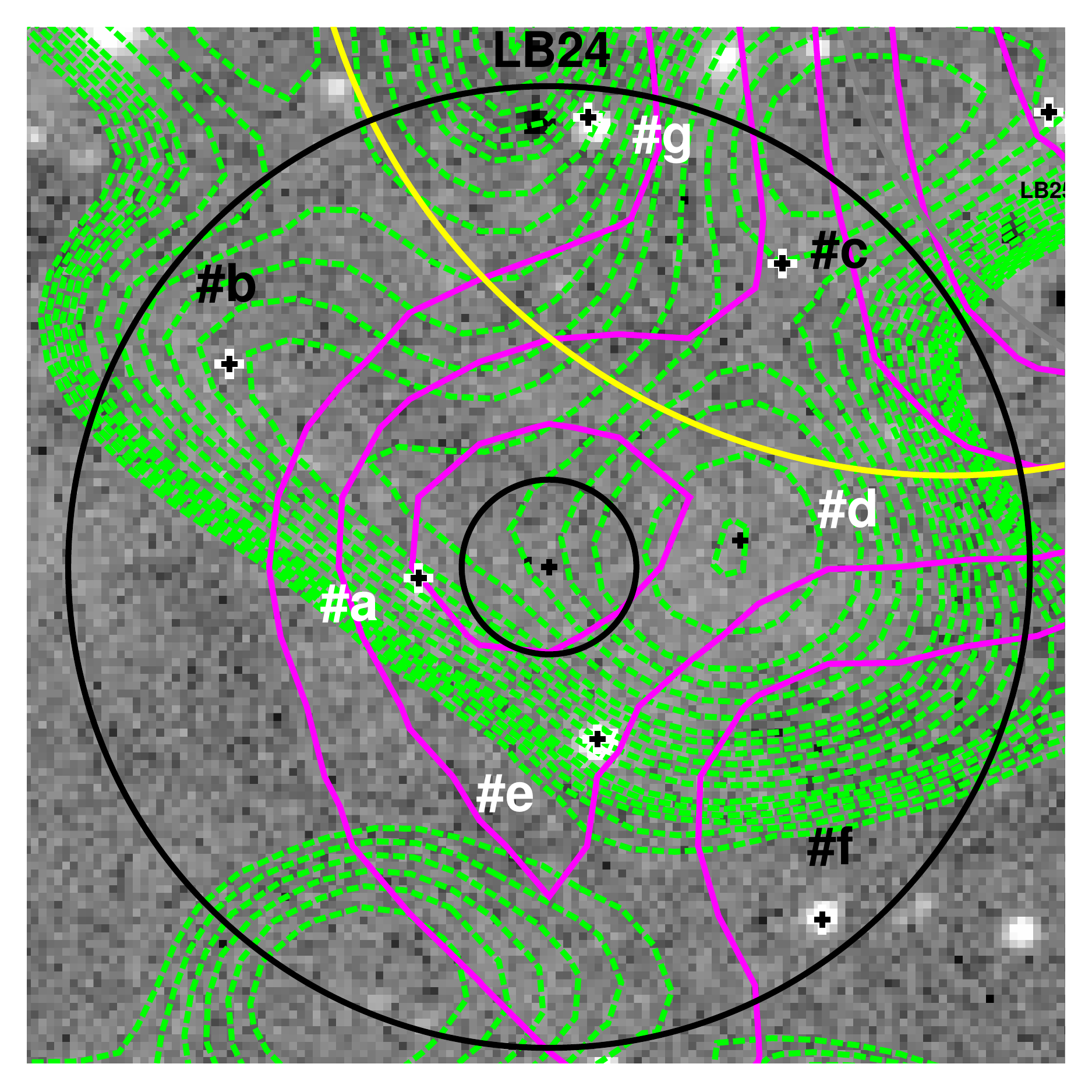}
\includegraphics[width=0.45\textwidth,scale=0.33]{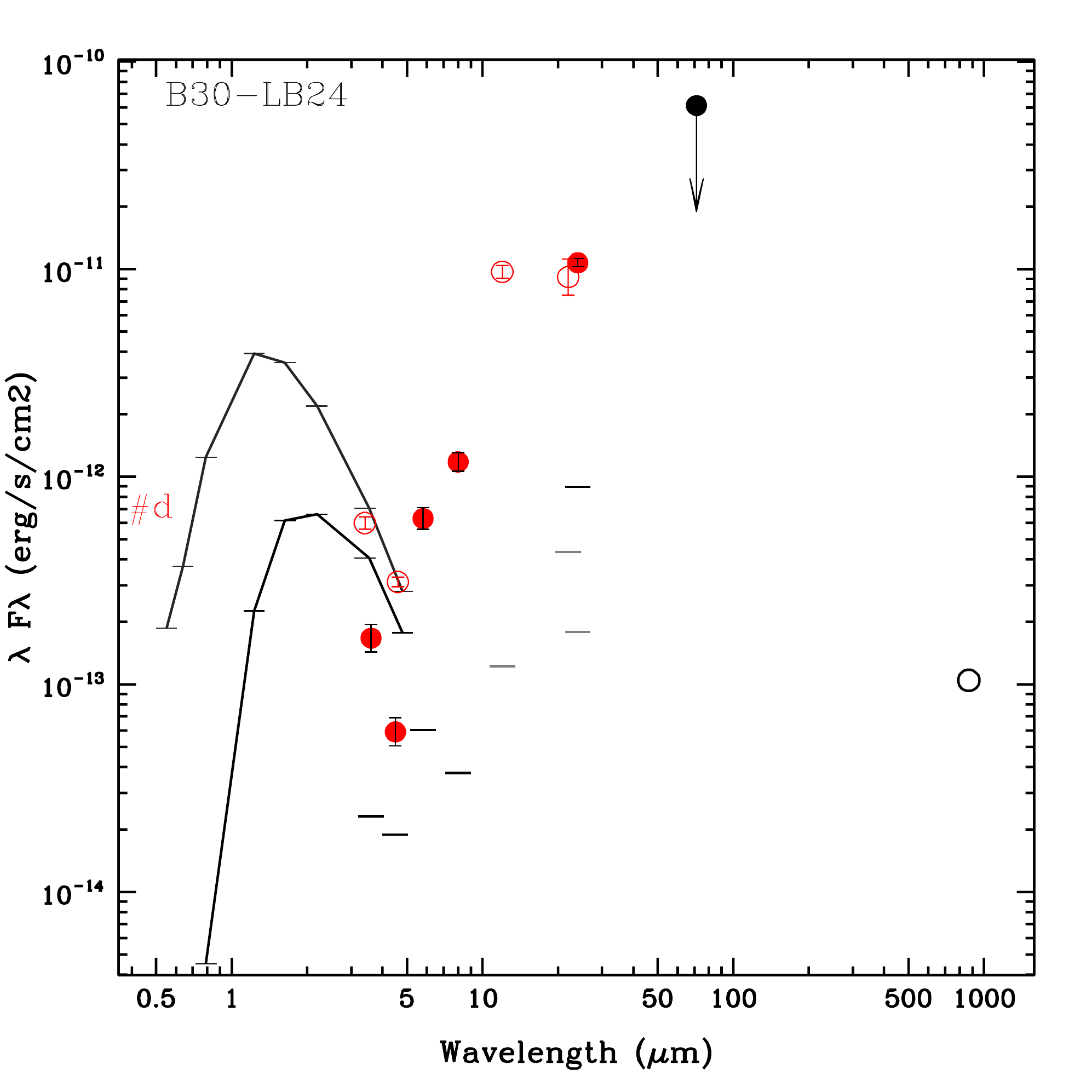} 
\caption{\label{FCh_groupB1} \label{SED_groupB1} \label{FCh_SED_groupB1} 
  Finding chart and   spectral energy distribution for B30-LB24, the only member within the  B1 group (upper limit at 70 $\mu$m  emission at 24 $\mu$m).
Symbols are as in Fig. \ref{FCh_groupA1} and  Fig. \ref{SED_groupA1}.
}
\end{figure*}

\begin{figure*}  
\center
%
\includegraphics[width=0.32\textwidth,scale=0.33]{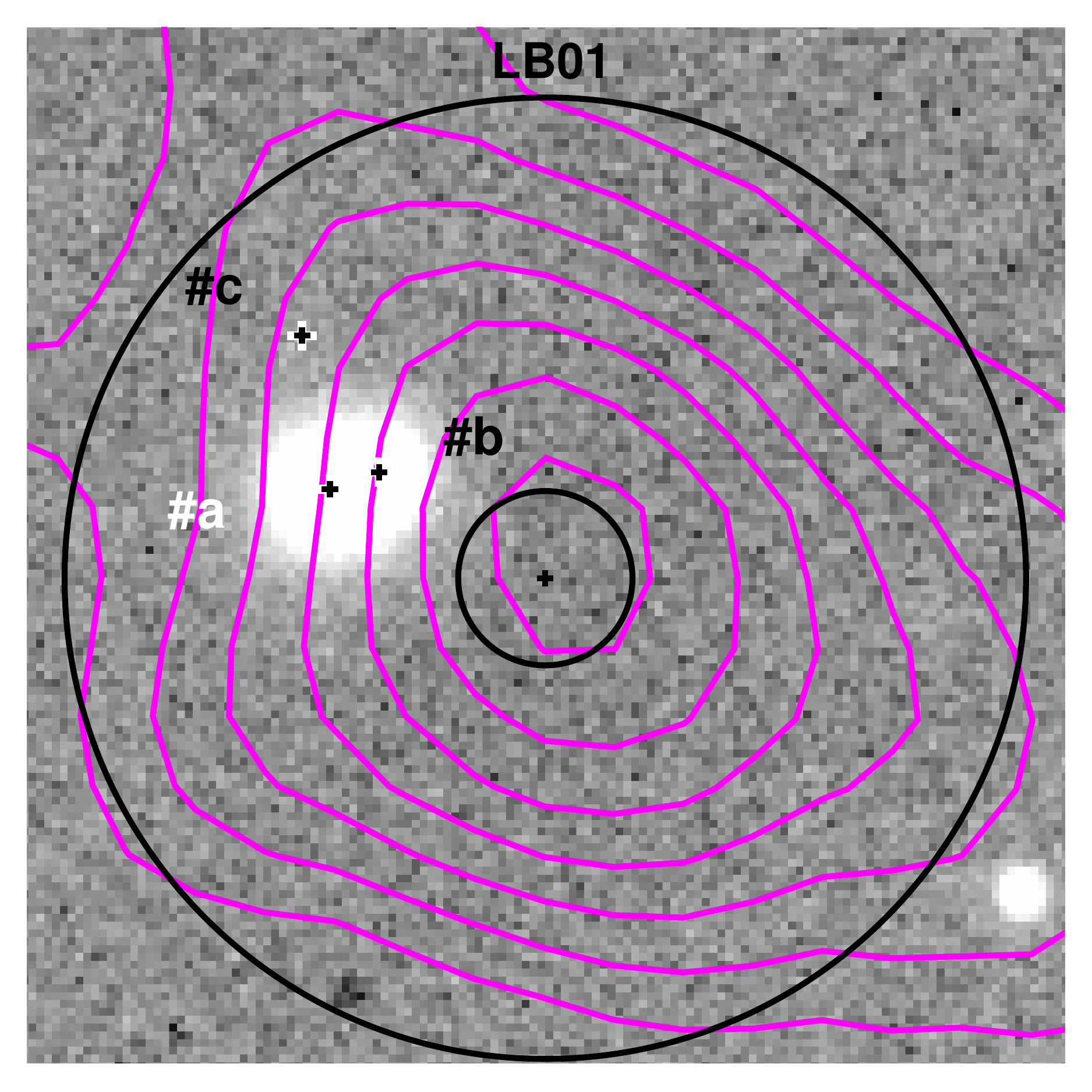}
\includegraphics[width=0.32\textwidth,scale=0.33]{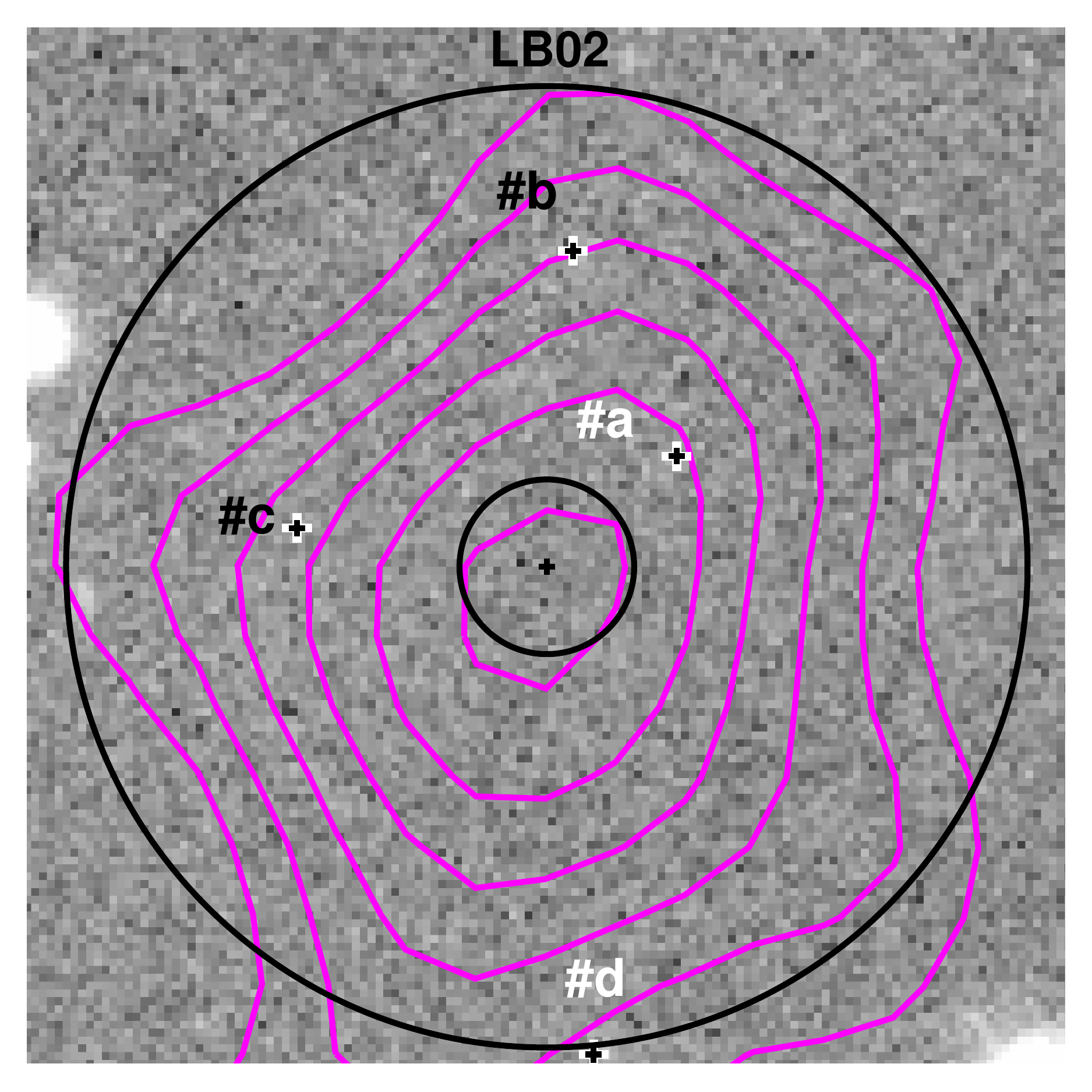}
\includegraphics[width=0.32\textwidth,scale=0.33]{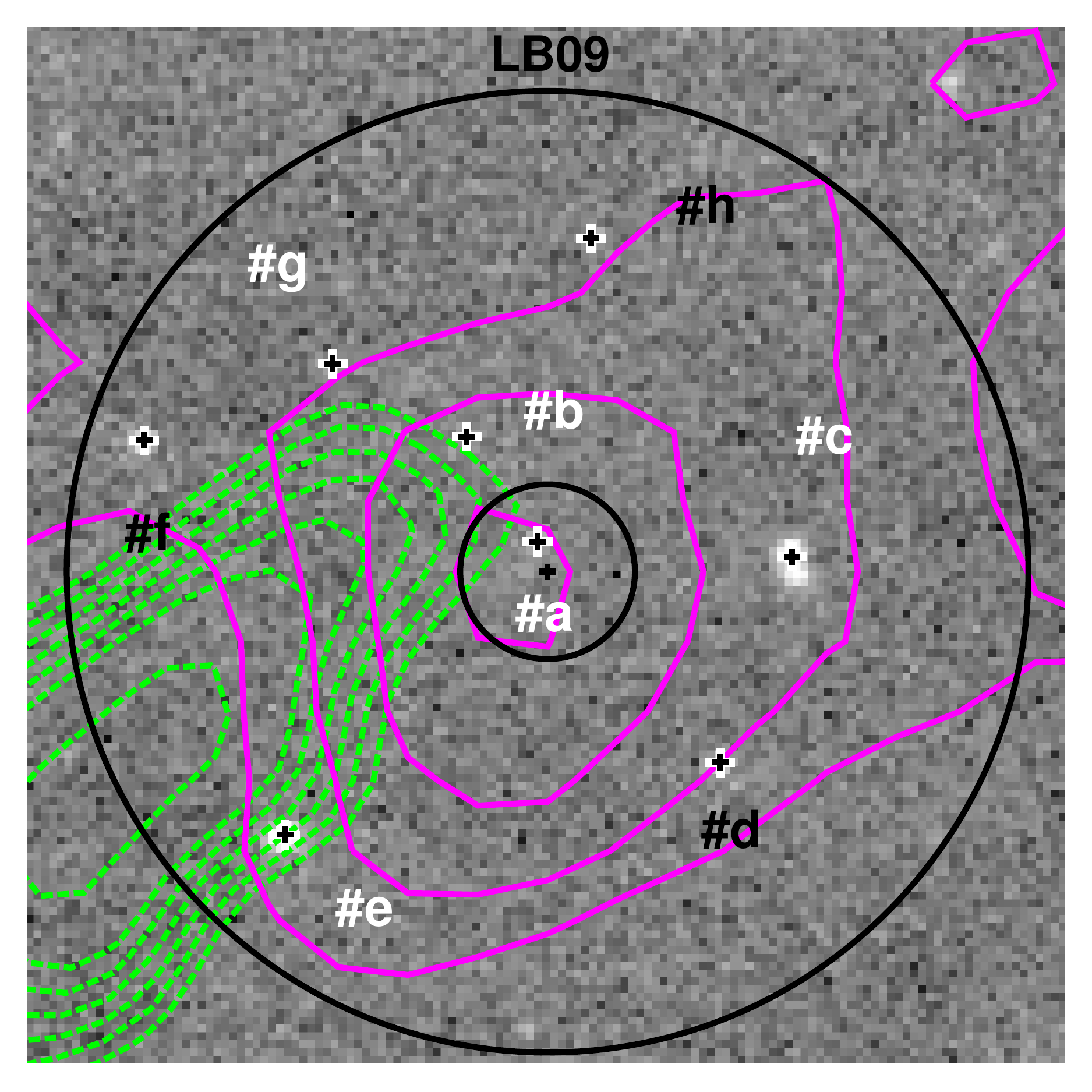}
\includegraphics[width=0.32\textwidth,scale=0.33]{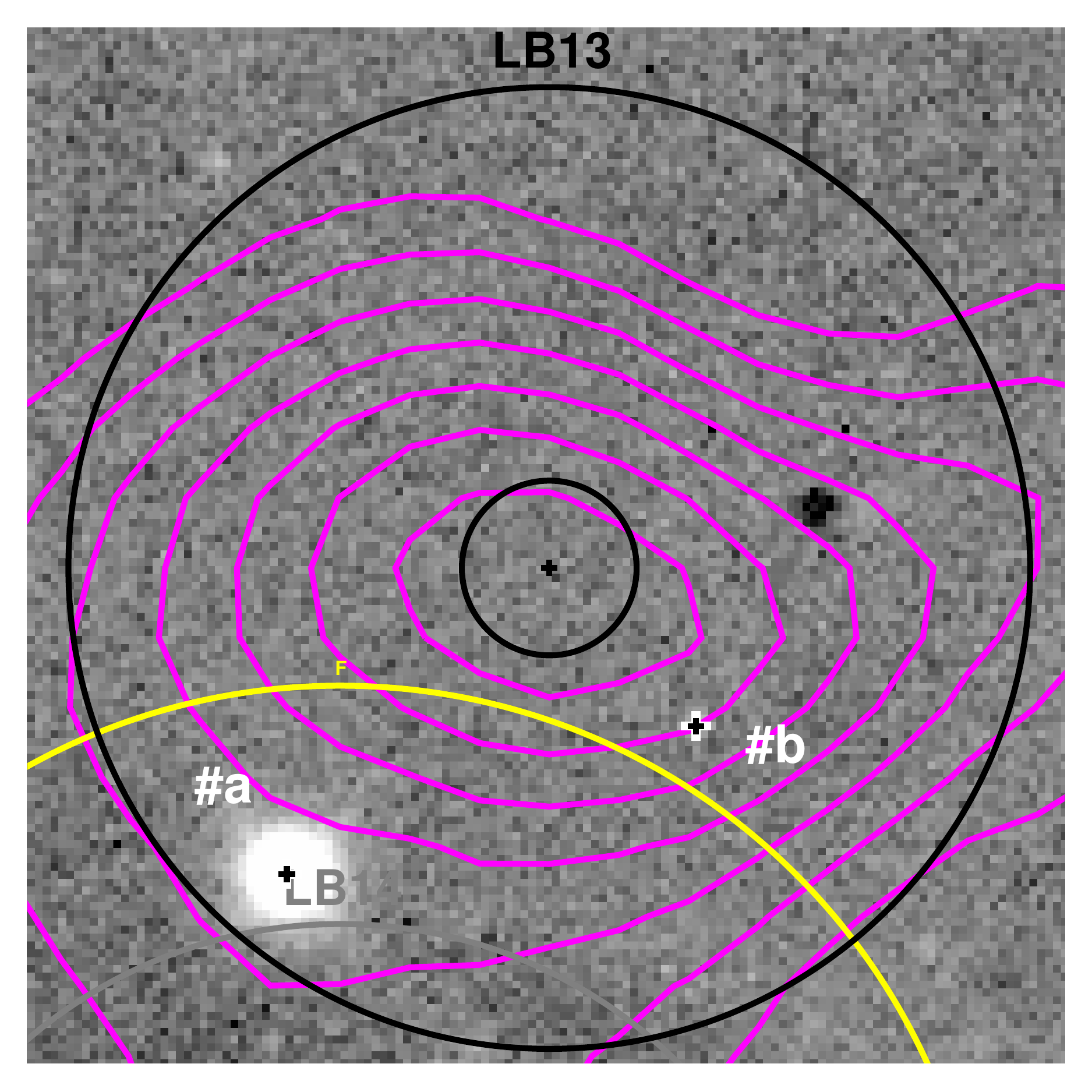}
\includegraphics[width=0.32\textwidth,scale=0.33]{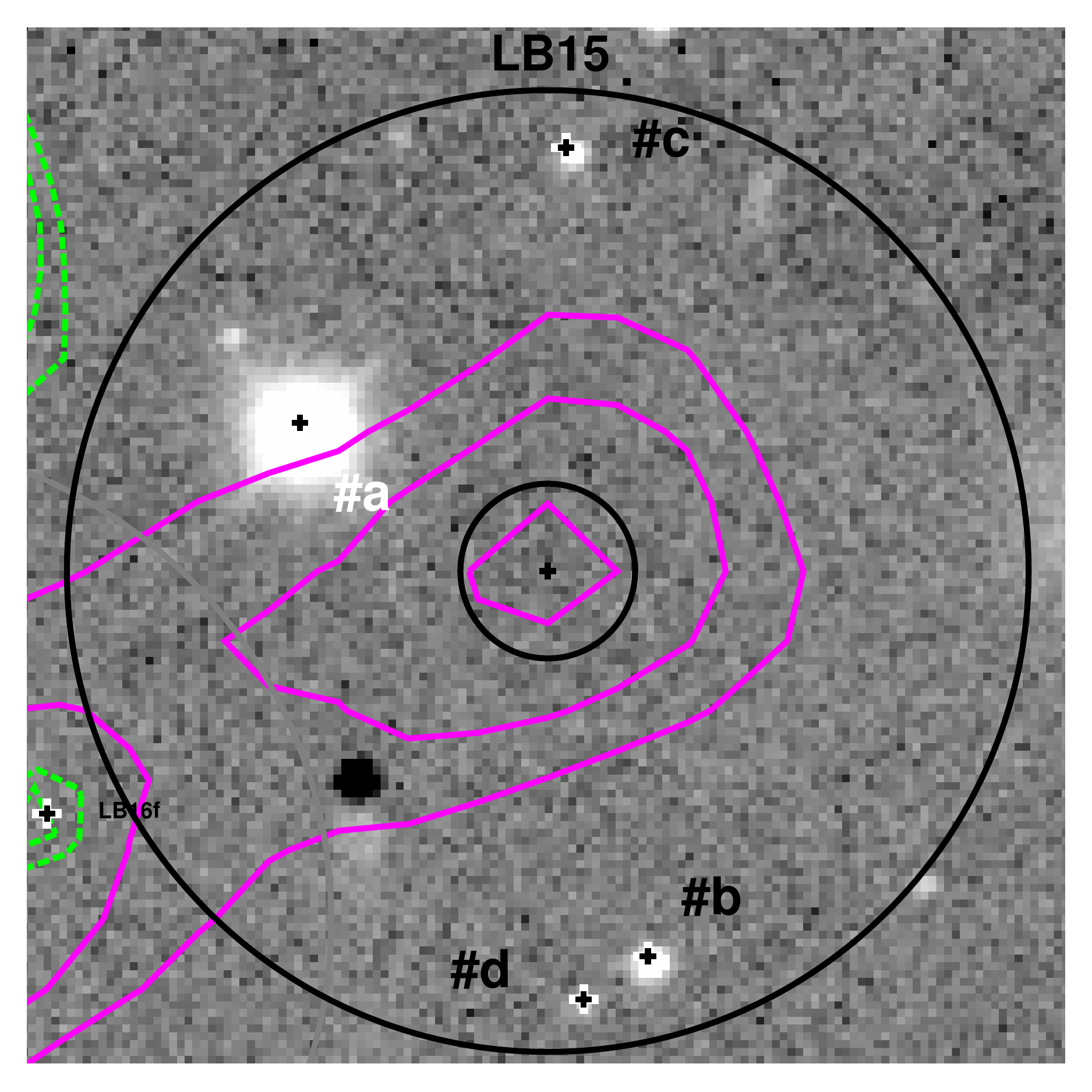}
\includegraphics[width=0.32\textwidth,scale=0.33]{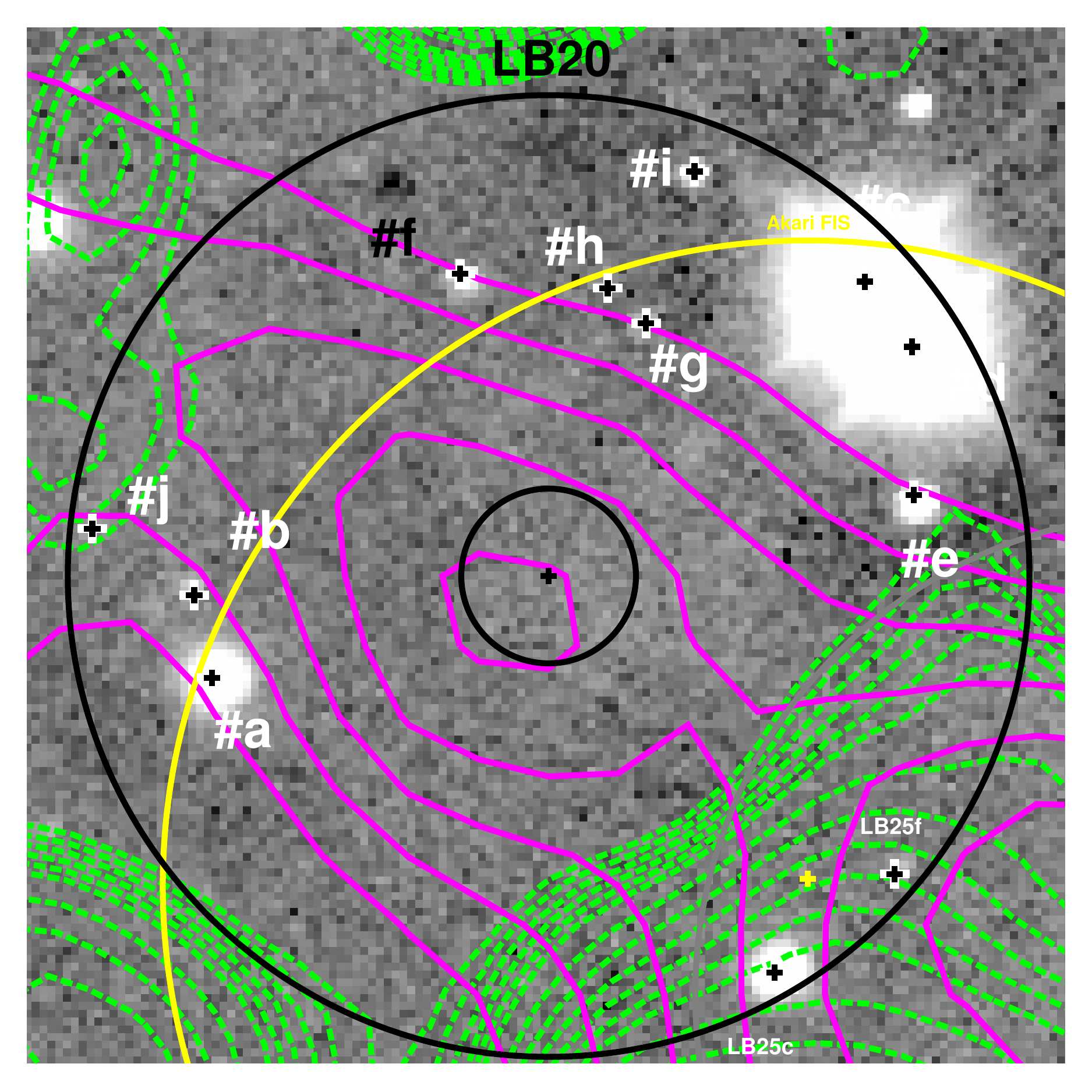}
\includegraphics[width=0.32\textwidth,scale=0.33]{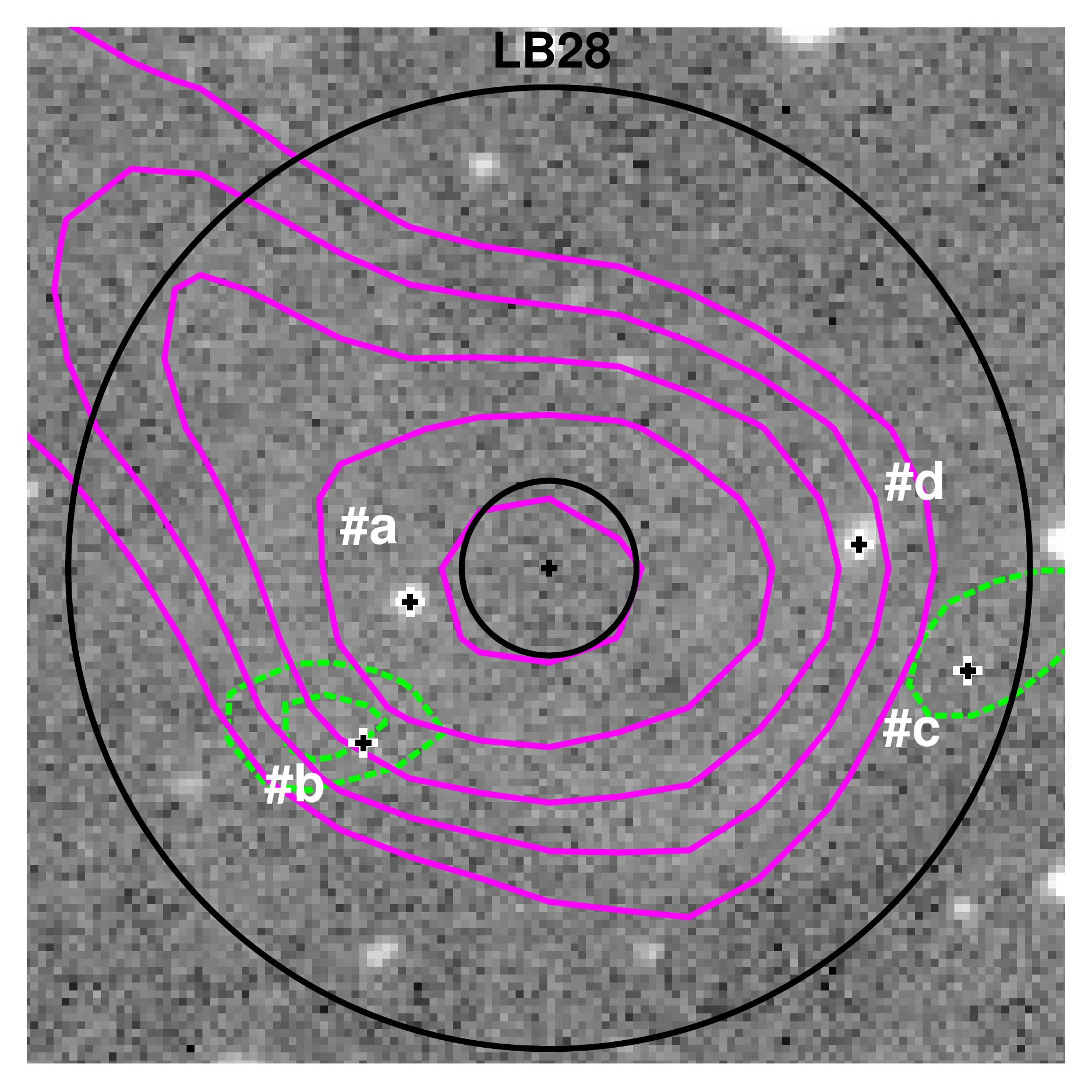}
\caption{\label{FCh_groupB2} 
  Finding charts for group B2, undetected both at  70 $\mu$m  and at 24 $\mu$m (upper limits in both cases).
Symbols as in Fig. \ref{FCh_groupA1}.
}
\end{figure*}

\begin{figure*}   
\center
%
\includegraphics[width=0.32\textwidth,scale=0.33]{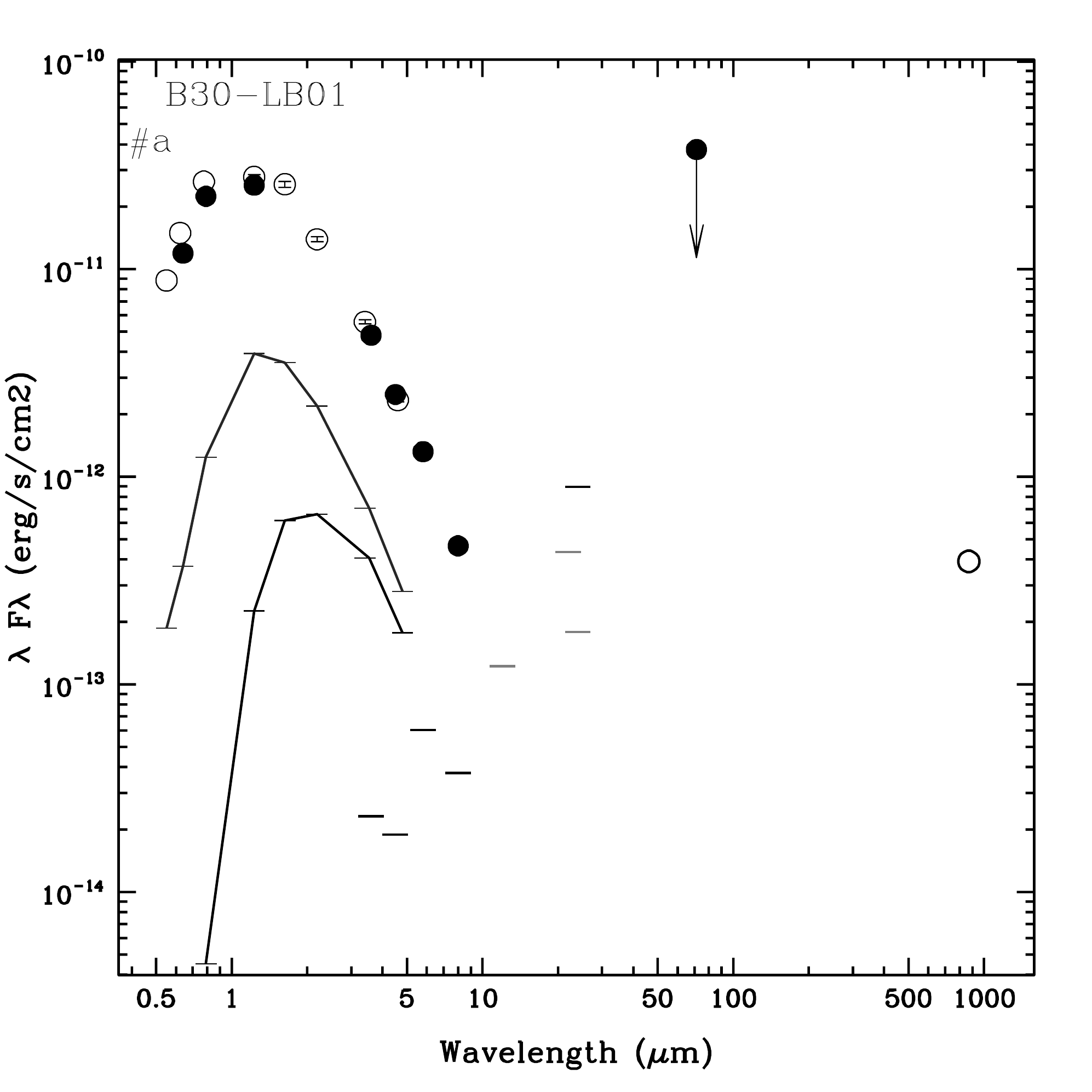} 
\includegraphics[width=0.32\textwidth,scale=0.33]{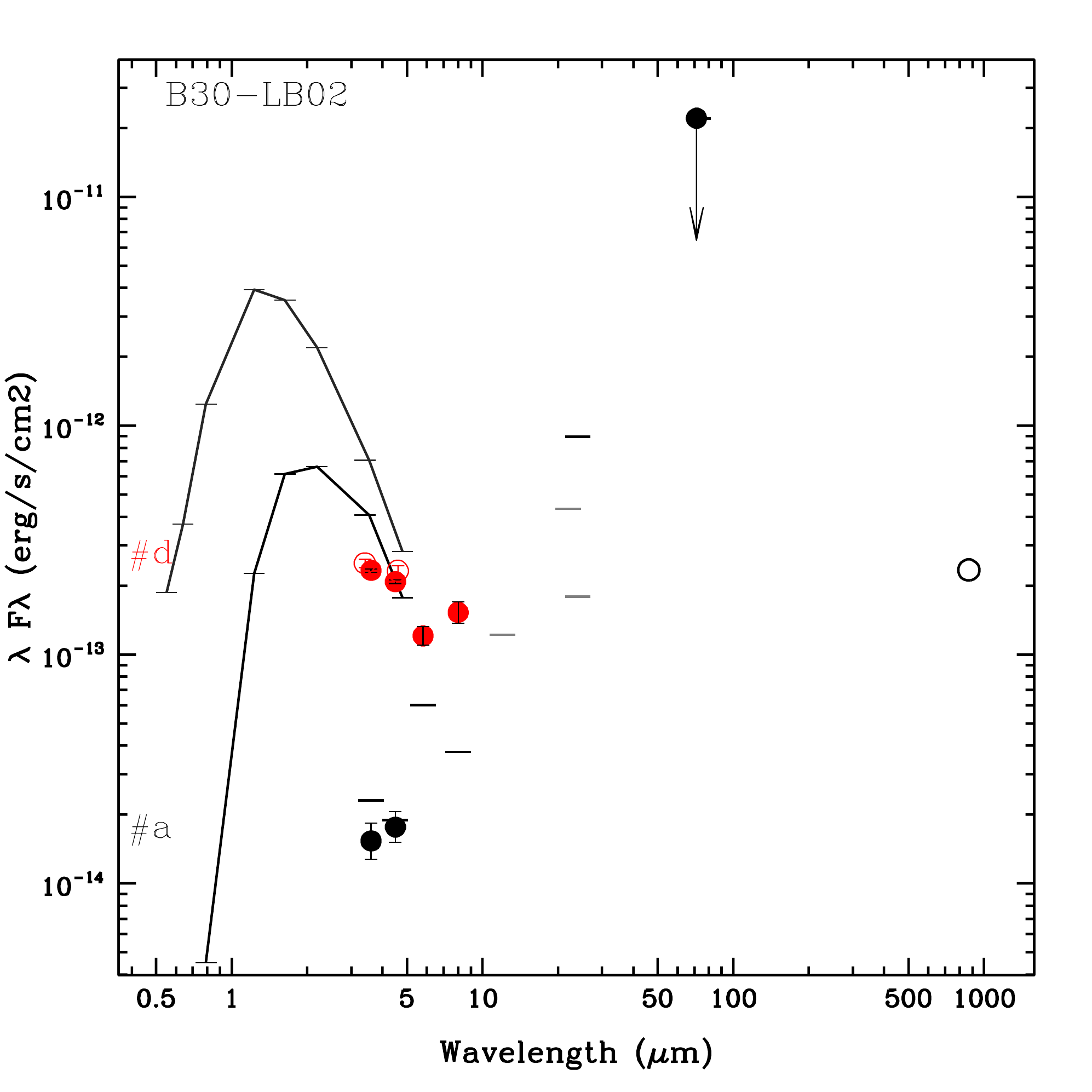} 
\includegraphics[width=0.32\textwidth,scale=0.33]{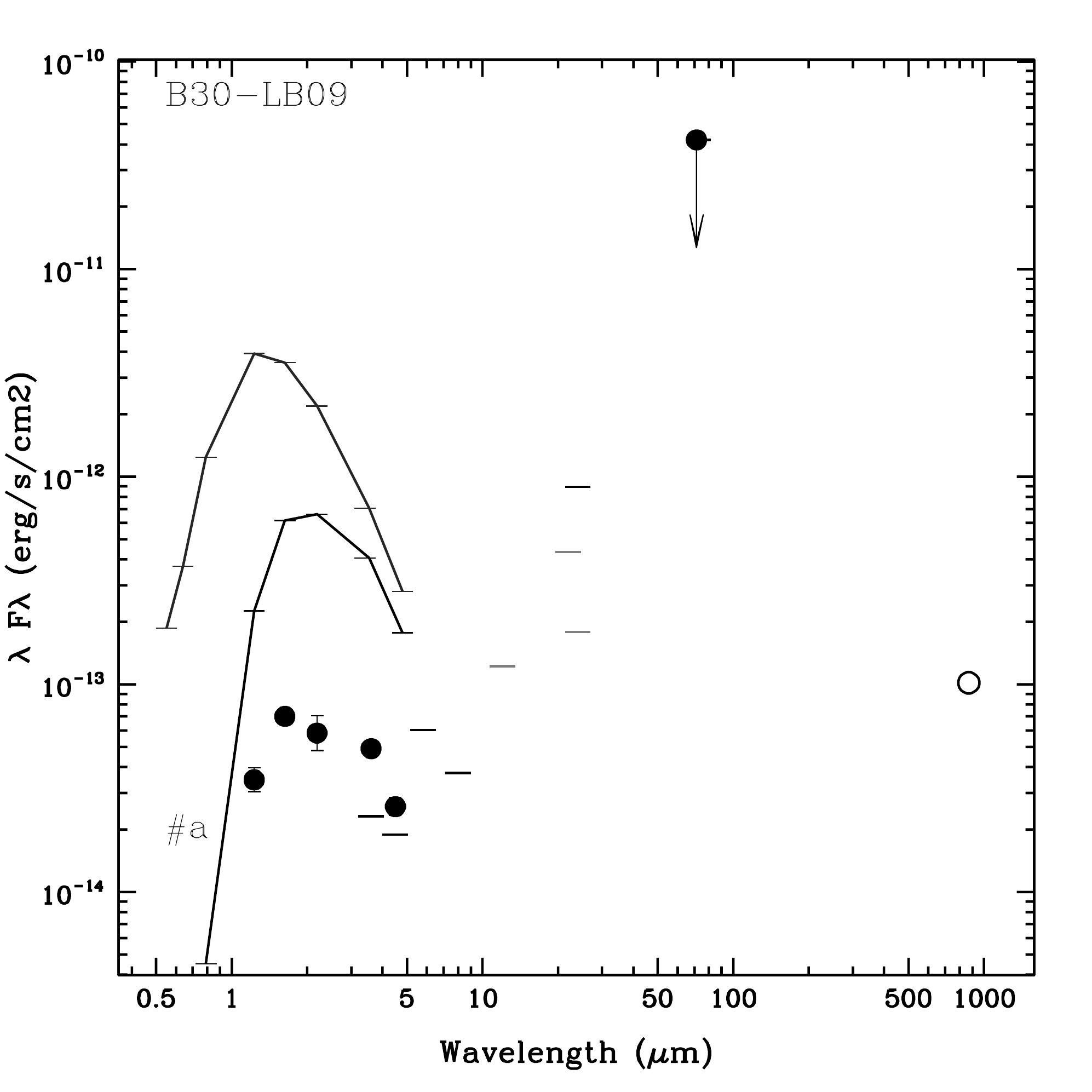} 
\includegraphics[width=0.32\textwidth,scale=0.33]{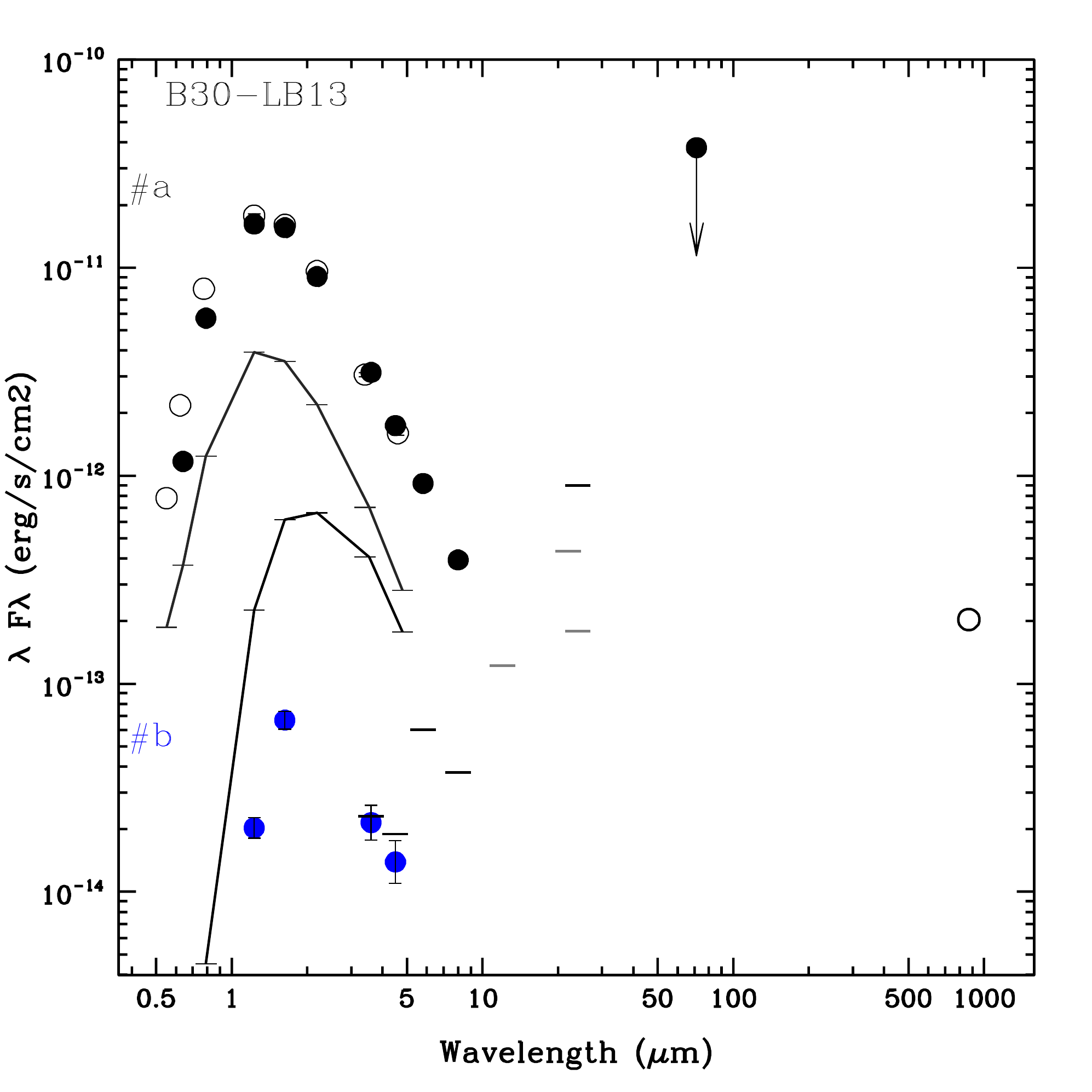} 
\includegraphics[width=0.32\textwidth,scale=0.33]{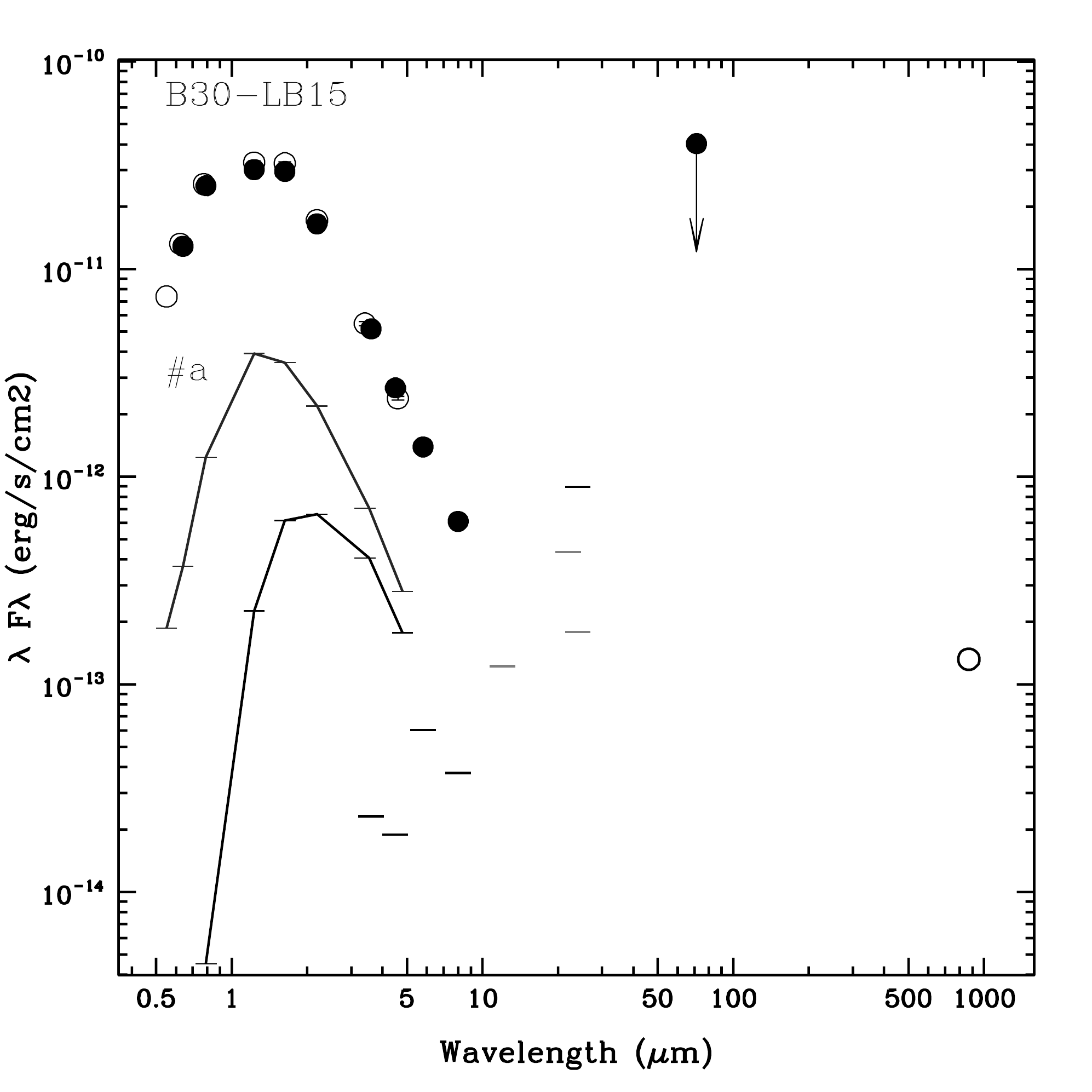} 
\includegraphics[width=0.32\textwidth,scale=0.33]{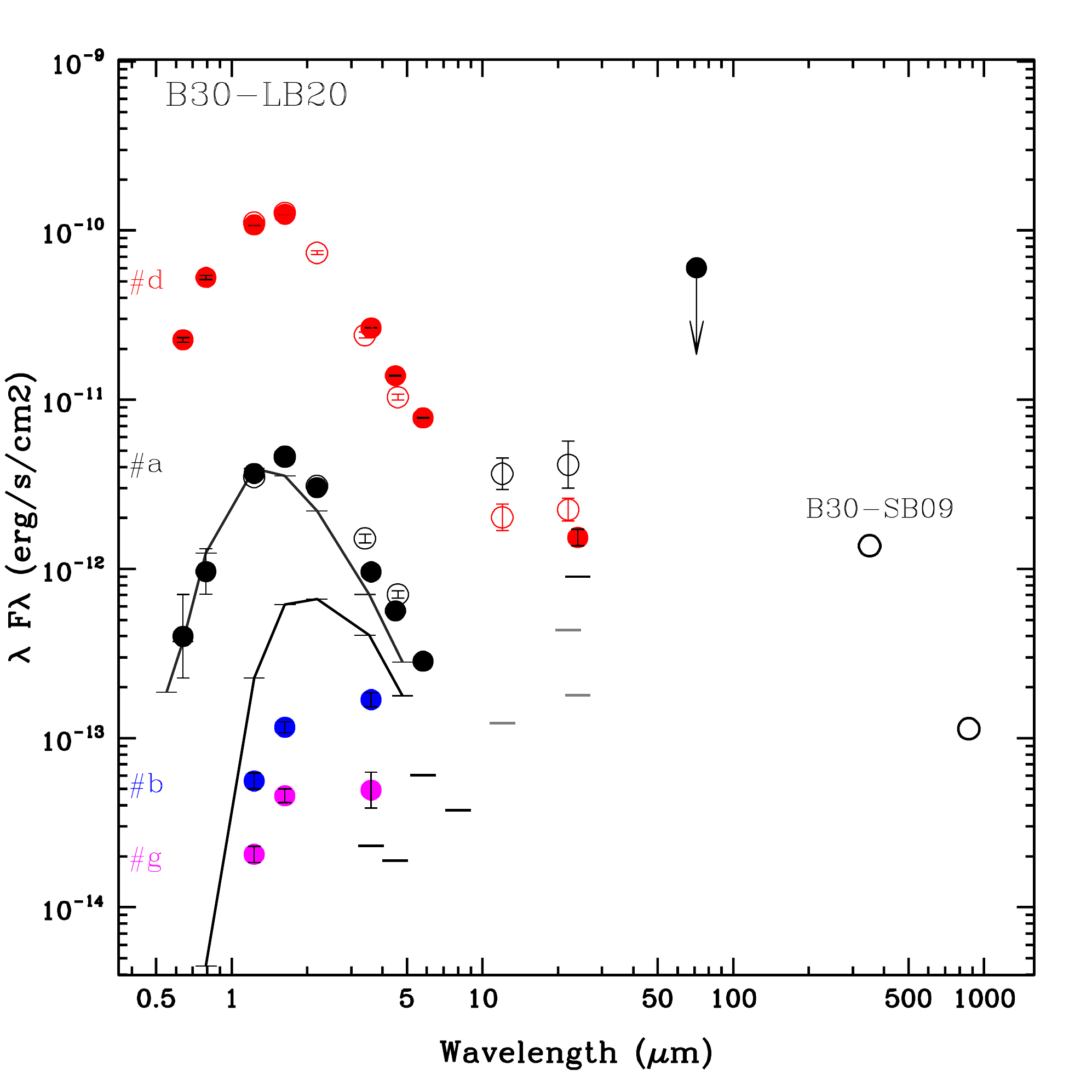} 
\includegraphics[width=0.32\textwidth,scale=0.33]{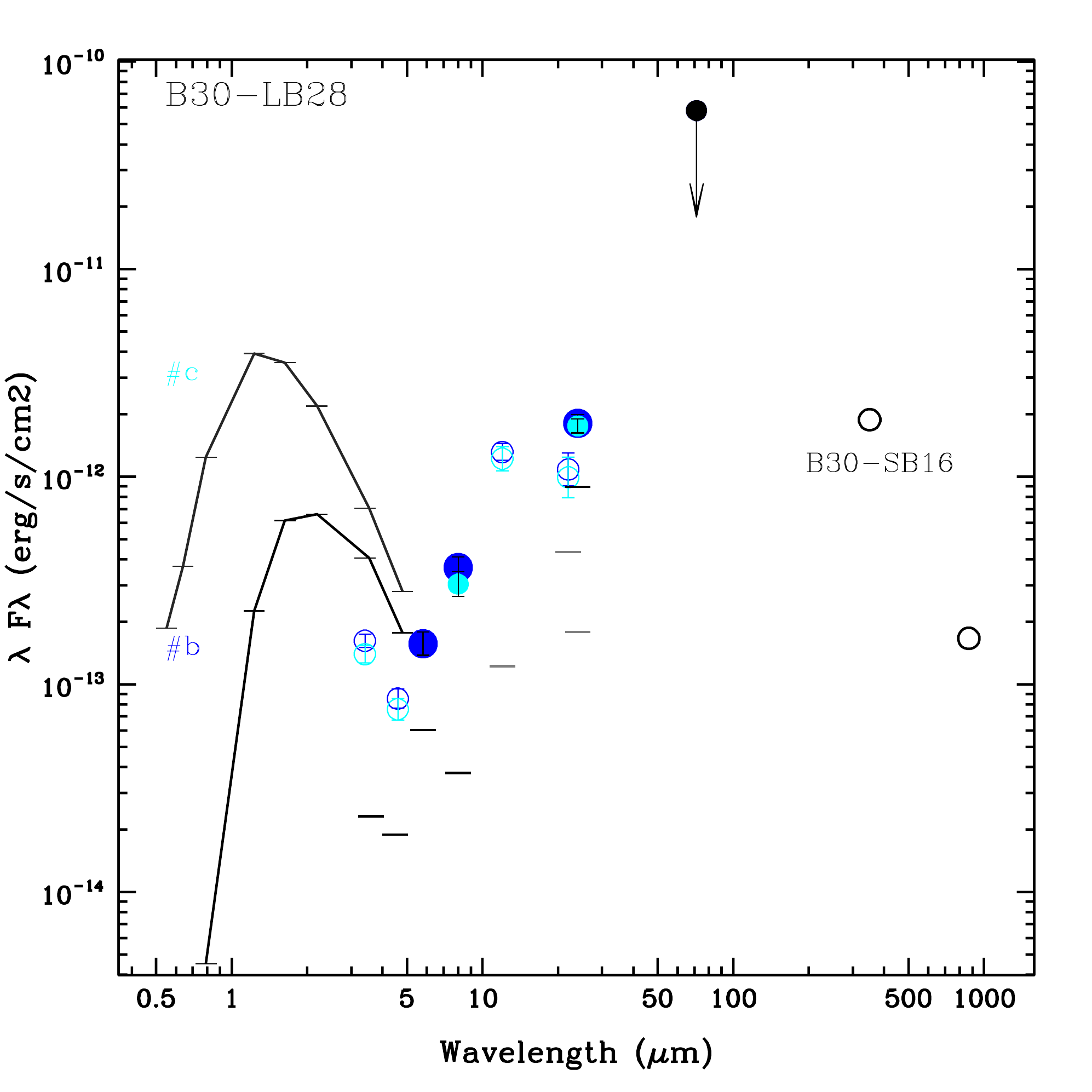} 
\caption{\label{SED_groupB2} 
  Spectral energy distribution for group B2  sources undetected  at 70 and  24 $\mu$m  (upper limits).
  We note that B30-LB20 and B30-LB28 have a possible extended emission at 24 $\mu$m.
Symbols as in Fig. \ref{SED_groupA1}.
}
\end{figure*}

\begin{figure*}  
\center
\includegraphics[width=0.32\textwidth,scale=0.33]{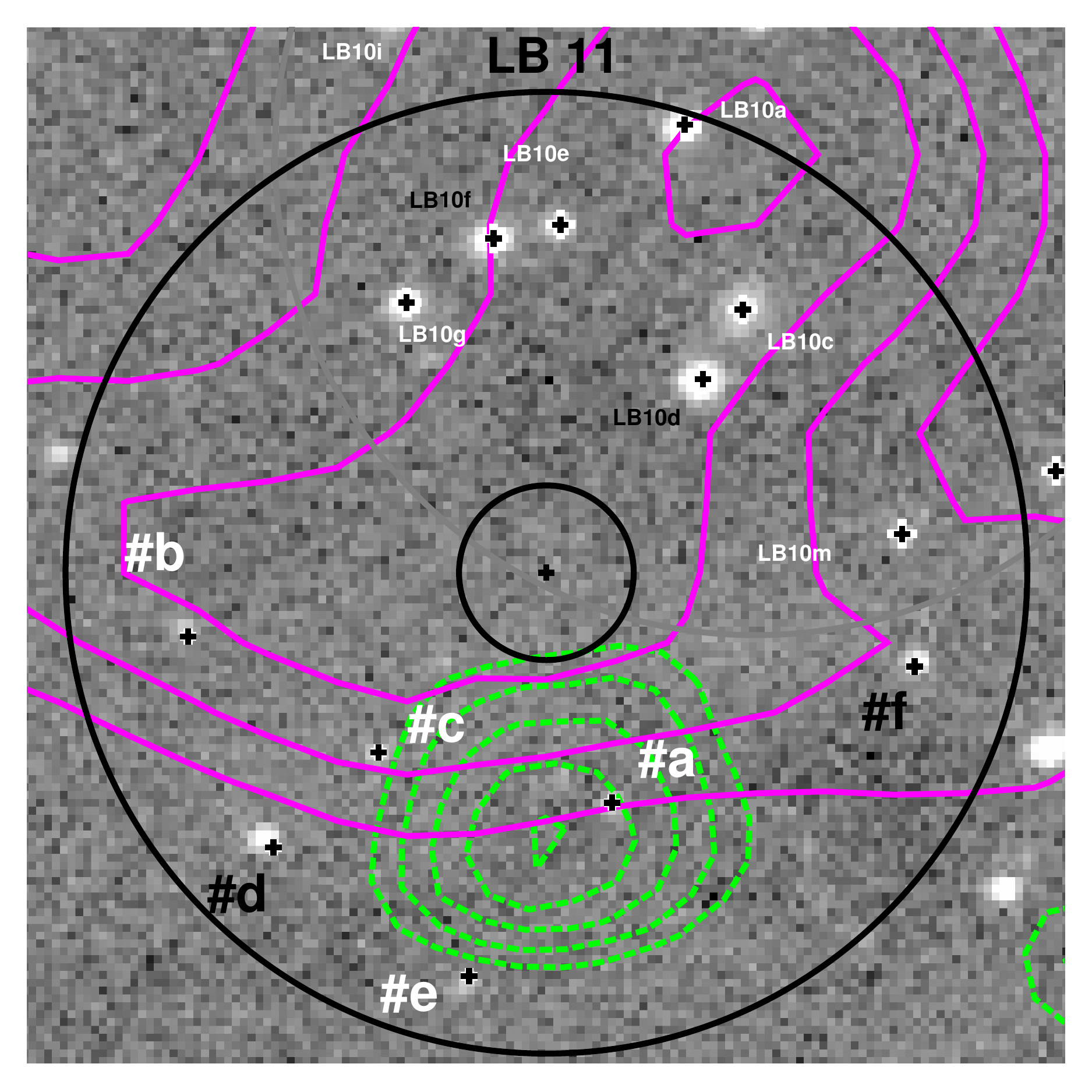}
\includegraphics[width=0.32\textwidth,scale=0.33]{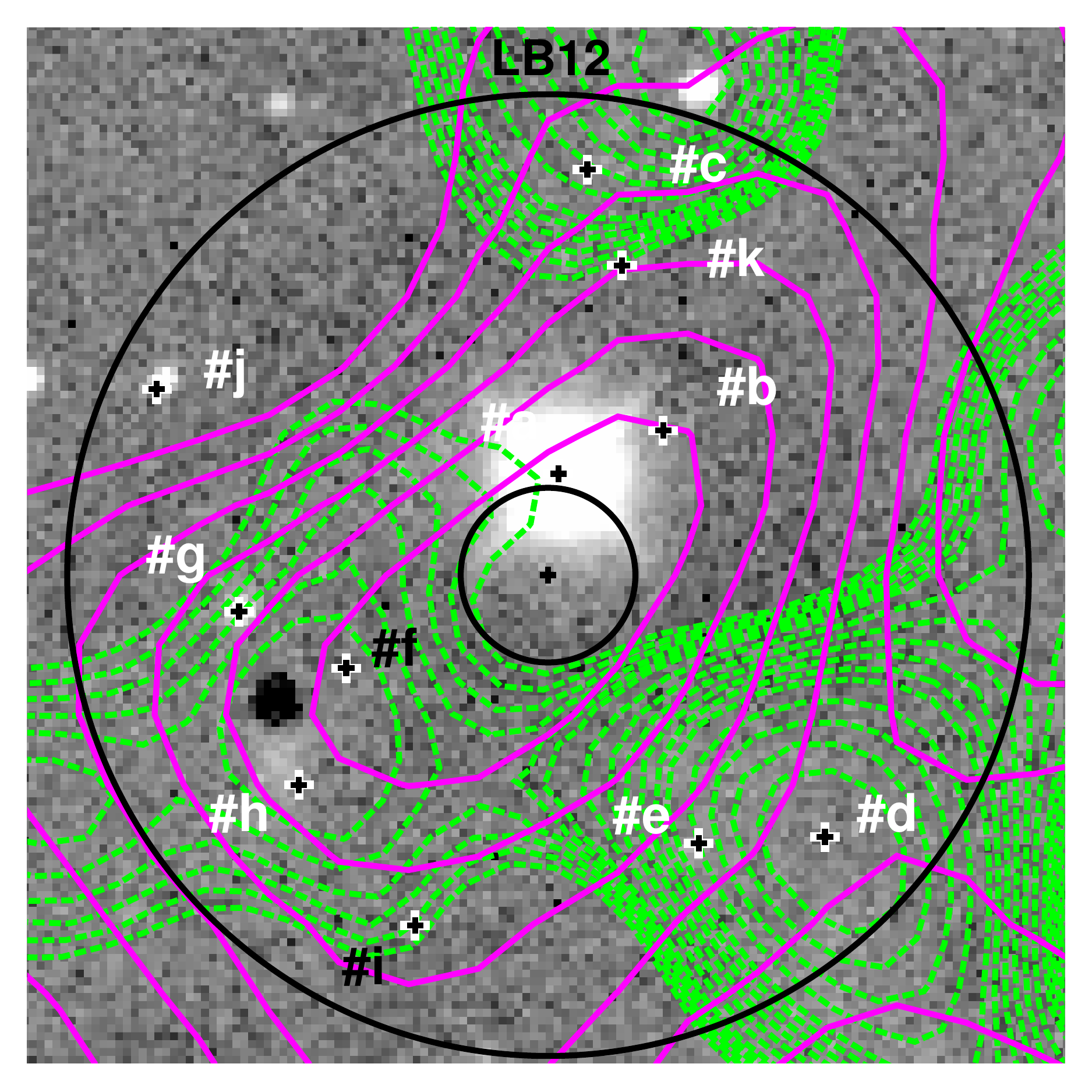}
\includegraphics[width=0.32\textwidth,scale=0.33]{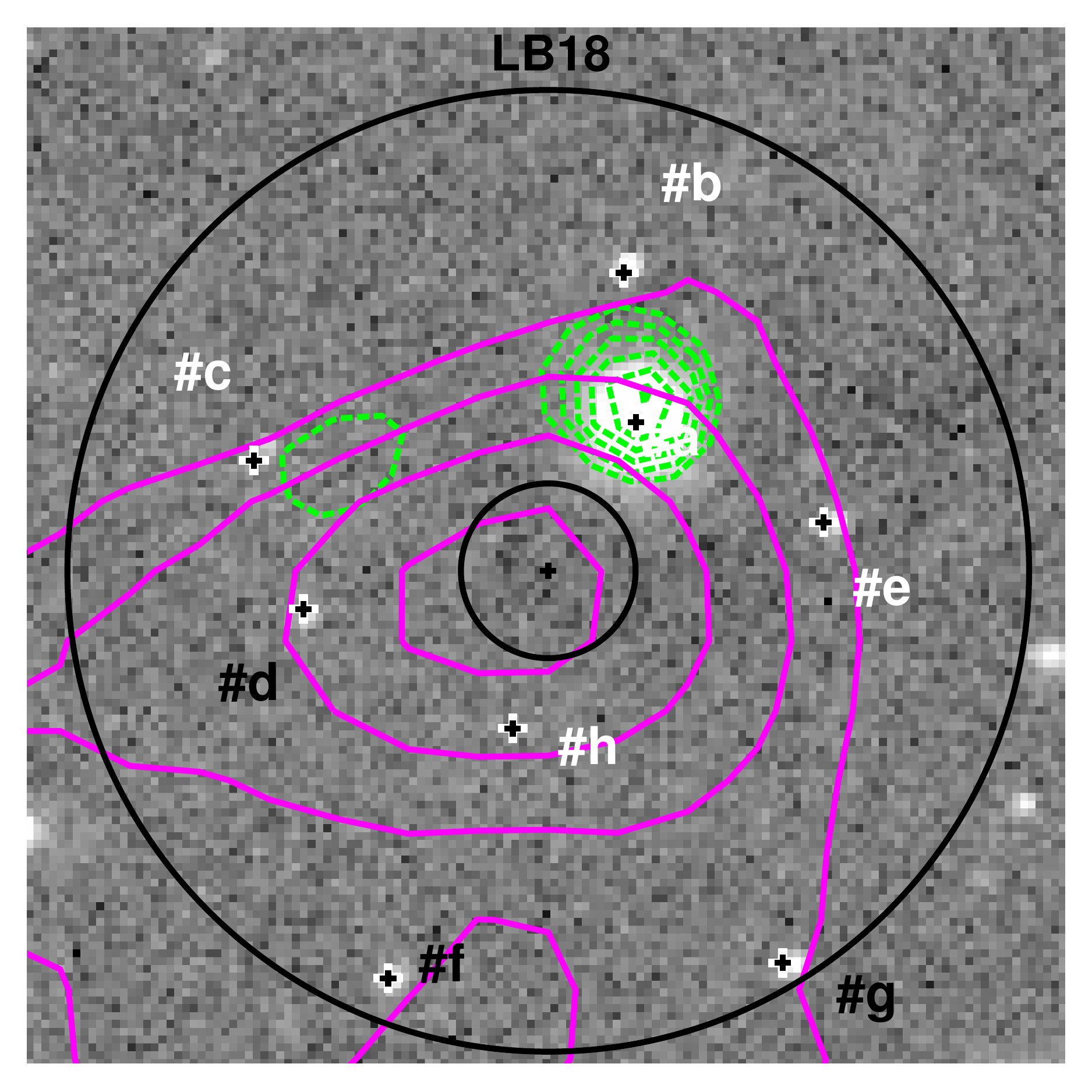}
\includegraphics[width=0.32\textwidth,scale=0.33]{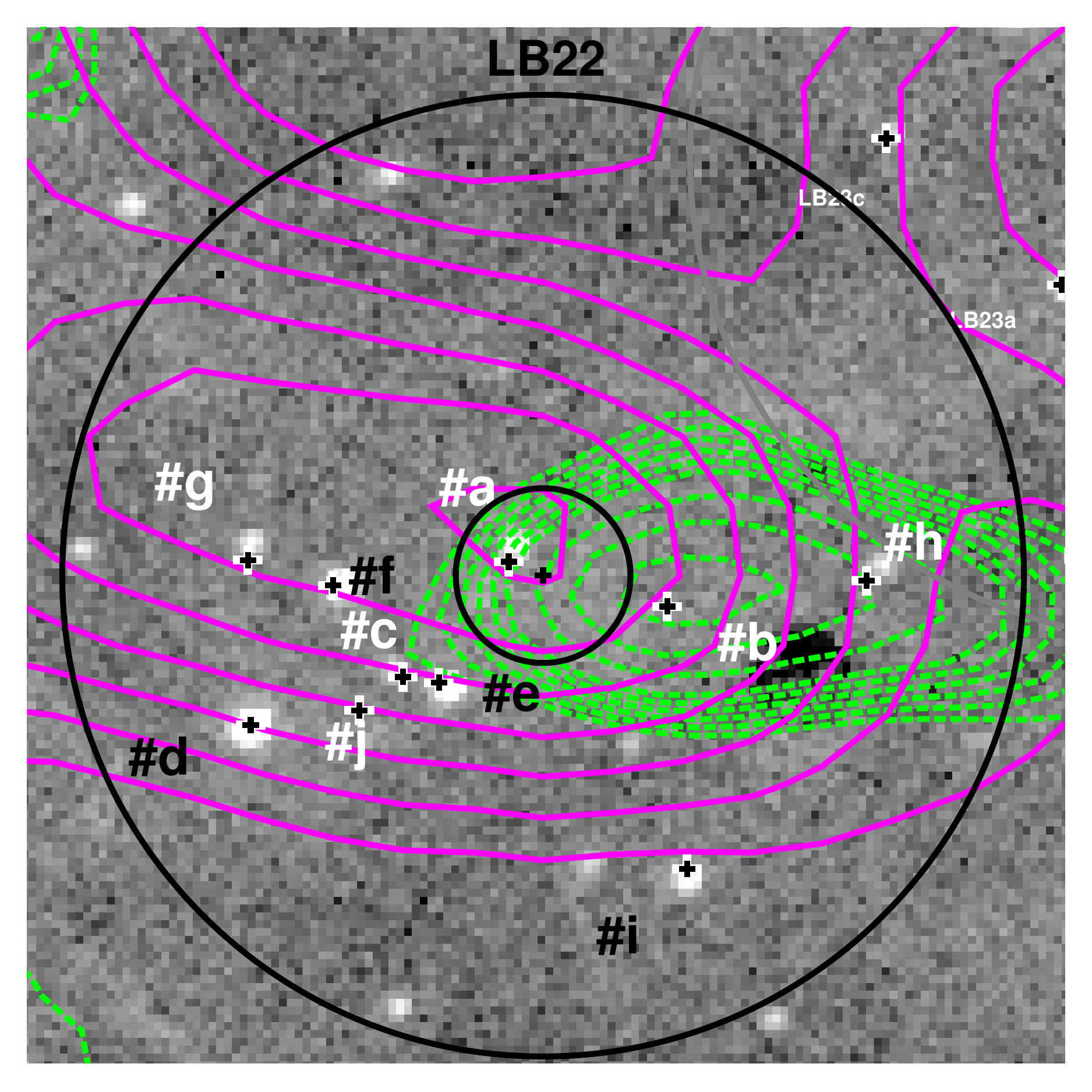}
\includegraphics[width=0.32\textwidth,scale=0.33]{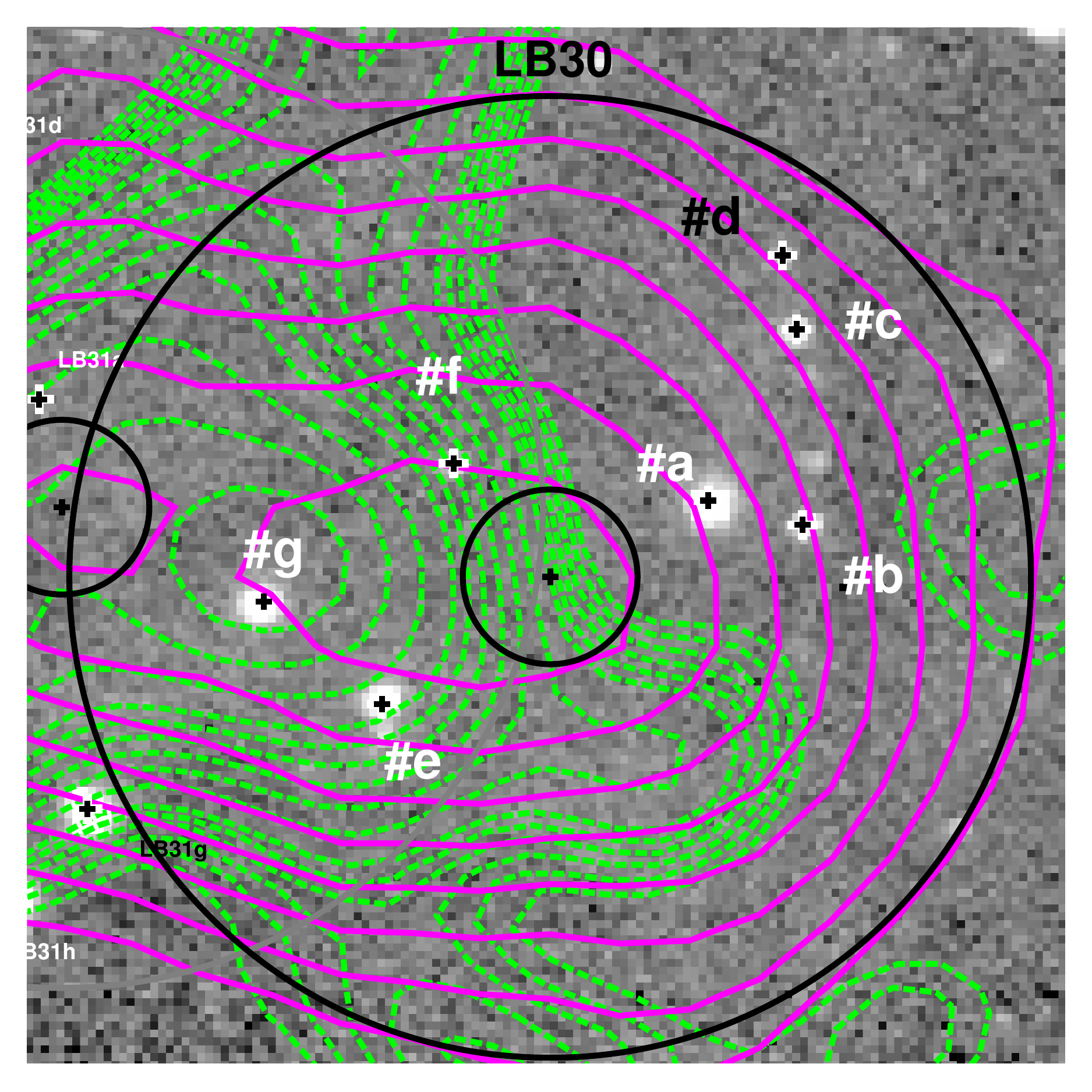}
\includegraphics[width=0.32\textwidth,scale=0.33]{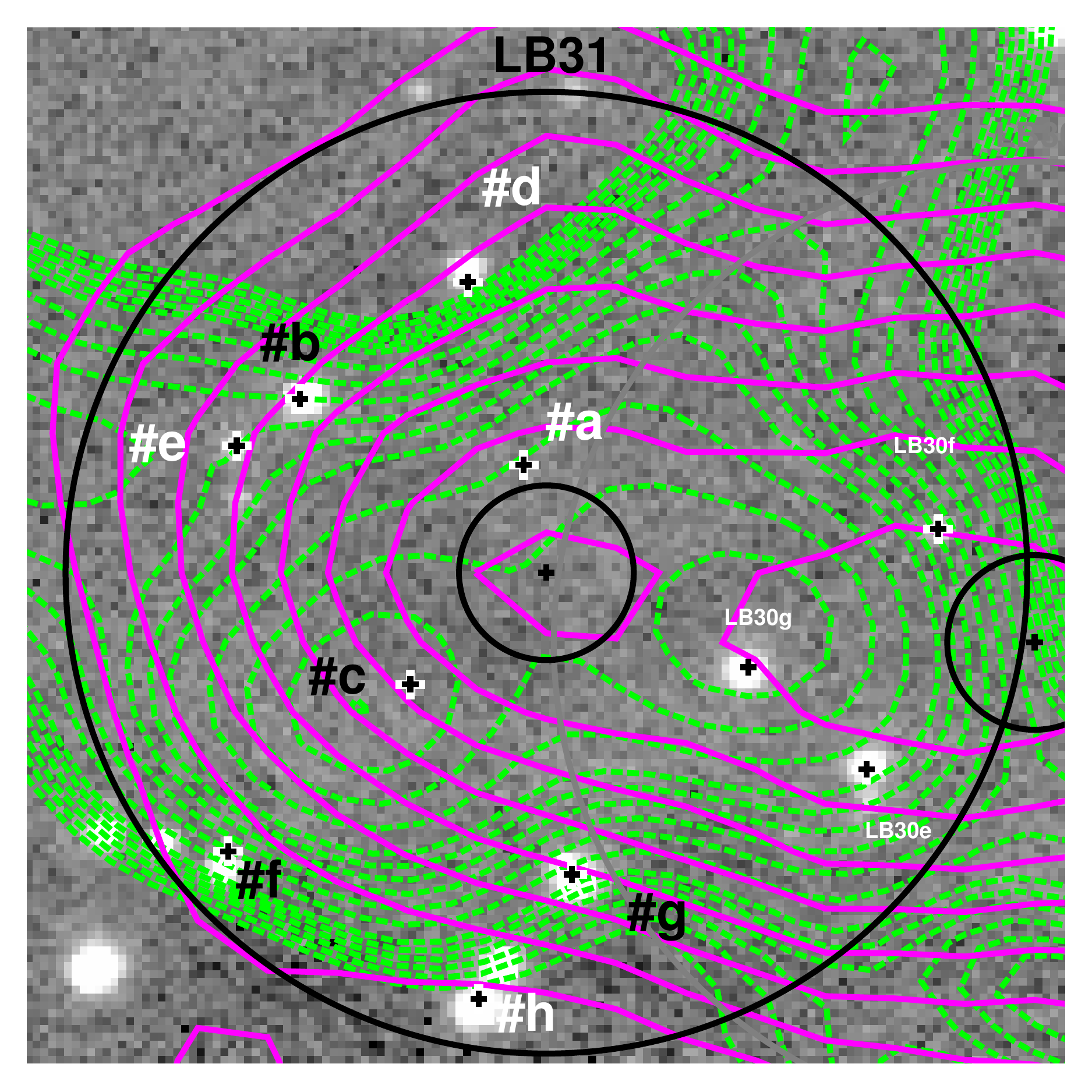}
\includegraphics[width=0.32\textwidth,scale=0.33]{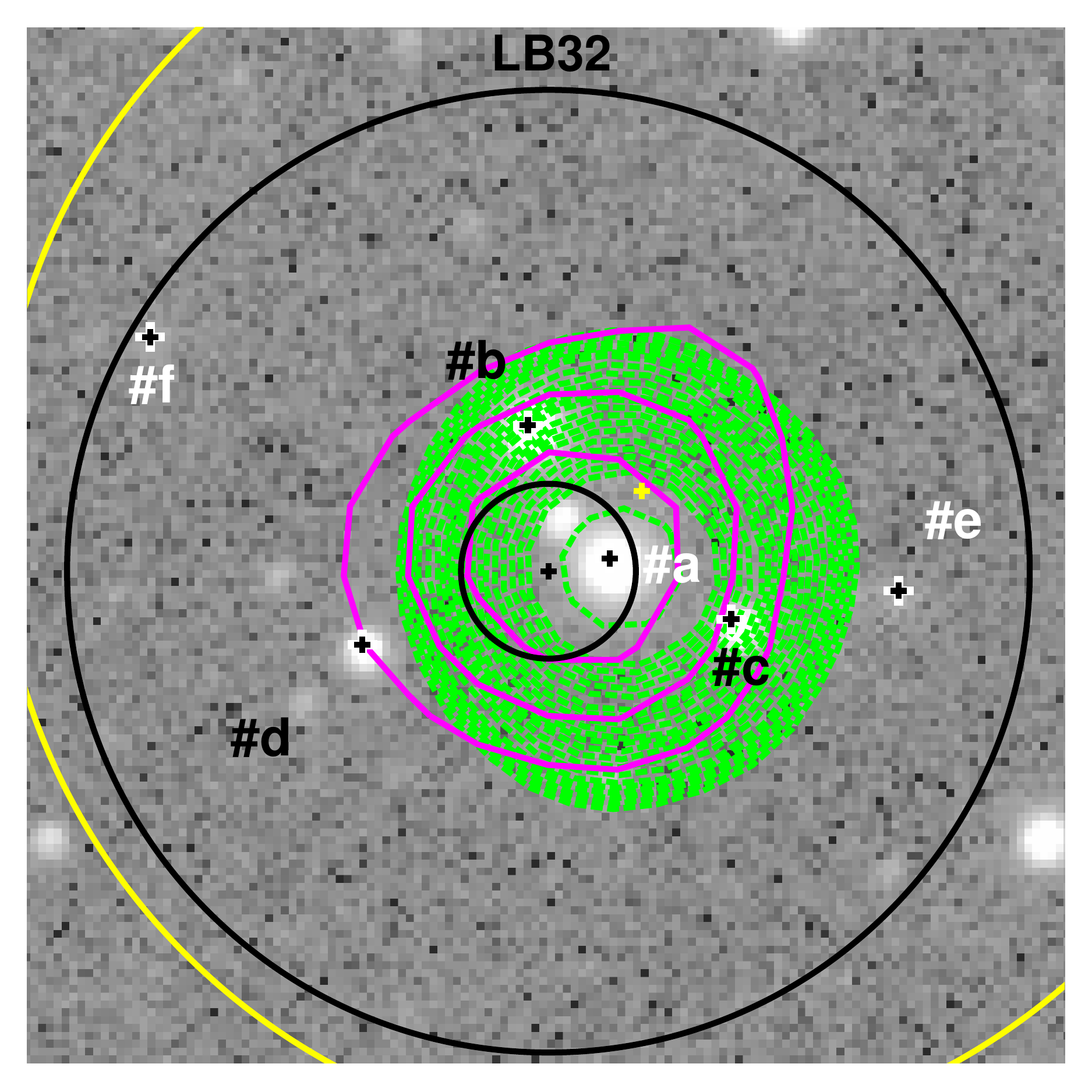}
\includegraphics[width=0.32\textwidth,scale=0.33]{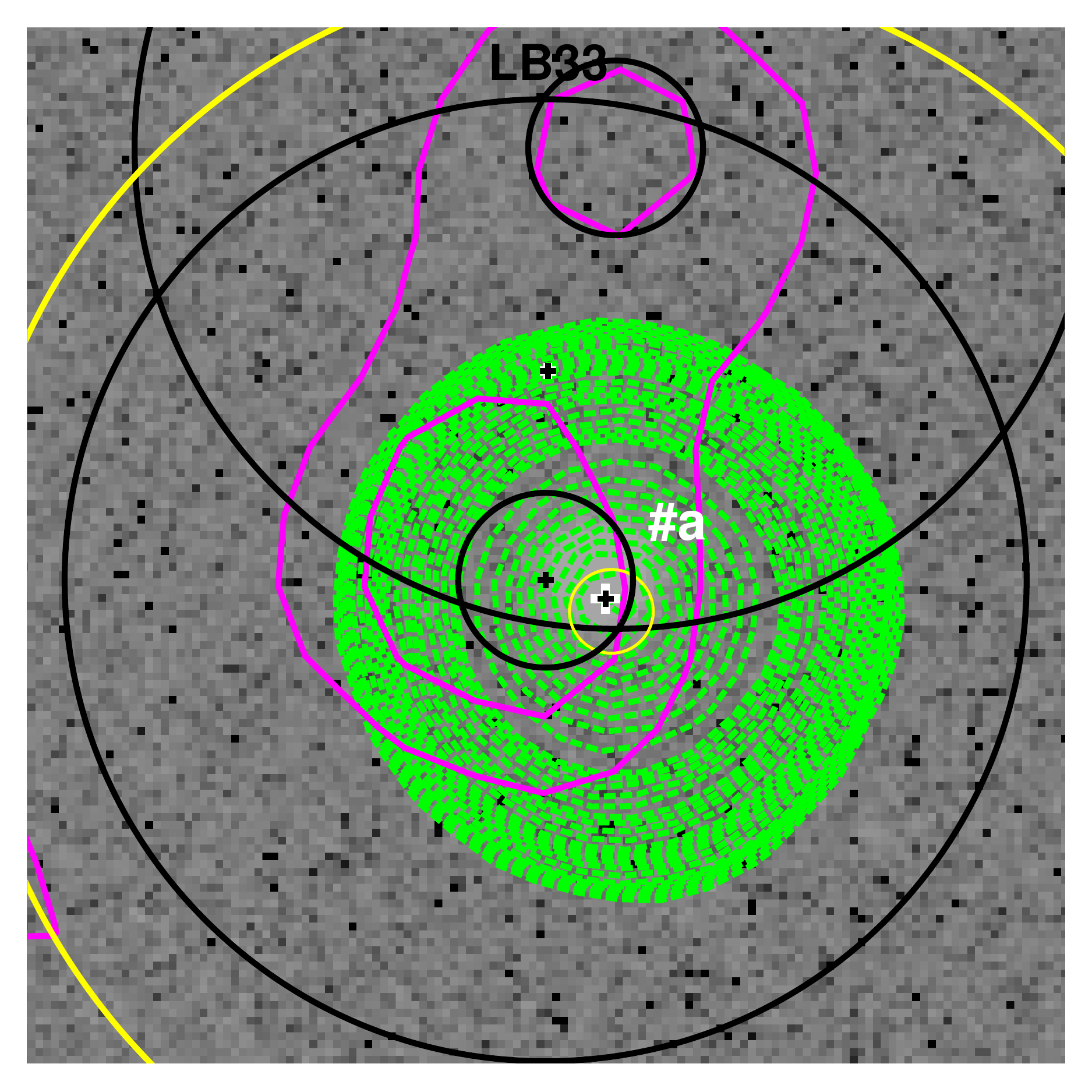}
\caption{\label{FCh_groupC1} 
Finding charts for group C1 or sources  outside the FOV of the MIPS M2 image.
 All have been detected  at 24 $\mu$m with MIPS.
Symbols as in Fig. \ref{FCh_groupA1}.
}
\end{figure*}

\begin{figure*}   
\center
%
\includegraphics[width=0.32\textwidth,scale=0.33]{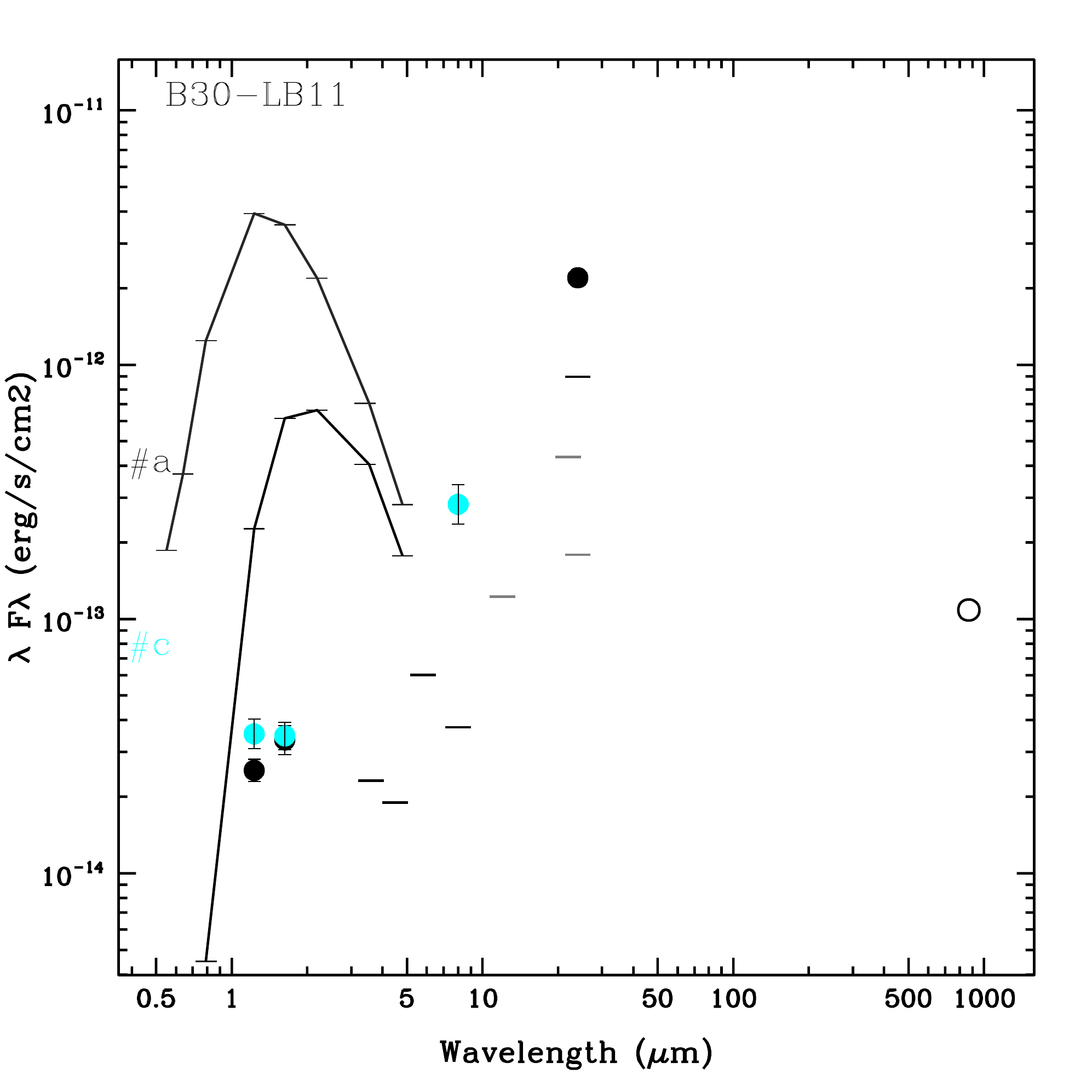} 
\includegraphics[width=0.32\textwidth,scale=0.33]{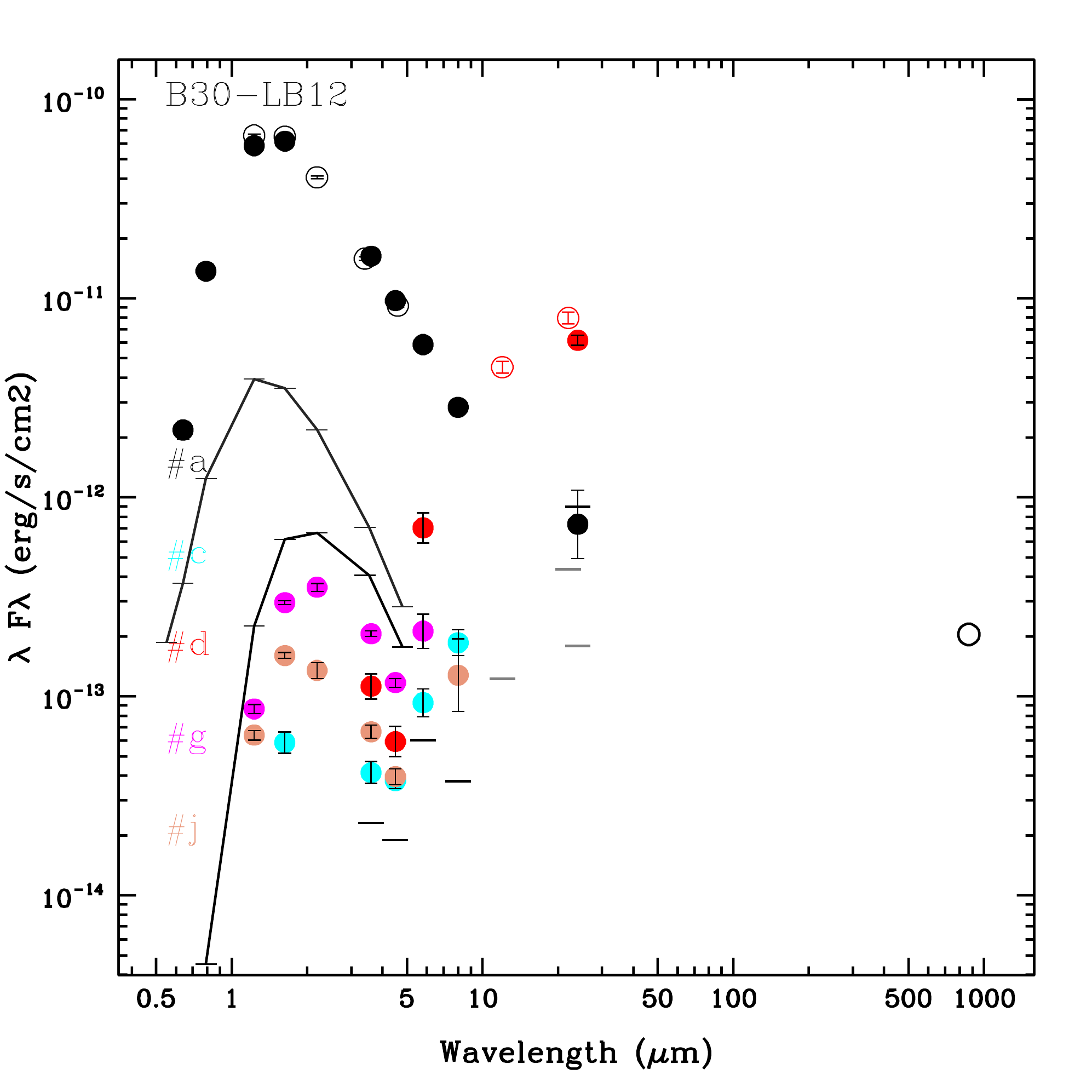} 
\includegraphics[width=0.32\textwidth,scale=0.33]{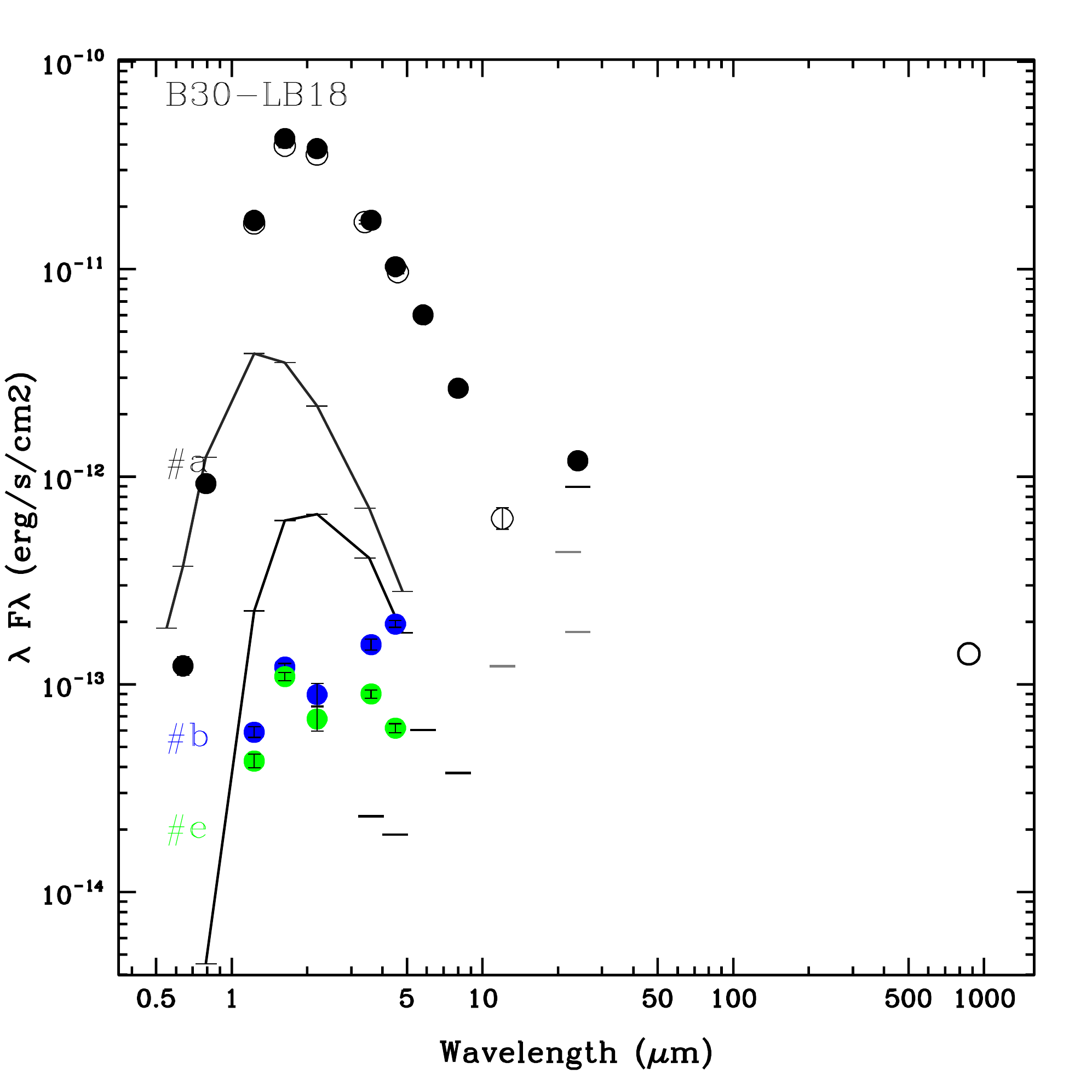} 
\includegraphics[width=0.32\textwidth,scale=0.33]{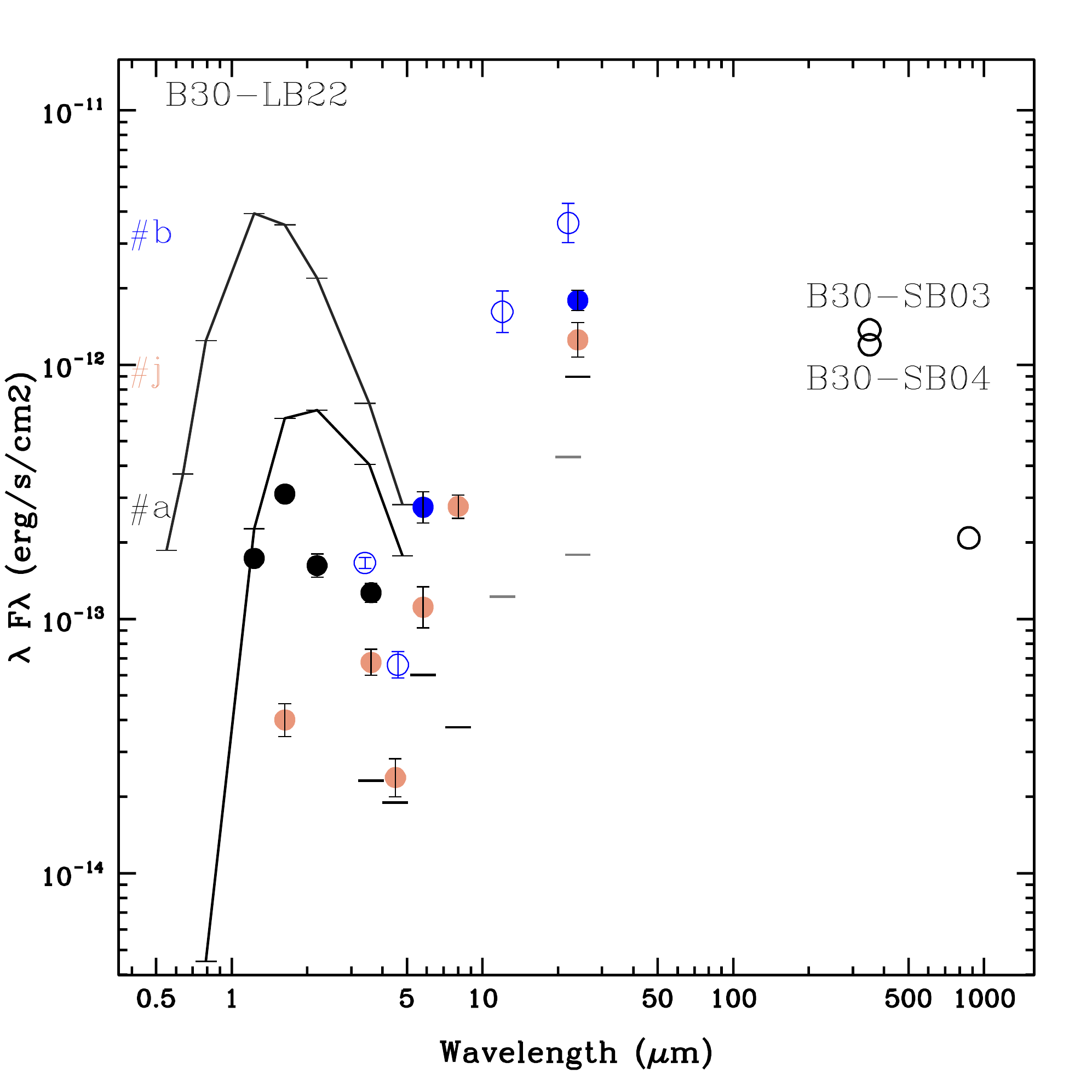} 
\includegraphics[width=0.32\textwidth,scale=0.33]{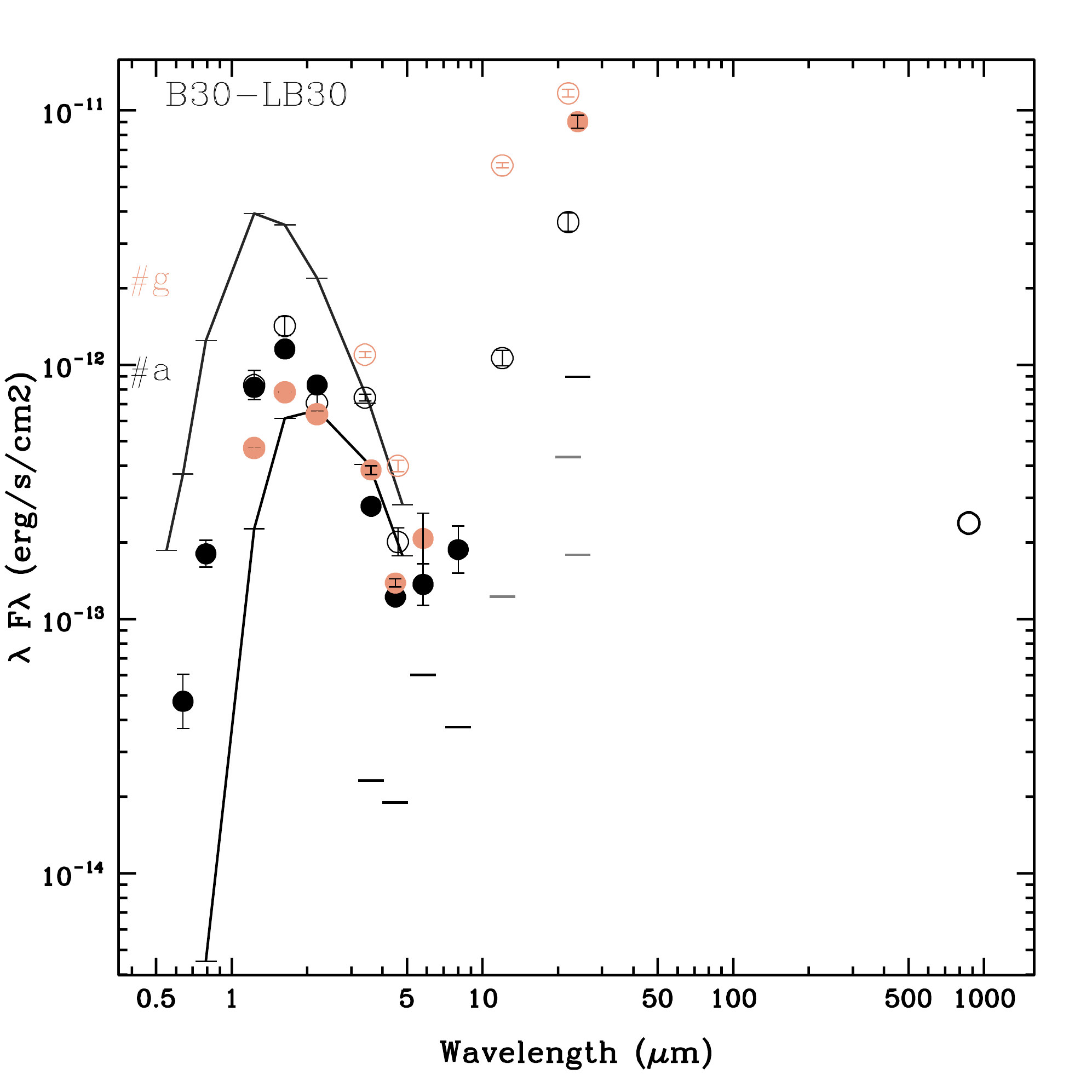} 
\includegraphics[width=0.32\textwidth,scale=0.33]{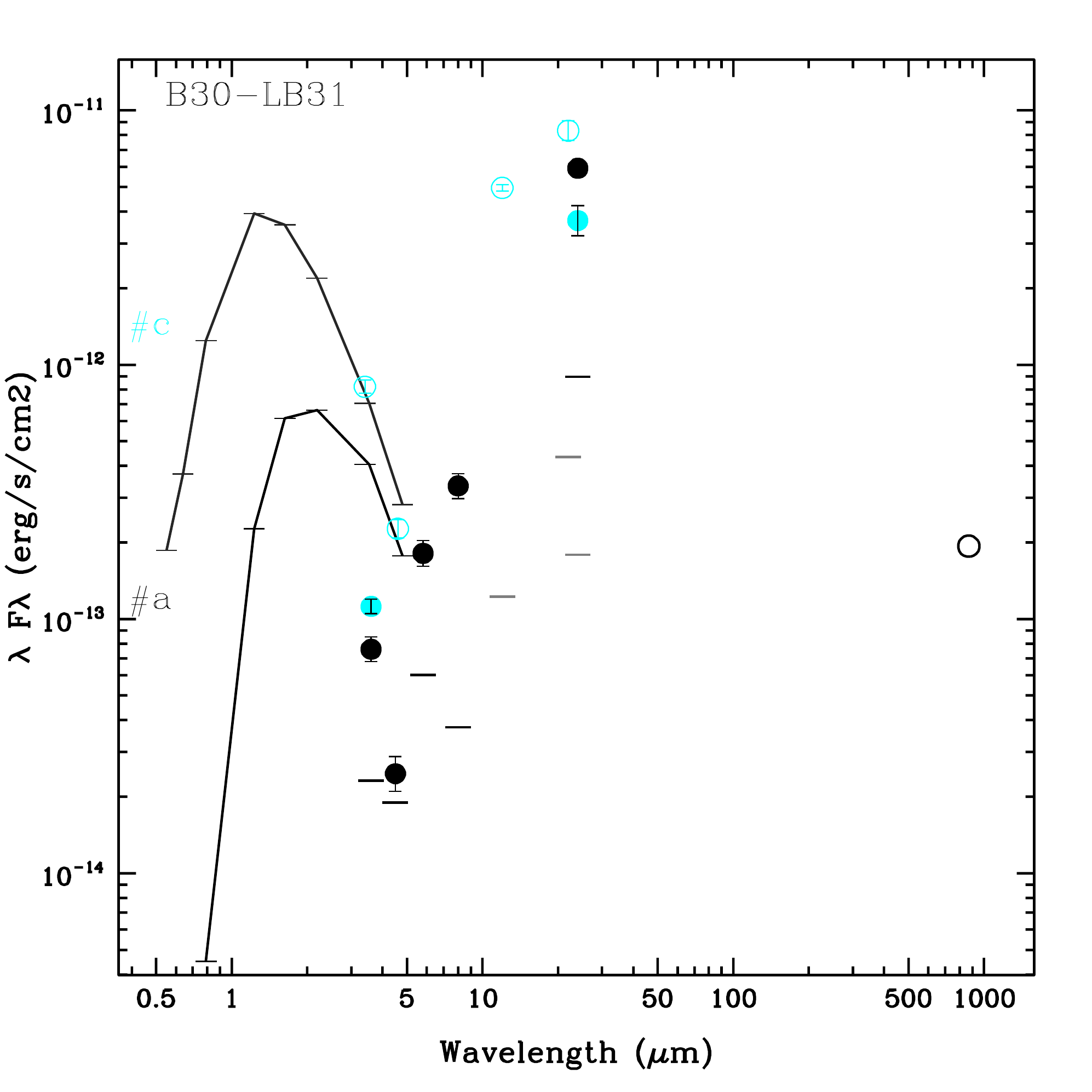} 
\includegraphics[width=0.32\textwidth,scale=0.33]{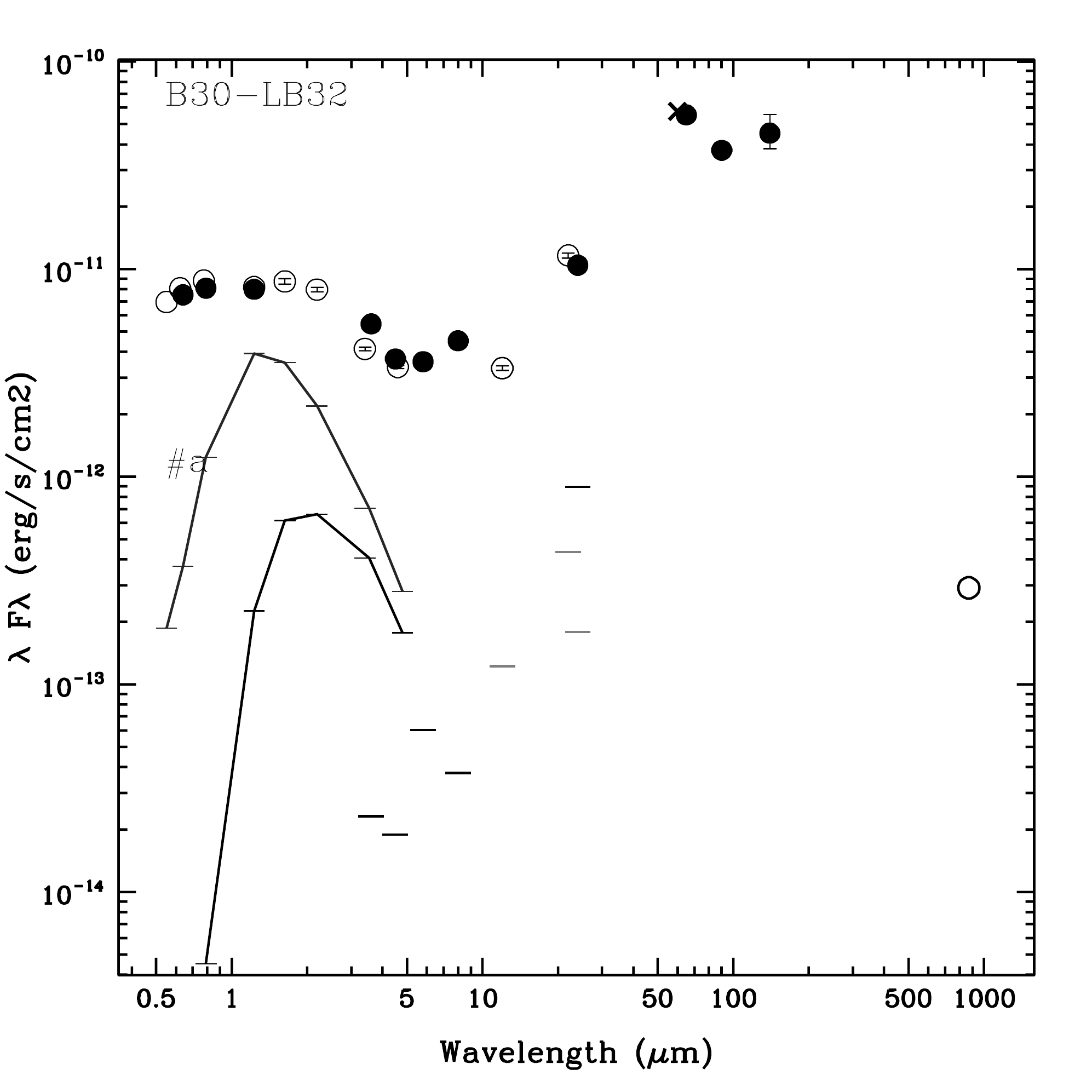} 
\includegraphics[width=0.32\textwidth,scale=0.33]{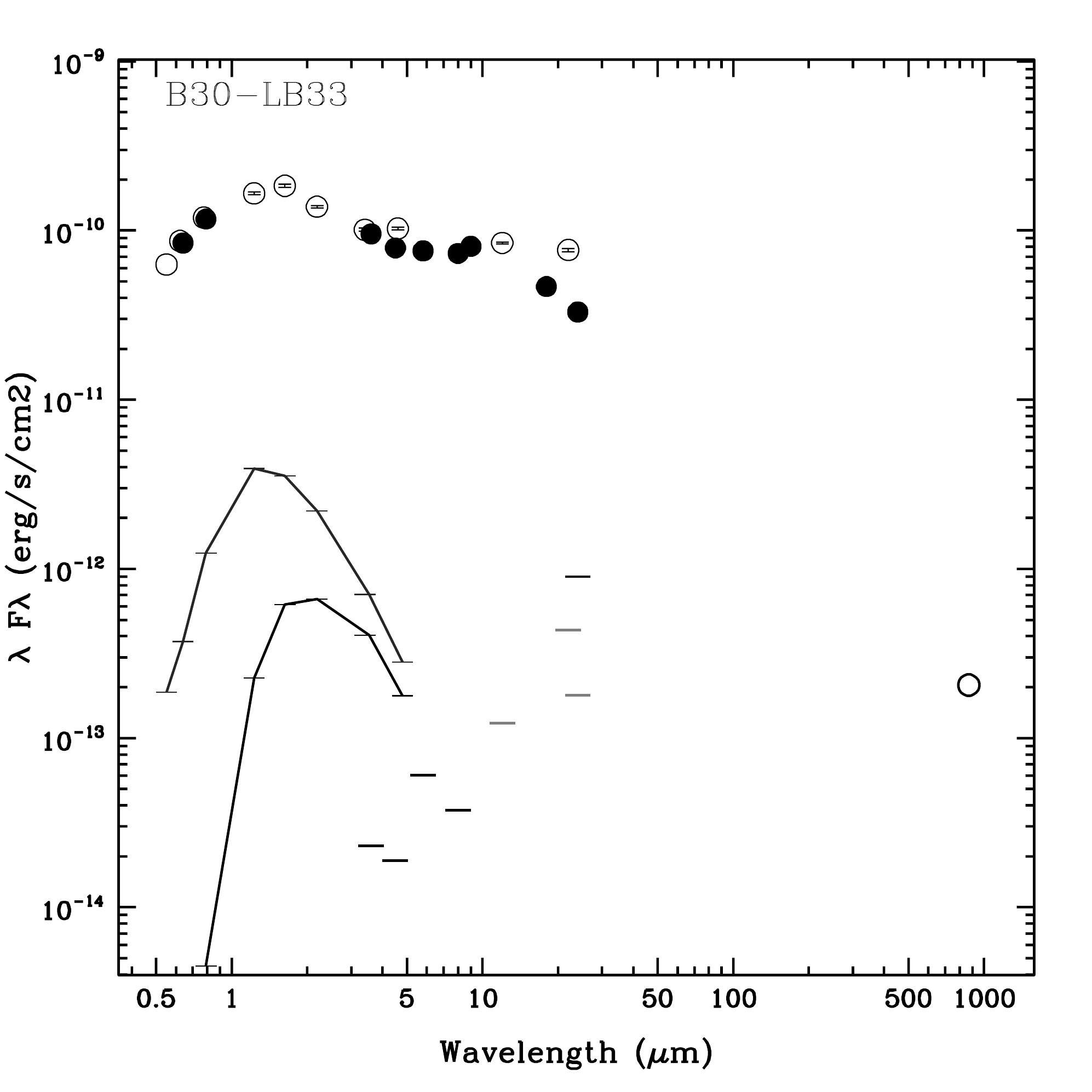} 
\caption{\label{SED_groupC1} 
  Spectral energy distribution for some relevant sources outside the FOV of the MIPS M2 image.
  All have been detected at  24 $\mu$m with MIPS.
Symbols as in Fig. \ref{SED_groupA1}.
}
\end{figure*}

\begin{figure*}   
\center
%
\includegraphics[width=0.32\textwidth,scale=0.33]{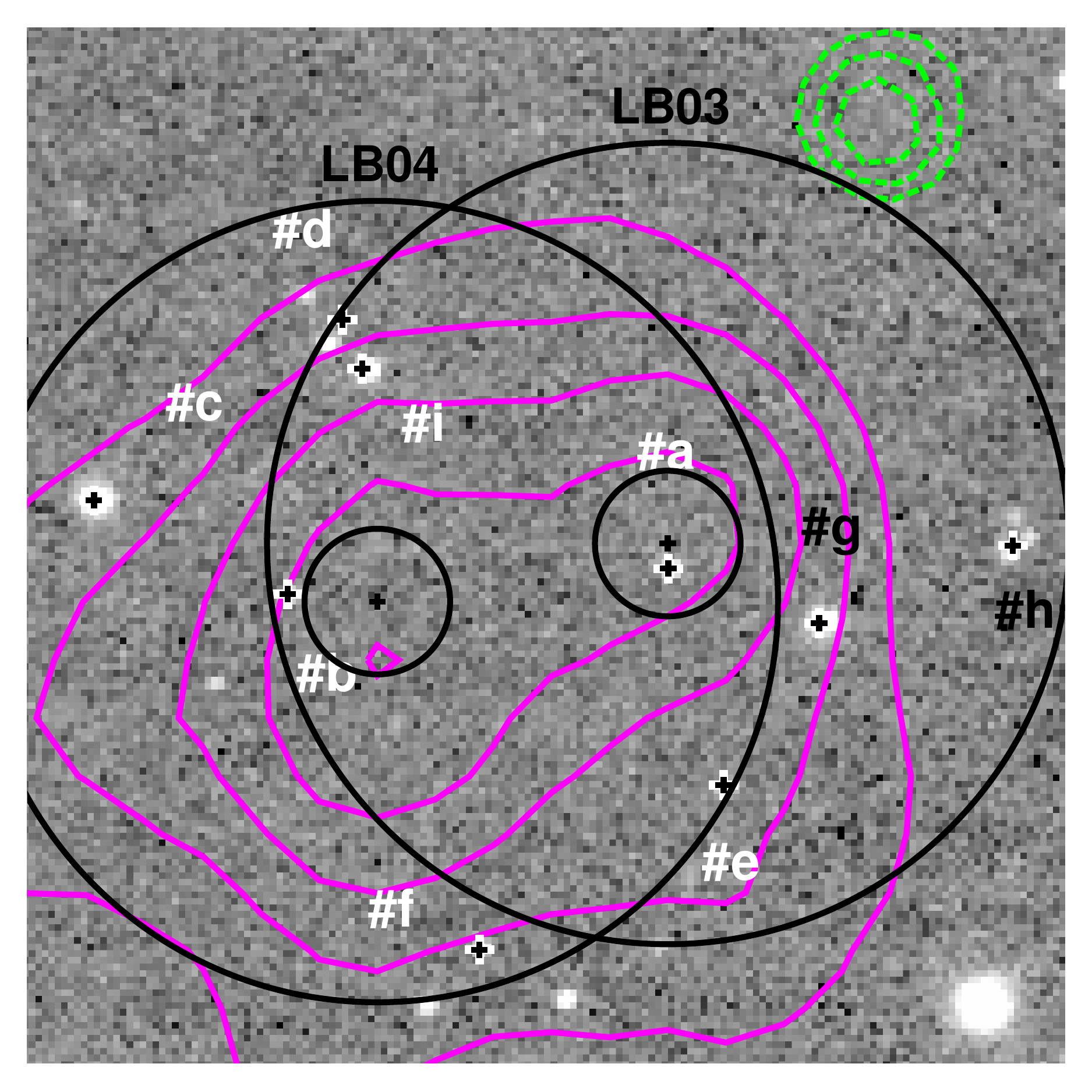}
\includegraphics[width=0.32\textwidth,scale=0.33]{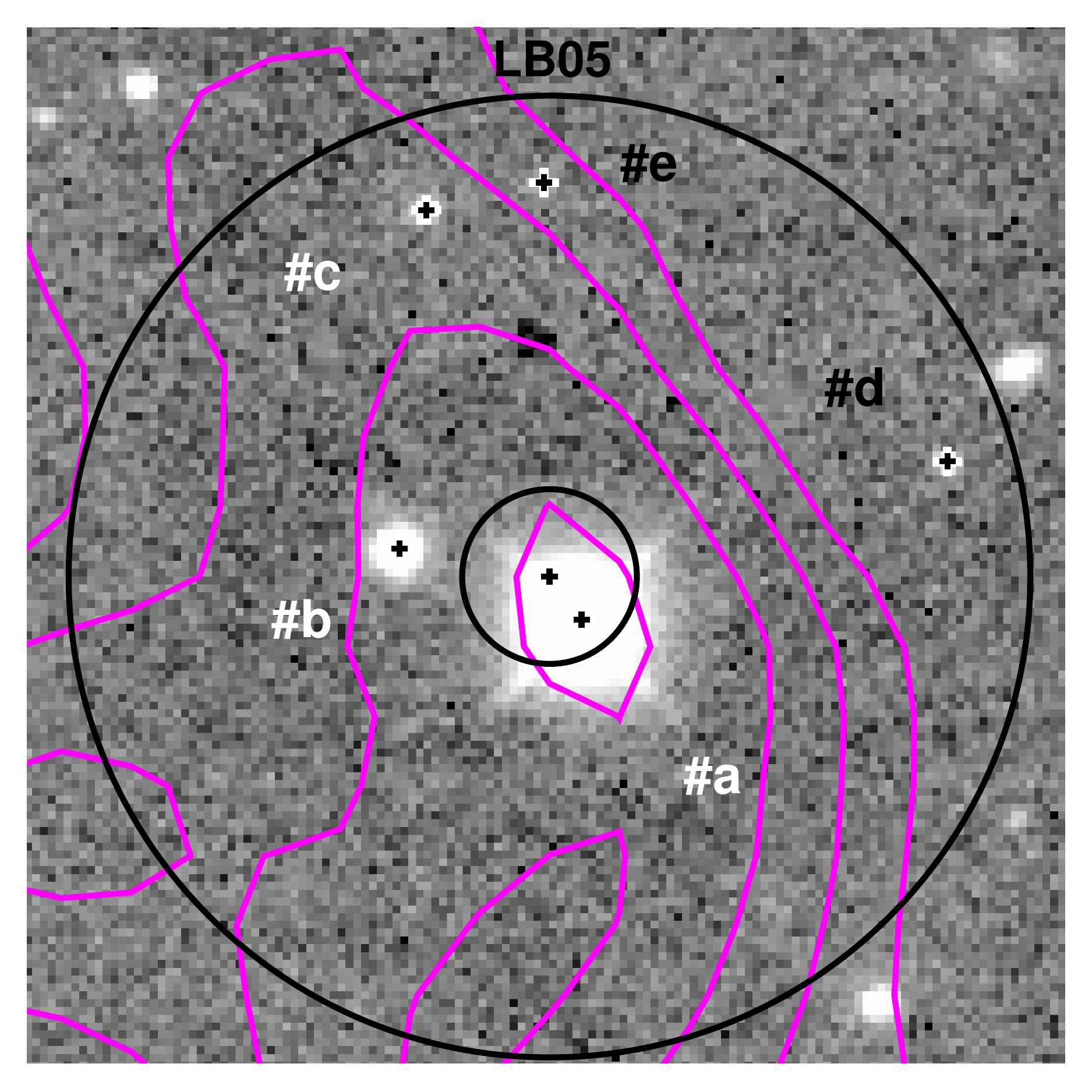}
\includegraphics[width=0.32\textwidth,scale=0.33]{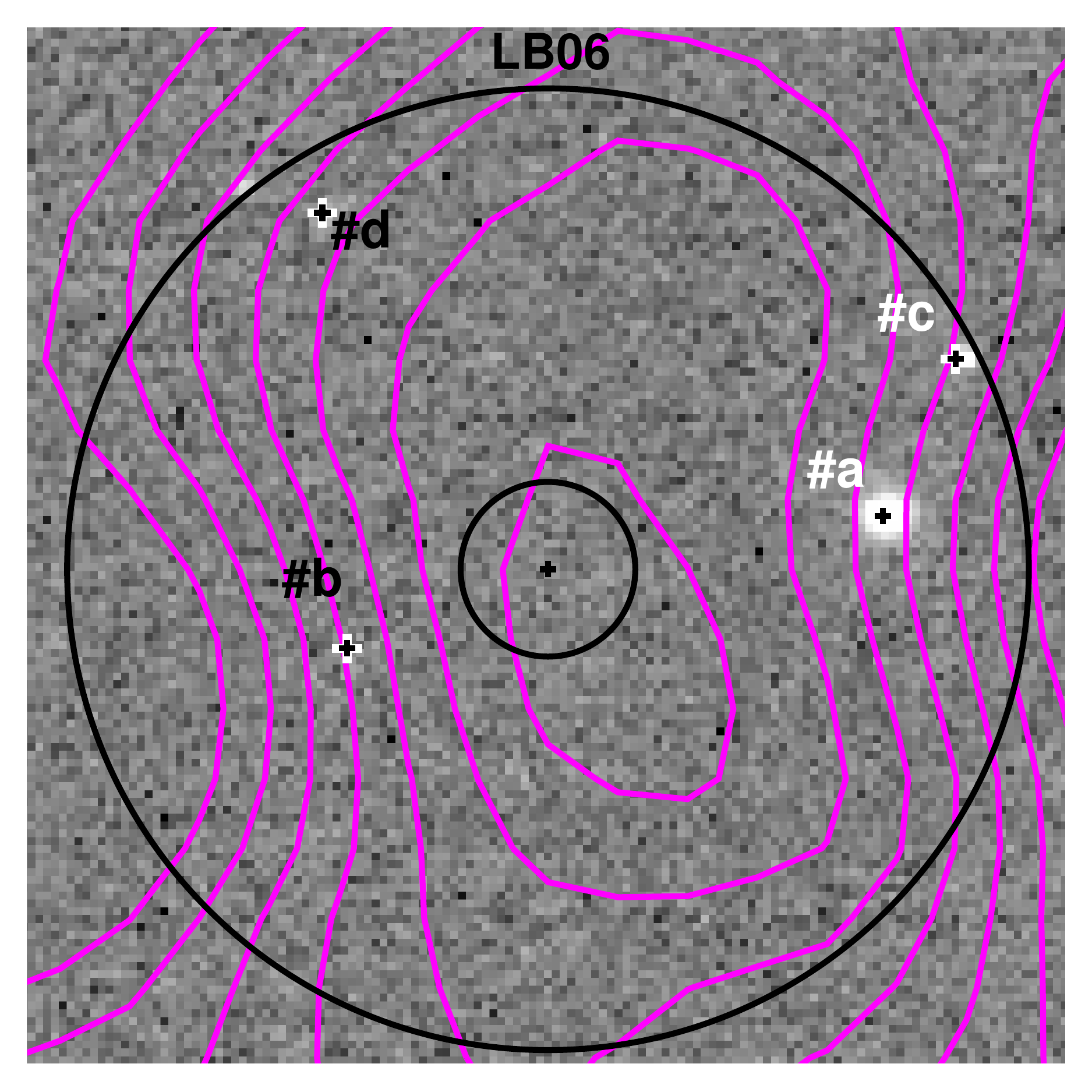}
\includegraphics[width=0.32\textwidth,scale=0.33]{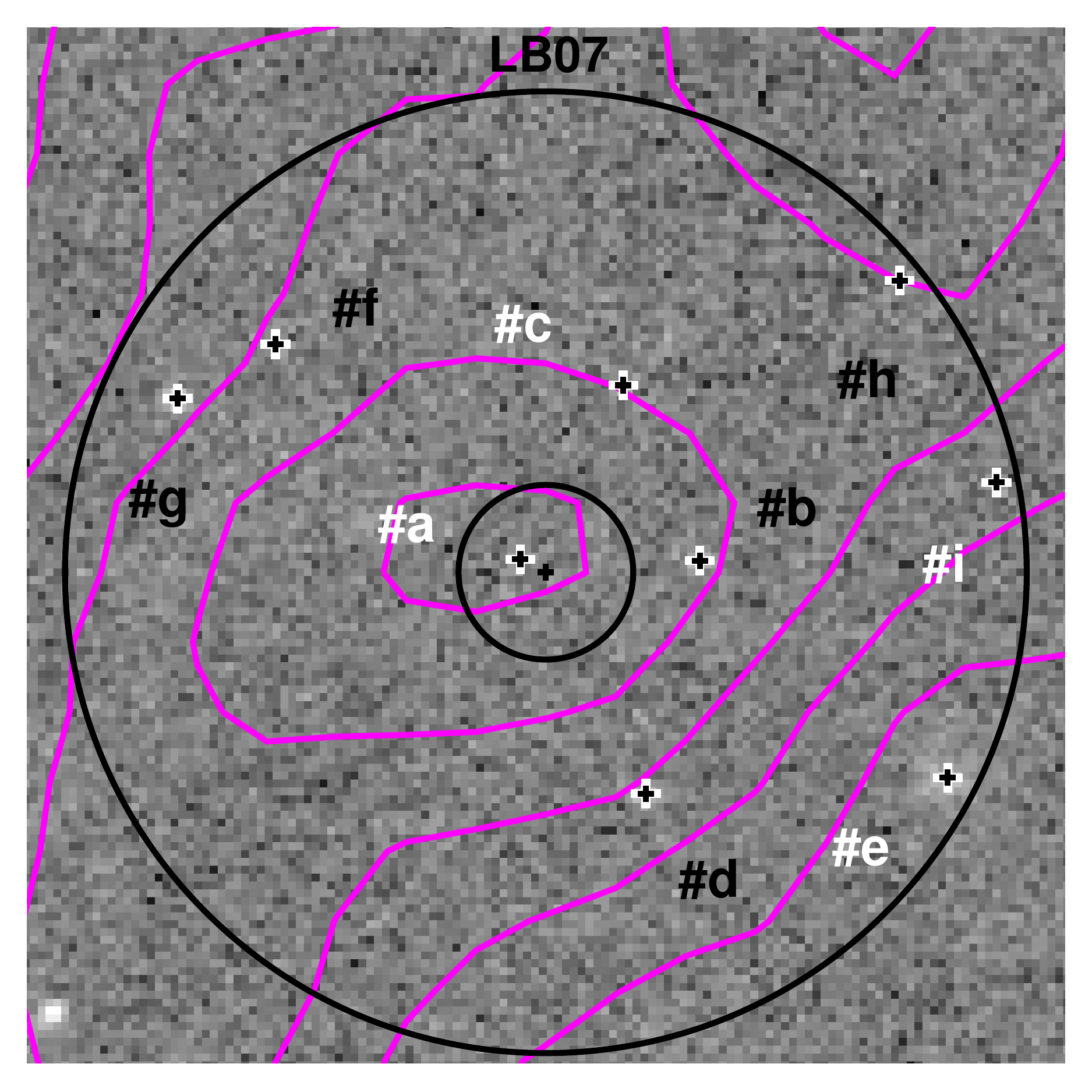}
\includegraphics[width=0.32\textwidth,scale=0.33]{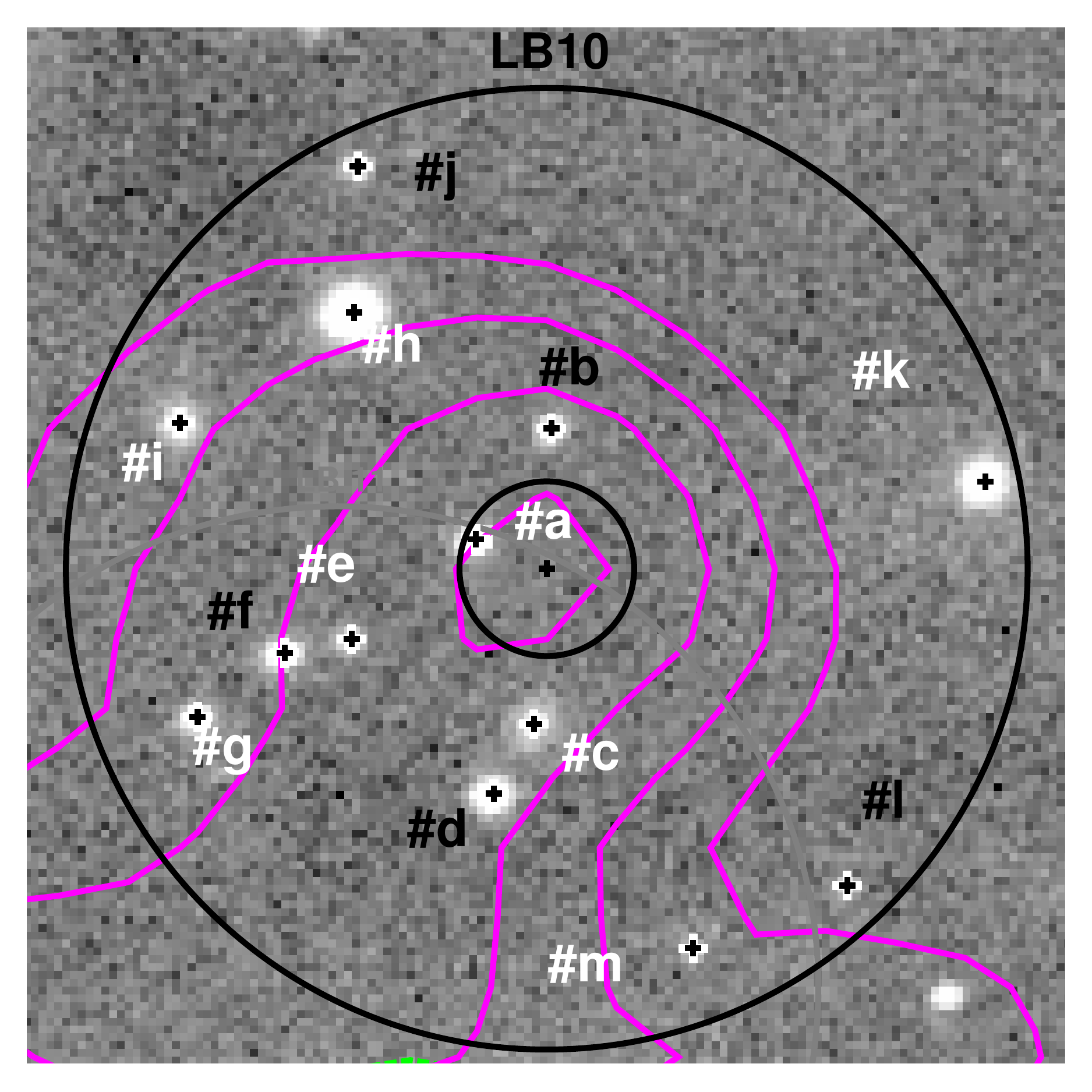}
\includegraphics[width=0.32\textwidth,scale=0.33]{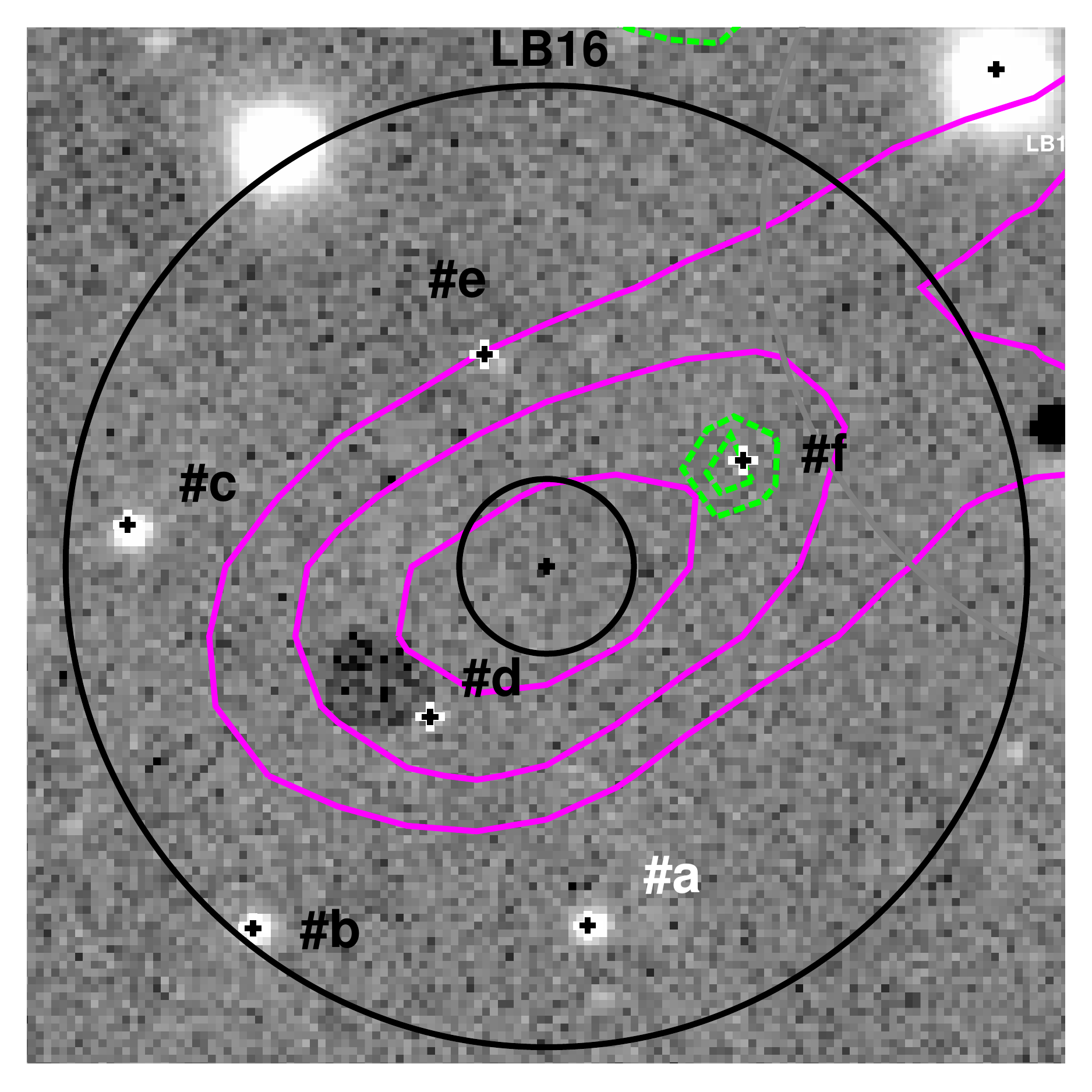}
\includegraphics[width=0.32\textwidth,scale=0.33]{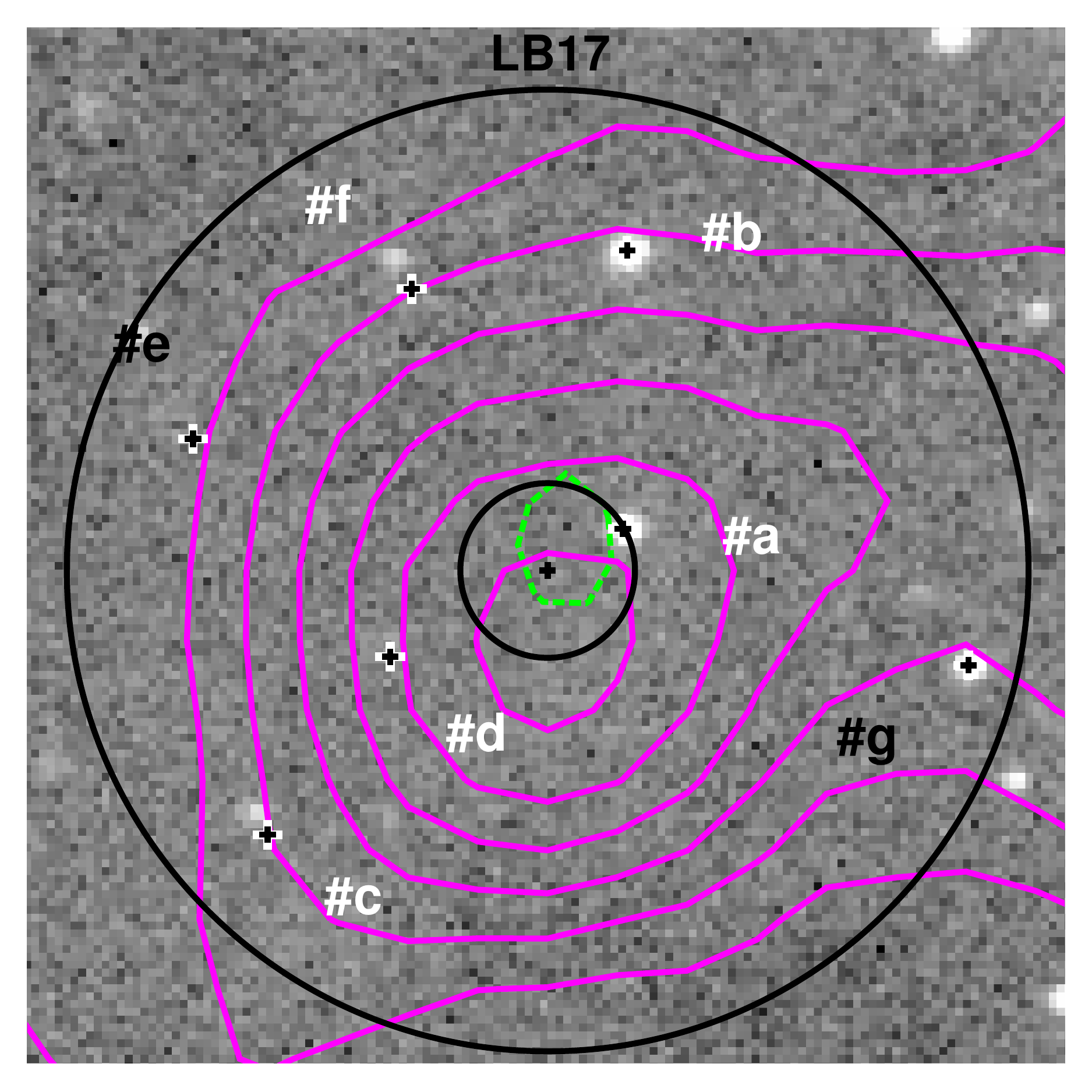}
\includegraphics[width=0.32\textwidth,scale=0.33]{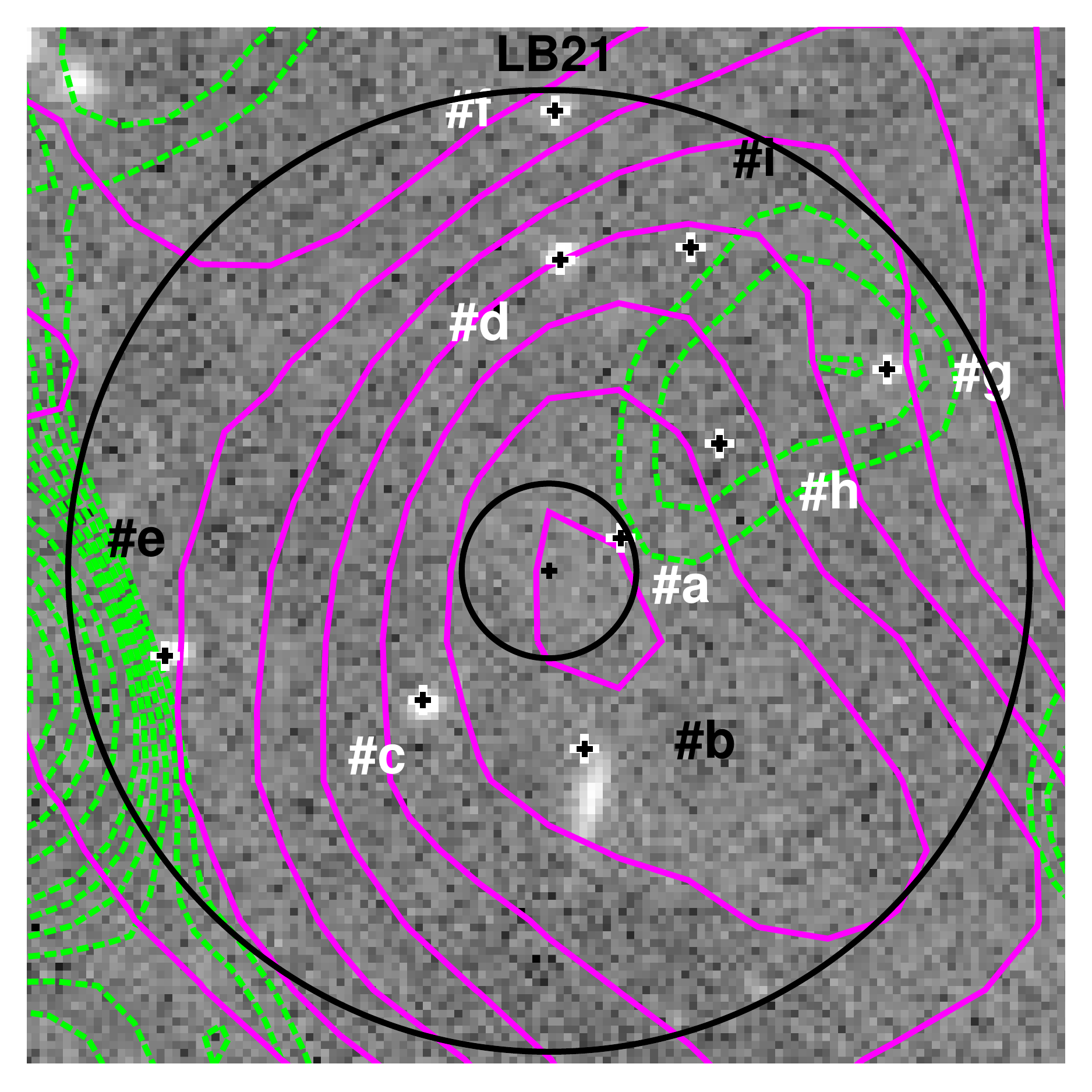}
\includegraphics[width=0.32\textwidth,scale=0.33]{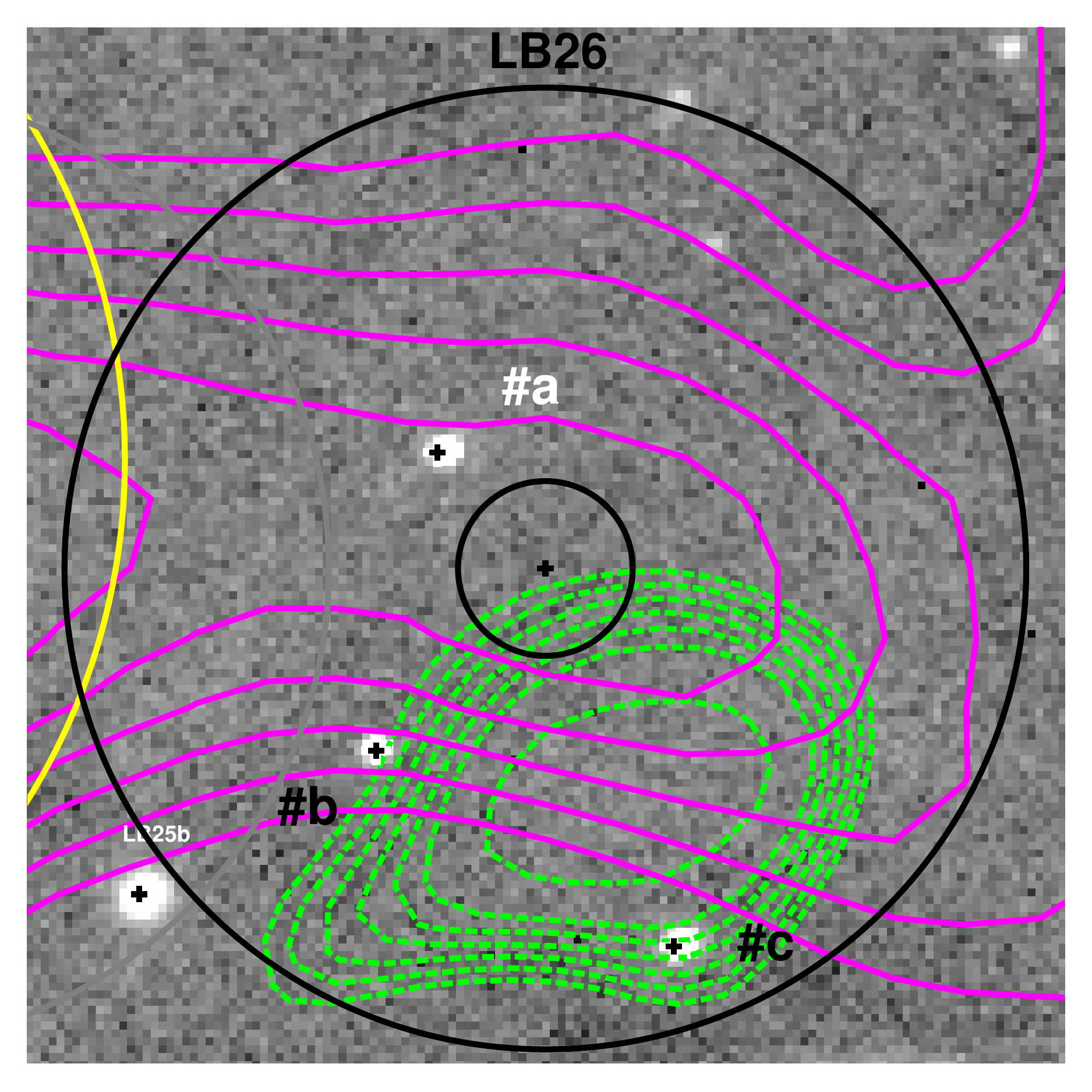}
\includegraphics[width=0.32\textwidth,scale=0.33]{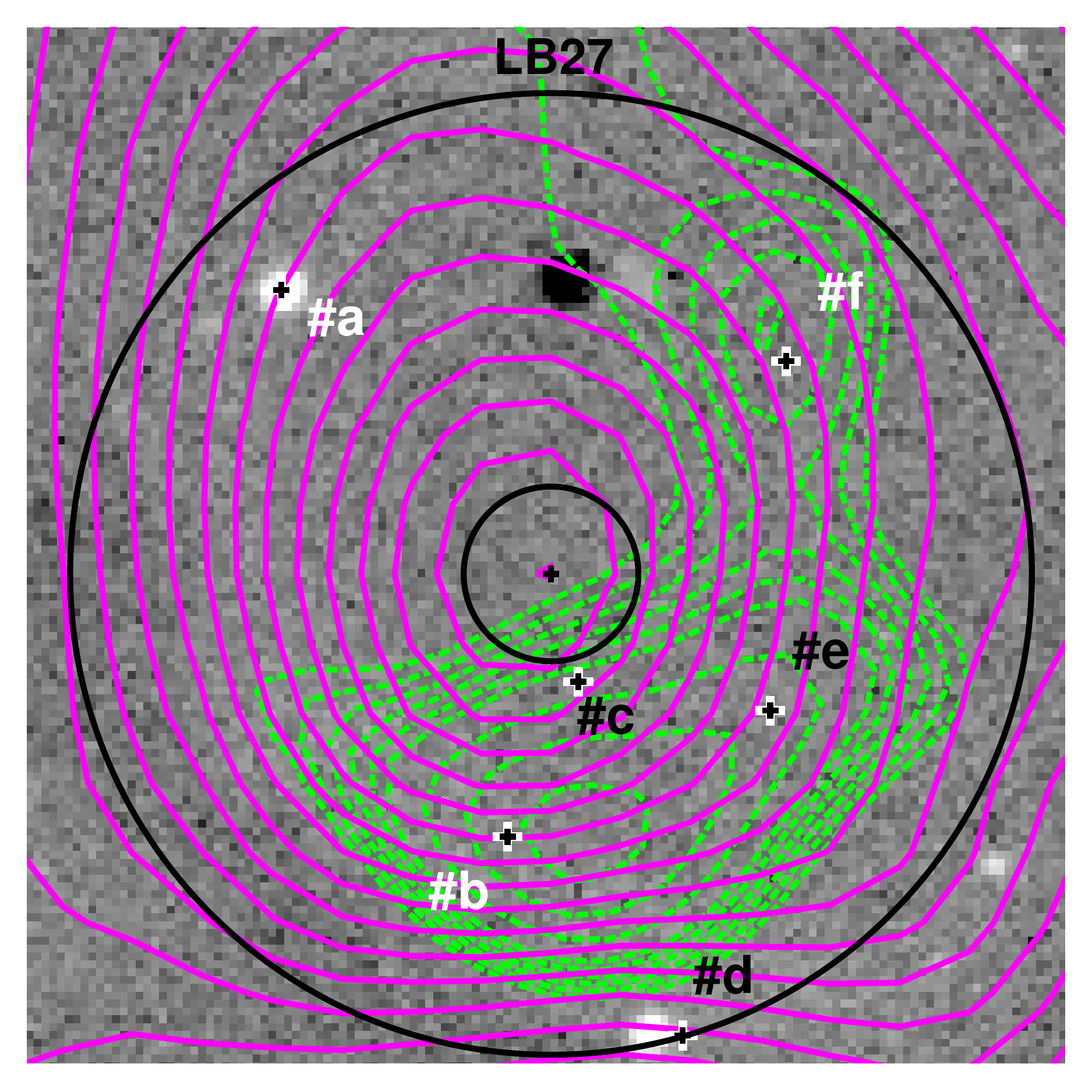}
\caption{\label{FCh_groupC2} 
  Finding charts for group C2 or sources  outside the FOV of the MIPS M2 image.
  None of them have been detected  24 $\mu$m with MIPS.
Symbols as in Fig. \ref{FCh_groupA1}.
We note that the feature displayed at 24 $\mu$m for B30-LB16f corresponds to an artifact.
 }
\end{figure*}

 On the other hand,
 {\bf B30-LB06},
 {\bf B30-LB16,} and
 {\bf B30-LB26}
 have not been detected with MIPS and there are no detectable optical nor near-IR sources within the 5 arcsec central peak.
 Several facts have to be taken into account.
 B30-LB16 has an envelope mass of 0.053  $M_\odot$. Counterpart \#a might also be substellar. There is a counterpart only
 detected at 24 $\mu$m (\#f), but it seems to be a artifact and we have not considered it for further discussions.
  Thus, we believe it is a good  candidate to be starless core.
 In the case of B30-LB06,
 the counterparts \#a and \#c seem to be substellar but they lack IR excess.
 The submm source is extended with masses of 0.210 and 0.077  $M_\odot$ (whole beam and peak, respectively).
 Again, it is starless core candidate.
 A similar situation appears for B30-LB26a, with a envelope mass of 0.058 $M_\odot$.
 However, this last one  might have a small IRAC excess and its submm emission is
 connected with B30-LB25 (in fact it is elongated into that direction), which has been detected both at 24 and at 70 $\mu$m
 (see Fig. \ref{FCh_groupA2} for the finding chart and Fig. \ref{SED_groupA2} for the SED).

\newpage
\clearpage

\newpage
\clearpage

\newpage
\clearpage

\newpage
\clearpage

\newpage
\clearpage

\newpage
\clearpage

\newpage
\clearpage

\newpage
\clearpage

\begin{figure*}   
\center
%
\includegraphics[width=0.32\textwidth,scale=0.33]{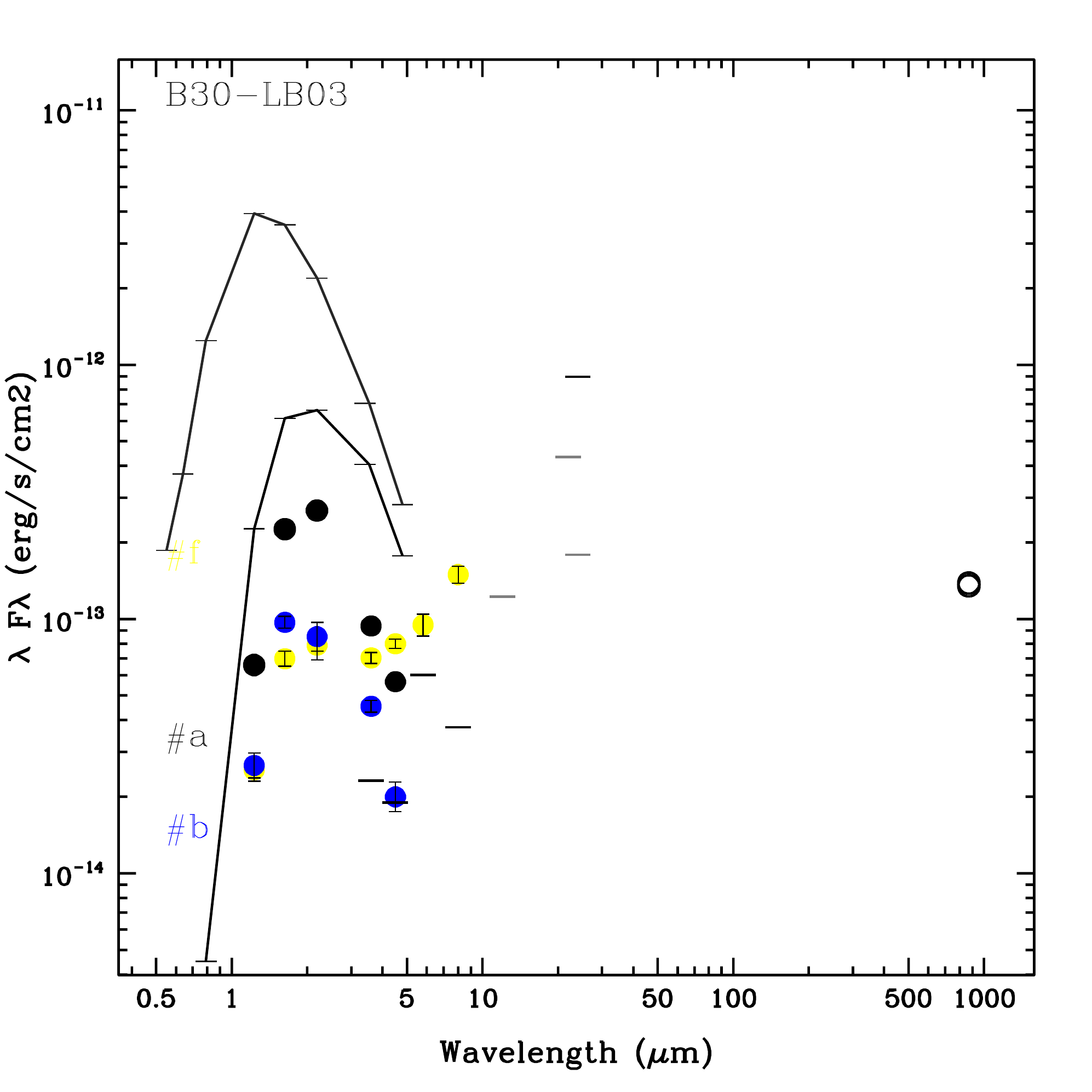} 
\includegraphics[width=0.32\textwidth,scale=0.33]{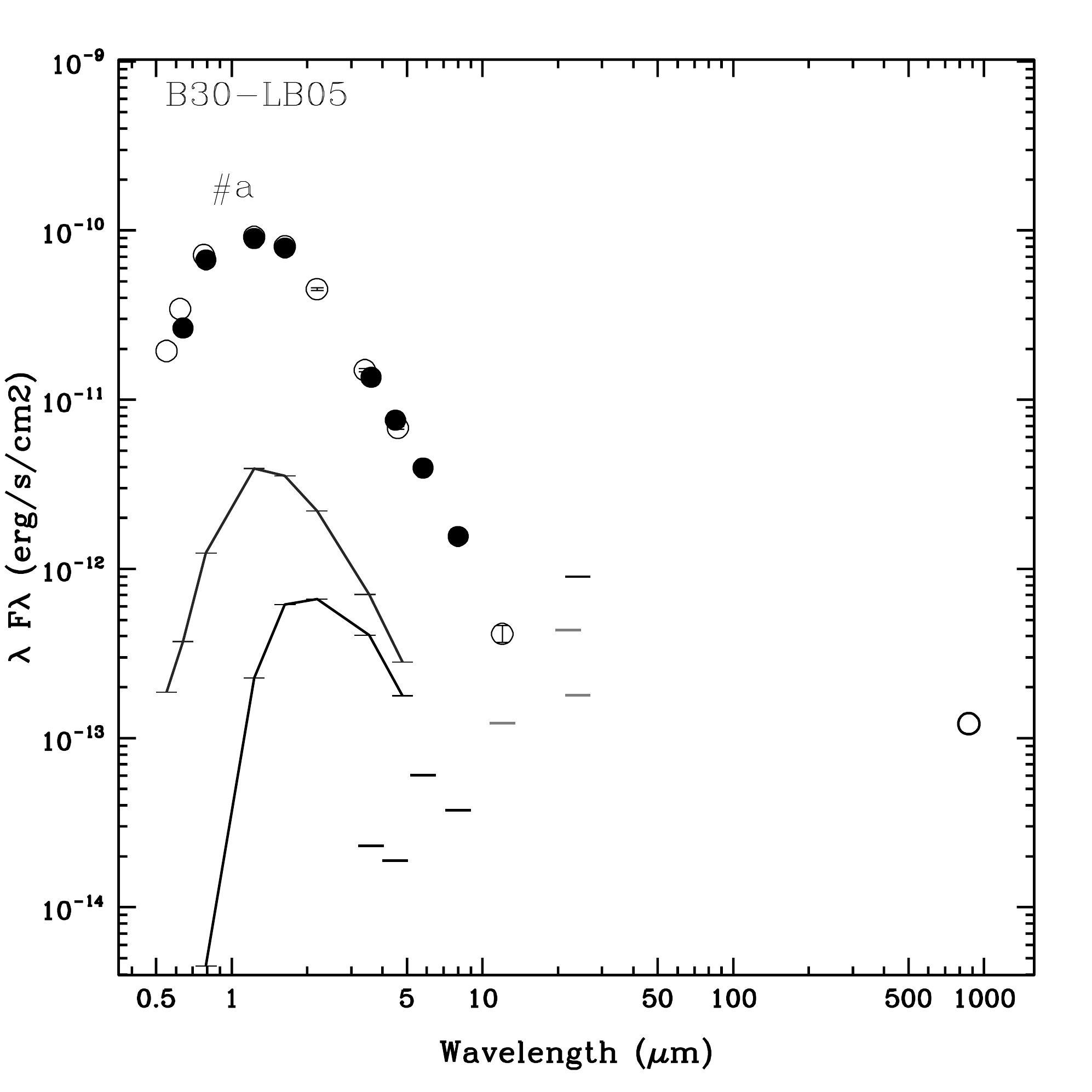} 
\includegraphics[width=0.32\textwidth,scale=0.33]{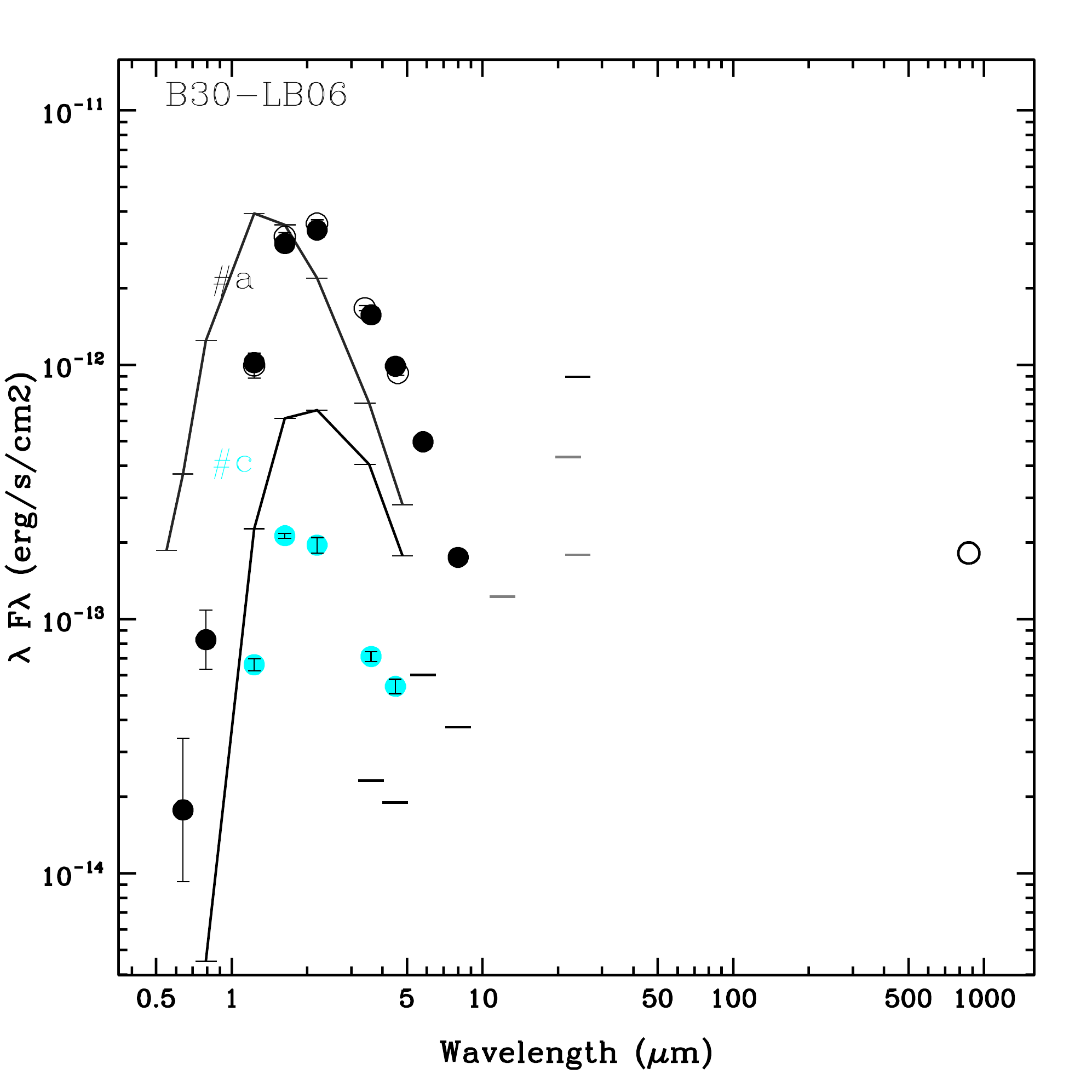} 
\includegraphics[width=0.32\textwidth,scale=0.33]{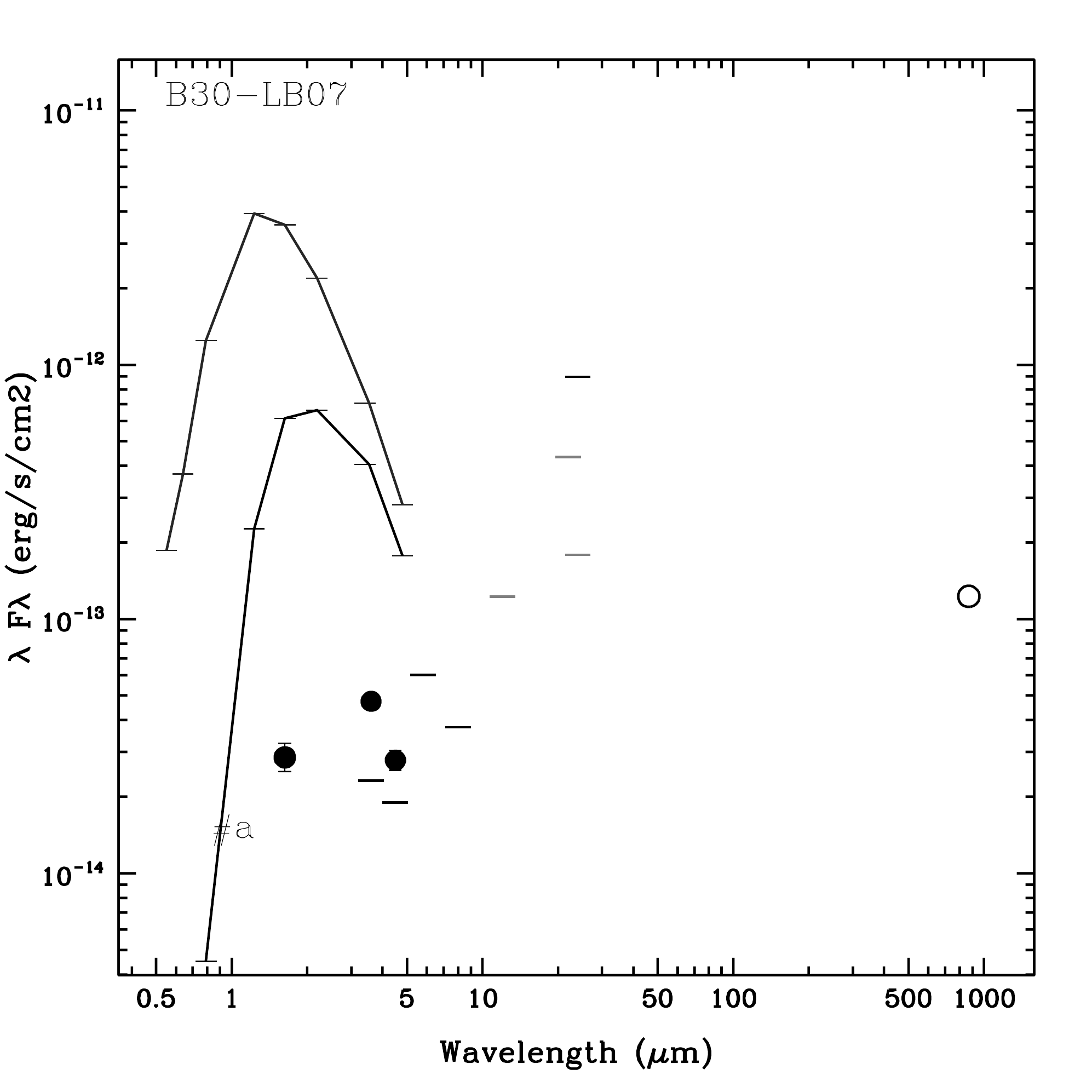} 
\includegraphics[width=0.32\textwidth,scale=0.33]{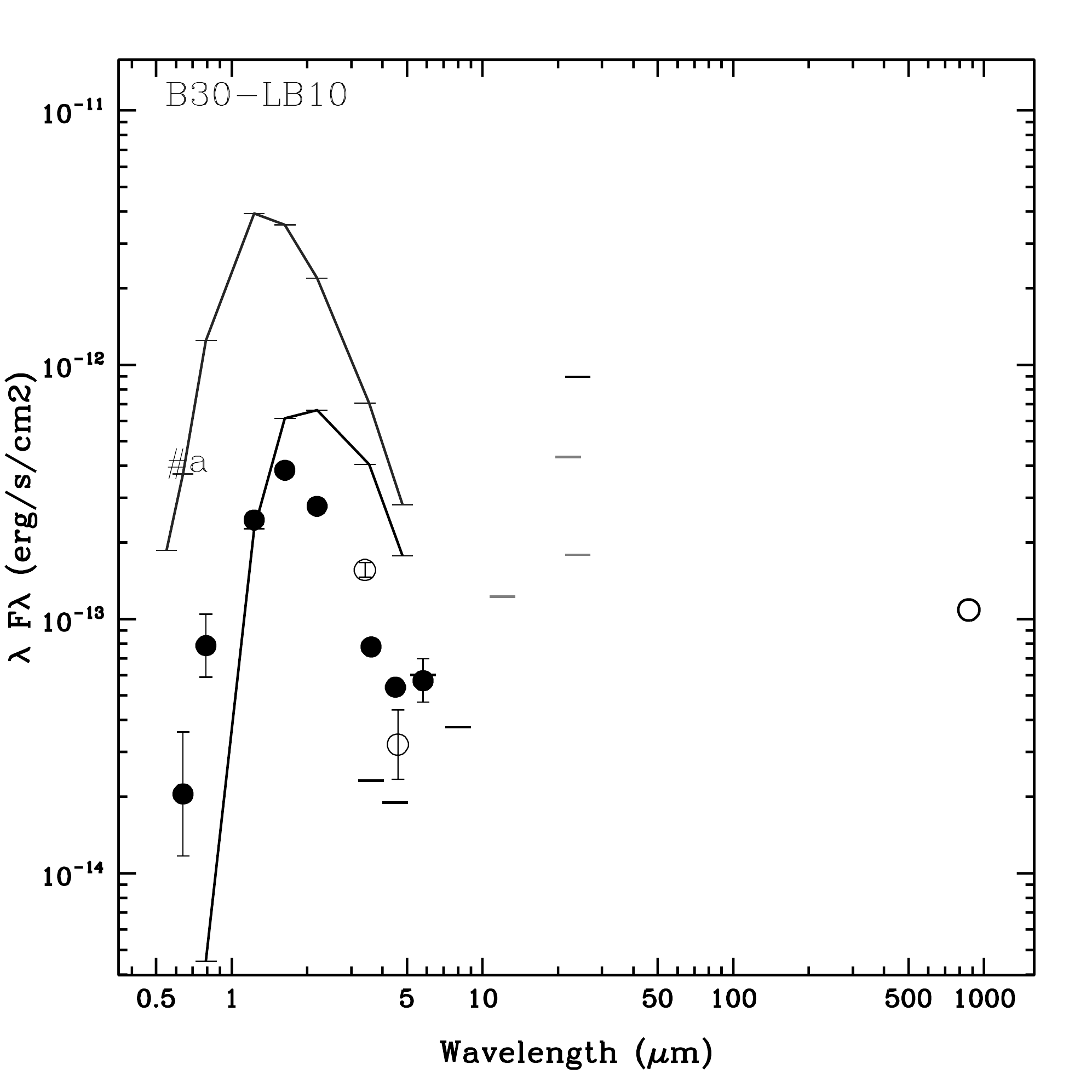} 
\includegraphics[width=0.32\textwidth,scale=0.33]{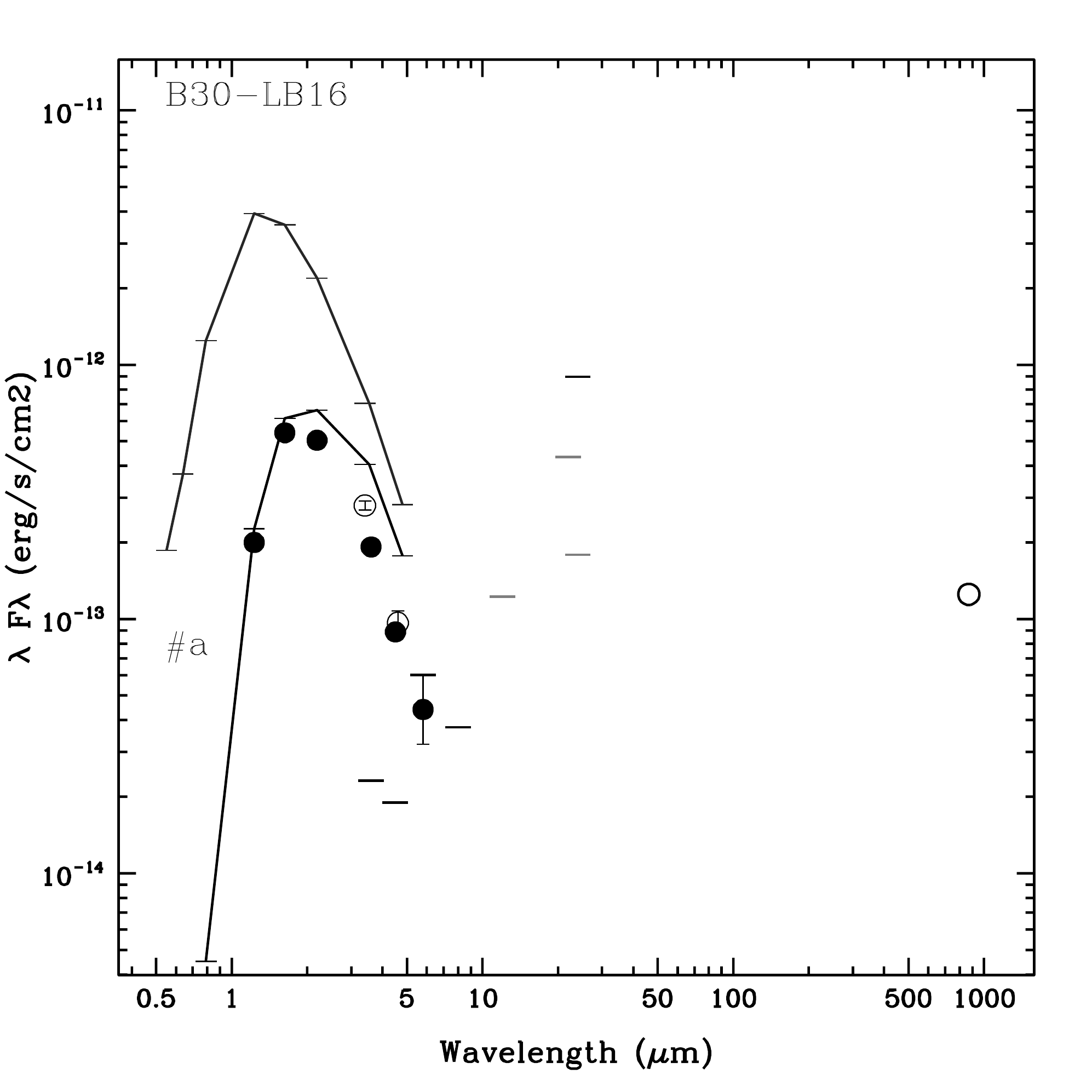} 
\includegraphics[width=0.32\textwidth,scale=0.33]{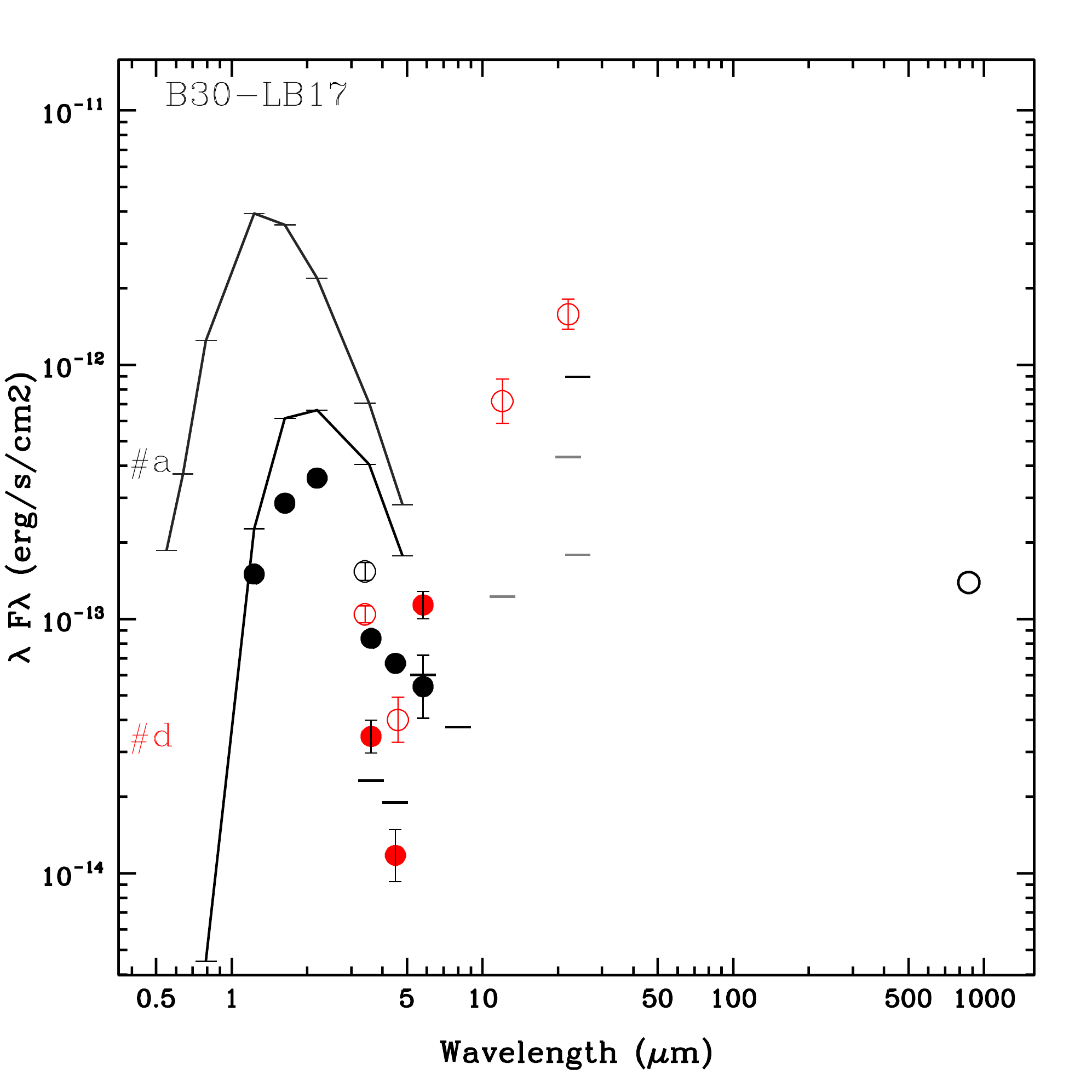} 
\includegraphics[width=0.32\textwidth,scale=0.33]{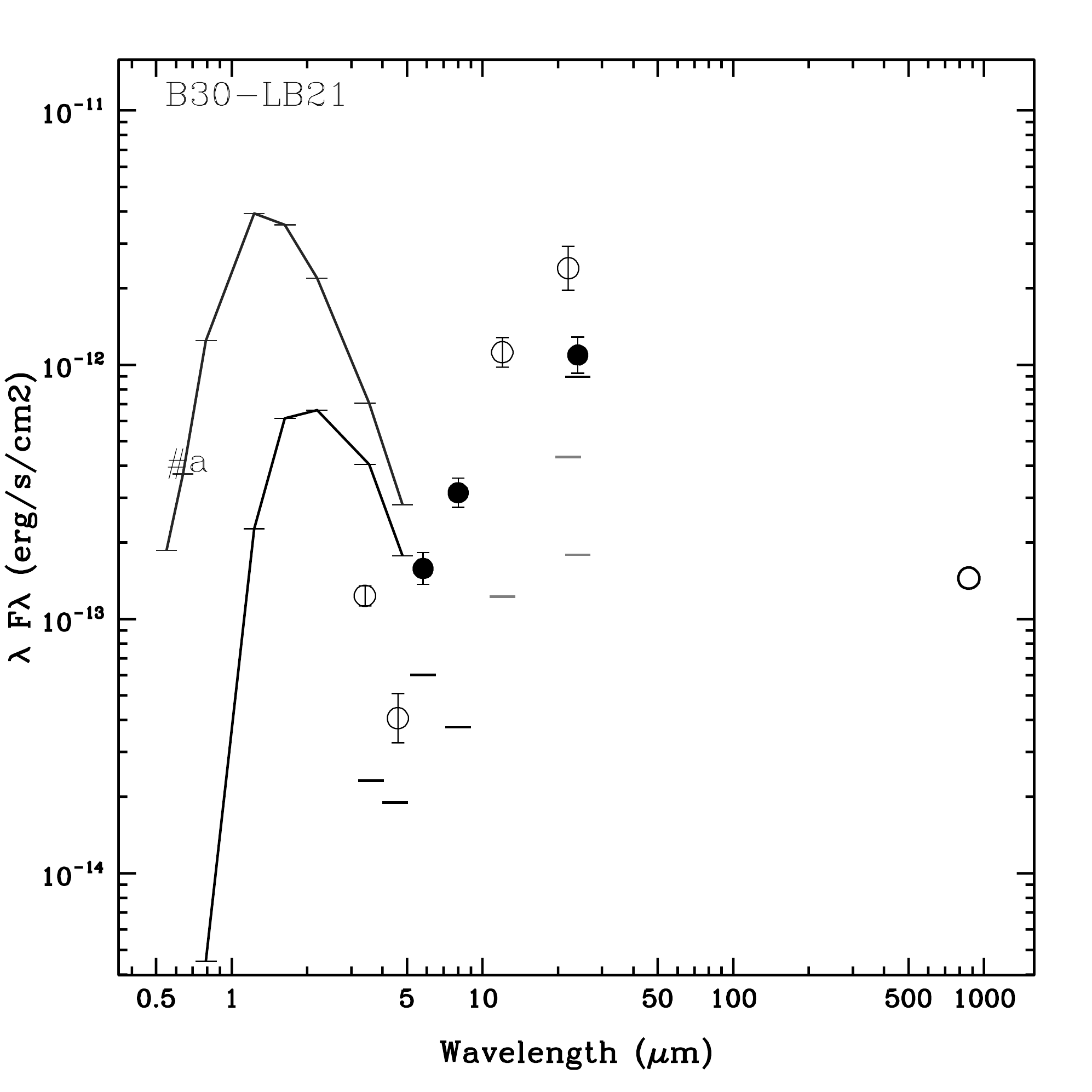} 
\includegraphics[width=0.32\textwidth,scale=0.33]{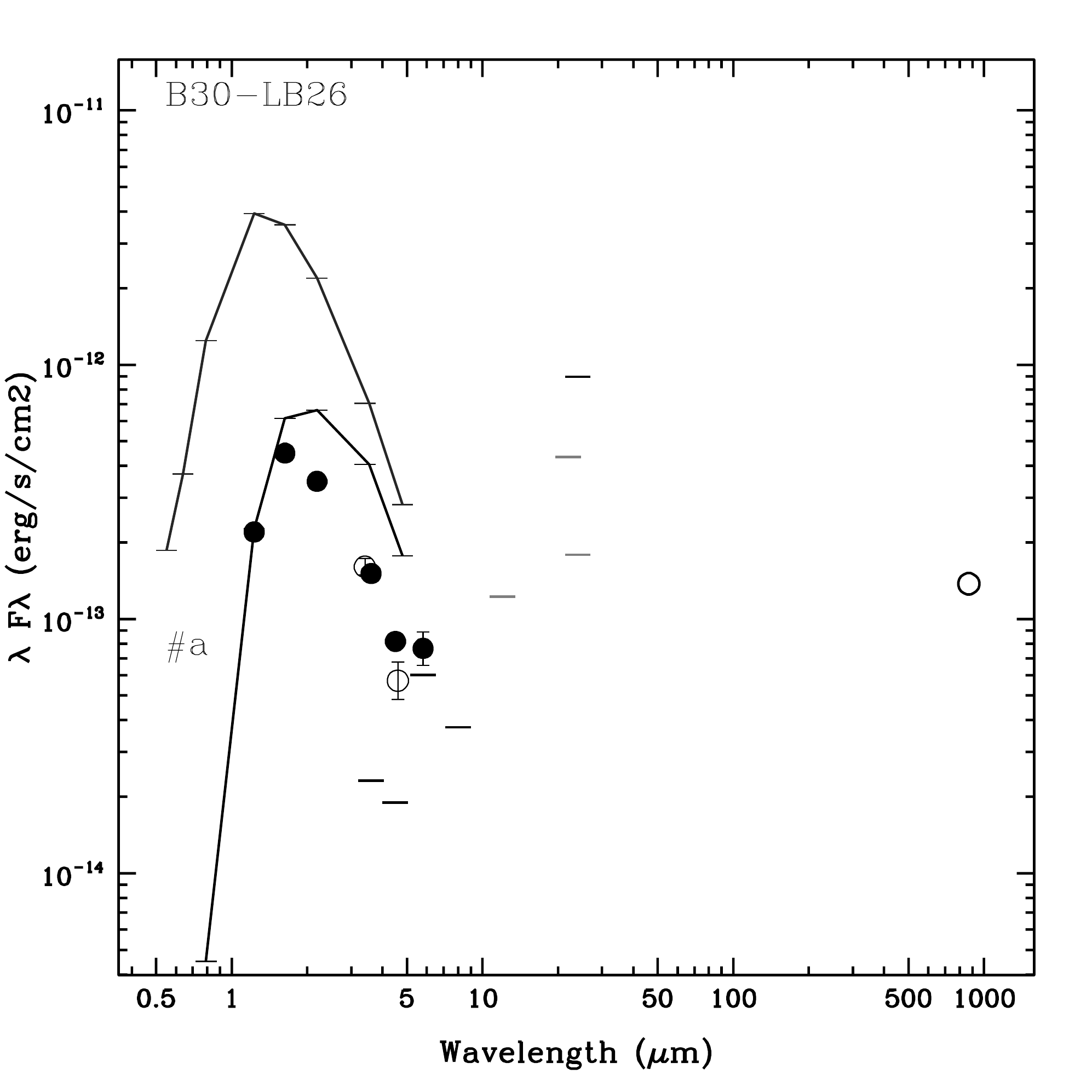} 
\includegraphics[width=0.32\textwidth,scale=0.33]{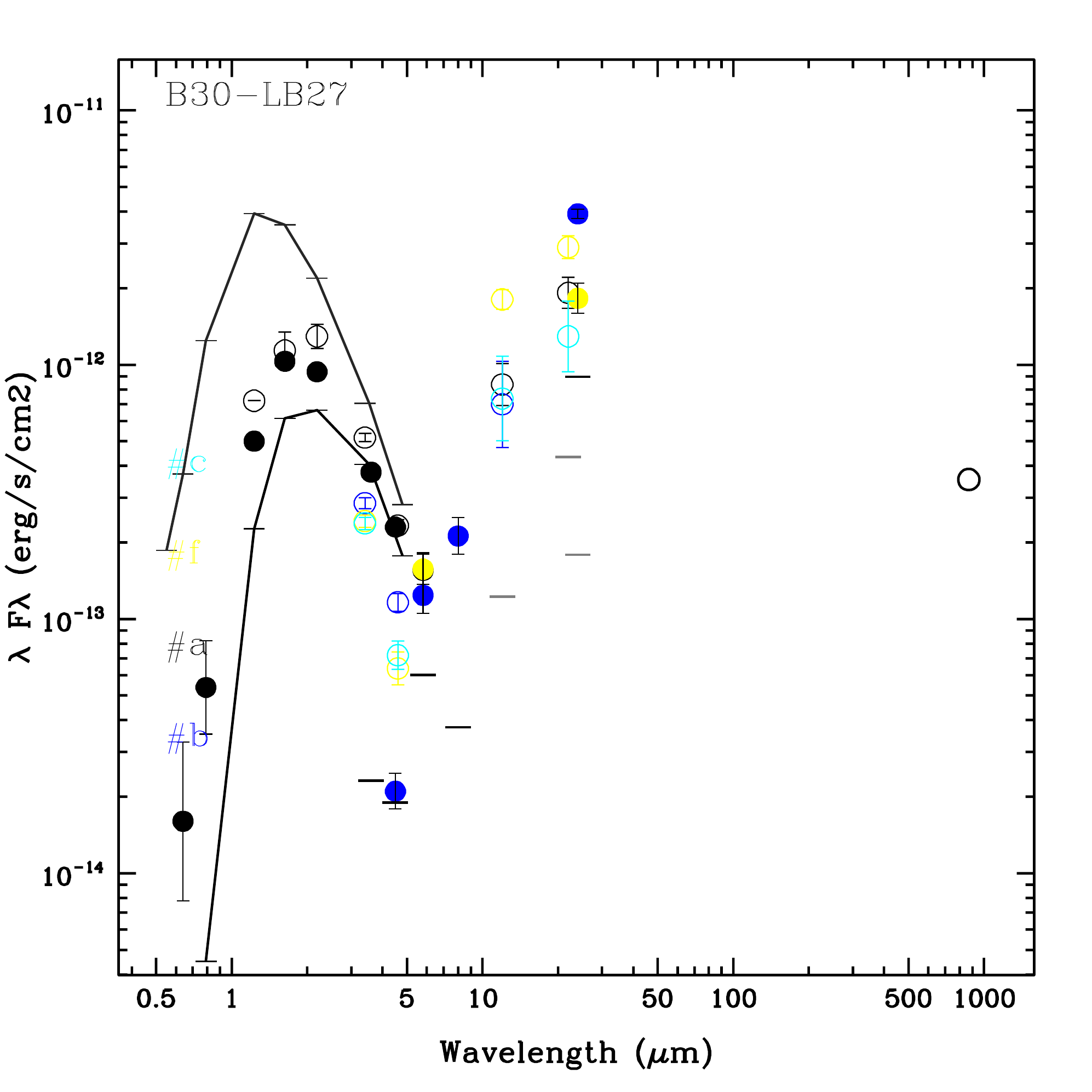} 
\caption{\label{SED_groupC2} 
  Spectral energy distribution for group C2 or sources  outside the FOV of the MIPS M2 image.
  None of them have been detected at 24 $\mu$m with MIPS.
  However, we note that B30-LB21 and B30-LB27 have a possible extended emission at 24 $\mu$m.
Symbols as in Fig. \ref{SED_groupA1}.
}
\end{figure*}

\newpage
\clearpage


\end{document}